\documentclass[12pt]{ociamthesis}  % default square logo 

\usepackage{amsmath}
\usepackage{amsthm}
\usepackage{amssymb}
\usepackage{mathtools}
\usepackage{stmaryrd}

\DeclareFontFamily{U}{mathb}{\hyphenchar\font45}
\DeclareFontShape{U}{mathb}{m}{n}{
  <5> <6> <7> <8> <9> <10> gen * mathb
  <10.95> mathb10 <12> <14.4> <17.28> <20.74> <24.88> mathb12
}{}
\DeclareSymbolFont{mathb}{U}{mathb}{m}{n}
\DeclareMathSymbol{\dotdiv}{2}{mathb}{"01}

\usepackage{fancybox}

\usepackage{tikz-cd}

\usepackage{tikz}

\usepackage{fancybox}
\usepackage{epigraph}
\usepackage{multicol}
\usepackage{import}
\usepackage{mVersion}
\usepackage{enumitem}
\usepackage{pgfornament}
\usepackage{hyperref}
\usepackage{graphicx}

\usepackage{natbib}
\newtheorem{thm}{Theorem}
\newtheorem{lem}{Lemma}
\newtheorem{cor}{Corollary}
\newtheorem{prop}{Proposition}

\newcommand{\parfunc}{\rightharpoonup}
\newcommand{\setcomp}[2]{\left\{ \, #1 \; \middle| \; #2 \, \right\}} 
\newcommand{\defeq}
  {\mathrel{\overset{\makebox[0pt]{\mbox{\normalfont\tiny def}}}{=}}}
\newcommand{\sem}[2]{\left\llbracket #1 \right\rrbracket^{#2}}
\newcommand{\myeq}{\stackrel{\mathclap{\mbox{\tiny def}}}{=}}

\newcommand{\ctxt}[2]{#1\mathbin{;}#2}
\newcommand{\ibox}[1]{\mathsf{box\;}#1}
\newcommand{\letbox}[3]{\mathsf{let\;box\;} #1 \Leftarrow #2 \mathsf{\;in\;} #3}
\newcommand{\fixbox}[2]{\mathsf{fix\;} #1 \mathsf{\;in\;box\;} #2}
\newcommand{\fixlob}[2]{\mathsf{fix\;} #1 \mathsf{\;in\;} #2}

\newcommand{\hash}{\mathbin{\#}}
\newcommand{\dom}{\text{dom} }

\newcommand{\vj}{\vdash_\mathcal{J}}
\newcommand{\vint}{\vdash_\text{int.}}
\newcommand{\vext}{\vdash_\text{ext.}}

\newcommand{\fv}[1]{\textsc{fv}\left(#1\right)}
\newcommand{\ufv}[1]{\textsc{fv}_0\left(#1\right)}
\newcommand{\bfv}[1]{\textsc{fv}_{\geq 1}\left(#1\right)}
\newcommand{\vars}[1]{\textsc{Vars}\left(#1\right)}

\newcommand{\red}{\mathrel{\longrightarrow}}
\newcommand{\redp}{\mathrel{\Longrightarrow}}
\newcommand{\redt}{\mathrel{\longrightarrow^\ast}}

\newcommand{\vct}[1]{\overrightarrow{#1}}

\newcommand{\bars}[1]{\left\lvert #1 \right\rvert}
\newcommand{\dbars}[1]{\left\| #1 \right\|}
\newcommand{\expo}{\looparrowright}
\newcommand{\oper}{\dashrightarrow}
\newcommand{\natexp}{\overset{\bullet}{\looparrowright}}

\newcommand{\mb}[1]{\mathbf{#1}}

\newlist{indproof}{itemize}{5}
\setlist[indproof]{%
  itemsep=5pt,  % space between items
  font={\sc}, % set the label font
  label={}
}
\newcommand{\case}[1]{\item{\sc Case({\normalfont #1})}.}

\theoremstyle{definition}
\newtheorem{defn}{Definition}
\newtheorem{example}{Example}

\usepackage{packages/prooftree}
\usepackage{packages/derivation}
\usepackage{packages/agda}

\DeclareUnicodeCharacter{"0393}{$\Gamma$}
\DeclareUnicodeCharacter{"0394}{$\Delta$}
\DeclareUnicodeCharacter{"2081}{1}
\DeclareUnicodeCharacter{"22A2}{\ensuremath{\vdash}}
\DeclareUnicodeCharacter{"00B7}{$\cdot$}
\DeclareUnicodeCharacter{"25A1}{$\Box$}

\usepackage[th]{optional}

\usepackage{tabularx}

\title{On the Semantics of Intensionality\\[1ex]
        and Intensional Recursion}

\author{G. A. Kavvos}
\college{St John's College}

\degree{Doctor of Philosophy}     %the degree
\degreedate{Trinity Term 2017}         %the degree date

\setVersion{1.0}

\begin{document}

% Increases the build counter.
\increaseBuild

%this baselineskip gives sufficient line spacing for an examiner to easily
%markup the thesis with comments
\baselineskip=18pt plus1pt

%set the number of sectioning levels that get number and appear in the contents
\setcounter{secnumdepth}{3}
\setcounter{tocdepth}{3}

\maketitle                  % create a title page from the preamble info
\begin{dedication}
  This thesis is dedicated to my father, John G. Kavvos
  (1948--2015), who taught me to persist in the face of all
  adversity.

%  \begin{figure*}
%    \centering
%    \includegraphics[width=\textwidth]{slides0027}
%  \end{figure*}
\end{dedication}
        % include a dedication.tex file
\begin{acknowledgements}
  A thesis in the making is a process of controlled deconstruction
  of its author's character. This fact in itself suffices to
  warrant a rich list of acknowledgees.

  First and foremost, I would like to thank my doctoral
  supervisor, Samson Abramsky, \emph{sine qua non}. Not only did
  he suggest the---admittedly unusual---topic of this thesis, but
  his unfailingly excellent advice and his unparalleled wit were
  vital ingredients in encouraging me to plough on, even when it
  seemed futile to do so. I shall never forget the time he quoted
  the experience of J. E. Littlewood on the development of chaos
  theory: \begin{quote}
    ``For something to do we went on and on at the thing with no
    earthly prospect of ``results''; suddenly the whole vista of
    the dramatic fine structure of solutions stared us in the
    face.''
    \end{quote}
  It was a pleasure to work for a few short years next to a
  scientist of his stature, who also happens to be a kind man with
  a wonderful sense of humour.

  I also want to thank my examiners, Luke Ong and Martin Hyland,
  who examined this work in detail, and provided many enlightening
  comments and suggestions.

  I am grateful to the UK Engineering and Physical Sciences
  Research Council (EPSRC) for their financial support. I am also
  greatly indebted to the Department of Computer Science, the
  European Cooperation in Science and Technology (COST) framework,
  and St John's College, for generously funding the trips involved
  in presenting my work to the wider community.

  I would also like to thank all those who reviewed parts of this
  work before its completion. These include quite a few anonymous
  reviewers; John Longley, who toiled over the first version of
  the manuscript on which this thesis is based; Neil Jones, with
  whom I have had many interesting discussions during and after
  his visits to Oxford, and who also read many parts of this
  thesis in detail; and my interim assessors within the
  department, Sam Staton and Hongseok Yang. This thesis could
  hardly have been completed without their support and
  encouragement.

  Many thanks to the fellow scientists around me: Kohei Kishida,
  for many interesting discussions, for providing help and advice
  at a moment's notice, and for translating and transliterating
  the title of \citep{Kiselyov2015}; Martin Berger, for
  inviting me to Sussex to present my work, and for the numerous
  exchanges that followed; and Sean Moss, who was eager to discuss
  all sorts of interesting categorical diversions. Many thanks to
  Geraint Jones for allowing me to use his macros for
  Eindhoven-style calculations. 
  
  I also want to thank my fellow students: Mario Alvarez, to whom
  I wish the best of luck in his ongoing attempt to `debunk' this
  thesis; Amar Hadzihasanovic, for many discussions on higher
  category theory and its applications; and Matthijs V\'ak\'ar and
  Norihiro Yamada, who were always eager to discuss categories,
  logic, and semantics.
  
  Finally, I would like to thank Andrew Ker, who was my tutor
  during my time as an undergraduate at University College, and my
  employer thereafter. I learned an enormous amount from him as a
  student, and even more whilst teaching under his supervision.
  I am indebted to him for his wise advice, and all the things
  that he has inadvertently taught me over a cup of coffee in the
  Senior Common Room. One would be hard-pressed to find better
  guidance or more generosity.
\end{acknowledgements}
   % include an acknowledgements.tex file
\begin{abstract}
  \small
  Intensionality is a phenomenon that occurs in logic and
  computation. In the most general sense, a function is
  intensional if it operates at a level finer than (extensional)
  equality. This is a familiar setting for computer scientists,
  who often study different programs or processes that are
  interchangeable, i.e. extensionally equal, even though they are
  not implemented in the same way, so intensionally distinct.
  Concomitant with intensionality is the phenomenon of intensional
  recursion, which refers to the ability of a program to have
  access to its own code. In computability theory, intensional
  recursion is enabled by Kleene's Second Recursion Theorem.

  This thesis is concerned with the crafting of a logical toolkit
  through which these phenomena can be studied. Our main
  contribution is a framework in which mathematical and
  computational constructions can be considered either
  \emph{extensionally}, i.e. as abstract values, or
  \emph{intensionally}, i.e. as fine-grained descriptions of their
  construction. Once this is achieved, it may be used to analyse
  intensional recursion.

  To begin, we turn to type theory. We construct a modal
  $\lambda$-calculus, called Intensional PCF, which supports
  non-functional operations at modal types. Moreover, by adding
  L\"ob's rule from provability logic to the calculus, we obtain a
  type-theoretic interpretation of intensional recursion. The
  combination of these two features is shown to be consistent
  through a confluence argument.

  Following that, we begin searching for a semantics for
  Intensional PCF. We argue that 1-category theory is not
  sufficient, and propose the use of P-categories instead.
  On top of this setting we introduce \emph{exposures}, which are
  P-categorical structures that function as abstractions
  of well-behaved intensional devices. We produce three
  examples of these structures, based on G\"odel numberings on
  Peano arithmetic, realizability theory, and homological algebra.
  
  The language of exposures leads us to a P-categorical analysis
  of intensional recursion, through the notion of
  \emph{intensional fixed points}. This, in turn, leads to
  abstract analogues of classic intensional results in logic
  and computability, such as G\"odel's Incompleteness Theorem,
  Tarski's Undefinability Theorem, and Rice's Theorem. We are thus
  led to the conclusion that exposures are a useful framework,
  which we propose as a solid basis for a \emph{theory of
  intensionality}.

  In the final chapters of the thesis we employ exposures to endow
  Intensional PCF with an appropriate semantics. It transpires
  that, when interpreted in the P-category of assemblies on the
  PCA $K_1$, the L\"ob rule can be interpreted as the type of
  Kleene's Second Recursion Theorem.
\end{abstract}
          % include the abstract

\begin{romanpages}          % start roman page numbering
\tableofcontents            % generate and include a table of contents
\listoffigures              % generate and include a list of figures

\begin{center}
  \pgfornament[scale=0.45]{84}
\end{center}

\begin{center}
  This is the arXiv version of this thesis (identical to build
  7049), and was compiled on \today.
\end{center}

\end{romanpages}            % end roman page numbering

%%% VERSION

%%% INTRODUCTION

\chapter{Introduction}
  \label{chap:intro}

  This thesis concerns the computational phenomenon of
intensionality.

\section{Intensionality}
  \label{sec:intensionality}

To be \emph{intensional} is to contain not only \emph{reference},
but also \emph{sense}. This philosophical distinction was drawn by
Frege; see \cite{Fitting2015}. An intensional sign denotes an
external \emph{referent}, yet inherently connotes more
information---its elusive \emph{sense}. The classic example is
that of the planet Venus, which may be referred to as either the
morning star, or the evening star.

In the most mathematically general sense, to be `intensional' is
to somehow operate at a level finer than some predetermined
`extensional' equality. Intensionality is omnipresent in
constructive mathematics, where the question of equality is
non-trivial, see e.g. \cite{Beeson1985}. An example that dates
back to the work of \cite{Bishop1967} on constructive analysis
is that of real numbers, and their construction as Cauchy
sequences of rationals: two different Cauchy sequences of
rationals may stand for the same real number, thus being
\emph{extensionally} equal, yet \emph{intensionally} distinct.

Most mainstream mathematics is extensional: we usually reason
about some underlying, `ideal' mathematical object, and not its
concrete descriptions. The latter is only a way to refer to the
former. In this light, intensionality is merely a nuisance, so
common set theories assume some \emph{axiom of extensionality}:
sets are identified by their members, and functions by their
graphs.  Glimpses of intensionality appear very rarely, and
usually only because we are interested in proving that some
extensionality axiom is independent from the rest of some logical
theory: see e.g. \cite{Streicher2013} for a recent example.

This is a difficult-to-work-in setting for Computer Science, where
intensionality is the norm. In fact, the present author believes
that it would be fair to say that many branches of computer
science are in essence the study of programs and processes seen
under some appropriate notion of equality. At one end of the
spectrum, correctness of programs is discussed in the context of
some relation of \emph{observational equivalence}, i.e.
indistinguishability of programs. Intermediately, complexity
theory requires a slightly stronger notion, which we could call
\emph{complexity equivalence}: this is observational equivalence
strengthened by some account of the resources the program
consumes. At the other extreme, computer viruses sometimes make
decisions based strictly on patterns of object code they
encounter, disregarding the actual function of what they are
infecting; one could say they operate up to
\emph{$\alpha$-equivalence}, i.e. syntactic identity.  Each of the
aforementioned notions of equality is more intensional than the
one preceding it, and each level is interesting in its own right.

Thus, depending on our point of view, there are always two ways in
which we can see a computational process. There is an extensional
level, which corresponds to \emph{what may be computed}. But there
is also an intensional level, corresponding to the \emph{programs
and processes that carry out the computation}. The shift between
these two viewpoints has been discussed by \cite{Moschovakis1993}
and \cite{Abramsky2014}: computational processes may be understood
by the ideal objects that they refer to (e.g. functions), or their
internal characteristics: length, structure, and, ultimately, the
\emph{algorithm} they embody.

This thesis concerns the difficulties that arise when we are
trapped between \emph{intension} and \emph{extension}, or
\emph{description} and \emph{behaviour}.

\section{Objectives}
  \label{sec:objectives}

The main objective of this thesis is to answer the following
question:
\begin{center}
  \shadowbox{
    \begin{minipage}{0.9\textwidth}
      \centering
      Is there a consistent, logical universe where the same
      mathematical objects can be viewed both as intension, and
      extension?
    \end{minipage}
  }
\end{center}

\noindent This research programme is a suggestion of
\cite{Abramsky2014}; in his words: \begin{quote}
  The notions of intensionality and extensionality carry
  symmetric-sounding names, but this apparent symmetry is
  misleading. Extensionality is enshrined in mathematically
  precise axioms with a clear conceptual meaning. Intensionality,
  by contrast, remains elusive. It is a ``loose baggy monster''
  into which all manner of notions may be stuffed, and a
  compelling and coherent general framework for intensional
  concepts is still to emerge.
\end{quote} Our discussion in the previous section hints at the
fact that the choice of what is extensional, and thereby what is
intensional is entirely up to us. There are many shades on the
spectrum between ideal mathematical object and full symbolic
description, and the choice of perspective depends on what we wish
to study. However, the design of a mathematical framework or
universe where both the extensions and intensions of our choice
co-exist harmoniously is the difficult and well-defined problem
with which this thesis is concerned.

The present author believes that such a framework will be
instrumental in making progress in providing answers to the
following questions: \begin{enumerate}
  \item
    What is the meaning of \emph{intensional programming}? Is
    there a logical interpretation of it?
  \item
    What is the meaning of \emph{reflective programming}? Is it
    possible to program reflectively in a consistent way?
  \item
    What is the meaning of \emph{intensional operations} in a
    \emph{higher-order setting}? How can we have a non-functional
    operation without resorting to first-order manipulations of
    G\"odel numbers?
  \item
    How can we add recursion to type theory?
\end{enumerate} The theory developed in this thesis fully
addresses the first two of these questions. Considered as a
toolkit, it is likely to be useful in answering the third one. The
fourth one is left for future work. Nevertheless, we shall now
briefly consider all of them.

\subsection{Intensional Programming}
  \label{sec:intprog}

In the realm of \emph{functional computation}, we can immediately
distinguish two paradigms: 
\begin{itemize}
  \item \textbf{The Extensional Paradigm}. It has been exactly 50
  years since Christopher Strachey articulated the notion of
  \emph{functions as first-class citizens} in his influential
  notes on programming \citep{Strachey2000}. In a purely
  functional world, a higher-order program can use a functional
  argument extensionally by evaluating it at a finite number of
  points: this leads to a form of \emph{continuity}, which is the
  basis of \emph{domain theory} \citep{Abramsky1994}.

  \item \textbf{The Intensional Paradigm}. This way of computing
  originated in \emph{computability theory} \citep{Cutland1980,
  Jones1997}: a program can compute with the \emph{source
  code}---or \emph{intension}---of another program as an argument.
  It can edit, optimise, call, or simulate the running of this
  code.
\end{itemize} Whereas the first paradigm led to a successful
research programme on the \emph{semantics of programming
languages}, the second is often reduced to \emph{symbolic
evaluation}. This is one of the reasons for which the intensional
paradigm has not reached the sophistication of its extensional
counterpart. Yet, the question remains: what can the intensional
paradigm contribute to programming? What is the additional
expressivity or programming power afforded by intensionality?

This is a theme that is often discussed by the \textsc{Lisp}
community. Indeed, certain dialects of \textsc{Lisp} are the
closest we have to a true paradigm of intensional programming,
both through \emph{\textsc{Lisp} macros} \citep{Graham1993} as
well the construct of \emph{quoting} \citep{Bawden1999}. Either of
these features can also be used for \emph{metaprogramming}, which
is the activity of writing programs that write other programs.

The notion of intensional programming is also central to the work
of the \emph{partial evaluation community}, which uses a rather
extreme form of intensionality that is better known under the name
\emph{programs-as-data}: see \cite{Jones1993, Jones1996,
Jones1997}. Their work uses insights from computability theory and
G\"odel numberings to build a metaprogramming-oriented methods for
the automatic generation of compilers, all based on the power of
the \emph{s-m-n theorem} of computability theory: see
\S\ref{sec:pareval} for more details and references. Even though
\textsc{Lisp}-inspired, and never seriously considering
non-functional operations, this is perhaps the closest anybody has
come to what we mean by intensional programming.

But, to this day, no satisfying theoretical account of
intensionality in programming has been produced. The present
author believes that this has to do with the fact that
\textsc{Lisp} is an unstructured, untyped language.  Hence, it is
not amenable to any kind of analysis, other than maybe that of the
untyped $\lambda$-calculus \citep{Barendregt1984}. See also
\cite{Wadler1987} for an early critique of \textsc{Scheme} in
programming methodology.

We shall introduce a new approach to intensional programming that
is fundamentally \emph{typed} in \S\ref{chap:ipcf}.

\subsection{Reflective Programming}
  \label{sec:reflprog}

For a very long time, programmers and theoreticians have sought to
understand \emph{computational reflection}, a concept introduced
by Brian Cantwell \cite{Smith1982, Smith1984}. Computational
reflection is an obscure idea that has been used in a range of
settings: see \cite{Malenfant1995} for a (slightly dated) survey.
In broad strokes, it refers to the ability of a program to
\emph{access its own code}, \emph{refer to its own description},
or \emph{examine its own internals}. 

The kind of reflection envisaged by Brian Cantwell Smith was to be
implemented by using \emph{reflective towers}. The idea is that a
program could be understood as running on the topmost level of an
infinity of interpreters. Reflective constructs then accept code
that is to be injected to the interpreter situated one level
below, hence having access to the complete state of the top-level
interpreter, all its registers and variables, and so on---to
infinity. This is a rather mysterious construction with unclear
semantics that have been the subject of investigation by
\cite{Friedman1984}, \cite{Wand1988}, and \cite{Danvy1988}. This
line of work eventually concluded with a theorem of
\cite{Wand1998}, which shows that there are no useful semantic
descriptions of the reflective tower. We will discuss Wand's
result in \S\ref{sec:wand}.

Despite its being poorly understood, computational reflection
seems to be a recurring and useful concept. Demystifying it is a
pressing problem, as many modern programming languages have
reflective or `introspective' facilities that lead to pernicious
bugs. The author suspects that a central theme of this thesis, the
notion of \emph{intensional recursion}, will prove fundamental in
obtaining a logical foundation for reflection. To quote
\cite{Polonsky2011}: \begin{quote}
  Of course, they [the questions posed by Smullyan] are only a
  sliver in the more global puzzle of understanding reflection as
  a distinct phenomenon. There is still lacking a general concept,
  an all-inclusive definition through which the common features of
  the constructions in G\"odel’s theorem, computability, number
  theory (systems of arithmetic), and set theory could be related.
  Finding such a concept remains a fascinating open problem.
\end{quote} We believe that the notion of \emph{intensional fixed
points}, to which we devote \S\ref{chap:irec}, is precisely an
abstract definition of (well-behaved) reflection. Our framework
thereby provides a candidate solution to the problem of Polonsky.

\subsection{Higher-Order Non-Functional Computation}
  \label{sec:honfc}

Over the past years, a new field of theoretical computer science
has emerged under the name of \emph{higher-order computability}.
As a subject, higher-order computability has its roots in the
1950s, but the work of \cite{Longley2000} has produced a unifying
account that overlaps significantly with the study of logic,
$\lambda$-calculus, category theory, realizability theory, and the
semantics of programming languages; see the recent book by
\cite{Longley2015}.

In classical computability theory \citep{Rogers1987, Cutland1980,
Odifreddi1992} there was a \emph{confluence of ideas}
\citep{Gandy1988} in the 1930s, culminating in the
\emph{Church-Turing thesis}, viz. that all `effectively
calculable' functions---whatever that might mean---are partial
recursive. In contrast, higher-order computability suffers from a
\emph{Church-Turing anti-thesis}: there are multiple notions of
computation at higher order, and some of them can be shown to be
strongly incompatible with each other. This situation generates a
lot of debate regarding which notion of higher-order computability
is `more natural' than the others: see the discussion in \citep[\S
1]{Longley2000}

Perhaps one of the most difficult challenges in higher-order
computability is to clarify what intensional, or
\emph{non-functional higher-order computation} is. This does not
only pertain to computation with the usual effects, like memory,
exceptions, or first-class continuations: such effects are
well-understood, either through \emph{game semantics}
\citep{Abramsky1996a, Abramsky1998}, or more abstractly through
the various theories of effects: see \citep{Moggi1989, Moggi1991,
Plotkin2004b, Hyland2007, Levy2003}. Instead, we consider more
general non-functional computation, where the non-functional
aspect arises from the ability of a device to read the
description or code of its higher-order argument.

A known example that is impossible to accommodate in an
extensional setting is the \emph{modulus of continuity
functional}. This is a type 3 functional, \[
  \Phi : \left((\mathbb{N} \rightarrow \mathbb{N}) \rightarrow
  \mathbb{N}\right) \rightarrow \mathbb{N}
\] Intuitively, when given a type 2 functional $F : (\mathbb{N}
\rightarrow \mathbb{N}) \rightarrow \mathbb{N}$, $\Phi(F)$ returns
an \emph{upper bound} $n$ on the range of values $F$ would examine of
any argument given to it, so that if $f, g : \mathbb{N}
\rightarrow \mathbb{N}$ agree on those values, i.e. \[
  \forall i \in \{0, \dots, n\}.\ f(i) = g(i)
\] then $Ff = Fg$. It can be shown that one cannot define a
modulus of continuity functional that is extensional. To compute
$\Phi(F)$ it is necessary to examine the internals of $F$:
we simulate its run on some function, and see what the maximum
argument it examines is. This can be computed in a language with
side-effects: we call $F(f_\spadesuit)$, where $f_\spadesuit$ uses
side-effects to record the maximum value at which it is called,
and $\Phi(F)$ then returns this recorded value. We can see that
this is highly dependent on the way $F$ is implemented, and thus
not extensional: see the blog post of Andrej \cite{Bauer2011}. The
question we want to ask in this setting is this: can we make the
modulus of continuity computable, perhaps by admitting it at
\emph{certain intensional types only}, but \emph{without}
generally violating neither extensionality nor freedom from side
effects?

The main ways to understand such computation are either through a
`highly intensional,' essentially untyped, first-order
formulation, which suffers from a lack of logical structure and
useful properties; or through computational effects, which is even
less clear in terms of higher-order computability. One way out of
this impasse has been suggested by Longley \cite[\S
6]{Longley2000}, and it employs realizability. Longley argues that
posing the question of whether a non-functional operation is
definable amounts to writing down a logical formula that specifies
it, and then examining whether that formula is \emph{realizable}:
see \citep{Longley1999} and the note \citep{Longley1999b}.

Whereas this is an informative approach, it does not smoothly lead
us toward the design of programming languages that harness these
non-functional powers. We propose, instead, the following research
programme: we should first use the type-theoretic techniques
proposed in this thesis to add intensionality to a
$\lambda$-calculus. Then, we shall be able to vary the intensional
operations available, and study the expressivity of the resulting
systems.

\subsection{Recursion in Type Theory}
  \label{sec:typethrec}

General recursive definitions are prohibited in most type
theories, including \emph{Martin-L\"of type theory
(\textsf{MLTT})} \citep{MartinLof1972, MartinLof1984,
Nordstrom1990}. This is not an arbitrary design choice:
\textsf{MLTT} features \emph{dependent types}, i.e. types depend
on terms. Thus, its terms ought to be strongly normalising, so
that the types themselves are.

In that way, \textsf{MLTT} is logically well-behaved, but not
ideal for programming. In fact, writing down the definitions of
many ordinary programs becomes a difficult exercise. The intuitive
reason is that every program we want to write must somehow contain
its own proof of termination. A lot of work has gone into
regaining the lost expressivity, from the results of
\cite{Constable1993} on \emph{partial types}, to the method of
\cite{Bove2005}, and all the way to the coinductive types of
\cite{Capretta2005} and the \emph{delay monad}.  There is also
recent work on the \emph{partiality monad} and higher inductive
types: see \cite{Altenkirch2017}.

In any case, working within \textsf{MLTT} has a fundamental
disadvantage: an old theorem of \cite{Blum1967} ensures that, no
matter what kind of `blow-up factor' we choose, we will be able to
write a program (in a partial, Turing-complete language) whose
shortest equivalent in a total programming language is larger by
the given factor: see the blog post of \cite{Harper2014}. Thus,
ultimately, there is no working around this theorem: at some point
we might have to add general recursion to type theory.

In the setting of \emph{simple types} adding recursion is
relatively easy. It was first done by \cite{Scott1993} and
\cite{Plotkin1977}, who defined the prototypical fixed point
language PCF, a simply-typed $\lambda$-calculus of booleans,
integers, and a fixed point combinator $\mathbf{Y} : (A
\rightarrow A) \rightarrow A$. There is a long and rich literature
on the models of PCF: the reader may consult any of
\citep{Plotkin1977, Gunter1992, Mitchell1996, Abramsky1996,
Hyland2000, Streicher2006, Longley1995}.

Merely throwing $\mathbf{Y}$ into the mixture is not advisable
in more expressive type theories, e.g. in System F or
\textsf{MLTT}. If sum or $\Sigma$ types are available (or
are definable, e.g. in System F), a theorem of \cite{Huwig1990}
guarantees that the resulting theory will be inconsistent: even
the existence of the coproduct $\mathbf{1} + \mathbf{1}$ is enough
to make a cartesian closed category with fixpoints degenerate to a
\emph{preorder}.

A short abstract by \cite{Plotkin1993} proposes that we forget the
cartesian setting, and work in either a (second-order)
\emph{intuitionistic linear type theory}, or a \emph{relevant type
theory}. The full manuscript for that abstract never appeared, but
the linear type system to which it refers (but without recursion)
was studied in detail by \cite{Barber1996}.

Further efforts went into defining a \emph{Linear PCF}, based on
these intuitions. Invariably, the type of $\mathbf{Y}$ or the rule
for recursion has one of the following forms:
\begin{align*}
  !(!A \multimap A) \multimap A 
    & \qquad \text{\cite{Brauner1995, Brauner1997b}} \\
  \begin{prooftree}
    !(!A \multimap A)
      \justifies
    A
  \end{prooftree} & \qquad\text{\cite{Maraist1995, Bierman2000}} \\
  \begin{prooftree}
    !(!A \multimap A)
      \justifies
    !A
  \end{prooftree} & \qquad\text{\cite{Maraist1995}}
\end{align*} We will shortly see that this pattern is not
accidental. We will develop a type-theoretic approach to
intensionality that will admit a slightly unusual kind of
recursion, namely \emph{intensional recursion}. Its type-theoretic
interpretation will be strongly reminiscent of the pattern of
these rules, but linearity will prove to be a red herring.

\section{Quoting is Impossible}

The first impression that one usually acquires regarding
intensional phenomena is that they can only spell trouble. After
all, encoding programs or logical formulae as data \emph{in the
same language} is the typical function of \emph{G\"odel
numbering}. Such constructs quickly lead to \emph{negative}
theorems that pinpoint the inherent limitations of logical
systems, e.g. G\"odel's \emph{First Incompleteness Theorem}
\citep{Smullyan1992}.

Our aim in this thesis is to tell a \emph{positive story}, but we
shall first recount the negative tales of the past. We shall not
engage in a foundational debate regarding G\"odel's theorems and
related results here, but a more in-depth discussion can be found
in \S \ref{chap:irec}, in the book of \cite{Smullyan1992}, and in
\cite[\S 2]{Girard2011}. Instead, we will focus on the negative
repercussions on programming.

\subsection{Tale 1: Quoting is Not Definable}

Let $\Lambda$ be the set of untyped $\lambda$-terms
\citep{Barendregt1984}, and let \[
  \ulcorner \cdot \urcorner : \Lambda \rightarrow \Lambda
\] be a map on $\lambda$-terms. The intention is that, for each
$\lambda$-term $M$, the term $\ulcorner M \urcorner$ represents
the \emph{program} $M$ as a \emph{datum} in the
$\lambda$-calculus. We call $\ulcorner M \urcorner$ the
\emph{quote of $M$}. The properties of such (external) quoting
functions $\ulcorner \cdot \urcorner : \Lambda \rightarrow
\Lambda$ have been systematically studied by \cite{Polonsky2011},
who also broadly surveys the sporadic literature on such
`meta-encodings' of the $\lambda$-calculus into itself.

In standard accounts, e.g. \citep{Barendregt1984}, the quote
function is a \emph{G\"odel numbering}, as known from the
literature on G\"odel's theorems. To each term $M$ one assigns a
number $\hash M$, and defines $\ulcorner M \urcorner$ to be the
Church numeral for $\hash M$.

A fundamental question then arises: \emph{is quoting internally
definable?} The answer is, of course, negative, as internal
definability of quoting would lead to inconsistency. The following
argument is due to \cite{Barendregt1991}: suppose there is a term
$\mathbf{Q} \in \Lambda$ such that \[
  \mathbf{Q}\,M =_\beta \ulcorner M \urcorner
\] It is a fact that Church numerals are in normal form. Hence,
both $\ulcorner \mathbf{I} \urcorner$ and $\ulcorner
\mathbf{I}\,\mathbf{I} \urcorner$ are in normal form, where
$\mathbf{I} \myeq \lambda x.\,x$ is the identity combinator.  We
have that \[
  \ulcorner \mathbf{I} \urcorner
    =_{\beta}
  \mathbf{Q}\,\mathbf{I}
    =_{\beta}
  \mathbf{Q}\,(\mathbf{I}\, \mathbf{I})
    =_{\beta}
  \ulcorner \mathbf{I}\, \mathbf{I} \urcorner
\] This amounts to equating two distinct normal forms. But it is
known that the $\lambda$-calculus is confluent, hence consistent!
It follows that such a term $\mathbf{Q}$ cannot exist.

It is very important to notice that the crux of the argument
essentially rests on the confusion between the extension $M$, the
intension $\ulcorner M \urcorner$, and the ability to pass
\emph{from extension to intension}. To obtain a consistent account
of intensionality we ought to forbid this possibility. However,
let us seize the opportunity to remark that the opposite direction
is not only attainable, but a result of historical importance for
computability theory \citep{Kleene1936, Kleene1981}:

\begin{thm}
  [\citet{Kleene1936}]
  \label{thm:lambdainterp}
  There exists a term $\mathbf{E} \in \Lambda^0$ such that \[
    \mathbf{E}\ \ulcorner M \urcorner
      \twoheadrightarrow_\beta
    M
  \] for all $M \in \Lambda^0$.
\end{thm} 

\noindent This is essentially the same result as the one obtained
by \cite{Turing1937}, but for the $\lambda$-calculus instead of
the Turing Machine. See \cite[\S 8.1.6]{Barendregt1984} for a
proof.

\subsection{Tale 2: Quoting Collapses Observational Equivalence}
  \label{sec:wand}

Let us suppose that the above result does not deter us in our
plans to add reflection to the $\lambda$-calculus. All else
failing, we can do so by postulating constants $\mathbf{eval}$ and
$\mathbf{fexpr}$, along with the following reductions:
\begin{align*}
  \mathbf{eval}\ \ulcorner M \urcorner
    &\red{} M \\
  (\mathbf{fexpr}\ V)\ M 
    &\red{} V\ \ulcorner M \urcorner
\end{align*} where $V$ is some notion of value (e.g. a weak head
normal form). Then $\mathbf{Q}$ would be definable as
$\mathbf{fexpr}\ (\lambda x.\,x)$.

One of the most interesting and well-studied notions in
$\lambda$-calculus is that of \emph{observational equivalence}.
Two terms $M, N$ are observationally equivalent, written $M \cong
N$, if they are interchangeable in all possible contexts, without
any `observable changes.' The notion of observable change is up
for debate, and the exploration of different options leads to
interesting variations---see e.g. \cite{Bloom1989} and
\cite{Abramsky1990b}. The usual choice is that we can
observe normal forms at ground type (i.e. different numerals and
boolean values), or---equivalently---termination of evaluation at
ground type. If $M =_\beta N$, then certainly $M \cong N$, but the
converse is not normally true: terms can be observationally
equivalent, but the theory of equality (which is only computably
enumerable) is not strong enough to show that.

In this context, \cite{Wand1998} showed the following theorem:
\begin{thm}[Wand]
  $M \cong N$ if and only if $M \equiv_\alpha N$
\end{thm} 

\noindent Wand's definition of observation is termination at a
weak head normal form, and his quoting function $\ulcorner \cdot
\urcorner$ is a Scott-Mogensen encoding: see \citep{Polonsky2011,
Mogensen1992}. However, the result is strong and general, and puts
the last nail in the coffin: there can be no semantic study of
functional programming languages that are so strongly reflective
that they can internally define quoting. Such a language can
internally distinguish any two terms in the language, as long as
they are not equal up to renaming. This result concludes the long
line of research on Smith's reflective towers that we mentioned in
\S\ref{sec:reflprog}.

\section{Intensionality and Types: Modality-as-Intension}
  \label{sec:modasint}

The two negative tales we have just told are, fortunately, not the
final word. If we take a close look at the existing literature,
e.g. at the chapters of \citep{Barendregt1984} that concern
computability and diagonal arguments, we may see that some
intensional operations are indeed definable. For example, there
are $\lambda$-terms $\textbf{gnum}, \textbf{app}$ and $\mathbf{E}$
such that \begin{align*}
  \textbf{gnum}\ \ulcorner M \urcorner
    &=_\beta
  \ulcorner \ulcorner M \urcorner \urcorner \\
  \textbf{app}\ \ulcorner M \urcorner\, \ulcorner N \urcorner
    &=_\beta
  \ulcorner M\,N \urcorner \\
  \mathbf{E}\ \ulcorner M \urcorner
    &=_\beta
  M
\end{align*} We mentioned $\mathbf{E}$ in the previous section,
but the other two might come as a surprise: they are indeed pure
operations on syntax. The pattern that seems to be at work here is
that the only true restriction is quoting, and that everything
else is admissible.

The standard way to enforce restriction in computation is to use
\emph{types}. Indeed, the main contribution of this thesis is a
detailed investigation and analysis of the following idea:
\begin{center}
  \shadowbox{
    \begin{minipage}{0.6\textwidth}
      \centering
      Intensionality adheres to a typing discipline.
    \end{minipage}
  }
\end{center} This is an old but not so well-known observation. To
the best of our knowledge, it was first formulated by Neil
Jones,\footnote{Personal communication.} and led to his work on
\emph{underbar types}: see \cite[\S 16.2]{Jones1993}.

\subsection{Modality-as-Intension}

Given a type $A$, let us write $\textsf{Code}(A)$ for the
\emph{type of code of type $A$}.  We must certainly have that, if
$M : A$, then $\ulcorner M \urcorner : \textsf{Code}(A)$. Thus, if
$\textbf{gnum}\ \ulcorner M \urcorner =_\beta \ulcorner \ulcorner
M \urcorner \urcorner$, then it must have the type \[
  \textbf{gnum} : \textsf{Code}(A) \rightarrow
  \textsf{Code}\left(\textsf{Code}(A)\right)
\] This looks familiar! If we drop all pretense and write $\Box
A$ for $\textsf{Code}(A)$, we obtain the following types:
\begin{align*}
  \textbf{gnum} &: \Box A \rightarrow \Box\Box A \\
  \textbf{app}  &: 
    \Box(A \rightarrow B) \rightarrow \Box A \rightarrow \Box B \\
  \textbf{E}    &: \Box A \rightarrow A
\end{align*} Surprisingly, there is an underlying Curry-Howard
correspondence: the types of these operations correspond to the
axioms \textsf{4}, \textsf{K} and \textsf{T} of the modal logic
\textsf{S4}. This connection to modal logic was drawn by
\cite{Davies1996, Davies2001a}, who used it to embed two-level
$\lambda$-calculi \citep{Nielson1992} in a modal
$\lambda$-calculus, and to perform binding-time analysis.

We will thus introduce and study the \emph{modality-as-intension}
interpretation. This is an idea that pervades \citep{Davies1996,
Davies2001a}, and is even mentioned in name in the conclusion of
\citep{Davies2001}. To quote: \begin{quote}
  One particularly fruitful interpretation of $\Box A$ is as the
  intensional type for expressions denoting elements of type $A$.
  Embedding types of this form in a programming language means
  that we can compute with expressions as well as values. The term
  $\ibox{M}$ quotes the expression $M$, and the construct
  $\letbox{u}{M}{N}$ binds $u$ to the expression computed by $M$
  and then computes the value of $N$.  The restrictions placed on
  the introduction rule for $\Box A$ mean that a term $\ibox{M}$
  can only refer to other expression variables $u$ but not value
  variables $x$. This is consistent with the intensional
  interpretation of $\Box A$, since we may not know an expression
  which denotes a given value and therefore cannot permit an
  arbitrary value as an expression.
\end{quote} The above excerpt refers to the modal type system
introduced and used by Davies and Pfenning, which is a
\emph{dual-context $\lambda$-calculus}, with judgements of the
form \[
  \ctxt{\Delta}{\Gamma} \vdash M : A
\] where $\Delta$ and $\Gamma$ are two ordinary but disjoint
contexts. The variables that occur in $\Delta$ are to be thought
of as \emph{modal variables}, or variables that carry
\emph{intensions} or \emph{codes}, whereas the variables in
$\Gamma$ are ordinary (intuitionistic) variables. In this system,
the evaluator $\mathbf{E} : \Box A \rightarrow A$ exists as a
variable rule, i.e. the ability to use a code variable as if it
were a value: \[
  \begin{prooftree}
    \justifies
      \ctxt{\Delta, u{:}A, \Delta'}{\Gamma} \vdash u : A
  \end{prooftree}
\] The canonical terms of modal type are of the form $\ibox{M}$,
and they largely mimic the G\"odel numbering $\ulcorner M
\urcorner$. The shape of the introduction rule guarantees that all
the variables that occur in the `boxed term' $M$ are code
variables, and not value variables; we write $\cdot$ for the empty
context: \[
  \begin{prooftree}
    \ctxt{\Delta}{\cdot} \vdash M : A
      \justifies
    \ctxt{\Delta}{\Gamma} \vdash \ibox{M} : \Box A
  \end{prooftree}
\] And, as mentioned above, there is also a $\letbox{u}{(-)}{(-)}$
construct that allows for substitution of quoted terms for code
variables: \[
  \begin{prooftree}
    \ctxt{\Delta}{\Gamma} \vdash M : A
      \quad
    \ctxt{\Delta, u{:}A}{\Gamma} \vdash N : C
      \justifies
    \ctxt{\Delta}{\Gamma} \vdash \letbox{u}{M}{N} : C
  \end{prooftree}
\] along with the reduction \[
  \letbox{u}{\ibox{M}}{N} \red{} N[M/u]
\] The $\letbox{u}{(-)}{(-)}$ construct secretly requires $\Box(A
\times B) \cong \Box A \times \Box B$, which is just enough 
to define the $\textbf{app} : \Box(A \rightarrow B) \rightarrow
\Box A \rightarrow \Box B$ constant.

Finally, the \textsf{4} axiom is inherent in the system: \[
  \begin{prooftree}
    \ctxt{\cdot}{x : \Box A} \vdash x : \Box A
      \quad
    \[
    \[
      \ctxt{u{:}A}{\cdot} \vdash u : A
        \justifies
      \ctxt{u{:}A}{\cdot} \vdash \ibox{u} : \Box A
    \]
      \justifies
    \ctxt{u{:}A}{x : \Box A} \vdash \ibox{\ibox{u}} : \Box A
    \]
      \justifies
    \ctxt{\cdot}{x : \Box A} \vdash \letbox{u}{x}{\ibox{\ibox{u}}} : \Box A
  \end{prooftree}
\] The author has previously studied such dual-context systems in
\citep{Kavvos2017b}.

\subsection{A Puzzle: Intensional Recursion}
  \label{sec:intrec}

We touched upon the subject of recursion in type theory in
\S\ref{sec:typethrec}. To obtain recursion in simple types, we
have to add a fixed point combinator $\mathbf{Y} : (A \rightarrow
A) \rightarrow A$, and obtain PCF. In contrast, recursion is
definable in the untyped setting. This is the conclusion of the
\emph{First Recursion Theorem} \cite[\S 2.1, \S 6.1]{Barendregt1984}:
\begin{thm}
  [First Recursion Theorem]
  $\forall f \in \Lambda.\
  \exists u \in \Lambda.\
  u = fu$.
\end{thm}
\begin{proof}
  Let \[
    \mathbf{Y} \myeq \lambda f. (\lambda x. f(xx))(\lambda x.  f(xx))
  \] Then $\mathbf{Y}f =_\beta f(\mathbf{Y}f)$.
\end{proof}

The FRT corresponds to \emph{extensional recursion}, which is what
most programming languages support. When defining a recursive
function definition, a programmer may make a finite number of
calls to the definiendum itself, in the same vein as our
description of the functional-extensional paradigm in
\S\ref{sec:intprog}. Operationally, this leads a function to
examine its own values at a finite set of points at which it
has---hopefully---already been defined. 

However, as \cite{Abramsky2014} notes, in the \emph{intensional
paradigm}, which we also described in \S\ref{sec:intprog}, a
stronger kind of recursion is attainable. Instead of merely
examining the result of a finite number of recursive calls, the
definiendum can recursively have access to a \emph{full copy of
its own source code}. This is embodied in the \emph{Second
Recursion Theorem (SRT)}, which was proved by \cite{Kleene1938}.
Here is a version of the SRT in the untyped $\lambda$-calculus
\cite[\S 6.5]{Barendregt1984}:
\begin{thm}
  [Second Recursion Theorem]
  $\forall f \in \Lambda.\
  \exists u \in \Lambda.\
    u = f \, \ulcorner u \urcorner$.
\end{thm}
\begin{proof}
  Given $f \in \Lambda$, set $u \myeq W\ulcorner W \urcorner$, where
  \[
    W \myeq \lambda x.\ f(\mathbf{app}\ x\ (\textbf{gnum}\ x))
  \] where $\mathbf{app}$ and $\mathbf{gnum}$ are as above. Then
  \begin{align*}
    u 
      &\equiv
    W\ulcorner W\urcorner \\
      &=_\beta
    f (\textbf{app}\ \ulcorner W \urcorner\
                     (\textbf{gnum}\ \ulcorner W \urcorner)) \\
      &=_\beta
    f (\textbf{app}\ \ulcorner W \urcorner\
                     \ulcorner \ulcorner W \urcorner \urcorner) \\
      &=_\beta
    f (\ulcorner W \ulcorner W \urcorner \urcorner) \\
      &\equiv
    f\ \ulcorner u \urcorner
  \end{align*}
\end{proof}

\noindent It is not hard to see that, using Kleene's interpreter
for the $\lambda$-calculus (Theorem \ref{thm:lambdainterp}), the
SRT implies the FRT. It is not at all evident whether the converse
holds. This is because the FRT is a theorem that concerns
higher-order computation, whereas the SRT is very much grounded on
first-order, diagonal constructions. The exact relationship
between these two theorems is the subject of \S\ref{chap:srtht}.

The point that we wish to make is that, in the presence of
intensional operations, the SRT affords us with a much stronger
kind of recursion. In fact, it allows for exactly the sort of
\emph{computational reflection} that we discussed in
\S\ref{sec:reflprog}.

Perhaps the greatest surprise to be found in this thesis is that
the SRT \emph{admits a type}.  Indeed, suppose that $u : A$. Then
certainly $\ulcorner u \urcorner : \Box A$, and it is forced that
\[
  f : \Box A \rightarrow A
\] The Curry-Howard reading of the SRT is then the following: for
every $f : \Box A \rightarrow A$, there exists a $u : A$ such that
$u = f \, \ulcorner u \urcorner$. This corresponds to
\emph{L\"ob's rule} from \emph{provability logic}
\citep{Boolos1994}: \[
  \begin{prooftree}
    \Box A \rightarrow A
      \justifies
    A
  \end{prooftree}
\] L\"ob's rule is equivalent to adding the G\"odel-L\"ob axiom,
\[
  \Box (\Box A \rightarrow A) \rightarrow \Box A
\] to a modal logic. One of the punchlines of this thesis will
be that \begin{center}
  \shadowbox{
    \begin{minipage}{0.9\textwidth}
      \centering
      The type of the Second Recursion Theorem is the
      G\"odel-L\"ob axiom.
    \end{minipage}
  }
\end{center}

\noindent Thus, to obtain reflective features, all we have to do
is add a version of L\"ob's rule to the Davies-Pfenning modal
$\lambda$-calculus. 

%Our choice of rule will be equivalent to the
%above, but proof-theoretically well-behaved---see
%\citep{Ursini1979a, Kavvos2017b}: \[
%  \begin{prooftree}
%    \Box A \rightarrow A
%      \justifies
%    \Box A
%  \end{prooftree}
%\] If we have a moment to think as follows, then it is intuitively
%evident why this version is computationally more well-behaved: it
%%says that for every $f : \Box A \rightarrow A$, there exists a
%$u : \Box A$ such that \[
%  \mathbf{E}\,u = f\,u
%\] so that the existential part of the statement provides an
%intensional object ($u : \Box A$), which contains `more
%information' than an extensional one (e.g. $u : A$).

\section{A Road Map}

The rest of this thesis consists of a thorough discussion of the
above observations, and the development of appropriate syntax and
categorical semantics that capture the aforementioned intuitions.

In \S\ref{chap:srtht}, we dive back to find the origins of
Kleene's two Recursion Theorems. As it happens, these correspond
to the two ways in which we can define a mathematical object by
recursion: one can either use diagonalisation, or some kind of
infinite (or transfinite) iteration. The origin of both of these
methods is lost in the mists of time, but both find their first
documented expressions in the work of \cite{Kleene1952,
Kleene1938}. \S\ref{chap:srtht} states and proves these two
theorems, and engages in a thorough discussion regarding their
similarities and differences. Whereas the SRT works by
diagonalising and is applicable to a first-order setting, the FRT
requires a higher-order perspective. In some cases both theorems
are applicable, but one is stronger than the other: the FRT always
produces least fixed points, but this is not always the case with
the SRT. Nevertheless, an old result by Hartley Rogers Jr.
bridges this gap. We also find the opportunity to engage in some
speculation regarding the possible applications of the reflective
features provided by the SRT.

In \S\ref{chap:ipcf}, we revisit the system of \cite{Davies2001}
that we described in \S\ref{sec:modasint}. After noting that it
does not feature any actual intensional operations, we add some to
the system. We also take the hint from \S\ref{sec:intrec}, and add
a form of L\"ob's rule as well. The result is a programming
language that is (a) \emph{intensional}, exactly in the sense
described in \S\ref{sec:intprog}, and (b) \emph{reflective}, in
the sense described in \S\ref{sec:reflprog}. A confluence proof
ensures that the resulting system, which we call \emph{Intensional
PCF}, is consistent. It is evident that the central device by
which everything comes together is the modal types, which separate
the two worlds of intension and extension, by way of containing
the former under the modality.

Following that, in \S\ref{chap:expo} we begin to seek categorical
semantics for intensionality. We argue that 1-category theory is
ill-suited for modelling intensionality. We are thus led to
consider the \emph{P-categories} of \v{C}ubri\'{c}, Dybjer and
Scott \citep{Cubric1998}, which are categories only up to a
\emph{partial equivalence relation} (PER). In this setting, we
introduce a new P-categorical construct, that of \emph{exposures}.
Exposures are very nearly functors, except that they do not
preserve the PERs of the P-category, but \emph{reflect} them
instead. Inspired by the categorical semantics of \textsf{S4}
\citep{Bierman2000a}, we begin to develop the theory of exposures.

To substantiate the discussion, \S\ref{chap:expoexamples} builds
on the previous chapter by presenting three concrete examples of
exposures. The first example shows that, when built on an
appropriate P-category, an exposure is really an abstraction of
the notion of a well-behaved intensional device, such as a
\emph{G\"{o}del numbering}. The second example is based on
\emph{realizability theory}; it is also the motivating example for
exposures, and is later used to show that Kleene's SRT is a form
of intensional recursion. The final example illustrates that
intensionality and exposures may occur outside logic and
computability, and is related to basic homological algebra.

Then, in \S\ref{chap:irec}, we put on our P-categorical spectacles
and examine intensional recursion. We find that it has a simple
formulation in terms of exposures. We then show that
classic theorems of logic that involve intensional recursion, such
as G\"odel's \emph{First Incompleteness Theorem}, Tarski's
\emph{Undefinability Theorem}, and \emph{Rice's Theorem}, acquire
concise, clear formulations in the unifying framework of
exposures. Moreover, our theory ensures that each logical device
or assumption involved in their proofs can be expressed in the
same algebraic manner. The chapter concludes by using exposures to
generalise a famous categorical fixed-point theorem of
\cite{Lawvere1969, Lawvere2006}, which roughly corresponds to a
restricted version of the FRT. The resulting Intensional Recursion
Theorem is a categorical analogue of Kleene's SRT.

At last, in \S\ref{chap:intsem1} and \S\ref{chap:intsem2}, we
bring Intensional PCF (\S\ref{chap:ipcf}) and exposures
(\S\ref{chap:expo}) together, by using the second to provide a
semantics for the first. Intensional PCF is too expressive for
this endeavour, so we discuss a restriction of it, which we call
Intensional PCF v2.0. We find that a sound semantics for
it consists of a \emph{cartesian closed P-category} equipped with
a \emph{product-preserving, idempotent comonadic exposure}. We
then discuss in which cases we may lift some of these new
restrictions of Intensional PCF v2.0. We conclude by showing that
$\mathfrak{Asm}(K_1)$ is a model of Intensional PCF v2.0, with the
intensional fixed points being interpreted by Kleene's SRT.

Finally, in \S\ref{chap:conc}, we conclude our investigation by
evaluating our contribution, and delineating a number of future
directions towards which this thesis seems to point.

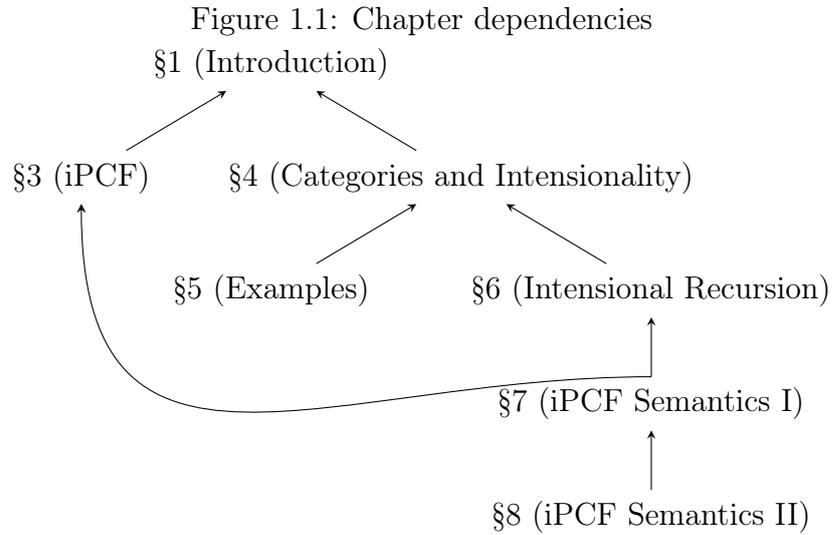
\begin{figure}[h]
  \caption{Chapter dependencies}
  \label{fig:chapdep}
  \centering
  \begin{tikzpicture}[stealth-]
    \node (is-root) {\S 1 (Introduction)}
      [sibling distance=5cm]
      child { node (ipcf) {\S 3 (iPCF)} }
      child {
        node {\S 4 (Categories and Intensionality)}
          [sibling distance=5cm]
          child { node (examples) {\S 5 (Examples)} }
          child { 
            node (irec) {\S 6 (Intensional Recursion)}
            [sibling distance=1.5cm]
            child {
              node (ipcfsem1) {\S 7 (iPCF Semantics I)}
              child { node (ipcfsem2) {\S 8 (iPCF Semantics II)} }
            }
          }
      };
    
      \draw (ipcf.south) edge[out=-90, in=180, looseness=1.4] (ipcfsem1.north);
  \end{tikzpicture}
\end{figure}

The figure presents a rough outline of the way the chapters of
this thesis depend on each other. \S 2 is not included in the
diagram, as it functionally independent of the developments in the
thesis. Nevertheless, we recommend that the reader consult it in
order to understand the origins and importance of intensional
recursion. The sequence \S1, \S4, \S6 may be read independently,
hence constituting the basics of our `theory of intensionality.'
Finally, the diagram does not capture two small further
dependencies: that of \S\ref{sec:fpexamples} on the examples
developed in \S 5, and that of the construction of IFPs in
$\mathfrak{Asm}(K_1)$ in \S\ref{sec:asmipcf} on the definition of
that P-category in \S\ref{sec:realexpo}.

\chapter[Kleene's Two Kinds of Recursion]
  {Kleene's Two Kinds of Recursion\footnote{A preprint is
  available as
  \href{https://arxiv.org/abs/1602.06220}{arXiv:1602.06220}}}
  \label{chap:srtht}

  It is well known that there are two ways to define a function by
recursion.

One way is through a \emph{diagonal construction}. This method
owes its popularity to Cantor, and forms the backbone of a large
number of classic diagonalisation theorems. Diagonal constructions
are a very concrete, syntactic, and computational method of
obtaining fixed points, which we in turn use to obtain recursion.

Another way is through a \emph{least fixed point} that is obtained
as a result of some kind of infinite (or even transfinite)
iteration. This kind of construction is more abstract and
mathematical in style. It is a very common trope in the study of
denotational semantics of programming languages, particularly
those based on \emph{domain theory} \citep{Abramsky1994}. The
origins of this lattice-theoretic argument are lost in the mists
of time, but see \citet{Lassez1982} for a historical exposition.

Both of these ways were famously used by Stephen C. Kleene
\citep{Kleene1981} to define functions by recursion in
computability theory. The least fixed point construction is the
basis of Kleene's \emph{First Recursion Theorem (FRT)}
\citep{Kleene1952}, whereas the diagonal construction is found at
the heart of his \emph{Second Recursion Theorem (SRT)}
\citep{Kleene1938}.

Nevertheless, it is not so well known that there is a slight
mismatch between these two theorems. This is mainly due to the
context in which they apply: the FRT is essentially a theorem
about \emph{computation at higher types}, whereas the SRT is a
\emph{first-order theorem} of a syntactic nature.

Modulo the above mismatch, it so happens that \emph{the SRT is
more general than the FRT}. Indeed, the SRT allows for a
computationally `stronger' kind of recursion---namely
\emph{intensional recursion}---whereas the FRT has a more
\emph{extensional} flavour. However, there is a close yet slightly
mysterious relationship between these two theorems, the
particulars of which we shall examine.

The SRT has numerous applications in computability, but it is
deafeningly absent in computer science. \citet{Abramsky2014} has
suggested further investigation, likening the SRT to ``the dog
that didn't bark'' in the Sherlock Holmes story, and has also
discussed several related issues.

In the sequel we shall investigate both of these types of
recursion, as well as their intricate relationship. In \S
\ref{sec:intensionalcomp} we discuss the notion of intensionality
and its relationship to computability; we state and prove the SRT,
and discuss the extra generality afforded by what we call
intensional recursion; and we sketch a number of speculative
applications of intensional recursion. Subsequently, in \S
\ref{sec:higher}, we move to the discussion of higher types: we
look at two slightly different notions of computation at higher
types, discuss their interaction, and prove the FRT for each one.
Finally, in \S\ref{sec:frtvssrt}, we investigate when each of the
recursion theorems applies, and when the resulting recursive
constructions match each other.

\section{Intensionality and Computability}
  \label{sec:intensionalcomp}

In loose philosophical terms, to be \emph{intensional} is to
contain not only \emph{reference}, but also \emph{sense}.  The
distinction between these two notions is due to Frege, see e.g.
\citet{Fitting2015}. An intensional sign denotes an external
\emph{referent}, yet inherently connotes more information---its
elusive \emph{sense}. The classic example is that of the planet
Venus, which may be referred to as either the morning star, or the
evening star.

Most mainstream mathematics is rather \emph{extensional}: we
normally reason about underlying, `ideal' mathematical objects,
and not their concrete descriptions; the latter are, in a way,
only there for our referential convenience. In most presentations
of set theory, for example, the \emph{axiom of extensionality}
equates any two sets whose members are the same.  Thus, in the
mathematical sense, \emph{to be intensional is to be finer than
some presupposed `extensional equality.'}
  
It is not difficult to argue that this setting is most inadequate
for Computer Science. On a very rough level, extensions correspond
to \emph{what may be computed}, whereas intensions correspond to
the \emph{programs and processes that carry out the computation},
see e.g. \citet{Moschovakis1993}. Once more, there is a
distinction to be made: programs may be understood by the ideal
objects that they refer to (e.g. functions), or their internal
characteristics: length, structure, and---ultimately---the
\emph{algorithm} they express.  

The former aspect---viz. the study of ideal objects behind
programs---is the domain of \emph{Computability Theory}
(previously known as \emph{Recursion Theory}), where the object of
study is `effectively computable' functions over the natural
numbers.  Computability Theory began with the `confluence of
ideas' \citep{Gandy1988} of multiple researchers in the late 1930s
in their attempt to characterise `automatic' or `mechanical'
calculability.  Remarkably, all roads led to Rome: different
notions were shown to coincide, leading to the identification of
the class of \emph{partial recursive functions}. Subsequent to
this fortuitous development, things took a decisive turn, as
further developments mostly concerned the study of the
\emph{incomputable}.\footnote{Harvey \citet{Friedman1998} once
made the following tongue-in-cheek recommendation to the
Foundations of Mathematics (FOM) mailing list: ``Why not rename
recursion theory as: noncomputability theory?  Maybe that would
make everybody happy.''}

So much for the extensional side. What about intensions? Here, we
encounter a diverse ecosystem. On one side, fixing the
\emph{Turing Machine} as one's model of computation leads to
\emph{Complexity Theory}, which attempts to classify algorithmic
problems with a view to identifying the exact resources that one
needs to solve a a problem. This aspect is largely reliant on more
\emph{combinatorial reasoning}. Alternatively, adopting the
$\lambda$-calculus as a point of reference leads to the study of
\emph{programming languages}, which includes---amongst other
things---type systems, semantics, program analysis, and program
logics. Here, the emphasis is on logical aspects. Finally, the
subject of concurrent and interactive computation unfolds as a
bewildering, obscure, and diverse landscape: see e.g.
\citet{Abramsky2006}.

\subsection{The Space of All Programming Languages}

It is, however, a curious state of affairs that standard
computability theory---as presented, for example, in the classic
textbooks of \citet{Rogers1987}, \citet{Cutland1980} and
\citet{Odifreddi1992}---begins with a small set of abstract
results that concern G\"odel numberings of partial recursive
functions. Even though they are the central pillar of an
extensional theory, these results have a decidedly intensional
flavour.

These results begin by putting some very mild conditions on
G\"odel numberings. Indeed, if one thinks of a G\"odel numbering
as a `programming language,' then these conditions comprise the
absolute minimum that intuitively needs to hold if that
programming language is `reasonable.' A clear presentation of this
part of the theory can be found in the classic textbook of
\citet[\S II.5]{Odifreddi1992}. A more modern account that is
informed by programming language theory is that of Neil
\citet{Jones1997}.

The story begins to unfold as soon as we encode
\emph{programs-as-data}, by assigning partial recursive functions
to natural numbers. Following tradition, we write $\phi$ for an
arbitrary numbering, and $\phi_p : \mathbb{N} \parfunc \mathbb{N}$
for the partial recursive function indexed by $p \in \mathbb{N}$
under the numbering $\phi$.

From a programming perspective, we may consider $p$ to be a
`program,' and $\phi : \mathbb{N} \rightarrow (\mathbb{N}
\parfunc{} \mathbb{N})$ a semantic function that maps programs to
the functions they compute. In practice, $p \in \mathbb{N}$ is
usually a G\"odel number that encodes the syntax of a Turing
Machine, or the instructions for a register machine, or even a
$\lambda$-term.

Let us write $e_1 \simeq e_2$ for \emph{Kleene equality}, viz.
to mean that expressions $e_1$ and $e_2$ are both undefined, or
both defined and equal in value. Of the numbering $\phi$ we shall
require the following conditions:

\begin{description}
  \item[Turing-Completeness]
    That for each partial recursive function $f$ there exists a
    $p$ such that $f = \phi_p$. 
  \item[Universal Function]
    That there is a program $U$ such that $\phi_U(x, y) \simeq
    \phi_x(y)$ for all $x, y \in \mathbb{N}$.
  \item[S-m-n]
    That there is a \emph{total} recursive function $S$ such that
    $\phi_{S(p, x)}(y) \simeq \phi_p(x, y)$ for all $x, y \in
    \mathbb{N}$.
\end{description}

That the first and second conditions are achievable was
popularised by \citet{Turing1937}, and the third is a result of
\citet{Kleene1938}. The first condition corresponds, by the
\emph{Church-Turing thesis}, to the fact that our programming
language is as expressive as possible (in extensional terms). The
second corresponds to the ability to write a
\emph{self-interpreter}, under suitable coding. And the third
allows us to computably `fix' an argument into the source code of
a two-argument program, i.e. that substitution is computable (c.f.
one-step $\beta$-reduction in the $\lambda$-calculus).

In logical terms, we may regard these as sanity conditions for
G\"odel numberings of the partial recursive functions. The `sane'
numberings that satisfy them are variously known as
\emph{acceptable numberings} \citep[Ex. 2.10]{Rogers1987},
\emph{acceptable programming systems} \citep[\S 3.1.1]{MY:1978},
or \emph{systems of indices} \citep[\S II.5.1]{Odifreddi1992}.

It was first shown by \citet{Rogers1958} that acceptable
numberings have very pleasant properties.

\begin{defn} 
  For numberings $\phi$ and $\psi$, define \[
    \phi \leq_{R} \psi
      \quad \text{ iff } \quad
    \exists t: \mathbb{N} \parfunc \mathbb{N} 
      \text{ total recursive}.\
    \forall p \in \mathbb{N}.\ \phi_{p} = \psi_{t(p)}
\] We then say that $\phi$ Rogers-reduces to $\psi$, and
$\leq_{R}$ is a preorder.
\end{defn}

Hence, thinking of $\phi$ and $\psi$ as different programming
languages,  $\phi \leq_{R} \psi$ just if every $\phi$-program may
be effectively translated---or \emph{compiled}---to a $\psi$
program. Then $\equiv_{R}\ \defeq\ \leq_{R} \cap \geq_{R}$ is an
equivalence relation.  Quotienting by it yields the \emph{Rogers
semilattice} under the extension of $\leq_{R}$ to the equivalence
classes (see \emph{op. cit}).
More specifically,
\begin{thm}
  \label{thm:indices}
  The following are equivalent:
    \begin{enumerate}
      \item
	$\psi$ is an acceptable numbering, as above.
      \item
	$\psi$ is a member of the unique top element of the Rogers
	semilattice.
      \item
	$\psi$ is an enumeration for which there is a universal
	function, and a total recursive $c: \mathbb{N} \parfunc
	\mathbb{N}$ such that \[
	  \psi_{c(i, j)} = \psi_{i} \circ \psi_{j}
      \]
  \end{enumerate}
\end{thm}

The first two equivalences are due to Rogers, see \emph{op. cit},
and \citet[\S II.5.3]{Odifreddi1992}, and the third is due to
\citet[Theorem 3.2]{MWY:1978}. Note that one should exercise great
caution with these equivalences, for their proofs liberally invoke
pairing tricks, loops, iterations, and other programming
constructs. Finally, the equivalence between (1) and (2) above was
strengthened by \citet{Rogers1958} to

\begin{cor}
  [Rogers' Isomorphism Theorem]
  Any two acceptable enumerations are recursively isomorphic.
\end{cor}

The possible numberings of the partial recursive functions,
whether acceptable or pathological, as well as the various forms
of SRTs that may or may not hold of them, have been investigated
by the school of John Case and his students: David
\citet{Riccardi1980, Riccardi1981}, James \citet{Royer1987} and,
more recently, Samuel Moelius III \citep{Case2007, Case2009,
Case2009a, Moelius2009}. For example, this community has shown
that there are numberings where certain known theorems, just as
the s-m-n theorem or Kleene's second recursion theorem, are not
`effective,' or simply do not hold. Earlier work on this front
seems to have concentrated on enumerations of subrecursive classes
of functions, see e.g. \citet{Royer1994}, whereas the later work
of Case and his students concentrated on the study of what they
called \emph{control structures}, i.e. constructs which provide
``a means of forming a composite program from given constituent
programs and/or data.''

\subsection{The Second Recursion Theorem}
  \label{sec:compsrt}

The central intensional result of elementary computability theory
is Kleene's \emph{Second Recursion Theorem (SRT)}, also first
shown in \citep{Kleene1938}.
\begin{thm}
  [Kleene]
  For any partial recursive $f: \mathbb{N} \times \mathbb{N}
  \parfunc \mathbb{N}$, there exists $e \in \mathbb{N}$ such that \[
    \forall y \in \mathbb{N}.\ \phi_e(y) \simeq f(e, y)
  \]
\end{thm}
\begin{proof}
  Consider the function defined by \begin{align*}
    &\delta_f : \mathbb{N} \times \mathbb{N} \parfunc{} \mathbb{N} \\
    &\delta_f(y, x) \simeq f(S(y, y), x)
  \end{align*} Since $f$ is partial recursive, simple arguments
  concerning the computability of composition and substitution
  yield that $\delta_f$ is partial recursive. Hence $\delta_f =
  \phi_p$ for some $p \in \mathbb{N}$. Consider $e \myeq S(p, p)$;
  then \[
    \phi_e(y) \simeq \phi_{S(p, p)}(y) \simeq \phi_p(p, y) \simeq
    \delta_f(p, y) \simeq f( S(p, p), y) \simeq f(e, y)
  \] The second Kleene equality follows by the s-m-n theorem, and
  the others are simply by definition or construction.
\end{proof}

In the above theorem, consider $f(x, y)$ as a function that treats
its first argument as code, and its second argument as data. The
equation $\phi_e(y) \simeq f(e, y)$ implies that $e$ is a program
which, when run on some data, will behave like $f$ with $e$ being
its first argument. In slogan form,

\begin{center}
  \shadowbox{
    We can always write a program in terms of its own source.
  }
\end{center}
  
Indeed, the trick has become standard: $f(e, y)$ is a `blueprint'
that specifies what to do with its own code $e$, and we take its
fixed point (c.f. the functionals in the untyped
$\lambda$-calculus to which we apply the $\mathbf{Y}$ combinator).
Moreover, in much the same way that the $\mathbf{Y}$ combinator is
itself a term of the untyped $\lambda$-calculus, it so happens
that the construction used in the proof of the SRT is itself
computable:

\begin{thm}
  [Constructive Kleene SRT]
  \label{thm:cksrt}
  There is a total recursive $h : \mathbb{N} \parfunc
  \mathbb{N}$ such that, for every $p \in \mathbb{N}$, \[
    \phi_{h(p)}(y) \simeq \phi_p(h(p), y)
  \] 
\end{thm}

\noindent That is: in the original statement, we may calculate a
code for $\delta_f$ from a code for $f$, and hence obtain $e$ in
an effective manner.

The Kleene SRT lies at the heart many proofs in computability,
especially those involving diagonalisation, or results pertaining
to fixed points, self-reference and the like. A smattering of more
theoretical applications of this `amazing' theorem has been
compiled by \citet{Moschovakis2010}.

The SRT is perhaps more familiar in the form popularised by
Hartley Rogers Jr. in his book \citep{Rogers1987}. We state a
slightly generalised version, and prove it from Kleene's version.
We write $e\downarrow$ to mean that the expression $e$ has a
defined value.

\begin{thm}[Rogers SRT]
  For partial recursive $f: \mathbb{N} \parfunc \mathbb{N}$,
  there is a $e \in \mathbb{N}$ such that, if $f(e)\downarrow$,
  then \[
    \phi_e = \phi_{f(e)}
  \] 
\end{thm}
\begin{proof}
  \citep[Lemma 14.3.7]{Jones1997} Define \[
    d_f(x, y) \simeq \phi_U(f(x), y)
  \] Again, by standard arguments, $d_f$ is partial recursive. By
  Kleene's SRT, there is a $e \in \mathbb{N}$ such that, for all $y
  \in \mathbb{N}$, \[
    \phi_e(y) \simeq d_f(e, y) \simeq \phi_U(f(e), y) \simeq
    \phi_{f(e)}(y)
  \] which is to say $\phi_e = \phi_{f(e)}$.
\end{proof} 

\noindent This result is equivalent to the previous formulation
\citep[Ex.  11-4]{Rogers1987}. We may summarise it in the
following slogan:

\begin{center}
  \shadowbox{
    \begin{minipage}{0.8\textwidth}
      \centering
      Every computable syntactic program transformation \\
      has a semantic fixed point.
    \end{minipage}
  }
\end{center}

Moreover, this version of the SRT comes in  a `constructive'
variant as well---see \citet[\S 11.2-II]{Rogers1987} or \citet[\S
11-3.1]{Cutland1980}: 

\begin{thm}
  [Constructive Rogers SRT] 
  \label{thm:fprt} 
  There is a total recursive $n:
  \mathbb{N} \parfunc \mathbb{N}$ such that, for any $z \in
  \mathbb{N}$ such that $\phi_z$ is total, \[
    \phi_{n(z)} = \phi_{\phi_z(n(z))}
  \]
\end{thm}

All of the above variants of the SRT are \emph{equivalent} under
the assumption that $\phi$ is an acceptable enumeration.
Nevertheless, if we relax the assumption of acceptability, there
are ways to compare and contrast them. \citet{Riccardi1980}
showed that there are enumerations of the partial recursive
functions for which there exist Rogers-type fixed points, but not
Kleene-type fixed points. In a rather technical section of his
thesis, \citet[\S 3]{Moelius2009} painstakingly compares the
various entailments between different formulations of the
recursion theorem, and concludes that Kleene's is the one that is
more natural and general.

No matter how enticing it may be, we shall not dwell on this
particular line of discussion. In fact, we shall avoid it as much
as possible, because it does not fit our view of programming. Any
numbering that is \emph{not} acceptable is somehow pathological:
by contraposition, it follows that either substitution is not
computable (and the s-m-n theorem does not hold), or that there is
no self-interpreter (universal function)---which, by
Turing-completeness, means that the interpreter as a function is
not computable. Things are even more subtle if the language is
\emph{not} Turing-complete: we then have a \emph{subrecursive}
indexing, as in \citep{Royer1994}, and this is not a path we would
like to tread presently.

\subsection{Intensional Recursion}

From the point of view of programming languages, the very general
and still rather strange application of the SRT is the ability to
define functions by \emph{intensional recursion}. This means that
a function can not only call itself on a finite set of points
during its execution---which is the well-known extensional
viewpoint---but it may also examine its own \emph{intension},
insofar as the source code for a program is a finite and complete
representation of it.

Indeed, the SRT is the only basic tool available in standard
(non-higher-order) accounts of computability that enables one to
make any `unrestricted' recursive definition whatsoever. For
example, if we define \[
  f(x, y) \simeq \begin{cases}
    1 &\text{if $y = 0$} \\
    y \cdot \phi_U(x, y-1) &\text{otherwise}
  \end{cases} 
\] and apply Kleene's SRT, we obtain a code $e \in \mathbb{N}$
such that $\phi_e$ is the factorial function.

However, the above use is slightly misleading, in that it is
\emph{extensional}: $x$ is only used as an argument to the
universal function $\phi_U$. Hence, the resulting behaviour does
not depend on the code $x$ itself, but only on the values of the
function $\phi_x$ for which it stands. The following definition
captures that phenomenon.

\begin{defn}
  A total recursive $f : \mathbb{N} \parfunc \mathbb{N}$ is
  \emph{extensional} just if \[
    \phi_a = \phi_b
      \quad \Longrightarrow \quad
    \phi_{f(a)} = \phi_{f(b)}
  \] for any $a, b \in \mathbb{N}$.
\end{defn}

That is, $f: \mathbb{N} \parfunc \mathbb{N}$ is extensional just
if the program transformation it effects depends solely on the
extension of the program being transformed. By a classic result of
\citet{Myhill1955}, such transformations correspond to a certain
class of functionals which we discuss in \S \ref{sec:effop}.

However, even if $f: \mathbb{N} \parfunc \mathbb{N}$ is
extensional, the paradigm of intensional recursion strictly
increases our expressive power. For example, suppose that we use
the SRT to produce a program $e$ satisfying a recursive definition
of the form \[
  \phi_e(x, y) \simeq\
    \dots \phi_U(e, g(x), y) \dots
\] for some $g : \mathbb{N} \parfunc \mathbb{N}$. 
We could then use the s-m-n function to replace this recursive
call by a specialisation of $e$ to $g(x)$, thereby obtaining
another program $e'$ of the form \[
  \phi_{e'}(x, y) \simeq\
    \dots \phi_{S(e', g(x))}(y) \dots
\] This is an equivalent program, in the sense that $\phi_e =
\phi_{e'}$. But if the s-m-n function $S(e,x)$ performs some
\emph{optimisation} based on the argument $x$---which it may
sometimes do---then $e'$ may provide a more efficient definition,
in that it the code for the recursive call may be optimised for
this particular recursive call. This line of thought is the
driving force of the \emph{partial evaluation} community: there,
the s-m-n function is called a \emph{specialiser} or a
\emph{partial evaluator}, and it is designed so that it optimises
the programs it is called to specialise; see \S\ref{sec:pareval}
for more details.

But the SRT also allows for recursive definitions which are
\emph{not functional}, in that the `blueprint' of which we take a
fixed point may not be extensional. For example, one may be as
daft as to define \[
  f(x) \simeq \begin{cases}
    x+1 \quad &\text{if $x$ is even} \\
    x-1 \quad &\text{if $x$ is odd}
  \end{cases}
\] This $f: \mathbb{N} \parfunc \mathbb{N}$ is total recursive,
but decidedly not extensional. However, we may still use Rogers'
theorem to obtain a fixed point $e \in \mathbb{N}$ such that
$\phi_e = \phi_{f(e)}$, and the resulting behaviour will depend on
the parity of $e$ (!). To this day, it is unclear what the use of
this is, except of course its underlying r\^{o}le in powerful
diagonal arguments---see e.g. \citet[\S 11.2]{Cutland1980}---as
well as many kinds of \emph{reflection}.

\subsection{Applications of Intensional Recursion}
  \label{sec:appirec}

\citet{Abramsky2014} observed that the SRT, as well
as other simple results on program codes, are strangely absent
from Computer Science. He comments: \begin{quote}
  ``This reflects the fact which we have already alluded
  to, that while Computer Science embraces wider notions of
  processes than computability theory, it has tended to refrain
  from studying intensional computation, despite its apparent
  expressive potential. This reluctance is probably linked to
  the fact that it has proved difficult enough to achieve
  software reliability even while remaining within the confines
  of the extensional paradigm. Nevertheless, it seems reasonable
  to suppose that understanding and harnessing intensional
  methods offers a challenge for computer science which it must
  eventually address.''
\end{quote} In many ways, we empathise with this programme.
Consequently, we catalogue some applications of such intensional
`results about program codes,' both within and on the fringes of
Computer Science, and then engage in some speculation regarding
various future directions.

\subsubsection*{Partial Evaluation}
  \label{sec:pareval}

Kleene's s-m-n theorem allows one to `specialise,' or `partially
evaluate' a certain program by fixing some of its arguments.
It may appear simple and innocuous, but this is deceptive: the
s-m-n theorem is an essential result in computability.

The power afforded by the s-m-n theorem bestowed considerable
success upon the \emph{partial evaluation} community, which began
with the work of \citet{Futamura1999} and \citet{Ershov1977,
Ershov1982} in the 1970s. Futamura observed that the ability to
write an interpreter for a language (i.e. a universal function),
as well as the ability to `specialise' an argument of a program
(s-m-n function) followed by source-level optimisation yields an
easy approach to generate a compiler, thus leading to the three
\emph{Futamura projections}. Writing $S = \phi_s$, where $S$ is
the s-m-n function, and $U$ for the program corresponding to the
universal function, we have:
\begin{align*}
  \textsf{target code}        &\myeq \phi_s(U, \textsf{source code}) \\
  \textsf{compiler}           &\myeq \phi_s(s, U) \\ 
  \textsf{compiler generator} &\myeq \phi_s(s, s)
\end{align*} One can then verify that these equations do yield the
desired behaviour, as shown in an elementary fashion by
\citet{Jones1997}. This led to a successful programme of automatic
generation of compilers, first realised in Copenhagen.  The
results are documented in \citet{Jones1996}, and the book of
\citet{Jones1993}.

Contrasting the simplicity of the s-m-n theorem with its success
in the partial evaluation community also led Jones to ponder
whether the SRT, which is a much more powerful result, could
have interesting practical applications. Quoting from
\citep{Jones1992}:

\begin{quote}
    ``While this theorem has many applications in mathematics, it
    is not yet clear whether it will be as constructively useful
    in computer science as the s-m-n theorem has turned out to
    be.''
\end{quote}

\noindent Taking this as a point of departure, he has posed two
significant questions:

\begin{itemize}
  \item
    What is the exact relationship between the First Recursion
    Theorem (FRT) and the Second Recursion Theorem (SRT)?
  \item 
    If one implements the SRT, how can the accumulating layers of
    self-interpretation be avoided?
\end{itemize}
Looking back at the literature on computability, we find that the
first question has been answered in a mostly satisfactory way.
Regarding the second question, some recent progress is documented
in \citet{Kiselyov2015}.

A series of experiments with the SRT and further discussion are
documented in \citet{Hansen1989}, and \citet{Jones1992, Jones2013}.

\subsubsection*{Abstract Computer Virology}
  \label{sec:abscompvir}

Computer viruses rely heavily on the ability to propagate their
own code, which is a kind of reflection. This was noticed by
\citet{Cohen1989}, who was the first to introduce the term
\emph{computer virus} alongside an early Turing Machine model of
viruses.  A few years later, his supervisor---Leonard
\citet{Adleman1990}---concocted his own model of viruses that is
based on elementary computability theory. In that model,
viruses are program transformations that `infect' ordinary
programs. This is also where the connection with the SRT was made
explicit, as Adleman invokes it to construct a program
that---under his own definition---is classified as a virus. He
also proved a result on his model that makes crucial use of the
SRT for a diagonal construction.

These were the two cornerstones that laid the foundation of
\emph{abstract computer virology}. In more recent years there have
been further developments, owing to the work of
\citet{Bonfante2005,Bonfante2006, Bonfante2007}, who discuss and
classify different types of viruses that correspond to multiple
variants of the SRT. See also \citet{Marion2012}.

\subsubsection*{Computational Reflection \& Reflective Towers}
  \label{sec:reflection}

The concept of \emph{computational reflection} was introduced in
the context of programming languages by Brian Cantwell
\citet{Smith1982, Smith1984} in his voluminous thesis.  The
underlying intuition is that a program may be considered to be
running on an interpreter, which interpreter itself is running on
another copy of this interpreter, and so on  A special construct
that makes use of this structure is available in the programming
language: it allows one to inject code in the interpreter that
lies one level below. Hence, a program has access not only to its
code, but the entire state of the interpreter that runs it, and
even the interpreter that runs the interpreter. This is embodied
by the language 3-LISP, introduced in \emph{op. cit.}

The resulting structure of \emph{reflective towers}
has captured the imagination of many, but---while couched in
colourful imagery---its construction is logically and
computationally mysterious. A series of publications
\citep{Friedman1984, Wand1988, Danvy1988} have made partial
attempts to explain this `tower' in more concrete terms.

Nevertheless, the concept of \emph{computational reflection} seems
to be a general recurring theme with broader scope, but is not
well-understood. For the state of affairs up to the mid-1990s, see
the short comparative survey \citet{Malenfant1995}. Demystifying
reflection is a pressing concern, as many modern programming
languages have reflective or `introspective' facilities, which are
infamous for pernicious bugs, and generally wreaking havoc. 

Some ideas regarding reflection seem to be experiencing a
resurgence of interest, mainly because of the appearance of a new
candidate foundation for reflection, namely Barry Jay's
\emph{factorisation calculus}: see \citet{Jay2011a},
\citet{Jay2011b}, and \citet{Jay2016}.

As the SRT is a result with fundamental connections to reflection,
we believe that a better understanding of intensional recursion
must be instrumental in laying a logical foundation for reflective
constructs.

\subsubsection*{Economics}

This section concerns a  speculative application of the SRT to
economic modelling. Historically, there has been a lot of
discussion---especially in the years following the Great
Recession---regarding the foundational principles of economics,
and whether these are an adequate substrate for the science. Some
of these critical approaches touch on the \emph{self-referential}
aspects that are ignored by (neoclassical) economic theory.
\citet{Winrich1984} offers an early example of these criticisms;
we give him the floor, for the argument is compelling:
  \begin{quote}
  In order for preferences to be complete the choice set must
  include preferences themselves! But, as soon as you allow
  preferences within the choice set you have preferences
  ``talking'' to preferences. In a world of preference
  self-reference we can, if you will, produce a neoclassical liar.
  As an example let our liar be a smoker. In such a situation it
  is not uncommon to hear a smoker say, ``I dislike my desire to
  smoke.'' What are we to make of such a statement? In the static
  framework of neoclassical choice theory this is a contradiction.
  But at the same time, it cannot be prevented. Not only is the
  ``act of smoking'' an element in the choice set, but the
  ``desire to smoke'' is itself an element in the choice set and
  also subject to discretion.

  Continuously expanding the choice set by including not only
  commodities but also social conditions in an attempt to explain
  invidious distinctions, other individuals in an attempt to
  explain altruism, and so forth, neoclassicism has been able to
  ``absorb'' heterodox attacks within the atomic individualistic
  perspective. However, the inclusion of choice itself is
  devastating to its own premise of consistency.

  Let me make the point clear.  Preference functions, preference
  orderings, or preference rankings do not exist. [...]
  \end{quote}
Indeed, changes to an economic process originating from within an
economic process have been rather difficult to model in a
reductionist fashion. Some early work that attempted to model the
`change of institutions' in game theory was carried out by
\citet{Vassilakis1989, Vassilakis1992}. Moreover, there have even
been approaches---even predating \emph{algorithmic game
theory}---that approach economics with computational feasibility
in mind, see e.g. the work of \citet{Velupillai2000} on
\emph{computable economics}. See also \citet{Blumensath2013} for a
recent attempt involving coalgebra.

\section{Extensional Recursion and the FRT}
  \label{sec:higher}

\subsection{Effective Operations}
\label{sec:effop}

Suppose that we have some $f: \mathbb{N} \parfunc \mathbb{N}$
that is total and extensional. It is not hard to see that---by
the definition of extensionality---$f$ uniquely induces a very
specific type of functional. Let us write $\mathcal{PR}$ for the
set of partial recursive functions.

\begin{defn}
  A functional $F: \mathcal{PR} \rightarrow \mathcal{PR}$ is an
  \emph{effective operation}, if there exists a total,
  extensional $f: \mathbb{N} \parfunc \mathbb{N}$ such that \[
    F(\phi_x) = \phi_{f(x)}
  \]
\end{defn} 

\noindent This is well-defined precisely because $f$ is
extensional. We are entering the realm of \emph{higher types}, by
computing functions from functions. Nevertheless, we are doing so
in a computable and finitary sense: functions may be infinite
objects, but this computation occurs, or \emph{is tracked}, on the
level of program codes.

\subsubsection*{The Myhill-Shepherdson Theorem}

Functionals such as the one defined above are perhaps one of the
most straightforward ways to define computability at higher order,
namely as effective code transformations. Surprisingly,
\citet{Myhill1955} showed that the same functionals can be defined
in a much more abstract manner that dispenses with code
transformations entirely. In fact, anyone familiar with the
domain-theoretic semantics of the $\lambda$-calculus will
recognise it immediately.  
  
We first need to discuss the simple order-theoretic structure of
the set $\mathcal{P}$ of unary partial functions: its elements may
indeed be ordered by \emph{subset inclusion}: \[
  \psi \subseteq \chi
    \quad \text{iff} \quad
  \forall x, y \in \mathbb{N}.\
    \psi(x) \simeq y \Longrightarrow \chi(x) \simeq y
\] This makes $\mathcal{P}$ into a $\omega$-complete partial order
($\omega$-cpo), in that least upper bounds of increasing chains
always exist, and they are unions. We can now make the following
definition.

\begin{defn}
  A functional $F: \mathcal{P} \rightarrow \mathcal{P}$ is
  \emph{effectively continuous} just if it satisfies the following
  properties:
  \begin{description}
    \item[Monotonicitity]
      $\psi \subseteq \chi\ \Longrightarrow\ F(\psi) \subseteq
      F(\chi)$

    \item[Continuity]
      For any increasing sequence of partial
      functions, \[
        f_0 \subseteq f_1 \subseteq f_2 \dots
      \] we have \[
        F\left(\bigsqcup_i f_i\right) = \bigsqcup_i F(f_i)
      \]

    \item[Effectivity on Finite Elements]
      Given an encoding $\hat{\cdot}$ of the graph of every finite
      function $\theta: \mathbb{N} \parfunc \mathbb{N}$ as a
      number $\hat{\theta} \in \mathbb{N}$, there is a partial
      recursive $g_F: \mathbb{N} \times \mathbb{N} \parfunc
      \mathbb{N}$ such that, for every finite $\theta: \mathbb{N}
      \parfunc \mathbb{N}$, and for all $x \in \mathbb{N}$, \[ 
        F(\theta)(x) \simeq g_F(\hat{\theta}, x)
      \]
  \end{description} 
\end{defn}

\noindent These are called \emph{recursive functionals} by
\citet{Cutland1980} and \citet{Rogers1987}: we beseech the reader
to exercise caution, as terminology varies \emph{wildly}.

Let us restate continuity, by using the following equivalent
formulation, for which see \citet[\S II.2.23]{Odifreddi1992}:

\begin{lem}
  [Compactness]
  $F: \mathcal{P} \rightarrow \mathcal{P}$ is continuous if and
  only if \[
    F(f)(x) \simeq y
      \quad \Longleftrightarrow \quad
    \exists \text{ finite } \theta \subseteq f.\
      F(\theta)(x) \simeq y
\] for all $x, y \in \mathbb{N}$.
\end{lem}

Putting these together, we see that effectively continuous
functionals are indeed very strongly \emph{computational} and
\emph{effective}: the value of $F(f)(x)$ only depends on a finite
part of the graph of $f$. In
fact, we can show that the behaviour of $F$ on finite elements
completely determines its behaviour on $f$:

\begin{lem}
  [Algebraicity]
  Let $F: \mathcal{P} \rightarrow \mathcal{P}$ be continuous.
  Then \[
    F(f) = \bigsqcup
\setcomp{ F(\theta) }
  { \theta \text{ finite } \land \theta \subseteq f }
  \]
\end{lem}
  
\noindent So, how do these functionals, which are computable in
finite approximations, relate to the aforementioned effective
operations, which are based on computations on indices? The answer
is astonishingly simple:

\begin{thm}
  [Myhill-Shepherdson]
  An effective operation $F_{\text{eff}} : \mathcal{PR}
  \rightarrow \mathcal{PR}$ can be uniquely extended to an
  effectively continuous functional $F: \mathcal{P} \rightarrow
  \mathcal{P}$ (with $F_{\text{eff}} \subseteq F$).

  Conversely, any effectively continuous functional, when
  restricted to the partial recursive functions $\mathcal{PR}$,
  is an effective operation.
\end{thm} 

\noindent See \citet[10-\S 2]{Cutland1980}, \citet[\S
15.3, XXIX]{Rogers1987}, or \citet[\S II.4.2]{Odifreddi1992} for
proofs.

\subsubsection*{A Turing Machine characterisation}

Our discussion of functionals began with extensional operations
on codes, as the most natural definition of higher-order
computation.

There is, however, an alternative, which some would argue is
even more simple: one may envisage the implementation of any
functional $F: \mathcal{P} \rightarrow \mathcal{P}$ as a Turing
Machine which has access to an \emph{oracle} for the argument of
the functional. During a computation, the machine may write a
number $x$ on a separate tape, and then enter a special state,
in order to query the oracle. The oracle then replaces $x$ with
$f(x)$ if $f$ is defined at $x$. If it is not, the oracle does
not respond, forcing the machine to eternally wait for an
answer, and the computation diverges.

Computation with oracles was first considered by
\citet{Turing1939}, leading to the intricate theory of
\emph{Turing reducibility} and \emph{relative computability}.
However, Turing-type oracles are \emph{fixed} in relative
computability, whilst we consider them as \emph{arguments to a
computation}. This shift in perspective, as well as the first
concrete results involving higher types, are due to
\citet{Kleene1952}.

In this context, subtle issues arise with
\emph{non-determinism}. It is well-known that non-deterministic
Turing Machines are equivalent to deterministic Turing machines
at the first order, but at higher order this is no longer true.
In fact, the following theorem was first shown---to the best of
our knowledge---by \citet[\S 3]{Moschovakis2010}, even though it
was simply labelled as the Myhill-Shepherdson theorem:
\begin{thm}
  [Moschovakis]
  \label{thm:ims}
  A functional $F: \mathcal{PR} \rightarrow \mathcal{PR}$ is an
  effective operation if and only if $F(f)$ is computable by a
  non-deterministic Turing machine with an oracle for $f$.
\end{thm} 
Great care has to be taken in combining non-determinism with
oracles: the machine should be designed so as to avoid the
presence of two halting branches in the same computation tree
with different candidate outputs. Similar restrictions occur in
defining what it means to non-deterministically compute a
polynomial time function---see e.g. \citet[\S 4.5.2]{Lewis1997}.
However, in this case, non-halting branches do not harm anyone;
in fact, they are in some sense necessary for the expressive
power afforded by this model of computation.

\subsubsection*{The First Recursion Theorem}

We have seen that effectively continuous functionals exhibit an
impressive amount of inherent order-theoretic structure. By the
Myhill-Shepherdson theorem, this structure emerges automatically
once a functional can be `realised' on codes by an extensional
operation on codes.

This order-theoretic structure is the basis of the proof of the
First Recursion Theorem (FRT). This fact was discovered by Dana
Scott, who was the first one to notice that the core argument in
the proof of the FRT applies to all so-called \emph{simple
types}.\footnote{The simple types of PCF, as defined by
\citet{Scott1993}, are generated by $\sigma\ ::= \mathbb{B} \;|\;
\mathbb{N} \;|\; \sigma \rightarrow \sigma$.  $\mathbb{B}$ is
supposed to connote the \emph{booleans}, and $\mathbb{N}$ the type
of natural numbers.} This discovery of Scott led to the
development of \emph{domain theory}, and the study of \textsf{PCF}
\citep{Plotkin1977}, both of which were the firstfruits of the
field of programming language semantics.  Indeed, Scott
acknowledged his debts; quoting from \citep{Scott1975}:
\begin{quote}
  ``[...] It is rather strange that the present model was not
  discovered earlier, for quite sufficient hints are to be found
  in the early paper of Myhill and Shepherdson and in Rogers' book
  (especially \S \S 9.7-9.8). These two sources introduce
  effective enumeration operators and indicate that there is a
  certain amount of algebra about that gives these operators a
  pleasant theory, but no one seemed ever to take the trouble to
  find out what it was.''
\end{quote}

\noindent The central result underlying the FRT is therefore
this:

\begin{thm}
  [The Fixpoint Theorem]
  Let $(D, \sqsubseteq)$ be a $\omega$-cpo with a least element
  $\bot \in D$, and let $F: D \rightarrow D$ be a continuous
  function. Then $F$ has a \emph{least fixed point}, defined
  explicitly by \[
    \mathbf{lfp}(F) = \bigsqcup_i F^i(\bot)
  \]
\end{thm} 

\noindent This is largely considered a folk theorem: its origins
are difficult to trace, and many variants of it have been proved
and used widely in Logic and Computer Science---see
\citet{Lassez1982} for a historical view. In \emph{op. cit} the
authors note that Kleene knew of it at least as early as 1938; see
also \citep{Kleene1981}. 

In fact, it is high time that we show to the reader how it
constitutes the first half of Kleene's FRT. But first, a caveat:
the version of the FRT that we will presently prove pertains to
effective operations (\`{a} la Myhill and Shepherdson), and a
proof similar to ours may be found in the book by \citet[\S
10-3]{Cutland1980}. The original statement, found in Kleene's book
\citep[\S 66]{Kleene1952}, concerns partial recursive functionals,
which we discuss in \S \ref{sec:prf}; see also \citet[\S
II.3.15]{Odifreddi1992}.

\begin{thm}
  [First Recursion Theorem]
  \label{thm:frt}
  Every effectively continuous functional $F: \mathcal{P}
  \rightarrow \mathcal{P}$ has a least fixed point
  $\mathbf{lfp}(F): \mathbb{N} \parfunc \mathbb{N}$, which is
  partial recursive.
  
  Furthermore, let $\phi_q : \mathbb{N} \parfunc \mathbb{N}$ be
  an extensional function that realises $F$ on the partial
  recursive functions, i.e. $F(\phi_x) = \phi_{\phi_q(x)}$.
  Then, a $p \in \mathbb{N}$ such that $\phi_p =
  \mathbf{lfp}(F)$ may be computed effectively from any such $q
  \in \mathbb{N}$.
\end{thm} \begin{proof} 
  The existence of the least fixed point follows from the fact
  $(\mathcal{P}, \subseteq)$ is a $\omega$-cpo, and the Fixpoint
  Theorem.

  By the Myhill--Shepherdson theorem, the functional $F$
  corresponds to some extensional $\phi_q: \mathbb{N} \parfunc
  \mathbb{N}$.  The proof of the Fixpoint Theorem constructs a
  chain, \[
    f_0 = \emptyset \subseteq f_1 \subseteq f_2 \dots
  \] where $f_{i+1} = F(f_i)$, and the least fixed point is
  $\mathbf{lfp}(F) = \bigsqcup_i f_i$. The $f_i$ may be
  construed as increasingly defined `approximations' to $f$. The
  key to this proof lies in using the extensional $\phi_q$ to
  obtain indices that \emph{track} each element of this chain:
  \begin{align*}
    p_0 &= \text{(some index for the nowhere defined function)} \\
    p_1 &= \phi_q(p_0) \\
    \vdots
  \end{align*} so that $p_{i+1} = \phi_q(p_i)$ and hence, by
  induction, $f_i = \phi_{p_i}$ for all $i \in \mathbb{N}$. Then, by
  Church's Thesis, we define $p \in \mathbb{N}$ by writing a program
  that performs the following steps: on input $n$,
  \begin{enumerate}
    \item Set $i := 0$.
    \item Begin a \emph{simulation} of the program $p_0$ running on $n$.
    \item Loop: \begin{enumerate}
      \item For each $j \leq i$, simulate one step of $p_j$ on
      $n$.
      \item If any of these simulations have halted and produced a
      value $m$, halt and output $m$.
      \item Otherwise, compute $p_{i+1} = \phi_q(p_i)$, and begin
      a simulation of $p_{i+1}$ on input $n$. Set $i := i+1$.
    \end{enumerate}
  \end{enumerate} Since
  the $f_i$'s are a chain, this program cannot accidentally
  produce two contradictory values.  But since $\mathsf{lfp}(F) =
  \bigsqcup_i f_i$, if $f(n) \simeq m$, then $f_i(n) \simeq m$ for
  some $i$, so that $f = \phi_p$.
\end{proof} 

In the books of both \citet{Cutland1980} and
\citet{Odifreddi1992}, the above proof is obtained after the
recursively enumerable sets are characterised as the $\Sigma_1^0$
sets in the arithmetical hierarchy. We prefer the more primitive,
`algorithmic version' above, because we can isolate the expressive
power needed.

So, what do expressive power do we use? The program in the proof
curiously calls for \emph{countable dovetailing}, i.e. the
simulation of a slowly expanding yet potentially infinite set of
computations, of which we perform a few steps at each stage.
This requires access to the code $q \in \mathbb{N}$ of the
relevant extensional function, so that we can run it to obtain the
rest of the codes $p_i$. Furthermore, we require a handle on a
number of simulations we spawn, so that we can pause them,
schedule some steps of each, and possibly even discard some.

It is worth remarking once more that the output of this procedure
is obviously deterministic, but there is inherent non-determinism
and parallelism in the method we use to compute it. This is very
much in line with Moschovakis' version of the Myhill--Shepherdson
theorem (Theorem \ref{thm:ims}).

\subsection{Partial Recursive Functionals and Pure Oracles}
\label{sec:prf}

We have only discussed effective operations up to this point,
and shown that they correspond to effectively continuous
functionals.

In our discussion leading to Theorem \ref{thm:ims} we also
mentioned a slightly different paradigm, that of \emph{oracle
computation}. Theorem \ref{thm:ims}, however, guaranteed that
non-deterministic oracle computation coincided with effective
continuity. If, in contrast, we adopt deterministic oracle
computation as our notion of higher-order computation, we are led
to a different notion of computable functional. This kind of
functional was first discussed by
\citet{Kleene1952}:\footnote{However, his definition was not
identical to ours. Kleene defined his functionals through an
equation calculus. In this framework, even if the semantics of
composition were understood to be strict, multiple (and possibly
inconsistent) defining equations were still allowed, leading to
the non-determinism observed by \citet{Platek1966}, which allowed
parallel-or to be computable. We use the definition that is
widely believed that Kleene really intended to use, and found in
later textbooks. That this is the common interpretation I learned
from John Longley, in personal communication.}

\begin{defn}
  A functional $F : \mathcal{P} \rightarrow \mathcal{P}$ is a
  \emph{partial recursive functional} if $F(f)$ can be obtained
  from $f : \mathbb{N} \parfunc \mathbb{N}$ and the initial
  functions by composition, primitive recursion, and
  minimalisation. If its domain is restricted to the set
  $\mathcal{F}$ of total functions, such a functional $F:
  \mathcal{F} \rightarrow \mathcal{P}$ is called a
  \emph{restricted partial recursive functional}.
\end{defn}

\noindent Thus, if $F(f) = g$, then we say that $g$ is partial
recursive \emph{uniformly in $f$}. 

An implementation of such a functional $F$ would resemble a
deterministic Turing machine with an oracle for its argument. But
notice that there is no non-determinism in this case, and hence
calls to the oracle have to happen in a predetermined way. As soon
as we decide to make a query at an undefined point, the
computation diverges: there is no other branch of the computation
to save the day! Informally, we may say that \emph{calls to the
oracle may not be dovetailed}. In effect, partial recursive
functionals deal with their arguments as \emph{pure extensions},
whereas effectively continuous functionals interact in a more
involved manner with the phenomenon of non-termination.

In the case of total inputs, the above connection was made
precise by Kleene:

\begin{thm}
  \citep[\S 68, XXVIII]{Kleene1952}
  A functional $F: \mathcal{F} \rightarrow \mathcal{P}$ is a
  restricted partial recursive functional if and only if it is
  computed by a deterministic Turing machine with an oracle.
\end{thm}

\noindent We are not aware of a plausible analogue of this
theorem for partial recursive functionals and deterministic Turing
machines.

The definition of partial recursive functionals has a lot of
undesirable consequences: see the discussion in the thesis of
\citet[p. 128-130]{Platek1966}. Thus, the definition is often
restricted to total inputs, for which the above characterisation
through Turing Machines exists. The underlying reason seems to be
that, for total inputs, we may \emph{enumerate} the graph of the
oracle, as no call to it will diverge.

Let us not forget this trivial but pleasant consequence:
\begin{lem}
  \label{lem:prfprec}
  Let $F: \mathcal{P} \rightarrow \mathcal{P}$ be a partial
  recursive functional. If $g \in \mathcal{PR}$, then $F(g) \in
  \mathcal{PR}$.
\end{lem}

It is in this setting that Kleene obtained the First Recursion
Theorem, which first appeared in \citet[\S 66]{Kleene1952}:
\begin{thm}
  [FRT for Partial Recursive Functionals] Let $F: \mathcal{P}
  \rightarrow \mathcal{P}$ be a partial recursive functional.
  Then $F$ admits a partial recursive least fixed point.
\end{thm} \begin{proof}
  See \citet[\S II.3.15]{Odifreddi1992}. As before, the existence
  of the least fixed point follows from the Fixpoint Theorem. The
  fact it is partial recursive follows from Lemma
  \ref{lem:prfprec}: we conclude by induction that all the $f_i$'s
  are partial recursive; as the least fixed point is $f =
  \bigsqcup_i f_i$, we have that \[
    f(x) \simeq y 
      \quad \Longleftrightarrow \quad
    \exists i \in \mathbb{N}.\ f_i(x) \simeq y
  \] As the $f_i$'s are partial recursive, the predicate on the
  RHS of this equivalence is recursively enumerable, and hence so
  is the graph of $f$.
\end{proof}

\noindent Notice that the proof was rather abstract, and that all
references to indices have disappeared completely.

The following was shown by Uspenskii and Nerode, see \citet[\S
II.3.19]{Odifreddi1992} for a proof:

\begin{thm}
  \label{thm:nerode}
  Every partial recursive functional is effectively continuous.
\end{thm}

\noindent In particular, if we restrict a partial recursive
functional to $\mathcal{PR}$, it is an effective operation. The
converse was shown to fail by Sasso---see \citet[\S
II.3.20]{Odifreddi1992}: \begin{thm} The functional \[
  F(f) = \lambda x. \begin{cases}
     0 &\text{if $f(2x) \simeq 0$ or $f(2x+1) \simeq 0$} \\
    \text{undefined} &\text{otherwise}
  \end{cases}
  \] is effectively continuous, but not partial recursive.
\end{thm} This clearly demonstrates, once more, that there is
inherent \emph{parallelism} or \emph{non-determinism} in
effective operations, whilst partial recursive functionals are
purely \emph{sequential}. In more detail, to compute the above
functional we would have to concurrently query the argument $f$
at two points by dovetailing the computations. A deterministic
Turing machine would have to either query $f$ at either $2x$ or
$2x+1$ first; if the first call were to an undefined point, it
would diverge and never examine the second. A non-deterministic
Turing machine would deal with the same difficulty by branching
at the point where a choice between $2x$ and $2x+1$ is to be
made.

\section{FRT vs. SRT}
  \label{sec:frtvssrt}

\subsection{Effective Operations and the SRT}
  \label{sec:effsrt}

Suppose we would like to construct a fixed point in a more
simplistic manner than the one employed in the proof of Theorem
\ref{thm:frt}. All we need to do is use the Myhill-Shepherdson
theorem to restrict an effectively continuous functional to an
effective operation and extract an extensional function from it,
followed by applying the SRT.

\begin{lem}
  \label{lem:srtfix}
  Given an effective operation $F: \mathcal{PR} \rightarrow
  \mathcal{PR}$ defined by an extensional $\phi_p: \mathbb{N}
  \parfunc \mathbb{N}$, we may effectively obtain a code for one
  of its fixed points from $p \in \mathbb{N}$.
\end{lem}

\begin{proof}
By Theorem \ref{thm:fprt}, that code is $n(p)$; for then, \[
    \phi_{n(p)} = \phi_{\phi_p (n(p))} = F(\phi_{n(p)})
  \] so that $\phi_{n(p)}$ is a fixed point of $F$.
\end{proof}

\noindent So far, so good; but what sort of fixed point have we
obtained? In particular, is it minimal? The following
construction, due to \citet[\S 11-XIII]{Rogers1987} demonstrates
that it is not.

\begin{thm}
  There is an extensional $\phi_m: \mathbb{N} \parfunc \mathbb{N}$
  such that the fixed point obtained by the SRT as in Lemma
  \ref{lem:srtfix} is not minimal.
\end{thm}

\begin{proof}
  Use Church's Thesis, Kleene's SRT, and the function $n$ from
  \ref{thm:fprt}  to define $m \in \mathbb{N}$ such that \[ 
    \phi_m(x) \simeq
      \begin{cases}
        x &\text{if } x \neq n(m) \\
        t &\text{if } x = n(m)
      \end{cases}
  \] where $t$ is an index for the constant zero function, i.e.
  $\phi_t(x) \simeq 0$ for all $x \in \mathbb{N}$. Observe that, as
  $n$ is total recursive, $\phi_m$ is total recursive.  It is also
  extensional. Essentially, $\phi_m$ asks: is the input my own
  Rogers fixed point? If yes, output code for the constantly zero
  function; otherwise, echo the input. Thus, if $x \neq n(m)$, we
  have that $\phi_m(x) \simeq x$, so that $\phi_{\phi_m(x)} \simeq
  \phi_x$. Otherwise, \[ 
    \phi_{\phi_m(n(m))} \simeq \phi_{n(m)}
  \] as $n(m)$ is a Rogers-style fixed point. In either case,
  $\phi_m$ is extensional, and defines the identity functional.
  The least fixed point of it is the empty function. However, the
  fixed point that results from the SRT has code $n(m)$, and \[
    \phi_m(n(m)) \simeq t
  \] so that $\phi_{n(m)} = \phi_t$, which is equal to the
  constantly zero function.
\end{proof}

The key aspect of this construction seems to be that, unlike
oracle computation, an extensional function is able to
syntactically inspect its input, thus creating a `singularity' at
one point. We maintain extensionality by arranging that the point
at which the `singularity' is to be found is---incidentally---the
extensional function's own fixed point! This is more evidence that
effective operations really hide something more than `pure
extension' under the hood.

\subsubsection*{The Standard Form}

Can this mend this situation? The answer is positive: any
extensional function can be rewritten in a `standard form,' which
guarantees that the SRT really defines a minimal fixed point. This
is the exact sense in which the SRT implies the FRT.

The construction is due to \citet[\S 11-XIV]{Rogers1987}. The
original statement is horribly complicated, and involves multiple
layers of enumeration; there is a lot of concurrency happening
here, and we cannot do much better than keep the description
informal.

For the following, we assume that there is also a standard way
to enumerate the graph of a partial recursive function given its
index. This may be done by dovetailing simulations of \[
  \phi_x(0), \; \phi_x(1), \dots
\] and emitting pairs $(i, \phi_x(i))$ as soon as the $i$th
simulation halts. We do not care about the exact details, but we
do care that the exact same construction is used throughout.

Thus, let there be a effective operation $F: \mathcal{PR}
\rightarrow \mathcal{PR}$, and let $f : \mathbb{N} \parfunc
\mathbb{N}$ be total and extensional, such that $F(\phi_x) =
\phi_{f(x)}$.  We will define a total and extensional $h_f:
\mathbb{N} \parfunc \mathbb{N}$, which is \emph{co-extensional}
with $f$, in the sense that \[
  \forall x \in \mathbb{N}.\ \phi_{f(x)} = \phi_{h_f(x)}
\] This $h_f$ will be in `standard' form. Moreover, we may
effectively compute an index for $h_f$ from an index for $f$.

We use Church's Thesis to define $h_f$ so that, on input $y$, it
outputs a program that performs the following instructions:
\begin{quote}
  $h_f(y) \simeq$ ``On input $x$, run the following processes in
  parallel: \begin{enumerate}
  \item
    One process enumerates the graphs of all finite functions
    $\mathbb{N} \parfunc{} \mathbb{N}$, encoded as numbers: \[
      \hat{\theta_0} = \emptyset, \; \hat{\theta_1}, \;
      \hat{\theta_2}, \; \dots
    \] This may be done in many ways, but it is necessary that
    we begin with the empty function (in order to include the
    covert base case in the strong induction of the following
    theorem).

  \item
    There is a total recursive $d: \mathbb{N} \parfunc \mathbb{N}$
    which turns a graph of a finite function into an index for
    that function (by writing code that simply checks if the
    input is in the graph, outputting the relevant value if so,
    and diverging otherwise). The second process receives the
    encoded graphs of the finite functions above, and enumerates
    codes, \[
      d(\hat{\theta_0}), \; d(\hat{\theta_1}), \;
      d(\hat{\theta_2}), \; \dots
    \] with $\phi_{d(\hat{\theta_i})} = \theta_i$.

  \item
    A third process receives messages from the second
    process, and applies $f$ to those codes, outputting
    \[
      f(d(\hat{\theta_0})), \; f(d(\hat{\theta_1})), \; \dots
    \] with $\phi_{f(d(\hat{\theta_i}))} =
    F(\phi_{d(\hat{\theta_i})}) = F(\theta_i)$. We thus obtain
    codes for all the applications of the effective operation $F$
    on all the finite functions. Beware: these functions
    $\phi_{f(d(\hat{\theta_i}))}$ may now be infinite!

  \item
    (This is the process where partiality enters the
    construction.) Enumerate, simultaneously, all the pairs in the
    graph of the function $\phi_y$, as well as the pairs in the
    graphs of $\phi_{f(d(\hat{\theta_i}))} = F(\theta_i)$.  This
    can be done using the method postulated above.

  \item
    As soon as we find some $t$ such that \[
      F(\theta_i)(x) \simeq t \quad \text{and} \quad \theta_i
      \subseteq \phi_y
    \] we halt and output that $t$. This may be done by
    periodically checking whether $x$ is defined in the
    enumeration of some $F(\theta_i)$, and then confirming that
    the entire graph of that $\theta_i$ is contained in the
    enumeration of $\phi_y$.''
  \end{enumerate}
\end{quote}

Notice that, in this construction, the code for $f$ may be
abstracted away. Using the s-m-n theorem, we can then effectively
produce code for it from any index of $f$.  Trivially, $h_f$ is
total. Furthermore, notice that we needed to enumerate the
$\phi_{f(d(\hat{\theta_i}))}$, for---in general---they will not be
finite functions.

By the compactness of $F$, which follows from the theorem of
Myhill and Shepherdson, we know that, $F(\phi_y)(x) \simeq t$ if
and only if $F(\theta_i)(x) \simeq y$ for some finite $\theta_i
\subseteq \phi_y$. This construction will always find such a
$\theta_i$ if there exists one. Hence $h_f$ defines the same
functional as $f$.

Now, using the SRT on $h_f$ will produce a minimal fixed point:

\begin{thm}
  \label{thm:rogsrt}
  If $\phi_v = h_f$, then $n(v)$ is a code for the least fixed
  point of the effective operation $F : \mathcal{PR} \rightarrow
  \mathcal{PR} $ defined by $f : \mathbb{N} \parfunc{} \mathbb{N}$.
\end{thm}
\begin{proof}
  We have that \[
    \phi_{n(v)} = \phi_{\phi_v(n(v))} = \phi_{h_f(n(v))}
  \] so that $n(v)$ behaves exactly as $h_f(y)$ would if $y$ were
  fixed to be its own code. That is to say, we can read
  $\phi_{n(v)}$ wherever $\phi_y$ is occurs in the definition of
  $h_f$, and the check \[
    \theta_i \subseteq \phi_y
  \] becomes \[ 
    \theta_i \subseteq \phi_{n(v)} 
  \] which is to say that the program checks whether each finite
  function is a subset of its own graph! By Lemma
  \ref{lem:srtfix}, this defines a fixed point for the functional
  $F$.

  To prove that this fixed point is least, we proceed by
  \emph{strong induction on the number of steps taken to enumerate
  the graph of $\phi_{n(v)}$}. That is, we shall show that if we
  begin enumerating $\phi_{n(v)}$ using our standard enumeration
  procedure on the code $n(v)$, all the pairs produced will belong
  to the least fixed point, hence $\phi_{n(v)} \subseteq
  \mathbf{lfp}(F)$, whence $\phi_{n(v)} = \mathbf{lfp}(F)$.

  Begin enumerating $\phi_{n(v)}$. This involves running the code
  $n(v)$. One of the sub-processes in that code involves
  enumerating $\phi_{n(v)}$ itself, using the same procedure as we
  are. Since this is a \emph{sub-computation} of our enumeration, it
  is always shorter in length. Hence, by the inductive hypothesis,
  we assume that the enumeration in the sub-computation produces
  the least fixed point. Hence, if the check \[
    \theta_i \subseteq \phi_{n(v)}
  \] succeeds, we know that \[
    \theta_i \subseteq \mathbf{lfp}(F)
  \] By monotonicity, it follows that \[
    F(\theta_i) \subseteq F(\mathbf{lfp}(F)) = \mathbf{lfp}(F)
  \] Hence, when the check $F(\theta_i)(x) \simeq t$ succeeds and
  the pair $(x, t)$ is output by the enumeration, we know it
  belongs to the least fixed point. It follows that every pair
  produced by the enumeration is in the least fixed point.
\end{proof}

\subsection{Partial Recursive Functionals and the SRT}

In contrast with effective operations, the situation is simpler
in the case of oracle computation: because partial recursive
functionals are decidedly extensional in their behaviour, the
problems that arose in the preceding section vanish. There is no
`inherent' parallelism in computing such a functional, and the
SRT immediately yields least fixed points.

The following theorem was shown by \citet[\S
II.3.16]{Odifreddi1992}, who wrongly attributes it to
Rogers:\footnote{Rogers instead sketched the proof to our Theorem
\ref{thm:rogsrt}, which strictly concerns effective operations.}
  
\begin{thm}[Odifreddi]
  Let $F: \mathcal{P} \rightarrow \mathcal{P}$ be a partial
  recursive functional, and define \[
    f(e, x) \simeq F(\phi_e)(x)
  \] Then $f$ is partial recursive. Moreover, there exists $q \in
  \mathbb{N}$ such that $f = \phi_q$, and the function $h:
  \mathbb{N} \parfunc \mathbb{N}$ of Theorem \ref{thm:cksrt}
  produces a code $h(q)$ such that $\phi_{h(q)} =
  \mathbf{lfp}(F)$.
\end{thm}

\begin{proof} 
  As $F$ is a partial recursive
  functional, it is also an effective operation on the
  partial recursive functions, by Theorem \ref{thm:nerode}. We
  define $q \in \mathbb{N}$ by Church's thesis: on input $(e, x)$,
  process the code of $e$ with the total extensional function
  associated to $F$ by Myhill-Shepherdson, and call the resulting
  code on $x$.

  Let $g = \phi_{h(q)}$. We have that \[
    g(x) \simeq \phi_{h(q)}(x) \simeq f(h(q), x) \simeq
    F(\phi_{h(q)})(x) \simeq F(g)(x)
  \] for any $x \in \mathbb{N}$, so that $g$ is a fixed
  point of $F$.

  It remains to show minimality. The proof is by \emph{strong induction
  on the length of computations of $F(g)$ on its arguments}.
  Suppose $F(g)(x) \simeq t$. $F$ is effectively continuous, so
  there exists a finite $\theta \subseteq g$ such that \[
    F(\theta)(x) \simeq t
  \] by compactness. Choose a minimal such $\theta$, and let \[
    \theta = \{ \, (x_1, y_1), \dots, (x_n, y_n) \, \}
  \] By construction, the $x_i$ are exactly the `questions' with
  which a Turing machine that computes $F$ would
  query the oracle, on input $x$.

  By using Kleene's SRT, we have replaced calls to the oracle
  by recursive calls to another copy of itself. It follows that
  the computation of each $F(g)(x_i) \simeq y_i$ is strictly
  shorter in length than the overall computation of $F(g)(x)
  \simeq y$. Hence, by the induction hypothesis,  $(x_i, y_i) \in
  \mathbf{lfp}(F)$ for all $i$, and $\theta \subseteq
  \mathbf{lfp}(F)$. 

  By monotonicity, $F(\theta) \subseteq \mathbf{lfp}(F)$, and
  since $F(\theta)(x) \simeq t$, we have that $(x, t) \in
  \mathbf{lfp}(F)$. Hence $g \subseteq \mathbf{lfp}(F)$, and as
  $g$ is also a fixed point, equality holds.
\end{proof}

\chapter[iPCF: An Intensional Programming Language]
  {\huge iPCF: An Intensional Programming Language\footnote{This chapter
  is based on the paper \citep{Kavvos2017d}, which was presented at
  the 7th workshop on Intuitionistic Modal Logics and Applications
  (IMLA 2017). A preprint is available as
  \href{https://arxiv.org/abs/1703.01288}{arXiv:1703.01288}}}

  \label{chap:ipcf}

  \opt{th}{
  This chapter concerns the elaboration of the
  modality-as-intension interpretation that we introduced in
  \S\ref{sec:modasint}.  Our starting point will be the
  Davies-Pfenning calculus for \textsf{S4}, which is a typed
  $\lambda$-calculus with \emph{modal types}. The intuitive
  meaning of the modal type $\Box A$ will be that of \emph{code},
  that---when evaluated---yields a value of type $A$. We wish to
  use this calculus for intensional and reflective programming.

  The Davies-Pfenning calculus already supports a notion of
  \emph{programs-as-data}: to each term $M : A$ that uses only
  `code' variables there corresponds a term $\ibox{M} : \Box A$
  that stands for the term $M$ considered as a datum. This is
  already considerably stronger than ordinary higher-order
  functional programming with `functions as first-class citizens,'
  as it also entails a kind of \emph{homoiconicity}, similar to
  the one present in dialects of \textsc{Lisp}. But we want to go
  even further than that: in \textsc{Lisp}, a program is able to
  process code by treating it as mere symbols, thereby
  disregarding its observable behaviour. 
  
  The true spirit of intensionality is the ability to support
  operations that are, according to the extensional viewpoint,
  \emph{non-functional}.  This was not the case in the work of
  Davies and Pfenning, who merely used their calculus for
  \emph{staged metaprogramming}, which did not require
  non-functional operations. In this chapter we shall mend this.
  We shall augment their calculus by adding \emph{intensional
  operations}, and \emph{intensional recursion}.  We shall call
  the resulting calculus \emph{Intensional PCF}, after the
  simply-typed $\lambda$-calculus with (extensional) fixed points
  studied by \cite{Scott1993} and \cite{Plotkin1977}.

  There has been some previous work on adding intensional
  operations to the Davies-Pfenning calculus.  A complicated
  system based on nominal techniques that fleshed out those ideas
  was presented by \cite{Nanevski2002}. The notions of intensional
  and extensional equality implicit in this system were studied
  using logical relations by Pfenning and Nanevski
  \cite{Nanevski2005}. However, none of these papers studied
  whether the induced equational systems are consistent. We show
  that, no matter the intensional mechanism at use, modalities
  enable consistent intensional programming.

  To our knowledge, this chapter presents (a) the first
  consistency proof for type-safe intensional programming, and (b)
  the first type-safe attempt at reflective programming.
}

\section{Introducing Intensional PCF}
  \label{sec:ipcfintro}

Intensional PCF (iPCF) is a typed $\lambda$-calculus with
modal types. As discussed before, the modal types work in
our favour by separating intension from extension, so that the
latter does not leak into the former. Given the logical flavour of
\opt{ipcf}{our previous work on intensionality
\cite{Kavvos2017a},}\opt{th}{our observations in
\S\ref{sec:modasint}} we shall model the types of iPCF after the
\emph{constructive modal logic \textsf{S4}}, in the dual-context
style pioneered by Pfenning and Davies \citep{Davies2001,
Davies2001a}. Let us seize this opportunity to remark that (a)
there are also other ways to capture \textsf{S4}, for which see
the survey \citep{Kavvos2016b}, and that (b) dual-context
formulations are not by any means limited to \textsf{S4}: they
began in the context of \emph{intuitionistic linear logic}, but
have recently been shown to also encompass other modal logics; see
\cite{Kavvos2017b}. 

iPCF is \emph{not} related to the language \textsf{Mini-ML} that
is introduced by \cite{Davies2001a}: that is a call-by-value,
ML-like language, with ordinary call-by-value fixed points. In
contrast, ours is a call-by-name language with a new kind of
fixed point, namely intensional fixed points. These fixed points
will afford the programmer the full power of \emph{intensional
recursion}.  In logical terms they correspond to throwing the
G\"odel-L\"ob axiom $\Box(\Box A \rightarrow A) \rightarrow \Box
A$ into \textsf{S4}. Modal logicians might object to this, as, in
conjunction with the \textsf{T} axiom $\Box A \rightarrow A$, it
will make every type inhabited. We remind them that a similar
situation occurs in PCF, where the $\mathbf{Y}_A : (A \rightarrow
A) \rightarrow A$ combinator allows one to write a term
$\mathbf{Y}_A(\lambda x{:}A.\ x)$ at every type $A$. As in the
study of PCF, we care less about the logic and more about the
underlying computation: \emph{it is the terms that matter, and the
types are only there to stop type errors from happening}.

The syntax and the typing rules of iPCF may be found in Figure
\ref{fig:ipcf}. These are largely the same as Pfenning and Davies'
\textsf{S4}, save the addition of some constants (drawn from PCF),
and a rule for intensional recursion. The introduction rule for
the modality restricts terms under a $\ibox{(-)}$ to those
containing only modal variables, i.e. variables carrying only
intensions or code, but never `live values:' \[
  \begin{prooftree}
    \ctxt{\Delta}{\cdot} \vdash M : A
      \justifies
    \ctxt{\Delta}{\Gamma} \vdash \ibox{M} : \Box A
  \end{prooftree}
\] There is also a rule for intensional recursion: \[
  \begin{prooftree}
    \ctxt{\Delta}{z : \Box A} \vdash M : A
      \justifies
    \ctxt{\Delta}{\Gamma} \vdash \fixlob{z}{M} : A
  \end{prooftree}
\] This will be coupled with the reduction $\fixlob{z}{M} \red{}
M[\ibox{(\fixlob{z}{M})}/z]$. This rule is actually just
\emph{L\"ob's rule} with a modal context, and including it in the
Hilbert system of a (classical or intuitionistic) modal logic is
equivalent to including the G\"odel-L\"ob axiom: see
\cite{Boolos1994} and \cite{Ursini1979a}. We recommend the survey
\cite{Litak2014} for a broad coverage of constructive modalities
with a provability-like flavour. Finally, let us record a fact
noticed by Samson Abramsky, which is that erasing the modality
from the types appearing in either L\"ob's rule or the
G\"odel-L\"ob axiom yields the type of $\mathbf{Y}_A : (A
\rightarrow A) \rightarrow A$, as a rule in the first case, or
axiomatically internalised as a constant in the second (both
variants exist in the literature: see \cite{Gunter1992} and
\cite{Mitchell1996}.)

\begin{figure}
  \caption{Syntax and Typing Rules for Intensional PCF}
  \label{fig:ipcf}
  \begin{align*}
  \textbf{Ground Types} \quad &
    G & ::=\quad &\textsf{Nat} \;|\; \textsf{Bool}
   \\ \\
  \textbf{Types} \quad &
    A, B & ::=\quad &G \;|\; A \rightarrow B \;|\; \Box A
   \\ \\
  \textbf{Terms} \quad &
    M, N & ::=\quad &x
      \;|\; \lambda x{:}A.\ M
      \;|\; M N
      \;|\; \ibox{M}
      \;|\; \letbox{u}{M}{N} \;| \\
   &      &    &\widehat{n}
      \;|\; \textsf{true}
      \;|\; \textsf{false}
      \;|\; \textsf{succ}
      \;|\; \textsf{pred}
      \;|\; \textsf{zero?}
      \;|\; \supset_G
      \;|\; \fixlob{z}{M}
    \\ \\
%  \textbf{Canonical Forms} \quad &
%    V & ::= \quad &\widehat{n}
%     \;|\; \textsf{true}
%     \;|\; \textsf{false}
%     \;|\; \lambda x{:}A.\ M
%     \;|\; \textsf{box } M
%   \\ \\
  \textbf{Contexts} \quad &
    \Gamma, \Delta & ::=\quad &\cdot \;|\; \Gamma, x: A
\end{align*}

\vfill

\renewcommand{\arraystretch}{3}

% TABLE FOR CONSTANTS
\begin{tabular}{c c}
  %% CONSTANT RULES
  $
    \begin{prooftree}
      \justifies
        \ctxt{\Delta}{\Gamma} \vdash \widehat{n} : \textsf{Nat}
    \end{prooftree}
  $
  
  &

  $   
    \begin{prooftree}
      \justifies
        \ctxt{\Delta}{\Gamma} \vdash b : \textsf{Bool}
      \using
        (b \in \{\textsf{true}, \textsf{false}\})
    \end{prooftree}
  $
  
  \\

  $ 
    \begin{prooftree}
      \justifies
        \ctxt{\Delta}{\Gamma} \vdash \textsf{zero?} : 
          \textsf{Nat} \rightarrow \textsf{Bool}
    \end{prooftree}
  $
  
  &

  $ 
    \begin{prooftree}
      \justifies
        \ctxt{\Delta}{\Gamma}
        \vdash f : \textsf{Nat} \rightarrow \textsf{Nat}
      \using
        (f \in \{\textsf{succ}, \textsf{pred}\})
    \end{prooftree}
  $
  
  \\

  \multicolumn{2}{c}{
  $
    \begin{prooftree}
      \justifies
        \ctxt{\Delta}{\Gamma} \vdash {\supset_G}  : 
        \textsf{Bool} \rightarrow G \rightarrow G \rightarrow G
    \end{prooftree}
  $
  }

  \\

  %% VARIABLE RULES

  $
    \begin{prooftree}
        \justifies
          \ctxt{\Delta}{\Gamma, x{:}A, \Gamma'} \vdash x:A
        \using
          {(\textsf{var})}
    \end{prooftree}
  $

  &

  $
    \begin{prooftree}
        \justifies
      \ctxt{\Delta, u{:} A, \Delta'}{\Gamma} \vdash u:A
        \using
      {(\Box\textsf{var})}
    \end{prooftree}
  $

  \\

  %% IMPLICATION RULES

  $
    \begin{prooftree}
      \ctxt{\Delta}{\Gamma}, x{:}A \vdash M : B
        \justifies
      \ctxt{\Delta}{\Gamma} \vdash \lambda x{:}A. \; M : A \rightarrow B
        \using
      {(\rightarrow\mathcal{I})}
    \end{prooftree}
  $

  &

  $
    \begin{prooftree}
      \ctxt{\Delta}{\Gamma} \vdash M : A \rightarrow B
        \quad
      \ctxt{\Delta}{\Gamma} \vdash N : A
        \justifies
      \ctxt{\Delta}{\Gamma} \vdash M N : B
        \using
      {(\rightarrow\mathcal{E})}
    \end{prooftree}
  $

  \\

  %% BOX RULES

  $
    \begin{prooftree}
      \ctxt{\Delta}{\cdot} \vdash M : A
        \justifies
      \ctxt{\Delta}{\Gamma} \vdash \ibox{M} : \Box A
        \using
      {(\Box\mathcal{I})}
    \end{prooftree}
  $

  &

  $
    \begin{prooftree}
      \ctxt{\Delta}{\Gamma} \vdash M : \Box A
        \quad\quad
      \ctxt{\Delta, u{:}A}{\Gamma} \vdash N : C
        \justifies
      \ctxt{\Delta}{\Gamma} \vdash \letbox{u}{M}{N} : C
        \using
      {(\Box\mathcal{E})}
    \end{prooftree}
  $

  \\

  \multicolumn{2}{c}{
    $
      \begin{prooftree}
        \ctxt{\Delta}{z : \Box A} \vdash M : A
          \justifies
        \ctxt{\Delta}{\Gamma} \vdash \fixlob{z}{M} : \Box A
          \using
        {(\Box\textsf{fix})}
      \end{prooftree}
    $
  }
\end{tabular}

\end{figure}

\section{Metatheory}
  \label{sec:ipcfmeta}

\opt{th}{

  This section concerns the basic metatheoretic properties of
  iPCF.  The expected structural rules are admissible. We also
  prove a theorem regarding the behaviour of free variables,
  similar to the ones in \citep{Kavvos2017b}, which demonstrates
  how the different layers of intension and extension are
  separated by the type system.

  \subsection{Structural Theorems \& Cut}
}

iPCF satisfies the expected basic results: structural and cut
rules are admissible. This is no surprise given its origin in the
well-behaved Davies-Pfenning calculus. We assume the typical
conventions for $\lambda$-calculi: terms are identified up to
$\alpha$-equivalence, for which we write $\equiv$, and
substitution $[\cdot / \cdot]$ is defined in the ordinary,
capture-avoiding manner. Bear in mind that we consider occurrences
of $u$ in $N$ to be bound in $\letbox{u}{M}{N}$.  Contexts
$\Gamma$, $\Delta$ are lists of type assignments $x : A$.
Furthermore, we shall assume that whenever we write a judgement
like $\ctxt{\Delta}{\Gamma} \vdash M : A$, then $\Delta$ and
$\Gamma$ are \emph{disjoint}, in the sense that $\vars{\Delta}
\cap \vars{\Gamma} = \emptyset$, where $\vars{x_1 : A_1, \dots,
x_n : A_n} \myeq \{x_1, \dots, x_n\}$.  We write $\Gamma, \Gamma'$
for the concatenation of disjoint contexts. Finally, we sometimes
write $\vdash M : A$ whenever $\ctxt{\cdot}{\cdot} \vdash M : A$.

\begin{thm}[Structural \& Cut]
  \label{thm:scut}
  The following rules are admissible in iPCF:
  \begin{multicols}{2}
  \begin{enumerate}
    \item (Weakening) \[
      \begin{prooftree}
        \ctxt{\Delta}{\Gamma, \Gamma'} \vdash M : A
          \justifies
        \ctxt{\Delta}{\Gamma, x{:}A, \Gamma'} \vdash M : A
      \end{prooftree}
    \]
    \item (Exchange) \[
      \begin{prooftree}
        \ctxt{\Delta}{\Gamma, x{:}A, y{:}B, \Gamma'} \vdash M : C
          \justifies
        \ctxt{\Delta}{\Gamma, y{:}B, x{:}A, \Gamma'} \vdash M : C
      \end{prooftree}
    \]
    \item (Contraction) \[
      \begin{prooftree}
        \ctxt{\Delta}{\Gamma, x{:}A, y{:}A, \Gamma'} \vdash M : A
          \justifies
        \ctxt{\Delta}{\Gamma, w{:}A, \Gamma'} \vdash M[w, w/x, y] : A
      \end{prooftree}
    \]
    \item (Cut) \[
      \begin{prooftree}
        \ctxt{\Delta}{\Gamma} \vdash N : A
          \qquad
        \ctxt{\Delta}{\Gamma, x{:}A, \Gamma'} \vdash M : A
          \justifies
        \ctxt{\Delta}{\Gamma, \Gamma'} \vdash M[N/x] : A
      \end{prooftree}
    \]
  \end{enumerate}
  \end{multicols}
\end{thm}

\opt{th}{
  \begin{proof}
    All by induction on the typing derivation of $M$. Verified in
    the proof assistant \textsc{Agda}: see Appendix
    \ref{sec:iPCF.agda}.
  \end{proof}
}

\begin{thm}[Modal Structural \& Cut]
  \label{thm:modalstruct}
  The following rules are admissible:
  \begin{multicols}{2}
  \begin{enumerate}
    \item (Modal Weakening) \[
      \begin{prooftree}
        \ctxt{\Delta, \Delta' }{\Gamma} \vdash M : C
          \justifies
        \ctxt{\Delta, u{:}A, \Delta'}{\Gamma} \vdash M : C
      \end{prooftree}
    \]
    \item (Modal Exchange) \[
      \begin{prooftree}
        \ctxt{\Delta, x{:}A, y{:}B, \Delta'}{\Gamma} \vdash M : C
          \justifies
        \ctxt{\Delta, y{:}B, x{:}A, \Delta'}{\Gamma} \vdash M : C
      \end{prooftree}
    \]
    \item (Modal Contraction) \[
      \begin{prooftree}
        \ctxt{\Delta, x{:}A, y{:}A, \Delta'}{\Gamma} \vdash M : C
          \justifies
        \ctxt{\Delta, w{:}A, \Delta'}{\Gamma} \vdash M[w, w/x, y] : C
      \end{prooftree}
    \]
    \item (Modal Cut) \[
      \begin{prooftree}
        \ctxt{\Delta}{\cdot} \vdash N :  A
          \quad
        \ctxt{\Delta, u{:}A, \Delta'}{\Gamma} \vdash M : C
          \justifies
        \ctxt{\Delta, \Delta'}{\Gamma} \vdash M[N/u] : C
      \end{prooftree}
    \]
  \end{enumerate}
  \end{multicols}
\end{thm}

\opt{th}{
  \begin{proof} 
    All by induction on the typing derivation of $M$. Verified in
    the proof assistant \textsc{Agda}: see Appendix
    \ref{sec:iPCF.agda}.
  \end{proof}
}

%\begin{thm}[Modal Dereliction]
%  \label{thm:dereliction}
%  The following rule is admissible: \[
%    \begin{prooftree}
%        \ctxt{\Delta}{\Gamma, \Gamma'} \vdash M : A
%      \justifies
%        \ctxt{\Delta, \Gamma}{\Gamma'} \vdash M : A
%    \end{prooftree}
%  \]
%\end{thm}

%\opt{th}{
%\begin{proof}
%  By induction on the derivation of $\ctxt{\Delta}{\Gamma, \Gamma'}
%  \vdash M : A$. Most cases are straightfoward, except
%  $(\textsf{var})$ and $(\Box\mathcal{I})$.
%
%  If the judgment holds by $(\textsf{var})$, then $M \equiv x$ for
%  some $(x : A) \in \Gamma, \Gamma'$. If $(x : A) \in \Gamma$, we
%  use $(\Box\textsf{var})$ to conclude that $\ctxt{\Delta,
%  \Gamma}{\Gamma'} \vdash x : A$. If $(x : A) \in \Gamma'$, then
%  another use of $(\textsf{var})$ suffices.
%
%  If the judgment holds by $(\Box\mathcal{I})$ then $M
%  \equiv \ibox{M'}$ and $A \equiv \Box A'$ for some $M', A'$ with
%  $\ctxt{\Delta}{\cdot} \vdash M' : A'$. Repeated use of weakening
%  for the modal context followed by an application of
%  $(\Box\mathcal{I})$ yields the result.
%\end{proof}
%}

\opt{th}{
\subsection{Free variables}

In this section we prove a theorem regarding the occurrences of
free variables in well-typed terms of iPCF. It turns out that, if
a variable occurs free under a $\ibox{(-)}$ construct, then it has
to be in the modal context. This is the property that enforces
that \emph{intensions can only depend on intensions}.

\begin{defn}[Free variables] \hfill
  \begin{enumerate}
    \item The \emph{free variables} $\fv{M}$ of a term
    $M$ are defined by induction on the structure of the term:
    \begin{align*}
      \fv{x}  &\myeq \{x\} \\
      \fv{MN} &\myeq \fv{M} \cup \fv{N} \\
      \fv{\lambda x{:}A.\ M} &\myeq \fv{M} - \{x\} \\
      \fv{\ibox{M}} &\myeq \fv{M} \\
      \fv{\fixlob{z}{M}} &\myeq \fv{M} - \{z\} \\
      \fv{\letbox{u}{M}{N}} &\myeq \fv{M} \cup \left(\fv{N} - \{u\}\right)
    \end{align*}

    \item The \emph{unboxed free variables} $\ufv{M}$ of a term
    are those that do \emph{not} occur under the scope of a
    $\ibox{(-)}$ or $\fixlob{z}{(-)}$ construct. They are formally
    defined by replacing the following clauses in the definition
    of $\fv{-}$:
    \begin{align*}
      \ufv{\ibox{M}} &\myeq \emptyset \\
      \ufv{\fixlob{z}{M}} &\myeq \emptyset
    \end{align*}

    \item The \emph{boxed free variables} $\bfv{M}$ of a term $M$
    are those that \emph{do} occur under the scope of a
    $\ibox{(-)}$ construct. They are formally defined by replacing
    the following clauses in the definition of $\fv{-}$:
    \begin{align*}
      \bfv{x} &\myeq \emptyset \\
      \bfv{\ibox{M}} &\myeq \fv{M} \\
      \bfv{\fixlob{z}{M}} &\myeq \fv{M} - \{z\}
    \end{align*}
  \end{enumerate}
\end{defn}

\begin{thm}[Free variables] \hfill
  \label{thm:freevar}
  \begin{enumerate}

    \item For every term $M$, $\fv{M} = \ufv{M} \cup \bfv{M}$.

    \item If and $\ctxt{\Delta}{\Gamma} \vdash M : A$, then
      \begin{align*}
	\ufv{M} &\subseteq \vars{\Gamma} \cup \vars{\Delta} \\
\bfv{M} &\subseteq \vars{\Delta}
     \end{align*}

  \end{enumerate}
\end{thm}

\begin{proof} \hfill
  \begin{enumerate}
    \item
      Trivial induction on $M$.

    \item
      By induction on the derivation of $\ctxt{\Delta}{\Gamma}
      \vdash M : A$. We show the case for $(\Box\mathcal{I})$; the
      first statement is trivial, so we show the second:
      \begin{derivation}
	\bfv{\ibox{M}}
	  \since{definition}
	\fv{M}
	  \since{(1)}
	\ufv{M} \cup \bfv{M}
	  \since[\subseteq]{IH, twice}
	\left(\vars{\Delta} \cup \vars{\cdot}\right) \cup \vars{\Delta}
	  \since{definition}
	\vars{\Delta}
      \end{derivation}
  \end{enumerate}
\end{proof}
}

\section{Consistency of Intensional Operations}
  \label{sec:ipcfconfl}

In this section we shall prove that the modal types of iPCF enable
us to consistently add intensional operations on the modal types.
These are \emph{non-functional operations on terms} which are not
ordinarily definable because they violate equality.  All we have
to do is assume them as constants at modal types, define their
behaviour by introducing a notion of reduction, and then
prove that the compatible closure of this notion of reduction is
confluent. A known corollary of confluence is that the equational
theory induced by the reduction is \emph{consistent}, i.e. does
not equate all terms.

There is a caveat involving extension flowing into intension. That
is: we need to exclude from consideration terms where a variable
bound by a $\lambda$ occurs under the scope of a $\ibox{(-)}$
construct. These will never be well-typed, but---since we discuss
types and reduction orthogonally---we also need to explicitly
exclude them here too.

\subsection{Adding intensionality}

\opt{ipcf}{Davies and Pfenning} \cite{Davies2001} suggested that
the $\Box$ modality can be used to signify intensionality. In
fact, in \citep{Davies1996, Davies2001a} they had prevented
reductions from happening under $\ibox{(-)}$ construct, `` [...]
since this would violate its intensional nature.'' But the truth
is that neither of these presentations included any genuinely
non-functional operations at modal types, and hence their only use
was for homogeneous staged metaprogramming. Adding intensional,
non-functional operations is a more difficult task. Intensional
operations are dependent on \emph{descriptions} and
\emph{intensions} rather than \emph{values} and \emph{extensions}.
Hence, unlike reduction and evaluation, they cannot be blind to
substitution. This is something that quickly came to light as soon
as \cite{Nanevski2002} attempted to extend the system of Davies
and Pfenning to allow `intensional code analysis' using nominal
techniques.

A similar task was also recently taken up by Gabbay and Nanevski
\citep{Gabbay2013}, who attempted to add a construct
$\textsf{is-app}$ to the system of Davies and Pfenning, along with
the reduction rules \begin{align*}
  \textsf{is-app}\ (\ibox{PQ}) &\red{} \textsf{true} \\
  \textsf{is-app}\ (\ibox{M})  &\red{} \textsf{false}
    \qquad \text{if $M$ is not of the form $PQ$}
\end{align*} The function computed by $\textsf{is-app}$ is truly
intensional, as it depends solely on the syntactic structure of
its argument: it merely checks if it syntactically is an
application or not. As such, it can be considered a
\emph{criterion of intensionality}, albeit an extreme one: its
definability conclusively confirms the presence of computation
up to syntax.

Gabbay and Nanevski tried to justify the inclusion of
$\textsf{is-app}$ by producing denotational semantics for modal
types in which the semantic domain $\sem{\Box A}{}$ directly
involves the actual closed terms of type $\Box A$. However,
something seems to have gone wrong with substitution. In fact, we
believe that their proof of soundness is wrong: it is not hard to
see that their semantics is not stable under the second of these
two reductions: take $M$ to be $u$, and let the semantic
environment map $u$ to an application $PQ$, and then notice that
this leads to $\sem{\textsf{true}}{} = \sem{\textsf{false}}{}$. We
can also see this in the fact that their notion of reduction is
\emph{not confluent}. Here is the relevant counterexample: we can
reduce like this: \[
  \letbox{u}{\ibox{(PQ)}}{\textsf{is-app}\ (\ibox{u})}
    \red{}
  \textsf{is-app}\ (\ibox{PQ}) 
    \red{}
  \textsf{true}
\] But we could have also reduced like that: \[
  \letbox{u}{\ibox{(PQ)}}{\textsf{is-app}\ (\ibox{u})}
    \red{}
  \letbox{u}{\ibox{(PQ)}}{\textsf{false}}
    \red{}
  \textsf{false}
\] This example is easy to find if one tries to plough through a
proof of confluence: it is very clearly \emph{not} the case that
$M \red{} N$ implies $M[P/u] \red{} N[P/u]$ if $u$ is under a
$\ibox{(-)}$, exactly because of the presence of intensional
operations such as $\textsf{is-app}$. 

Perhaps the following idea is more workable: let us limit
intensional operations to a chosen set of functions $f :
\mathcal{T}(A) \rightarrow \mathcal{T}(B)$ from terms of type $A$
to terms of type $B$, and then represent them in the language by a
constant $\tilde f$, such that $\tilde f(\ibox{M}) \red{}
\ibox{f(M)}$. This set of functions would then be chosen so that
they satisfy some sanity conditions. Since we want to have a
\textsf{let} construct that allows us to substitute code for modal
variables, the following general situation will occur: if $N
\red{} N'$, we have
%  \[
%    \begin{tikzcd}
%      & \letbox{u}{\ibox{M}}{N}
%	  \arrow[dr]
%	  \arrow[dl]
%      &  \\
%      N[M/u]
%      & 
%      &
%      \letbox{u}{\ibox{M}}{N'}
%	\arrow[d] \\
%      &
%      &
%      N'[M/u]
%    \end{tikzcd}
%  \]
  \[
    \letbox{u}{\ibox{M}}{N}
      \red{}
    N[M/u]
  \] but also \[
    \letbox{u}{\ibox{M}}{N}
      \red{}
    \letbox{u}{\ibox{M}}{N'}
      \red{}
    N'[M/u]
  \]

\noindent Thus, in order to have confluence, we need $N[M/u]
\red{} N'[M/u]$. This will only be the case for reductions of the
form $\tilde f(\ibox{M}) \rightarrow \ibox{f(M)}$ if $f(N[M/u])
\equiv f(N)[M/u]$, i.e. if $f$ is \emph{substitutive}. But then a
simple naturality argument gives that $f(N) \equiv f(u[N/u])
\equiv f(u)[N/u]$, and hence $\tilde f$ is already definable by \[
  \lambda x : \Box A.\ \letbox{u}{x}{\ibox{f(u)}}
\] so such a `substitutive' function is not intensional after all.

In fact, the only truly intensional operations we can add to our
calculus will be those acting on \emph{closed} terms. We will see
that this circumvents the problems that arise when intensionality
interacts with substitution. Hence, we will limit intensional
operations to the following set:

\begin{defn}[Intensional operations]
  Let $\mathcal{T}(A)$ be the set of ($\alpha$-equivalence classes
  of) closed terms such that $\ctxt{\cdot}{\cdot} \vdash M : A$.
  Then, the set of \emph{intensional operations}, $\mathcal{F}(A,
  B)$, is defined to be the set of all functions $f :
  \mathcal{T}(A) \rightarrow
  \mathcal{T}(B)$.
\end{defn}

\noindent We will include all of these intensional operations $f :
\mathcal{T}(A) \rightarrow \mathcal{T}(B)$ in our calculus, as
constants: \[
  \ctxt{\Delta}{\Gamma} \vdash \tilde f : \Box A \rightarrow \Box B
\] with reduction rule $\tilde f(\ibox{M}) \rightarrow
\ibox{f(M)}$, under the proviso that $M$ is closed.  Of course,
these also includes operations on terms that might \emph{not be
computable}. However, we are interested in proving consistency of
intensional operations in the most general setting. The questions
of which intensional operations are computable, and which
primitives can and should be used to express them, are both still
open.

\subsection{Reduction and Confluence}

\begin{figure}
  \caption{Reduction for Intensional PCF}
  \label{fig:ipcfbeta}
  \renewcommand{\arraystretch}{4}

\begin{center}
\begin{tabular}{c c}
  %% BETA
  $
    \begin{prooftree}
        \justifies
      (\lambda x{:}A.\ M)N \red{} M[N/x]
        \using
      {(\red{}\beta)}
    \end{prooftree}
  $

  &

  $
    \begin{prooftree}
      M \red{} N
        \justifies
    \lambda x{:}A.\ M \red{} \lambda x{:}A.\ N
        \using
      {(\textsf{cong}_\lambda)}
    \end{prooftree}
  $

  \\

  %% APPLICATION CONGRUENCE

  $
    \begin{prooftree}
      M \red{} N
        \justifies
      MP \red{} NP
        \using
      {(\textsf{app}_1)}
    \end{prooftree}
  $

  &

  $
    \begin{prooftree}
      P \red{} Q
        \justifies
      MP \red{} MQ
        \using
      {(\textsf{app}_2)}
    \end{prooftree}
  $

  \\

  \multicolumn{2}{c}{
    $
      \begin{prooftree}
          \justifies
        \letbox{u}{\ibox{M}}{N} \red{} N[M/u]
          \using
        {(\Box\beta)}
      \end{prooftree}
    $
  }

  \\

  \multicolumn{2}{c}{
    $
      \begin{prooftree}
          \justifies
        \fixlob{z}{M} \red{} M[\ibox{(\fixlob{z}{M})}/z]
          \using
        {(\Box\textsf{fix})}
      \end{prooftree}
    $
  }
  
  \\

  \multicolumn{2}{c}{
    $
      \begin{prooftree}
        \text{$M$ closed} 
          \justifies
        \tilde f(\ibox{M}) \red{} \ibox{f(M)}
          \using
        {(\Box\textsf{int})}
      \end{prooftree}
    $
  }

  \\

  %% LET BOX CONGRUENCE

  \multicolumn{2}{c}{
    $
      \begin{prooftree}
        M \red{} N
          \justifies
        \letbox{u}{M}{P} \red{} \letbox{u}{N}{P}
          \using
        {(\textsf{let-cong}_1)}
      \end{prooftree}
    $
  }

  \\

  \multicolumn{2}{c}{
    $
      \begin{prooftree}
        P \red{} Q
          \justifies
        \letbox{u}{M}{P} \red{} \letbox{u}{M}{Q}
          \using
        {(\textsf{let-cong}_2)}
      \end{prooftree}
    $
  }

  \\

  $ 
    \begin{prooftree}
        \justifies
      \textsf{zero?}\ \widehat{0} \red{} \textsf{true}
        \using
      {(\textsf{zero?}_1)}
    \end{prooftree}
  $

  & 

  $
    \begin{prooftree}
        \justifies
      \textsf{zero?}\ \widehat{n+1} \red{} \textsf{false}
        \using
      {(\textsf{zero?}_2)}
    \end{prooftree}
  $

  \\

  $
    \begin{prooftree}
        \justifies
      \textsf{succ}\ \widehat{n} \red{} \widehat{n+1}
        \using
      {(\textsf{succ})}
    \end{prooftree}
  $

  & 

  $
    \begin{prooftree}
        \justifies
      \textsf{pred}\ \widehat{n} \red{} \widehat{n\dotdiv 1}
        \using
      {(\textsf{pred})}
    \end{prooftree}
  $

  \\

  $
    \begin{prooftree}
        \justifies
      \supset_G\ \textsf{true}\ M\ N \red{} M
        \using
      {(\supset_1)}
    \end{prooftree}
  $

  & 

  $
    \begin{prooftree}
        \justifies
      \supset_G\ \textsf{false}\ M\ N \red{} N
        \using
      {(\supset_2)}
    \end{prooftree}
  $
\end{tabular}
\end{center}

\end{figure}

\begin{figure}
  \caption{Equational Theory for Intensional PCF}
  \label{fig:ipcfeq}
  \renewcommand{\arraystretch}{4}

\begin{tabular}{c c}

%  \textbf{Equivalence Relation} & \\ \\
%
%  \multicolumn{2}{c}{
%    $ \begin{prooftree}
%        \ctxt{\Delta}{\Gamma} \vdash M : A
%          \justifies
%        \ctxt{\Delta}{\Gamma} \vdash M = M : A
%      \end{prooftree} $
%  } \\ \\
%
%  $ \begin{prooftree}
%      \ctxt{\Delta}{\Gamma} \vdash M = N : A
%        \justifies
%      \ctxt{\Delta}{\Gamma} \vdash N = M : A
%    \end{prooftree} $
%  &
%  $ \begin{prooftree}
%      \ctxt{\Delta}{\Gamma} \vdash P = Q : A
%        \qquad
%      \ctxt{\Delta}{\Gamma} \vdash Q = R : A
%        \justifies
%      \ctxt{\Delta}{\Gamma} \vdash P = R : A
%    \end{prooftree} $
%

  \textbf{Function Spaces} & \\

  \multicolumn{2}{c}{
    $
      \begin{prooftree}
        \ctxt{\Delta}{\Gamma} \vdash N : A
          \qquad
        \ctxt{\Delta}{\Gamma, x{:}A, \Gamma'} \vdash M : B
          \justifies
        \ctxt{\Delta}{\Gamma} \vdash (\lambda x{:}A. M)\,N = M[N/x] : B
          \using
        {(\rightarrow\beta)}
      \end{prooftree}
    $
  }

  \\

%  $
%    \begin{prooftree}
%      \ctxt{\Delta}{\Gamma} \vdash M : A \rightarrow B
%        \qquad
%      x \not\in \text{fv}(M)
%        \justifies
%      \ctxt{\Delta}{\Gamma} \vdash M = \lambda x{:}A. Mx : A \rightarrow B
%        \using
%      {(\rightarrow\eta)}
%    \end{prooftree}
%  $

%  $ \begin{prooftree}
%      \ctxt{\Delta}{\Gamma, x{:}A} \vdash M = N : B
%        \justifies
%      \ctxt{\Delta}{\Gamma} \vdash \lambda x{:}A. M = \lambda x{:}A. N :
%      A \rightarrow B
%        \using
%      {(\rightarrow\textsf{cong}_1)}
%    \end{prooftree} $

%  $ \begin{prooftree}
%      \ctxt{\Delta}{\Gamma} \vdash M = P : A \rightarrow B
%        \qquad
%      \ctxt{\Delta}{\Gamma} \vdash N = Q : A
%        \justifies
%      \ctxt{\Delta}{\Gamma} \vdash MN = PQ : B
%        \using
%      {(\rightarrow\textsf{cong}_2)}
%    \end{prooftree} $

  \textbf{Modality} & \\

  \multicolumn{2}{c}{
    $
      \begin{prooftree}
        \ctxt{\Delta}{\cdot} \vdash M : A
          \qquad
        \ctxt{\Delta, u : A}{\Gamma} \vdash N : C
          \justifies
        \ctxt{\Delta}{\Gamma}
            \vdash \letbox{u}{\ibox{M}}{N} = N[M/x] : C
          \using
        {(\Box\beta)}
      \end{prooftree}
    $
  }
    
  \\

%  \multicolumn{2}{c}{
%  $ \begin{prooftree}
%      \ctxt{\Delta}{\Gamma} \vdash M : \Box A
%        \justifies
%      \ctxt{\Delta}{\Gamma} \vdash \letbox{u}{M}{\ibox{u}} = M : \Box A
%        \using
%      {(\Box\eta)}
%    \end{prooftree} $
%  }
%
%  \\

  \multicolumn{2}{c}{
    $
      \begin{prooftree}
        \ctxt{\Delta}{z : \Box A} \vdash M : A
          \justifies
        \ctxt{\Delta}{\Gamma} \vdash 
            \fixlob{z}{M} = M[\ibox{(\fixlob{z}{M})}/z] : A
          \using
          {(\Box\textsf{fix})}
      \end{prooftree}
    $
  }

  \\

  \multicolumn{2}{c}{
    $
      \begin{prooftree}
        \ctxt{\cdot}{\cdot} \vdash M : A
          \quad
        f \in \mathcal{F}(A, B)
          \justifies
        \ctxt{\Delta}{\Gamma} \vdash 
            \tilde f (\ibox{M}) = \ibox{f(M)} : \Box B
          \using
        {(\Box\textsf{int})}
      \end{prooftree}
    $
  }

  \\

  \multicolumn{2}{c}{
    $
      \begin{prooftree}
        \ctxt{\Delta}{\Gamma} \vdash M = N : \Box A
          \qquad
        \ctxt{\Delta}{\Gamma} \vdash P = Q : C
          \justifies
        \ctxt{\Delta}{\Gamma} \vdash
            \letbox{u}{M}{P} = \letbox{u}{N}{Q} : B
          \using
        {(\Box\textsf{let-cong})}
      \end{prooftree}
    $
  }

  \\

  \multicolumn{2}{c}{
    \begin{minipage}{\textwidth}
      \textbf{Remark}. In addition to the above, one should also
      include (a) rules that ensure that equality is an equivalence
      relation, (b) congruence rules for $\lambda$-abstraction and
      application, and (c) rules corresponding to the behaviour of
      constants, as in Figure \ref{fig:ipcfbeta}. 
    \end{minipage}
  }

\end{tabular}

\end{figure}

We introduce a notion of reduction for iPCF, which we present in
Figure \ref{fig:ipcfbeta}. Unlike many studies of PCF-inspired
languages, we do not consider a reduction strategy but ordinary
`non-deterministic' $\beta$-reduction. We do so because are trying
to show consistency of the induced equational theory.

The equational theory induced by this notion of reduction is the
one alluded to in the previous section: it is a symmetric version
of it, annotated with types. It can be found in Figure
\ref{fig:ipcfeq}. Note the fact that, like in the work of Davies
and Pfenning, we do \emph{not} include the congruence rule for the
modality: \[
  \begin{prooftree}
      \ctxt{\Delta}{\cdot} \vdash M = N : A
    \justifies
      \ctxt{\Delta}{\Gamma} \vdash \ibox{M} = \ibox{N} : \Box A
    \using
      {(\Box\textsf{cong})}
  \end{prooftree}
\] In fact, the very absence of this rule is what will allow modal
types to become intensional. Otherwise, the only new rules are
intensional recursion, embodied by the rule $(\Box \mathsf{fix})$,
and intensional operations, exemplified by the rule
$(\Box\mathsf{int})$.

We note that it seems perfectly reasonable to think that we should
allow reductions under $\textsf{fix}$, i.e. admit the rule \[
  \begin{prooftree}
      M \red{} N
    \justifies
      \fixlob{z}{M} \red{} \fixlob{z}{N}
  \end{prooftree}
\] as $M$ and $N$ are expected to be of type $A$, which need not
be modal. However, the reduction $\fixlob{z}{M} \red{}
M[\ibox{(\fixlob{z}{M})}/z]$ `freezes' $M$ under an occurrence of
$\ibox{(-)}$, so that no further reductions can take place within
it. Thus, the above rule would violate the intensional nature of
boxes. We were likewise compelled to define $\ufv{\fixlob{z}{M}}
\myeq \emptyset$ in the previous section: we should already
consider $M$ to be intensional, or under a box.

We can now show that 

\begin{thm}
  \label{thm:conf}
  The reduction relation $\red{}$ is confluent.
\end{thm}

We will use a variant of the proof in \citep{Kavvos2017b}, i.e.
the method of \emph{parallel reduction}. This kind of proof was
originally discovered by Tait and Martin-L\"of, and is nicely
documented in \cite{Takahashi1995}. Because of the intensional
nature of our $\ibox{(-)}$ constructs, ours will be more nuanced
and fiddly than any in \emph{op. cit.} The method is this: we will
introduce a second notion of reduction, \[
  \redp\ \subseteq \Lambda \times \Lambda
\] which we will `sandwich' between reduction proper and its
transitive closure: \[
  \red{}\ \subseteq\ \redp\ \subseteq\ \redt
\] We will then show that $\redp$ has the diamond property. By the
above inclusions, the transitive closure $\redp^\ast$ of $\redp$
is then equal to $\redt$, and hence $\red$ is Church-Rosser.

In fact, we will follow \cite{Takahashi1995} in doing something
better: we will define for each term $M$ its \emph{complete
development}, $M^\star$. The complete development is intuitively
defined by `unrolling' all the redexes of $M$ at once. We will
then show that if $M \redp N$, then $N \redp M^\star$. $M^\star$
will then suffice to close the diamond: 
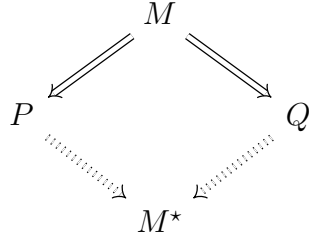
\[
  \begin{tikzcd}
    & M
	\arrow[dr, Rightarrow]
	\arrow[dl, Rightarrow]
    &  \\
    P
      \arrow[dr, Rightarrow, dotted]
    & 
    & Q 
      \arrow[dl, Rightarrow, dotted] \\
    & M^\star
  \end{tikzcd}
\]

\begin{figure}
  \centering
  \caption{Parallel Reduction}
  \renewcommand{\arraystretch}{3}

\begin{center}
\begin{tabular}{c c}

  %% BETA

  $
    \begin{prooftree}
        \justifies
      M \redp{}M
        \using
      {(\textsf{refl})}
    \end{prooftree}
  $

  &

  $
    \begin{prooftree}
      M \redp{}N
        \quad\quad
      P \redp{}Q
        \justifies
      (\lambda x{:}A.\ M)P \redp{}N[Q/x]
        \using
      {(\rightarrow\beta)}
    \end{prooftree}
  $

  \\

  $
    \begin{prooftree}
      M \redp{}N
        \justifies
      \lambda x{:}A.\ M \redp{}\lambda x{:}A.\ N
        \using
      {(\textsf{cong}_\lambda)}
    \end{prooftree}
  $

  &

  $
    \begin{prooftree}
      M \redp{}N
        \quad\quad
      P \redp{}Q
        \justifies
      MP \redp{}NQ
        \using
      {(\textsf{app})}
    \end{prooftree}
  $

  \\

  $
    \begin{prooftree}
      P \redp{} P'
        \justifies
      \supset_G\ \textsf{true}\ P\ Q \redp{} P'
        \using
      {(\supset_1)}
    \end{prooftree}
  $

  & 

  $
    \begin{prooftree}
      Q \redp{} Q'
        \justifies
      \supset_G\ \textsf{false}\ P\ Q \redp{} Q'
        \using
      {(\supset_2)}
    \end{prooftree}
  $

  \\

  \multicolumn{2}{c}{
    $ \begin{prooftree}
      M \redp{} N
        \justifies
      \letbox{u}{\ibox{P}}{M} \redp{} N[P/u]
        \using
      {(\Box\beta)}
    \end{prooftree} $
  }

  \\

  \multicolumn{2}{c}{
    $
      \begin{prooftree}
	M \redp{} M'
          \justifies
        \fixlob{z}{M} \redp{} M'[\ibox{(\fixlob{z}{M})}/z]
          \using
        {(\Box\textsf{fix})}
      \end{prooftree}
    $
  }

  \\ 

  \multicolumn{2}{c}{
    $
      \begin{prooftree}
        \text{$M$ closed}
          \justifies
        \tilde f(\ibox{M}) \redp{}\ibox{f(M)}	
          \using
        {(\Box\textsf{int})}
      \end{prooftree}
    $
  }

  \\

  \multicolumn{2}{c}{
    $
      \begin{prooftree}
        M \redp{}N
          \quad\quad
        P \redp{}Q
          \justifies
        \letbox{u}{M}{P} \redp{} \letbox{u}{N}{Q}
          \using
        {(\Box\textsf{let-cong})}
      \end{prooftree}
    $
  }

  \\

  \multicolumn{2}{c}{
    \begin{minipage}{\textwidth}
      \textbf{Remark}. In addition to the above, one should also
      include rules for the constants, but these are merely
      restatements of the rules in Figure \ref{fig:ipcfbeta}.
    \end{minipage}
  }

\end{tabular}
\end{center}

  \label{fig:parallel}
\end{figure}

The parallel reduction $\redp$ is defined in Figure
\ref{fig:parallel}. Instead of the axiom $(\textsf{refl})$ we
would more commonly have an axiom for variables, $x \redp{} x$,
and $M \redp{} M$ would be derivable. However, we do not have a
congruence rule neither for $\ibox{(-)}$ nor for L\"ob's rule, so
that possibility would be precluded. We are thus forced to include
$M \redp{} M$, which slightly complicates the lemmas that follow.

The main lemma that usually underpins the confluence proof is
this: if $M \redp{} N$ and $P \redp{} Q$, $M[P/x] \redp{} N[Q/x]$.
However, this is intuitively wrong: no reductions should happen
under boxes, so this should only hold if we are substituting for a
variable \emph{not} occurring under boxes.  Hence, this lemma
splits into three different ones:
\begin{itemize}
  \item
    $P \redp{} Q$ implies $M[P/x] \redp{} M[Q/x]$, if $x$ does not
    occur under boxes: this is the price to pay for replacing the
    variable axiom with (\textsf{refl}).
  \item
    $M \redp{} N$ implies $M[P/u] \redp{} N[P/u]$, even if $u$ is
    under a box.
  \item
    If $x$ does not occur under boxes, $M \redp{} N$ and $P
    \redp{} Q$ indeed imply $M[P/x] \redp{} N[Q/x]$
\end{itemize} But let us proceed with the proof.

\begin{lem}
  \label{lem:substint}
  If $M \redp{} N$ then $M[P/u] \redp{} N[P/u]$.
\end{lem}
\begin{proof}
  By induction on the generation of $M \redp{} N$. Most cases
  trivially follow, or consist of simple invocations of the IH. In
  the case of $(\rightarrow\beta)$, the known substitution lemma
  suffices. Let us look at the cases involving boxes.
  \begin{indproof}
    \case{$\Box\beta$}
      Then $M \redp{} N$ is $\letbox{v}{\ibox{R}}{S} \redp{}
      S'[R/v]$ with $S \redp{} S'$. By the IH, we have that
      $S[P/u] \redp{} S'[P/u]$, so \[
	\letbox{v}{\ibox{R[P/u]}}{S[P/u]} \redp S'[P/u][R[P/u]/v]
      \] and this last is $\alpha$-equivalent to $S'[R/v][P/u]$ by
      the substitution lemma.
    \case{$\Box\textsf{fix}$}
      A similar application of the substitution lemma.
    \case{$\Box\textsf{int}$}
      Then $M \redp{} N$ is $\tilde f(\ibox{Q}) \redp{}
      \ibox{f(Q)}$. Hence \[
        \left(\tilde f(\ibox{Q})\right)[P/u]
          \equiv
        \tilde f(\ibox{Q})
          \redp{}
        \ibox{f(Q)}
          \equiv
        \left(\ibox{f(Q)}\right)[P/u]
    \] simply because both $Q$ and $f(Q)$ are closed.
  \end{indproof}
\end{proof}

\begin{lem} 
  \label{lem:substredp}
  If $P \redp Q$ and $x \not\in \bfv{M}$, then $M[P/x] \redp
  M[Q/x]$.
\end{lem}
\begin{proof}
  By induction on the term $M$. The only non-trivial cases are
  those for $M$ a variable, $\ibox{M'}$ or $\fixlob{z}{M'}$. In
  the first case, depending on which variable $M$ is, use either
  $(\textsf{refl})$, or the assumption $P \redp Q$. In the latter
  two, $(\ibox{M'})[P/x] \equiv \ibox{M'} \equiv (\ibox{M'})[Q/x]$
  as $x$ does not occur under a box, so use $(\textsf{refl})$, and
  similarly for $\fixlob{z}{M'}$.
\end{proof}

\begin{lem} 
  \label{lem:redp}
  If $M \redp N$, $P \redp Q$, and $x \not\in \bfv{M}$, then \[
    M[P/x] \redp N[Q/x]
  \]
\end{lem}

\begin{proof}
  By induction on the generation of $M \redp N$. The cases for
  most congruence rules and constants follow trivially, or from
  the IH. We prove the rest.
  \begin{indproof}
    \case{$\textsf{refl}$}
      Then $M \redp N$ is actually $M \redp M$, so we use Lemma
      \ref{lem:substredp} to infer $M[P/x] \redp M[Q/x]$. 

    \case{$\Box\textsf{int}$}
      Then $M \redp N$ is actually $\tilde f(\ibox{M}) \redp
      \ibox{f(M)}$. But $M$ and $f(M)$ are closed, so
      $\left(\tilde f(\ibox{M})\right)[P/x] \equiv \tilde
      f(\ibox{M}) \redp \ibox{f(M)} \equiv
      \left(\ibox{f(M)}\right)[Q/x]$.

    \case{$\supset_i$}
      Then $M \redp N$ is $\supset_G\ \textsf{true}\ M\ N \redp
      M'$ with $M \redp M'$. By the IH, $M[P/x] \redp M'[Q/x]$, so
      \[
	\supset_G\ \textsf{true}\ M[P/x]\ N[P/x]
	  \redp
	M'[Q/x]
      \] by a single use of $(\supset_1)$. The case for
      $\textsf{false}$ is similar.

    \case{$\rightarrow \beta$}
      Then $(\lambda x'{:}A.\ M)N \redp N'[M'/x']$, where $M \redp
      M'$ and $N \redp N'$. Then \[
	\left((\lambda x'{:}A.\ M)N\right)[P/x]
	  \equiv
	(\lambda x'{:}A.\ M[P/x])(N[P/x])
      \] But, by the IH, $M[P/x] \redp M'[Q/x]$ and $N[P/x] \redp
      N'[Q/x]$. So by $(\rightarrow \beta)$ we have \[
        (\lambda x'{:}A.\ M[P/x])(N[P/x])
          \redp
        M'[Q/x]\left[N'[Q/x]/x'\right]
      \] But this last is $\alpha$-equivalent to
      $\left(M'[N'/x']\right)\left[Q/x\right]$ by the substitution
      lemma.

    \case{$\Box\beta$}
      Then $\letbox{u'}{\ibox{M}}{N} \redp N'[M/u']$ where $N
      \redp N'$. By assumption, we have that $x \not\in \fv{M}$
      and $x \not\in \bfv{N}$. Hence, we have by the IH that
      $N[P/x] \redp N'[Q/x]$, so by applying $(\Box\beta)$ we get
      \begin{align*}
	(\letbox{u'}{\ibox{M}}{N})[P/x]
	  \equiv\
	    &\letbox{u'}{\ibox{M[P/x]}}{N[P/x]} \\
	  \equiv\
	    &\letbox{u'}{\ibox{M}}{N[P/x]} \\
	  \redp\
	    &N'[Q/x][M/u']
      \end{align*} But this last is $\alpha$-equivalent to
      $N'[M/u'][Q/x]$, by the substitution lemma and the fact that
      $x$ does not occur in $M$.

    \case{$\Box\textsf{fix}$}
      Then $\fixlob{z}{M} \redp{} M'[\ibox{(\fixlob{z}{M})}/z]$,
      with $M \redp{} M'$. As $x \not\in \bfv{\fixlob{z}{M}}$, we
      have that $x \not\in \fv{M}$, and by Lemma
      \ref{lem:varmon}, $x \not\in \fv{M'}$ either, so \[
        (\fixlob{z}{M})[P/x] \equiv \fixlob{z}{M}
      \] and\[
        M'[\fixlob{z}{M}/z][Q/x]
          \equiv
        M'[Q/x][\fixlob{z}{M[Q/x]}/z]
	  \equiv
	M'[\fixlob{z}{M}/z]
      \] Thus, a single use of $(\Box\textsf{fix})$ suffices.
  \end{indproof}
\end{proof}

We now pull the following definition out of the hat:
\begin{defn}[Complete development]
  The \emph{complete development} $M^\star$ of a term $M$ is
  defined by the following clauses:
    \begin{align*}
      x^\star
        &\myeq x \\
      c^\star
        &\myeq c \qquad (c \in 
	  \{\tilde f, \widehat{n},
	    \textsf{zero?}, \dots \}) \\
      \left(\lambda x{:}A.\ M\right)^\star
        &\myeq \lambda x{:}A.\ M^\star \\
      \left(\tilde{f}(\ibox{M})\right)^\star
	&\myeq \ibox{f(M)} \qquad \text{if $M$ is closed} \\
      \left(\left(\lambda x{:}A.\ M\right)N\right)^\star
        &\myeq M^\star[N^\star/x] \\
      \left(\supset_G\ \textsf{true}\ M\ N\right)^\star
        &\myeq M^\star \\
      \left(\supset_G\ \textsf{false}\ M\ N\right)^\star
        &\myeq N^\star \\
      \left(MN\right)^\star
        &\myeq M^\star N^\star \\
      \left(\ibox{M}\right)^\star
        &\myeq \ibox{M} \\
      \left(\letbox{u}{\ibox{M}}{N}\right)^\star
        &\myeq N^\star[M/u] \\
      \left(\letbox{u}{M}{N}\right)^\star
        &\myeq \letbox{u}{M^\star}{N^\star} \\
      \left(\fixlob{z}{M}\right)^\star
        &\myeq M^\star[\ibox{(\fixlob{z}{M})}/z]
    \end{align*}
\end{defn}

\noindent We need the following two technical results as well.

\begin{lem}
  \label{lem:reflstar}
  $M \redp{} M^\star$
\end{lem}
\begin{proof}
  By induction on the term $M$. Most cases follow immediately by
  (\textsf{refl}), or by the IH and an application of the relevant
  rule. The case for $\ibox{M}$ follows by $(\textsf{refl})$, the
  case for $\fixlob{z}{M}$ follows by $(\Box\textsf{fix})$, and
  the case for $\tilde f(\ibox{M})$ by $(\Box\textsf{int})$.
\end{proof}

\begin{lem}[BFV antimonotonicity]
  \label{lem:varmon}
  If $M \redp{} N$ then $\bfv{N} \subseteq \bfv{M}$.
\end{lem}
\begin{proof}
  By induction on $M \redp{} N$.
\end{proof}

\noindent And here is the main result:

\begin{thm}
  \label{thm:mainredp}
  If $M \redp P$, then $P \redp M^\star$.
\end{thm}

\begin{proof}
  By induction on the generation of $M \redp P$. The case of
  $(\textsf{refl})$ follows by Lemma \ref{lem:reflstar}, and the
  cases of congruence rules follow from the IH. We show the rest.
  \begin{indproof}
    \case{$\rightarrow\beta$}
      Then we have $(\lambda x{:}A.\ M)N \redp M'[N'/x]$, with $M
      \redp M'$ and $N \redp N'$. By the IH, $M' \redp M^\star$
      and $N' \redp N^\star$. We have that $x \not\in \bfv{M}$,
      so by Lemma \ref{lem:varmon} we get that $x \not\in
      \bfv{M'}$. Hence, by Lemma \ref{lem:redp} we get $M'[N'/x]
      \redp M^\star[N^\star/x] \equiv \left(\left(\lambda x{:}A.\
      M\right)N\right)^\star$.

    \case{$\Box\beta$}
      Then we have \[
	\letbox{u}{\ibox{M}}{N} \redp N'[M/u]
      \] where $N \redp N'$. By the IH, $N' \redp N^\star$, so it
      follows that \[
	N'[M/u] \redp N^\star[M/u] \equiv \left(
	\letbox{u}{\ibox{M}}{N}\right)^\star
      \] by Lemma \ref{lem:substint}.

    \case{$\Box\textsf{fix}$}
      Then we have \[
	\fixlob{z}{M}
	  \redp{}
	M'[\ibox{(\fixlob{z}{M})}/z]
      \] where $M \redp{} M'$. By the IH, $M' \redp{} M^\star$.
      Hence \[
	M'[\ibox{(\fixlob{z}{M})}/z]
	  \redp{}
	M^\star[\ibox{(\fixlob{z}{M})}/z]
	  \equiv
	\left(\fixlob{z}{M}\right)^\star
      \] by Lemma \ref{lem:substint}.
    \case{$\Box\textsf{int}$}
      Similar.
  \end{indproof}
\end{proof}

\noindent As a result,

\begin{cor}
  The equational theory of iPCF (Figure \ref{fig:ipcfeq}) is
  consistent.
\end{cor}

\section{Some important terms}
  \label{sec:ipcfterms}

Let us look at the kinds of terms we can write in iPCF.

\begin{description}

\item[From the axioms of \textsf{S4}]

First, we can write a term corresponding to axiom \textsf{K}, the
\emph{normality axiom} of modal logics: \[
  \mathsf{ax_K} \myeq
    \lambda f : \Box (A \rightarrow B). \;
    \lambda x : \Box A . \;
      \letbox{g}{f}{\letbox{y}{x}{\ibox{(g\,y)}}}
\]
Then $\vdash \mathsf{ax_K} : \Box(A \rightarrow B) \rightarrow
(\Box A \rightarrow \Box B)$. An intensional reading of this is
the following: any function given as code can be transformed into
an \emph{effective operation} that maps code of type $A$ to code
of type $B$.

The rest of the axioms correspond to evaluating and quoting.
Axiom \textsf{T} takes code to value, or intension to extension:
\[
  \vdash \mathsf{eval}_A \myeq
    \lambda x : \Box A. \; \letbox{y}{x}{y}
      : \Box A \rightarrow A
\] and axiom \textsf{4} quotes code into code-for-code: \[
  \vdash \mathsf{quote}_A \myeq
    \lambda x : \Box A. \;
      \letbox{y}{x}{\ibox{\left(\ibox{y}\right)}}
    : \Box A \rightarrow \Box \Box A
\] 

\item[Undefined]

The combination of $\textsf{eval}$ and intensional fixed points
leads to non-termination, in a style reminiscent of the term
$(\lambda x.\,xx)(\lambda x.\,xx)$ of the untyped
$\lambda$-calculus. Let \[
  \Omega_A \myeq \fixlob{z}{(\textsf{eval}_A\,z)}
\] Then $\vdash \Omega_A : A$, and \[
  \Omega_A
    \red{}\
  \textsf{eval}_A\;\left(\ibox{\Omega_A}\right)\
    \redt{}\
  \Omega_A
\] 

\item[The G\"odel-L\"ob axiom: intensional fixed points]

Since $(\Box\textsf{fix})$ is L\"ob's rule, we expect to be able
to write down a term corresponding to the G\"odel-L\"ob axiom of
provability logic. We can, and it is an \emph{intensional
fixed-point combinator}: \[
  \mathbb{Y}_A \myeq
    \lambda x : \Box (\Box A \rightarrow A). \;
      \letbox{f}{x}{
          \ibox{\left(\fixlob{z}{f\,z}\right)}
      }
\] and $\vdash \mathbb{Y}_A : \Box(\Box A \rightarrow A)
\rightarrow \Box A$. We observe that \[
  \mathbb{Y}_A(\ibox{M})
    \redt{}
  \ibox{\left(\fixlob{z}{(M\, z)}\right)}
\]

Notice that, in this term, the modal variable $f$ occurs free
under a $\fixlob{z}{(-)}$ construct. This will prove important in
\S\ref{chap:intsem2}, where this occurrence will be prohibited.

\item[Extensional Fixed Points]

Perhaps surprisingly, the ordinary PCF $\mathbf{Y}$ combinator is
also definable in the iPCF.  Let \[
  \mathbf{Y}_A \myeq
    \fixlob{z}{
      \lambda f : A \rightarrow A.\
        f (\textsf{eval}\ z\  f)
    }
\] Then $\vdash \mathbf{Y}_A : (A \rightarrow A) \rightarrow A$, so that
\begin{align*}
  \mathbf{Y}_A
    \redt{}\
  &\lambda f : A \rightarrow A.\ f (\textsf{eval}\ (\ibox{\mathbf{Y}_A})\ f)) \\
    \redt{}\
  &\lambda f : A \rightarrow A.\ f (\mathbf{Y}_A\, f)
\end{align*}

Notice that, in this term, the modal variable $z$ occurs
free under a $\lambda$-abstraction. This will prove important in
\S\ref{chap:intsem2}, where this occurrence will be prohibited.

\end{description}

\section{Two intensional examples}
  \label{sec:ipcfexamples}

No discussion of an intensional language with intensional
recursion would be complete without examples that use these two
novel features. Our first example uses intensionality, albeit in a
functional way, and is drawn from the study of PCF and issues
related to sequential vs. parallel (but not concurrent)
computation. Our second example uses intensional recursion, so it
is slightly more adventurous: it is a computer virus.

\subsection{`Parallel or' by dovetailing}

In \citep{Plotkin1977} Gordon Plotkin proved the following
theorem: there is no term $\textsf{por} : \textsf{Bool}
\rightarrow \textsf{Bool} \rightarrow \textsf{Bool}$ of PCF such
that $\textsf{por}\ \textsf{true}\ M \twoheadrightarrow_\beta
\textsf{true}$ and  $\textsf{por}\ M\ \textsf{true}
\twoheadrightarrow_\beta \textsf{true}$ for any $\vdash M :
\textsf{Bool}$, whilst $\textsf{por}\ \textsf{false}\
\textsf{false} \twoheadrightarrow_\beta \textsf{false}$.
Intuitively, the problem is that $\textsf{por}$ has to first
examine one of its two arguments, and this can be troublesome if
that argument is non-terminating. It follows that the
\emph{parallel or} function is not definable in PCF. In order to
regain the property of so-called \emph{full abstraction} for the
\emph{Scott model} of PCF, a constant denoting this function has
to be manually added to PCF, and endowed with the above rather
clunky operational semantics. See \citep{Plotkin1977, Gunter1992,
Mitchell1996, Streicher2006}.

However, the parallel or function is a computable \emph{partial
recursive functional} \citep{Streicher2006, Longley2015}. The way
to prove that is intuitively the following: given two closed terms
$M, N : \textsf{Bool}$, take turns in $\beta$-reducing each one
for a one step: this is called \emph{dovetailing}. If at any point
one of the two terms reduces to \textsf{true}, then output
\textsf{true}. But if at any point both reduce to \textsf{false},
then output \textsf{false}.

This procedure is not definable in PCF because a candidate term
$\textsf{por}$ does not have access to a code for its argument,
but can only inspect its value. However, in iPCF we can use the
modality to obtain access to code, and intensional operations to
implement reduction. Suppose we pick a reduction strategy
$\red{}_r$. Then, let us include a constant $\textsf{tick} :
\Box\textsf{Bool} \rightarrow \Box\textsf{Bool}$ that implements
one step of this reduction strategy on closed terms: \[
  \begin{prooftree}
    M \red_r N, \text{ $M, N$ closed}
      \justifies
    \textsf{tick}\ (\ibox{M}) \red{} \ibox{N}
  \end{prooftree}
\] Also, let us include a constant $\textsf{done?} :
\Box\textsf{Bool} \rightarrow \textsf{Bool}$, which tells us if a
closed term under a box is a normal form: \[
  \begin{prooftree}
    \text{$M$ closed, normal}
      \justifies
    \textsf{done?}\ (\ibox{M}) \red{} \textsf{true}
  \end{prooftree}
    \qquad
  \begin{prooftree}
    \text{$M$ closed, not normal}
      \justifies
    \textsf{done?}\ (\ibox{M}) \red{} \textsf{false}
  \end{prooftree}
\] It is not hard to see that these two intensional operations can
be subsumed under our previous scheme for introducing intensional
operations: our proof still applies, yielding a consistent
equational system.

The above argument is now implemented by the following term:
\begin{align*}
  \textsf{por} :\equiv\ 
    \mathbf{Y}(
      \lambda &\textsf{por}. \; 
      \lambda x : \Box \textsf{Bool}. \; \lambda y : \Box \textsf{Bool}. \\
    &\supset_\textsf{Bool}
      (\textsf{done?} \; x)\
      &&(\textsf{lor} \; (\textsf{eval} \; x) (\textsf{eval} \; y)) \\
    & &&(\supset_\textsf{Bool} (\textsf{done?} \; y)\
      &&(\textsf{ror} \; (\textsf{eval} \; x) (\textsf{eval} \; y)) \\
    & && &&(\textsf{por} \; (\textsf{tick} \; x)
                            (\textsf{tick} \; y)))
\end{align*} where $\textsf{lor}, \textsf{ror} : \textsf{Bool}
\rightarrow \textsf{Bool} \rightarrow \textsf{Bool}$ are terms
defining the left-strict and right-strict versions of the `or'
connective respectively. Notice that the type of this term is
$\Box\textsf{Bool} \rightarrow \Box\textsf{Bool} \rightarrow
\textsf{Bool}$: we require \emph{intensional access} to the terms
of boolean type in order to define this function!

\subsection{A computer virus}

\emph{Abstract computer virology} is the study of formalisms that
model computer viruses. There are many ways to formalise viruses.
We will use the model of \opt{ipcf}{Adleman} \cite{Adleman1990},
where files can be interpreted either as data, or as functions. We
introduce a data type $F$ of files, and two constants \[
  \textsf{in} : \Box (F \rightarrow F) \rightarrow F
    \quad \text{ and} \quad
  \textsf{out} : F \rightarrow \Box (F \rightarrow F)
\] If $F$ is a file, then $\textsf{out}\ F$ is that file
interpreted as a program, and similarly for $\textsf{in}$. We
ask that $\textsf{out}\ (\textsf{in}\ M) \red{} M$,
making $\Box (F \rightarrow F)$ a retract of
$F$.\opt{th}{\footnote{Actually, in \S\ref{sec:buildipwps} and
\S\ref{sec:asmipcf} we will see it is very easy to construct
examples for the apparently more natural situation where
$\textsf{in} : \Box(\Box F \rightarrow F) \rightarrow F$,
$\textsf{out}: F \rightarrow (\Box F \rightarrow F)$, and
$\textsf{out}\ (\textsf{in}\ M) \red{}\ \textsf{eval}_{F
\rightarrow F}\ M $. Nevertheless, our setup is slightly more
well-adapted to virology.}} This might seem the same as the
situation where $F \rightarrow F$ is a retract of $F$, which
yields models of the (untyped) $\lambda$-calculus, and is not
trivial to construct \cite[\S 5.4]{Barendregt1984}. However, in
our case it is not nearly as worrying: $\Box(F \rightarrow F)$ is
populated by programs and codes, not by actual functions. Under
this interpretation, the pair $(\textsf{in}, \textsf{out})$
corresponds to a kind of G\"odel numbering---especially if $F$ is
$\mathbb{N}$.

Now, in Adleman's model, a \emph{virus} is a given by its infected
form, which either \emph{injures}, \emph{infects}, or
\emph{imitates} other programs. The details are unimportant in the
present discussion, save from the fact that the virus needs to
have access to code that it can use to infect other executables.
One can hence construct such a virus from its \emph{infection
routine}, by using Kleene's SRT. Let us model it by a term \[
  \vdash \textsf{infect} : \Box (F \rightarrow F) 
    \rightarrow F \rightarrow F
\] which accepts a piece of viral code and an executable file, and
it returns either the file itself, or a version infected with the
viral code. We can then define a term \[
  \vdash \textsf{virus} 
    \myeq \fixlob{z}{(\textsf{infect}\ z)} :
      F \rightarrow F
\] so that \[
  \textsf{virus}\ 
    \redt{}\ 
  \textsf{infect}\ (\ibox{\textsf{virus}})
\] which is a program that is ready to infect its input with its
own code.

\section{Open Questions}
  \label{sec:ipcfq}

We have achieved the desideratum of an intensional programming
language, with intensional recursion. There are two main questions
that result from this development. 

Firstly, does there exist a good set of \emph{intensional
primitives} from which all others are definable? Is there perhaps
\emph{more than one such set}, hence providing us with a choice of
programming primitives?

Secondly, what is the exact kind of programming power that we have
unleashed?  Does it lead to interesting programs that we have not
been able to write before? We have outlined some speculative
applications for intensional recursion in
\opt{th}{\S\ref{sec:appirec}}\opt{ipcf}{\cite{Kavvos2016a}}. Is
iPCF a useful tool when it comes to attacking these?

We discuss some more aspects of iPCF in the concluding chapter
(\S\ref{sec:expripcf}).

\chapter[Categories and Intensionality]
  {Categories and Intensionality\footnote{A preliminary form of
  the results in this chapter was first published as
  \citep{Kavvos2017a}, which is available at Springer:
  \url{https://dx.doi.org/10.1007/978-3-662-54458-7_32}}}
  \label{chap:expo}

  We turn now to the discussion of the categorical modelling of
intensionality.

We have discussed the importance of intensionality for Computer
Science in \S\ref{sec:intensionality}. One might therefore ask why
the concept has not led to many exciting developments.
\cite{Abramsky2014} suggests that it may simply be that the
extensional paradigm is already sufficiently challenging:
\begin{quote}
  `` [...] while Computer Science embraces wider notions of
  processes than computability theory, it has tended to refrain
  from studying intensional computation, despite its apparent
  expressive potential. This reluctance is probably linked to the
  fact that it has proved difficult enough to achieve software
  reliability even while remaining within the confines of the
  extensional paradigm.  Nevertheless, it seems reasonable to
  suppose that understanding and harnessing intensional methods
  offers a challenge for computer science which it must eventually
  address.''
\end{quote} But we believe that there is also a deeper reason:
once we step outside extensionality, there are no rules: one might
even say that `anything goes:' this is what Abramsky means by the
phrase ``loose baggy monster'' (quoted at the start of
\S\ref{sec:objectives}). A natural reaction to this state of
affairs is to turn to category theory in an attempt to find some
structure that can put things into perspective, or simply provide
a language for studying the interplay between the extensional and
the intensional.

Surprisingly, very little has been said about intensionality in
the categorical domain. \cite{Lawvere1969, Lawvere2006} observes
that there are categories which are \emph{not well-pointed},
hence---in some sense---`intensional.' But if we only allow for
slightly more generality, this intensionality vanishes: there is
only one notion of equality.

We shall take the hint regarding the relationship between modal
logic and intensionality from our discussion in
\S\ref{sec:modasint}. We already know that there is a deep
connection between logic, type theory, and category theory, namely
the \emph{Curry-Howard-Lambek correspondence}. This makes it
likely that attempting to transport the reading of
modality-as-intension to the realm of category theory could lay
the foundation for a basic theory of intensionality. We hence
revisit the appropriate categorical semantics of type theories in
the spirit of \textsf{S4}, which was introduced by
\cite{Bierman2000a}. We find that it is not appropriate for our
purposes, and we argue to that effect in \S\ref{sec:catint}.

Taking all of these points into account, one can only surmise that
there has to be a radical shift in our perspective: \emph{we need
to step outside classical 1-category theory}. Fortunately, this
necessary groundwork has been laid by \cite{Cubric1998} and their
discovery of \emph{P-category theory}. We introduce P-category
theory in \S\ref{sec:pcats}, and discuss its use in modelling
intensionality.

All that remains is to introduce a new concept that ties
intensionality and extensionality together under the same roof.
This is the notion of an \emph{exposure}, which we introduce and
study in \S\ref{sec:expo}. 

\section{Categories are not intensional}
  \label{sec:catint}

At the outset, things look promising: let there be a category
$\mathcal{C}$ with a terminal object $\mathbf{1}$. Arrows of type
$x : \mathbf{1} \rightarrow C$ are called \emph{points of the
object $C$}. An arrow $f : C \rightarrow A$ introduces a map \[
  x : \mathbf{1} \rightarrow C
    \quad\longmapsto\quad
  f \circ x : \mathbf{1} \rightarrow A
\] from points of $C$ to points of $A$. In this setting,
\cite{Lawvere1969, Lawvere2006} observes that we may have two
distinct arrows $f, g : C \rightarrow A$ that induce the same map,
i.e. such that \[
  \forall x : \mathbf{1} \rightarrow A.\
    f \circ x = g \circ x
\] all whilst $f \neq g$. In this case, we say that the category
$\mathcal{C}$ \emph{does not have enough points}, or \emph{is not
well-pointed}.

Nevertheless, lack of enough points is not enough to have
`intensionality,' and the discussion in \citep[\S 2.3]{Awodey2010}
provides the necessary intuition. Up to now, we have focused on
points $x : \mathbf{1} \rightarrow C$. These can be construed as
\emph{`tests'}, in the sense that we can look at each $f \circ x$
and infer some information about the arrow $f : C \rightarrow A$,
e.g. its value at some `argument' $x : \mathbf{1} \rightarrow C$.
However, we can conceive of more involved test objects $T$, and
significantly more comprehensive `tests' of type $x : T
\rightarrow C$ which we can call \emph{generalised points}.
Arrows $f : C \rightarrow A$ are completely distinguishable if
given arbitrary generalised points.  The argument is
dumbfoundingly trivial: the most thorough `test object' is $C$
itself, and for that one there's the generalised point $id_C : C
\rightarrow C$ with the unfortunate property that $f \circ id_C =
f \neq g = g \circ id_C$.

We can therefore draw the conclusion that \emph{categories cannot
model intensionality}: there cannot be two distinct arrows $f$ and
$g$ that are indistinguishable within the category. Of course, we
can always quotient $\mathcal{C}$ by some compatible equivalence
relation $\sim$ to obtain $\mathcal{C}\mathbin{/}\sim$. Then the
intensional universe would be $\mathcal{C}$, and its extensional
version would be $\mathcal{C}\mathbin{/}\sim$. But that would
entail that we are no longer operating within a single
mathematical universe, i.e. a single category!

\subsection{Intension, Modality, and Categories}
  \label{sec:intmodcat}

We thus return to the \emph{modality-as-intention} interpretation
of \S\ref{sec:modasint} to look for the answer. From a
categorical perspective, all is well with the intuitions we have
developed there and in \S\ref{chap:ipcf}, save the punchline: the
categorical semantics of the \textsf{S4} box modality, due to
\cite{Bierman2000a} and \cite{Kobayashi1997}, specifies that \[
  \Box : \mathcal{C} \longrightarrow \mathcal{C}
\] is part of a \emph{monoidal comonad} $(\Box, \epsilon, \delta)$
on a cartesian closed category  $\mathcal{C}$. Let us define some
of these notions for the sake of completeness.

\begin{defn}
  A \emph{comonad} $(Q, \epsilon, \delta)$ consists of an
  endofunctor \[
    Q: \mathcal{C} \longrightarrow \mathcal{C}
  \] and two natural transformations, \[
    \epsilon : Q \Rightarrow \mathsf{Id}_\mathcal{C},
      \quad
    \delta : Q \Rightarrow Q^2
  \] such that the following diagrams commute: \[
    \begin{tikzcd}
      QA
        \arrow[r, "\delta_A"]
        \arrow[d, "\delta_A", swap]
      & Q^2 A 
        \arrow[d, "\delta_{QA}"] \\
      Q^2 A 
        \arrow[r, "Q\delta_A", swap] 
      & Q^3 A
    \end{tikzcd} \quad
    \begin{tikzcd}
      QA
        \arrow[r, "\delta_A"]
        \arrow[d, "\delta_A", swap]
        \arrow[rd, "id_A", swap]
      & Q^2 A 
        \arrow[d, "\epsilon_{QA}"] \\
      Q^2 A
        \arrow[r, "Q\epsilon_A", swap]
      & QA
    \end{tikzcd}
  \]
\end{defn}

\noindent In the cartesian case, \emph{monoidality} requires the
provision of  morphisms \begin{align*}
  m_{A, B} &: QA \times QB \rightarrow Q(A \times B) \\
  m_0      &: \mathbf{1} \rightarrow Q\mathbf{1}
\end{align*} natural in each pair of objects $A, B \in
\mathcal{C}$, which must also make certain diagrams commute. We
will be particularly interested in the case where the functor is
\emph{strong monoidal}, i.e. $m_{A, B}$ and $m_0$ are natural
\emph{isomorphisms}. It has been shown in the technical report
\citep{Kavvos2017c} that this is the same as a
\emph{product-preserving functor}.

The transformations $\delta : Q \Rightarrow Q^2$ and $\epsilon : Q
\Rightarrow \mathsf{Id}_\mathcal{C}$ are \emph{monoidal} if the
following diagrams commute: \[
    \begin{tikzcd}
        QA \times QB
        \arrow[r, "\epsilon_A \times \epsilon_B",]
        \arrow[d, "m_{A, B}", swap]
      & A \times B
        \arrow[d, equal] \\
      Q(A \times B)
        \arrow[r, "\epsilon_{A \times B}", swap]
      & A \times B
    \end{tikzcd} \quad \begin{tikzcd}
      \mathbf{1}
        \arrow[rd, equal]
        \arrow[d, "m_0", swap]
      & \\
      Q\mathbf{1}
        \arrow[r, "\epsilon_\mathbf{1}", swap]
      & \mathbf{1}
    \end{tikzcd}
  \] \[
    \begin{tikzcd}
        QA \times QB
        \arrow[r, "\delta_A \times \delta_B",]
        \arrow[dd, "m_{A, B}", swap]
      & Q^2 A \times Q^2 B
        \arrow[d, "m_{QA, QB}"] \\
      & Q(QA \times QB) 
        \arrow[d, "Qm_{A, B}"] \\
      Q(A \times B)
        \arrow[r, "\delta_{A \times B}", swap]
      & Q^2(A \times B)
    \end{tikzcd} \quad \begin{tikzcd}
      \mathbf{1}
        \arrow[rd, "m_0"]
        \arrow[dd, "m_0", swap]
      & & \\
      & Q\mathbf{1}
        \arrow[rd, "Q(m_0)"] 
      & \\
      Q\mathbf{1}
        \arrow[rr, "\delta_\mathbf{1}", swap]
      & & Q^2\mathbf{1}
    \end{tikzcd}
\]

\noindent Please refer to \cite[\S XI.2]{MacLane1978} or \cite[\S
5]{Mellies2009} for more details, and the missing commuting
diagrams.

Now, as $\Box : \mathcal{C} \longrightarrow \mathcal{C}$ is a
functor, it will unfortunately trivially satisfy \[
  f = g
    \quad\Longrightarrow\quad
  \Box f = \Box g
\] and will hence necessarily validate the congruence rule for the
modality: \[
  \begin{prooftree}
    \ctxt{\Delta}{\cdot} \vdash M = N : A
      \justifies
    \ctxt{\Delta}{\Gamma} \vdash \ibox{M} = \ibox{N} : \Box A
      \using
    {(\Box\textsf{cong})}
  \end{prooftree}
\] This does not conform to the `no reductions under $\ibox{(-)}$'
restriction, and hence disrupts the intensional nature of the
modal types in iPCF.

As if this were not enough, we will now present another argument
that provides the last straw for the monoidal comonad
interpretation. Intuitively, if $\Box A$ is to represent code of
type $A$, then there should be many more points $\mathbf{1}
\rightarrow \Box A$ than points $\mathbf{1} \rightarrow A$: there
is more than one expression corresponding to the same value in any
interesting logical system. Under a very mild assumption, this
desideratum fails. To show that, suppose we have a monoidal
comonad $(\Box, \epsilon, \delta)$. Furthermore, suppose the
components of $\delta : Q \Rightarrow Q^2$ satisfy the following
definition:
\begin{defn}
  The component $\delta_A : \Box A \rightarrow \Box^2 A$ is
  \emph{reasonable quoting device at $A$} just if the following
  diagram commutes: \[
    \begin{tikzcd}
      \mathbf{1}
	\arrow[r, "a"]
	\arrow[d, "m_0", swap]
      & \Box A 
	\arrow[d, "\delta_A"] \\
      \Box \mathbf{1}
	\arrow[r, "\Box a", swap]
      & \Box^2 A
  \end{tikzcd}
  \] 
\end{defn}

\noindent This equation may be type-theoretically expressed in
iPCF as the following equation for any $\vdash M : \Box A$: \[
  \vdash \letbox{u}{M}{\ibox{(\ibox{u})}} = \ibox{M} : \Box\Box A
\]

\noindent Then,

\begin{prop}
  \label{prop:degeneracy}
  If each component $\delta_A : \Box A \rightarrow \Box^2 A$ of a
  monoidal comonad $(\Box, \epsilon, \delta)$ is a reasonable
  quoting device, then there is a natural isomorphism \[
    \mathcal{C}(\mathbf{1}, -)
      \cong \mathcal{C}(\mathbf{1}, \Box(-))
  \]
\end{prop}

\begin{proof}
  We can define maps \begin{align*}
  \textsf{in} :
    \mathcal{C}(\mathbf{1}, A) &\rightarrow \mathcal{C}(\mathbf{1}, \Box A) \\
    x                          &\mapsto     \Box x \circ m_0 \\
  \textsf{out} :
    \mathcal{C}(\mathbf{1}, \Box A) &\rightarrow \mathcal{C}(\mathbf{1}, A) \\
    a                               &\mapsto     \epsilon_A \circ a
\end{align*} and then calculate: \[
    \textsf{out}\left(\textsf{in}(x)\right)
    = \epsilon_A \circ \Box x \circ m_0
    = x \circ \epsilon_\mathbf{1} \circ m_0
    = x
  \] where the last equality is because $\mathbf{1}$ is a terminal
  object. Similarly, \[
    \textsf{in}\left(\textsf{out}(a)\right)
      = \Box(\epsilon_A \circ a) \circ m_0
      = \Box\epsilon_A \circ \Box a \circ m_0
      = \Box\epsilon_A \circ \delta_A \circ a
      = a
  \] where we have only used our `reasonable' condition,
  and one of the comonad laws. Naturality of the isomorphism
  follows by functoriality of $\Box$ and naturality of $\epsilon$.
\end{proof}

\noindent Hence, in these circumstances, there are no more codes
than values!

This definition of a reasonable quoting device is slightly
mysterious. In fact, it follows from the commutation of the
following more general diagram, for each $f : QA \rightarrow QB$:
\[
  \begin{tikzcd}
    QA
      \arrow[r, "f"]
      \arrow[d, "\delta_A", swap]
    & QB
      \arrow[d, "\delta_B"] \\
    Q^2 A
      \arrow[r, "Qf", swap]
    & Q^2 B
  \end{tikzcd}
\] Indeed, given $a : \mathbf{1} \rightarrow QA$, consider $a
\circ m_0^{-1} : Q\mathbf{1} \rightarrow QA$. Commutation of the
diagram them yields \[
  \delta_A \circ a \circ m_0^{-1}
    = 
  Q(a \circ m_0^{-1}) \circ \delta_\mathbf{1}
\] Taking the $m_0$ to the other side, we have \[
  \delta_A \circ a
    =
  Qa \circ Qm_0^{-1} \circ \delta_\mathbf{1} \circ m_0
    =
  Qa \circ Qm_0^{-1} \circ Qm_0 \circ m_0
    =
  Qa \circ m_0
\] by the monoidality of $\delta$. In turn, commutation of this
diagram for any $f : A \rightarrow B$ corresponds to the comonad
being \emph{idempotent}.

\begin{thm}[Idempotence]
  \label{thm:idem}
  Given a comonad $(Q, \epsilon, \delta)$, the following are equivalent:
  \begin{enumerate}
    \item
      $\delta : Q \Rightarrow Q^2$ is an isomorphism.
    \item
      $\delta \circ \epsilon_Q : Q^2 \Rightarrow Q^2$ is the identity
      natural transformation.
    \item
      For all $f : QA \rightarrow QB$, $Qf \circ \delta_A =
      \delta_B \circ f$.
  \end{enumerate} If any one of these holds, we say that $(Q,
  \epsilon, \delta)$ is \emph{idempotent}.
\end{thm}
\begin{proof} 
  We prove $(2) \Rightarrow (3) \Rightarrow (1) \Rightarrow (2)$.
  \begin{indproof}
    \case{$2 \Rightarrow 3$}
      We have \begin{derivation}
	  Qf \circ \delta_A
	\since{by (2)}
	  \delta_B \circ \epsilon_{QB} \circ Qf \circ \delta_A
	\since{naturality of $\epsilon$}
	  \delta_B \circ f \circ \epsilon_{QA} \circ \delta_A
	\since{comonad equation}
	  \delta_B \circ f
      \end{derivation}
    \case{$3 \Rightarrow 1$}
      We already know that $\epsilon_Q \circ \delta = \mathsf{Id}$
      from the comonad equations, so it remains to prove that
      $\delta \circ \epsilon_Q$ is the identity natural
      transformation on $Q$.
        \begin{derivation}
            \delta_A \circ \epsilon_{QA}
          \since{by (3)}
            Q\epsilon_{QA} \circ \delta_{QA}
          \since{comonad equation}
            id_{QA}
        \end{derivation}
            Hence $\delta^{-1} = \epsilon_Q$.
      \case{$1 \Rightarrow 2$}
        We have
          \begin{derivation}
              \delta_A \circ \epsilon_{QA}
            \since{by (1)}
              \delta_A \circ \epsilon_{QA} \circ \delta_A \circ \delta_A^{-1}
            \since{comonad equation}
              \delta_A \circ \delta_A^{-1}
            \since{$\delta$ iso}
              id_{Q^2 A}
          \end{derivation}
  \end{indproof}
\end{proof}

\noindent The behaviour of code will often be \emph{idempotent} in
the above sense: once something is quoted code, it is in the realm
of syntax, and more layers of quoting do not change its quality as
sense. Thus, the above argument delivers a fatal blow to the
monoidal comonad approach.

\subsection{PERs and P-categories}
  \label{sec:pcats}

The way out of this seeming impasse is to step outside 1-category
theory. We shall use the notion of \emph{P-category}, which was
introduced by \cite{Cubric1998} precisely so the authors could
deal with a form of intensionality.\footnote{For the record, the
gist of \citep{Cubric1998} is that the Yoneda embedding on the
categorical term model of typed $\lambda$-calculus is a key
ingredient in \emph{normalisation by evaluation}---with the
proviso that terms are not strictly identified up to $\beta\eta$
equality!}

\emph{P-category theory} is essentially category theory \emph{up
to a partial equivalence relation}.

\begin{defn}
  A \emph{partial equivalence relation (PER)} is a symmetric and
  transitive relation.
\end{defn}

Partial equivalence relations were introduced to Theoretical
Computer Science by Turing in an unpublished manuscript, and
brought to the study of semantics by \cite{Girard1972} and
\cite{Scott1975}. Broadly speaking, we can view an equivalence
relation on a set as a \emph{notion of equality} for that set.
However, partial equivalence relations might not be reflexive.
Elements that are not related to themselves can be understood as
\emph{not being well-defined}. We can, for example, define a PER
$\sim$ between sequences of rationals as follows: $\{x_i\} \sim
\{y_i\}$ just if both sequences are Cauchy and converge to the
same real number. Sequences that are \emph{not Cauchy} cannot be
real numbers.

\begin{defn}
  A \emph{P-set} is a pair $A = (\bars{A}, \sim_A)$ consisting of
  a set $\bars{A}$ and a PER $\sim_A$ on $A$.
\end{defn}

\noindent We will formally distinguish between \emph{elements} $x
\in \bars{A}$ of the P-set $A$, and \emph{points} $x \in A$ of
$A$: for the latter we will also require that $x \sim_A x$, i.e.
that they be well-defined. Given a P-set $A$, its \emph{domain}
$\dom(A)$ is the set of its points.

The notion of \emph{operation} will be instrumental in the
development of our theory. An operation is a transformation
between the elements of P-sets that is not functional, in that it
need not respect the PERs on P-sets.

\begin{defn}[Operation]
  Given two P-sets $A = (\bars{A}, \sim_A)$ and $B = (\bars{B},
  \sim_B)$, an \emph{operation}, written \[
    f : A \dashrightarrow B
  \] is a function $f : \bars{A} \rightarrow \bars{B}$ such that
  $x \sim_A x$ implies $f(x) \sim_B f(x)$.
\end{defn}

\noindent I.e. an operation takes elements to elements, but when
given a point (well-defined element) also returns a point. Some
operations are more well-behaved:

\begin{defn}
  An operation $f : A \dashrightarrow B$ is a \emph{P-function},
  written \[
    f : A \rightarrow B
  \] just if $x \sim_A y$ implies $f(x) \sim_B f(y)$.
\end{defn}

\noindent We will later see that P-sets and P-functions form a
cartesian closed P-category (Theorem \ref{thm:psetsccc}). The main
ingredients needed for that theorem are the subject of the first
three of the following examples.

\begin{example} \hfill
  \begin{enumerate}
    \item

      The P-set $\mathbf{1}$ is defined to be $(\{\ast\}, \{(\ast,
      \ast)\})$.

    \item

      Given P-sets $A = (\bars{A}, \sim_A)$ and $B = (\bars{B},
      \sim_B)$, we define the P-set $A \times B$ by $\bars{A
      \times B} \myeq \bars{A} \times \bars{B}$ and $(a, b)
      \sim_{A \times B} (a', b')$ just if $a \sim_A a'$ and $b
      \sim_B b'$.

    \item

      Given P-sets $A = (\bars{A}, \sim_A)$ and $B = (\bars{B},
      \sim_B)$, the P-functions $f : A \rightarrow B$ form a P-set
      $B^A = (\bars{B^A}, \sim_{B^A})$ as follows: $\bars{B^A}
      \myeq \setcomp{ f }{ f : A \dashrightarrow B }$, and $f
      \sim_{B^A} g$ just if $a \sim_A a'$ implies $f(a) \sim_B
      g(a')$.

    \item

      Given P-sets $A = (\bars{A}, \sim_A)$ and $B = (\bars{B},
      \sim_B)$, the P-operations $f : A \dashrightarrow B$ form a
      P-set $A \Yright B = (\bars{B^A}, \sim_{A \Yright B})$ with
      the same underlying set $\bars{B^A}$, but with $f \sim_{A
      \Yright B} g$ just if $f = g$.
  \end{enumerate}
\end{example}

In the definition of the P-set of P-functions, it is evident that
all operations $f : A \oper{} B$ are present as elements in
$\bars{B^A}$. However, they are only in $\dom(B^A)$ if they are
P-functions. This pattern of `junk' being present amongst elements
is characteristic when it comes to PERs, and it is the reason we
need the notion of points.  A \emph{P-point} of the P-set $A$
would be a P-function $x : \mathbf{1} \rightarrow A$, which is
determined by the point $x(*) \in \dom(A)$. We will systematically
confuse a P-point $x : \mathbf{1} \rightarrow A$ with the point
$x(\ast) \in \dom(A)$. % If $x \not\sim_A x$, we may variably call
% it `the code $x$,' `the element $x$,' or `the construction $x$.'

We can finally define what it means to be a 

\begin{defn}[P-category]
  A \emph{P-category} $(\mathfrak{C}, \sim)$ consists of:
  \begin{itemize}
    \item 
      a set of objects $ob(\mathfrak{C})$;
    \item 
      for any two objects $A, B \in ob(\mathfrak{C})$, a P-set
      $\mathfrak{C}(A, B) = \left(\bars{\mathfrak{C}(A, B)},
      \sim_{\mathfrak{C}(A, B)}\right)$;
    \item
      for each object $A \in ob(\mathfrak{C})$, a point $id_A \in
      \mathfrak{C}(A, A)$;
    \item
      for any three objects $A, B, C \in ob(\mathfrak{C})$, a
      P-function \[
        c_{A, B, C} :
          \mathfrak{C}(A, B) \times \mathfrak{C}(B, C)
            \rightarrow
          \mathfrak{C}(A, C)
      \] for which we write $g \circ f \myeq c_{A, B, C}(f, g)$,
  \end{itemize}
  such that, for any point $f \in \mathfrak{C}(A, B)$ we have
  \begin{align*}
    f \circ id_A &\sim_{\mathfrak{C}(A, B)} f \\
    id_B \circ f &\sim_{\mathfrak{C}(A, B)} f
  \end{align*}
  and for any $f \in \mathfrak{C}(A, B)$, $g \in \mathfrak{C}(B,
  C)$ and $h \in \mathfrak{C}(C, D)$, we have \begin{align*}
    h \circ (g \circ f)
      \sim_{\mathfrak{C}(A, D)}
	(h \circ g) \circ f
  \end{align*}
\end{defn}

In a P-category we have an ordinary set of objects, but only a
P-set of morphisms. We will say that $f$ is \emph{an arrow of
$\mathfrak{C}$ with domain $A$ and codomain $B$}, and write $f : A
\rightarrow B$, only whenever $f$ is a well-defined morphism, i.e.
$f \in \dom(\mathfrak{C}(A, B))$.

Furthermore, we will variously refer to P-categories by Fraktur
letters $\mathfrak{B}, \mathfrak{C}, \dots$, without mentioning
the family of relations $\{\sim_{\mathfrak{C}(A, B)}\}_{A, B \in
\mathfrak{C}}$. Sometimes we might use the same sets of morphisms
$\bars{\mathfrak{C}(A, B)}$, but with a different PER; we will
then indicate which PER we are using, by writing e.g.
$(\mathfrak{C}, \sim)$ or $(\mathfrak{C}, \doteq)$.  When the
types of two morphisms $f$ and $g$ are evident, we will write $f
\sim g$ without further ado. 

We can think of the morphisms as intensional, and of the PER on
them as describing extensional equality. That composition is a
P-function encodes the requirement that \emph{composition respects
extensional equality}: if $f \sim f'$ and $g \sim g'$, then $g
\circ f \sim g' \circ f'$.

As is expected, P-categories come with associated notions of
functor and natural transformation.

\begin{defn}[P-functor]
  Let $\mathfrak{C}, \mathfrak{D}$ be P-categories. A
  \emph{P-functor} $F : \mathfrak{C} \longrightarrow \mathfrak{D}$
  from $\mathfrak{C}$ to $\mathfrak{D}$ consists of a map assigning
  an object $FX \in \mathfrak{D}$ for each object $X \in
  \mathfrak{C}$, and a family of P-functions, \[
    F_{A, B} : \mathfrak{C}(A, B) \rightarrow \mathfrak{D}(FA, FB)
  \] for each pair of objects $A, B \in \mathfrak{C}$, such that
  \begin{align*}
    F(id_A) &\sim id_{FA} \\
    F(g \circ f) &\sim Fg \circ Ff
  \end{align*} for all pairs of arrows $f : A \rightarrow B$ and
  $g : B \rightarrow C$.
\end{defn}

\begin{defn}[P-natural transformation]
  Let $F, G : \mathfrak{C} \longrightarrow \mathfrak{D}$ be
  P-functors. A \emph{P-natural transformation} $\theta : F
  \Rightarrow G$ consists of an arrow $\theta_A : FA \rightarrow
  GA$ in $\mathfrak{D}$ for each $A \in \mathfrak{C}$, such that
  for each $f : A \rightarrow B$ we have \[
    \theta_B \circ Ff \sim Gf \circ \theta_A
  \]
\end{defn}

Finally, the definitions of finite products and exponentials carry
over smoothly. The various components of the definitions are
required to be unique with respect to their universal property,
but only up to the PERs.

\begin{defn}[Terminal object]
  Let $\mathfrak{C}$ be a P-category. An object $\mathbf{1} \in
  \mathfrak{C}$ is \emph{terminal} just if for every $C \in
  \mathfrak{C}$ there is an arrow \[
    {!}_C : C \rightarrow \mathbf{1}
  \] such that ${!}_C \sim {!}_C$, and for every arrow $h : C
  \rightarrow \mathbf{1}$ we have $h \sim {!}_C$.
\end{defn}

\begin{defn}[Binary Product]
  Let $\mathfrak{C}$ be a P-category, and let $A, B \in
  \mathfrak{C}$. The object $A \times B \in \mathfrak{C}$ is a
  \emph{product} of $A$ and $B$ if there are arrows \[
    A \xleftarrow{\pi_1} A \times B \xrightarrow{\pi_2} B
  \] and for any object $C \in \mathfrak{C}$, there is a
  P-function \[
    \langle \cdot, \cdot \rangle :
      \mathfrak{C}(C, A) \times \mathfrak{C}(C, B)
        \rightarrow
      \mathfrak{C}(C, A \times B)
  \] such that for every $f : C \rightarrow A$ and $g : C
  \rightarrow B$ we have \begin{align*}
    \pi_1 \circ \langle f, g \rangle &\sim f \\
    \pi_2 \circ \langle f, g \rangle &\sim g
  \end{align*} and, for any $h : C \rightarrow A \times B$, we
  have $\langle \pi_1 \circ h, \pi_2 \circ h \rangle \sim h$.
\end{defn}

\noindent Of course, the usual calculational rules of products
still hold, e.g. \begin{align*}
  \langle f, g \rangle \circ h
    &\sim
  \langle f \circ h, g \circ h\rangle \\
  (f \times g) \circ \langle h, k \rangle
    &\sim
  \langle f \circ h, g \circ k \rangle
\end{align*} and so on.

\begin{defn}
  A \emph{cartesian P-category} is a P-category that has a
  terminal object and binary products.
\end{defn}

\begin{defn}[Exponential]
  Let $\mathfrak{C}$ be a cartesian P-category, and let $A, B \in
  \mathfrak{C}$. The object $B^A \in \mathfrak{C}$ is an
  \emph{exponential} of $A$ and $B$ just if there is an arrow \[
    \textsf{ev}_{A, B} : B^A \times A \rightarrow B
  \] such that, for each $C \in \mathfrak{C}$, there is a
  P-function \[
    \lambda_C(-) :
      \mathfrak{C}(C \times A, B)
	\rightarrow
      \mathfrak{C}(C, B^A)
  \] such that, for any $h : C \times A \rightarrow B$ and $k : C
  \rightarrow B^A$, \begin{align*}
    \textsf{ev}_{A, B} \circ (\lambda_C(h) \times id_A)
      &\sim
    h \\
    \lambda_C\left(\textsf{ev}_{A, B} \circ (k \times id_A)\right)
      &\sim
    k
  \end{align*}
\end{defn}

\begin{defn}[P-ccc]
  A \emph{cartesian closed P-category}, or \emph{P-ccc}, is a
  P-category that has a terminal object, binary products, and an
  exponential for each pair of objects.
\end{defn}

\begin{thm}[\cite{Cubric1998}]
  \label{thm:psetsccc}
  The P-category $\mathfrak{PSet}$ of P-sets and P-functions is a
  P-ccc.
\end{thm}

\noindent We warn the reader that we might be lax with the prefix
``P-'' in the rest of this thesis, as the overwhelming majority of
it will solely concern P-categories.

\section{Exposures}
  \label{sec:expo}

In this section we introduce a new P-categorical notion, that of
an \emph{exposure}. The aim of exposures is to P-categorically
capture the modality-as-intension interpretation.

An exposure is \emph{almost a functor}: it is a map of objects and
arrows of a P-category into another, and it preserves identities
and composition. It is \emph{not} a functor because it does not
preserve the PERs on the hom-sets of the source P-category.
Instead, it \emph{reflects} them. In that sense, it may
\emph{expose} the structure of a particular arrow by uncovering
its inner workings, irrespective of the extensional equality
represented by the PER. The inner workings are then represented as
a well-defined arrow of some, possibly the same, P-category.

\begin{defn}
  An exposure $Q : (\mathfrak{B}, \sim) \expo{} (\mathfrak{C},
  \sim)$ consists of
  \begin{enumerate}[label=(\roman{*})]
    \item
      a map assigning to each object $A \in \mathfrak{B}$ an object
      $QA \in \mathfrak{C}$, and
    \item
      for each $A, B \in \mathfrak{C}$, an operation \[
        Q_{A, B} : \mathfrak{B}(A, B) \dashrightarrow \mathfrak{C}(QA, QB)
      \] for which we simply write $Qf$ when the source and target
      of $f$ are known
  \end{enumerate}
  such that
  \begin{enumerate}[label=(\roman{*})]
    \item 
      $Q(id_A) \sim id_{QA}$;
    \item
      $Q(g \circ f) \sim Qg \circ Qf$, for any arrows $f : A
      \rightarrow B$ and $g : B \rightarrow C$, and
    \item
      $Q_{A, B}$ reflects PERs: if $Qf \sim Qg$ then $f \sim g$.
  \end{enumerate}
\end{defn}

\noindent Like functors, exposures compose: it suffices to use
reflection of PERs twice. There is a close relationship between
functors and exposures. In fact, if exposures were functors, they
would be faithful functors.

\begin{lem}
  \label{lem:funcexpofunc}
  A (P-)functor $Q : (\mathfrak{B}, \sim) \longrightarrow
  (\mathfrak{C}, \sim)$ is an exposure if and only if it is
  (P-)faithful.
\end{lem}
\begin{proof}
  If $Q : (\mathfrak{B}, \sim) \longrightarrow (\mathfrak{C},
  \sim)$ is also an exposure, then the morphism map $Q_{A, B} :
  \mathfrak{B}(A, B) \rightarrow \mathfrak{C}(QA, QB)$ is a
  P-function which also reflects PERs, hence $Q$ is a (P-)faithful
  (P-)functor. The converse is similar.
\end{proof}

\noindent Thus, the identity functor is an exposure
$\mathsf{Id}_\mathfrak{B} : (\mathfrak{B}, \sim) \expo{}
(\mathfrak{B}, \sim)$. This lemma also enables us to compose an
exposure $Q : (\mathfrak{B}, \sim) \expo{} (\mathfrak{C}, \sim)$
with a faithful functor in either direction: if $F :
(\mathfrak{A}, \sim) \longrightarrow (\mathfrak{B}, \sim)$ is
faithful then we can define the exposure $Q \circ F :
(\mathfrak{A}, \sim) \expo{} (\mathfrak{C}, \sim)$ by
\begin{align*}
  (Q \circ F)(A) &\myeq Q(FA) \\
  (Q \circ F)_{A, B} &: \mathfrak{A}(A, B) \dashrightarrow \mathfrak{C}(Q(FA), Q(FB)) \\
  (Q \circ F)_{A, B} &\myeq f \mapsto Q_{A, B}(F_{A, B}(f))
\end{align*} and similarly for pre-composition.

The notion of natural transformations also naturally carries over
to the setting of exposures:

\begin{defn}
  A \emph{natural transformation of exposures} $t : F \natexp G$
  between two exposures $F, G: \mathfrak{B} \expo{} \mathfrak{C}$
  consists of an arrow $t_A : FA \rightarrow GA$ of $\mathfrak{C}$
  for each object $A \in \mathfrak{B}$, such that, for every arrow
  $f : A \rightarrow B$ of $\mathfrak{B}$, the following diagram
  commutes up to $\sim$: \[
    \begin{tikzcd}
      FA
        \arrow[r, "Ff"]
        \arrow[d, "t_A", swap]
      & FB
        \arrow[d, "t_B"] \\
      GA
        \arrow[r, "Gf", swap]
      & GB
    \end{tikzcd}
  \] 
\end{defn}

\noindent Nevertheless, we must not be cavalier when adopting
practices from 1-category theory. For example, we cannot
arbitrarily compose natural transformations of exposures with
other exposures. Let $t : F \natexp{} G$ be a natural
transformation between two exposures $F, G : \mathfrak{B} \expo{}
\mathfrak{C}$.  If $R : \mathfrak{A} \expo{} \mathfrak{B}$ is an
exposure, then we can define \[
  (t_R)_A \myeq t_{RA} : FRA \rightarrow GRA
\] These components form a natural transformation $t_R : FR
\natexp{} GR$. The defining diagram \[
  \begin{tikzcd}
    FRA
      \arrow[r, "FRf"]
      \arrow[d, "t_{RA}", swap]
    & FRB
      \arrow[d, "t_{RB}"] \\
    GRA
      \arrow[r, "GRf", swap]
    & GRB
  \end{tikzcd}
\] commutes, as it is the naturality square of $t : F \natexp{} G$
at $Rf : RA \rightarrow RB$. But if we instead have a P-functor $L
: \mathfrak{C} \longrightarrow \mathfrak{D}$, we can define
\begin{align*}
  &Lt : LF \natexp{} LG \\
  &(Lt)_A \myeq L(t_A) : LFA \rightarrow LGA
\end{align*} We must have that $L$ be a P-functor for this
to be natural: the diagram we want is commutative only if \[
  L(t_B) \circ LFf \sim L(t_B \circ Ff) \sim L(Gf \circ t_A) \sim
  LGf \circ L(t_A)
\] Whereas the first and last step would hold if $L$ were merely
an exposure, the middle step requires that we can reason
equationally `under $L$,' which can only happen if $L$ preserves
the PERs. 

\subsection{Intensional Equality}

As exposures give a handle on the internal structure of arrows,
they can be used to define intensional equality: if the images of
two arrows under the same exposure $Q$ are extensionally equal,
then the arrows have the same implementation, so they are
intensionally equal. This is an exact interpretation of the slogan
of \cite{Abramsky2014}: \emph{intensions become extensions}.

\begin{defn}[Intensional Equality]
  Let $Q : (\mathfrak{B}, \sim) \expo{} (\mathfrak{C}, \sim)$ be an
  exposure. Two arrows $f, g : A \rightarrow B$ of $\mathfrak{B}$ are
  \emph{intensionally equal (up to $Q$)}, written \[
    f \approx^{A,B}_{Q} g
  \] just if $Qf \sim Qg$.
\end{defn}

\noindent We often drop the source and target superscripts, and
merely write $f \approx_Q g$. The fact that exposures
\emph{reflect} PERs guarantees that

\begin{lem}
  \label{lem:inttoext}
  Intensional equality implies extensional equality.
\end{lem}

\begin{proof}
  $f \approx_{Q} g : A \rightarrow B$ is $Qf \sim Qg$, which
  implies $f \sim g$.
\end{proof}

\noindent In some cases the converse is true, but not for general
arrows $A \rightarrow B$: we often need some restrictions on $A$,
perhaps that it is an \emph{intensional context} i.e. of the form
$\prod_{i=1}^n QA_i$, or that it is simply a point. In that case,
we use the following definition.\footnote{Not to be confused with
Voevodsky's univalence axiom. When examining this thesis, Martin
Hyland strongly recommended that the name be changed. This will
most likely happen before subsequent publications.}

\begin{defn}
  \label{def:univalent}
  Let $Q : (\mathfrak{B}, \sim) \expo{} (\mathfrak{C}, \sim)$ be
  an exposure.
  \begin{enumerate}
    \item
      Let $\mathcal{A}$ be a class of objects of $\mathfrak{B}$.
      An object $U \in \mathfrak{B}$ is
      \emph{$\mathcal{A}$-univalent (up to $Q$)} just if
      extensional equality implies intensional equality for arrows
      with domain in $\mathcal{A}$ and codomain $U$, i.e. for
      every $f, g : A \rightarrow U$ with $A \in \mathcal{A}$, \[
        f \sim g
          \quad\Longrightarrow\quad
        f \approx_{Q} g
      \]
    \item
      If $\mathfrak{B}$ has a terminal object $\mathbf{1}$, and
      $U$ is $\{\mathbf{1}\}$-univalent, we say that $U$ is
      \emph{point-univalent}.
  \end{enumerate}
\end{defn}

Intensional equality is a PER, as we have defined it through $Q$
and extensional equality, which is a PER itself. Replacing $\sim$
with $\approx_Q$ yields another P-category, which is possibly
`more intensional' than $(\mathfrak{B}, \sim)$.

\begin{defn}
  \label{def:xray}
  Given an exposure $Q : (\mathfrak{B}, \sim) \expo{}
  (\mathfrak{C}, \sim)$, we define the \emph{x-ray category of
  $(\mathfrak{B}, \sim)$ up to $Q$} by replacing the hom-P-sets
  with \[
    (\bars{\mathfrak{B}(A, B)}, \approx^{A,B}_Q)
  \] We denote the x-ray category by $(\mathfrak{B}, \approx_Q)$.
  \end{defn}

\noindent To show that this is a valid definition, we have to
check that composition respects intensional equality, and that the
necessary axioms hold. If $f \approx_Q k$ and $g \approx_Q h$,
then \[
  Q(g \circ f) \sim Qg \circ Qf \sim Qh \circ Qk \sim Q(h \circ k)
\] because exposures preserve composition; hence $\circ$ is
indeed a well-defined P-function \[
  \circ :
    (\mathfrak{B}, \approx_Q)(A, B)
      \times 
      (\mathfrak{B}, \approx_Q)(B, C)
      \rightarrow
    (\mathfrak{B}, \approx_Q)(A, C)
\] Similarly, the PER $\approx_Q$ hereditarily satisfies
associativity of composition, and---as $Q$ preserves
identities---also satisfies the identity laws.

\subsection{Cartesian and Product-Preserving Exposures}
  \label{sec:cartexp}

Bare exposures offer no promises or guarantees regarding
intensional equality. For example, it is not a given that $\pi_1
\circ \langle f, g \rangle \approx_{Q} f$. However, from a certain
viewpoint one may argue there is no grand intensional content in
projecting a component: it is merely a structural operation and
not much more. Requiring this of an exposure strengthens the
notion of intensional equality, and further reinforces the point
that exposures offer a stratified and modular view of equality.

\begin{defn}
  Let $\mathfrak{B}$ be a cartesian P-category, and let
  $Q : \mathfrak{B} \expo{} \mathfrak{C}$ be an exposure. We say
  that $Q$ is \emph{cartesian} just if for any arrows $f : C
  \rightarrow A$, $g : C \rightarrow B$, $h : C \rightarrow A
  \times B$, and $k : D \rightarrow \mathbf{1}$ we have
  \begin{align*}
    \pi_1 \circ \langle f, g \rangle             &\approx_Q f \\
    \pi_2 \circ \langle f, g \rangle             &\approx_Q g \\
    \langle \pi_1 \circ h, \pi_2 \circ h \rangle &\approx_Q h \\
                                               k &\approx_Q {!}_D 
  \end{align*}
\end{defn} 

\noindent However, this is not enough to formally regain standard
equations like $\langle f, g \rangle \circ h \approx_{Q} \langle f
\circ h, g \circ h \rangle$. This is because we cannot be certain
that the pairing function $\langle \cdot, \cdot \rangle$ preserves
the intensional equality $\approx_Q$. We need something quite a
bit stronger, which is to ask for full extensional preservation of
products.

\begin{defn}
  A cartesian exposure $Q : \mathfrak{B} \expo{} \mathfrak{C}$ of a
  cartesian P-category $\mathfrak{B}$ in a cartesian P-category
  $\mathfrak{C}$ is \emph{product-preserving} whenever the
  canonical arrows \begin{align*}
    \langle Q\pi_1, Q\pi_2 \rangle 
      &: Q(A \times B) \xrightarrow{\cong} QA \times QB \\
    {!}_{Q\mathbf{1}} 
      &: Q\mathbf{1}   \xrightarrow{\cong} \mathbf{1}
  \end{align*} are (P-)isomorphisms. We write $m_{A, B} : QA
  \times QB \rightarrow Q(A \times B)$ and $m_0 : \mathbf{1}
  \rightarrow Q\mathbf{1}$ for their inverses.
\end{defn}

\noindent In essence, the isomorphism $Q(A \times B)
\xrightarrow{\cong} QA \times QB$ says that \emph{code for pairs
is a pair of codes}, and vice versa.  Amongst the exposures, the
product-preserving are the only ones that interact harmoniously
with the product structure. In fact, preservation of products
forces the pairing function to preserve intensional quality, thus
regaining all the standard equations pertaining to products up to
$\approx_Q$. To show all that, we first need the following
proposition.

\begin{prop}
  \label{prop:prodpresangle}
  In the above setting, the following diagram commutes up to
  $\sim$: \[
    \begin{tikzcd}
      QC
	\arrow[r, "\langle Qf{,} Qg \rangle"]
	\arrow[dr, "Q\langle f{,} g \rangle", swap]
      & QA \times QB
	\arrow[d, "m_{A, B}"]
      \\
      & Q(A \times B)
    \end{tikzcd}
  \] i.e. $m_{A, B} \circ \langle Qf, Qg \rangle \sim Q\langle f,
  g \rangle$ for $f : C \rightarrow A$ and $g : C \rightarrow B$.
\end{prop}

\begin{proof}
  We compute
  \begin{derivation}
    \langle Q\pi_1, Q\pi_2 \rangle \circ Q\langle f, g \rangle
      \since[\sim]{naturality}
    \langle Q\pi_1 \circ Q\langle f, g \rangle, Q\pi_2 \circ
	Q\langle f, g \rangle \rangle
      \since[\sim]{$Q$ is an exposure}
    \langle Q(\pi_1 \circ \langle f, g \rangle), Q(\pi_2 \circ
	\langle f, g \rangle) \rangle
      \since[\sim]{$Q$ is a cartesian exposure}
    \langle Qf, Qg \rangle
  \end{derivation}
  and hence $m_{A, B} \circ \langle Qf, Qg \rangle \myeq \langle
  Q\pi_1, Q\pi_2\rangle^{-1} \circ \langle Qf, Qg \rangle \sim
  Q\langle f, g \rangle$.
\end{proof}

\noindent One can easily easy to compute that the $m_{A, B}$'s
satisfy a naturality property, similar to the one for strong
monoidal categories. That is,

\begin{prop}
  \label{prop:mnat}
  In the above setting, the following diagram commutes up to
  $\sim$: \[
    \begin{tikzcd}
      QC \times QD
	\arrow[r, "Qf \times Qg"]
	\arrow[d, "m_{C, D}", swap]
      & QA \times QB
        \arrow[d, "m_{A, B}"] \\
      Q(C \times D)
	\arrow[r, "Q(f \times g)", swap]
      & Q(A \times B)
    \end{tikzcd}
  \] i.e. $m_{A, B} \circ (Qf \times Qg) \sim Q(f \times g) \circ
  m_{C, D}$ for $f : C \rightarrow A$ and $g : D \rightarrow B$.
\end{prop}

\begin{proof}
  Simple calculation as above, using the inverses of both $m_{A,
  B}$ and $m_{C, D}$, as well as the fact that $Q : \mathfrak{B}
  \expo{} \mathfrak{C}$ is cartesian.
\end{proof}

\noindent In monoidal 1-category theory this would simply be a
natural isomorphism between the functors $Q(- \times -)$ and $Q(-)
\times Q(-)$. However, the product functor $- \times -$ is not
necessarily an exposure: we may have $f \times g \sim h \times k$,
yet it may be that $f \not\sim h$ or $g \not\sim k$. However, if
the category is \emph{connected} (all hom-P-sets are nonempty),
then the projections are epic, and hence $- \times -$ is
\emph{faithful}. By Lemma \ref{lem:funcexpofunc}, this would then
allow us to compose it with $Q$ to make an exposure. But since
this is not the case in general, we do not.  However, the
requisite naturality property follows from the fact the $m$'s are
the inverses of canonical arrows, so we are not seriously
hampered.

We can also show that the following relationship holds
between the projection arrows and their `exposed' version, given
product-preservation:

\begin{prop}
  \label{prop:prodpresproj}
  In the above setting, let \[
    A \xleftarrow{\pi^{A, B}_1} 
      A \times B
    \xrightarrow{\pi^{A, B}_2} B
  \] and \[
    QA \xleftarrow{\pi^{QA, QB}_1}
      QA \times QB
    \xrightarrow{\pi^{QA, QB}_2} QB
  \] be product diagrams in $\mathfrak{B}$ and $\mathfrak{C}$
  respectively. Then \[
    Q\pi^{A, B}_i \circ m_{A, B} \sim \pi^{QA, QB}_i
  \]
\end{prop}

\begin{proof}
  $\pi_i \circ m^{-1}_{A, B} \myeq \pi_i \circ \langle Q\pi_1,
  Q\pi_2 \rangle \sim Q\pi_i$
\end{proof} 

\noindent The product-preserving structure of $Q : \mathfrak{B}
\expo{} \mathfrak{C}$ then suffices to guarantee that taking the
mediating morphism $\langle f, g\rangle$ preserves not just
extensional equality in $f$ and $g$, as it does by the definition
of products in P-categories, but also intensional equality. Hence,
$\langle \cdot, \cdot \rangle$ also induces products in the x-ray
category $(\mathfrak{B}, \approx_Q)$. 

\begin{prop}
  \label{prop:presint}
  If $Q : \mathfrak{B} \expo{} \mathfrak{C}$ is a
  product-preserving exposure, then the function \[
    \langle \cdot, \cdot \rangle :
      \mathfrak{B}(C, A) \times \mathfrak{B}(C, B)
	\rightarrow
      \mathfrak{B}(C, A \times B)
  \] preserves intensional equality $\approx_Q$, and is thus a
  function \[
    \langle \cdot, \cdot \rangle :
      (\mathfrak{B}, \approx_Q)(C, A) \times (\mathfrak{B}, \approx_Q)(C, B)
	\rightarrow
      (\mathfrak{B}, \approx_Q)(C, A \times B)
  \]
\end{prop}
\begin{proof}
  If $f \approx_Q h : C \rightarrow A$ and $g \approx_Q k : C
  \rightarrow B$, then \begin{derivation}
      Q\langle f, g \rangle
    \since[\sim]{ Proposition \ref{prop:prodpresangle} }
      m \circ \langle Qf, Qg \rangle
    \since[\sim]{ assumptions }
      m \circ \langle Qh, Qk \rangle
    \since[\sim]{ Proposition \ref{prop:prodpresangle} }
      Q\langle h, k \rangle
  \end{derivation} and hence $\langle f, g \rangle \approx_Q
  \langle h, k \rangle$.
\end{proof}

\noindent Note that in the above proof we used the `monoidality'
$m_{A,B}$ to `shift' the exposure $Q$ exactly where we want it to
be to use the assumptions $f \approx_Q h$ and $g \approx_Q k$. 

This result also implies that the standard equations for products
hold up to $\approx_Q$.\footnote{In fact, even if we exclude 
$\langle \pi_1 \circ h, \pi_2 \circ h \rangle \approx_Q h$ from
the definition of a cartesian exposure, we can then regain it
through product-preservation: in this sense, product-preservation
is a strong extensionality principle that even implies the
`$\eta$-rule' for the x-ray category. One might even entertain the
idea that they can show the `$\beta$-rule' $\pi_1 \circ \langle f,
g \rangle \approx_Q f$ simply by the existence of the isomorphism
$m_{A, B}$, and without explicitly assuming $Q$ to be cartesian,
thus ostensibly reducing cartesian exposures to product-preserving
ones. But the derivation of this requires $Q\pi_1 \circ m \sim
\pi_1$, which seems to only follow if the exposure is cartesian,
making the apparently simple argument circular.}

\begin{lem} \hfill
  \begin{enumerate}
    \item
      If $Q : \mathfrak{B} \expo{} \mathfrak{C}$ is a
      product-preserving exposure, then \[
        \langle f, g \rangle \circ h
          \approx_{Q}
        \langle f \circ h, g \circ h \rangle
      \] for $f : C \rightarrow A$, $g : C \rightarrow B$, and $h
      : D \rightarrow C$.
    \item
      If $Q$ is a product-preserving exposure, then \[
        (f \times g) \circ \langle h, k \rangle
          \approx_{Q}
        \langle f \circ h, g \circ k \rangle
      \] for $f : C \rightarrow A$, $g: D \rightarrow B$, $h : E
      \rightarrow C$, $k : F \rightarrow D$.
  \end{enumerate}
\end{lem}

\subsection{Comonadic Exposures}
  
We can now revisit the failed categorical approach to
modality-as-intension that we discussed in \S\ref{sec:intmodcat}.
It turns out that all the categorical equipment used for strong
monoidal (= product-preserving) comonads have direct analogues in
exposures. Throughout the rest of this section we fix a
product-preserving endoexposure $Q : \mathfrak{B} \expo{}
\mathfrak{B}$.

If we have an interpreter that maps code to values at each type,
then we can present it as a (well-behaved) natural transformation
from our selected exposure to the identity exposure.

\begin{defn}
  An \emph{evaluator} is a transformation of exposures, \[
    \epsilon : Q \natexp{} \textsf{Id}_{\mathfrak{B}}
  \] such that the following diagrams commute up to $\sim$: \[
    \begin{tikzcd}
        QA \times QB
        \arrow[r, "\epsilon_A \times \epsilon_B",]
        \arrow[d, "m_{A, B}", swap]
      & A \times B
        \arrow[d, equal] \\
      Q(A \times B)
        \arrow[r, "\epsilon_{A \times B}", swap]
      & A \times B
    \end{tikzcd} \quad \begin{tikzcd}
      \mathbf{1}
        \arrow[rd, equal]
        \arrow[d, "m_0", swap]
      & \\
      Q\mathbf{1}
        \arrow[r, "\epsilon_\mathbf{1}", swap]
      & \mathbf{1}
    \end{tikzcd}
  \]
\end{defn}

\noindent How about quoting, then? Given a point $a : \mathbf{1}
\rightarrow A$, we define its \emph{quote} to be the point \[
  Qa \circ m_0 : \mathbf{1} \rightarrow QA
\] The fact that $\epsilon : Q \natexp{} \textsf{Id}_\mathfrak{B}$
lets us then calculate that \[
      \epsilon_A \circ Qa \circ m_0
  \sim
      a \circ \epsilon_\mathbf{1} \circ m_0
  \sim
      a
\] So, post-composing a component of an evaluator to a quoted point
yields back the point itself! The naturality is there to guarantee
that the evaluator is defined `in the same way' at all objects.
The two diagrams that are required to commute as part of the
definition would---in the context of monoidal functors---ensure
that $\epsilon$ is a \emph{monoidal} natural transformation. In
the setting of exposures they are not only necessary in the final
step of the above calculation, but they also ensure that the
evaluators are compatible with products.\footnote{Nevertheless,
notice that---since we are in a cartesian setting and $\mathbf{1}$
is a terminal object---the second diagram commutes automatically.}

\begin{defn} 
  A \emph{quoter} is a transformation of exposures, \[
    \delta : Q \natexp Q^2
  \] for which the following diagrams commute up to $\sim$: \[
    \begin{tikzcd}
        QA \times QB
        \arrow[r, "\delta_A \times \delta_B",]
        \arrow[dd, "m_{A, B}", swap]
      & Q^2 A \times Q^2 B
        \arrow[d, "m_{QA, QB}"] \\
      & Q(QA \times QB) 
        \arrow[d, "Qm_{A, B}"] \\
      Q(A \times B)
        \arrow[r, "\delta_{A \times B}", swap]
      & Q^2(A \times B)
    \end{tikzcd} \quad \begin{tikzcd}
      \mathbf{1}
        \arrow[rd, "m_0"]
        \arrow[dd, "m_0", swap]
      & & \\
      & Q\mathbf{1}
        \arrow[rd, "Qm_0"] 
      & \\
      Q\mathbf{1}
        \arrow[rr, "\delta_\mathbf{1}", swap]
      & & Q^2\mathbf{1}
    \end{tikzcd}
  \]
\end{defn}

\noindent If we post-compose a component of a quoter to a quoted
point, we get \[
    \delta_A \circ Qa \circ m_0
  \sim
    Q^2 a \circ \delta_\mathbf{1} \circ m_0
  \sim
    Q^2 a \circ Qm_0 \circ m_0
  \sim
  Q(Qa \circ m_0) \circ m_0
\] So a quoter maps a quoted point to its doubly quoted version.
In this instance, the diagram that would correspond to the
transformation being monoidal is crucial in obtaining this
pattern.

All of these ingredients then combine to form a comonadic
exposure.

\begin{defn}
  A \emph{comonadic exposure} $(Q, \epsilon, \delta)$
  consists of an endoexposure \[
    Q: (\mathfrak{B}, \sim)
      \looparrowright (\mathfrak{B}, \sim)\text{,}
  \] along with an evaluator $\epsilon : Q \natexp{}
  \mathsf{Id}_\mathfrak{B}$, and a quoter $\delta : Q \natexp{}
  Q^2$, such that the following diagrams commute up to $\sim$: \[
    \begin{tikzcd}
      QA
        \arrow[r, "\delta_A"]
        \arrow[d, "\delta_A", swap]
      & Q^2 A 
        \arrow[d, "\delta_{QA}"] \\
      Q^2 A 
        \arrow[r, "Q\delta_A", swap] 
      & Q^3 A
    \end{tikzcd} \quad
    \begin{tikzcd}
      QA
        \arrow[r, "\delta_A"]
        \arrow[d, "\delta_A", swap]
        \arrow[rd, "id_{QA}", swap]
      & Q^2 A 
        \arrow[d, "\epsilon_{QA}"] \\
      Q^2 A
        \arrow[r, "Q\epsilon_A", swap]
      & QA
    \end{tikzcd}
  \]
\end{defn}

\noindent Comonadic exposures are the analogue of
product-preserving (= strong monoidal) comonads in the categorical
semantics of \textsf{S4}. They will prove instrumental in our
analysis of intensional recursion (\S\ref{chap:irec}), and in the
semantics of iPCF (\S\ref{chap:intsem1}, \S\ref{chap:intsem2}).

\subsection{Idempotent Comonadic Exposures}
  \label{sec:idempotent}

If the components of $\delta : Q \natexp{} Q^2$ are isomorphisms,
we shall call the comonadic exposure $(Q, \epsilon, \delta)$
\emph{idempotent}. 

As we discussed before, if one is to take the interpretation of
$Q$ as `code' seriously, then there are clearly two `regions' of
data: that of \emph{static code}, always found under an occurrence
of $Q$, and that of \emph{dynamic data}. Intuitively, the notion
of `code of code of $A$,' namely $Q(QA)$, should be the same as
`code of $A$.' If something is code, it is already
\emph{intensional} in a maximal sense: it can certainly be taken
`one level up' ($Q(QA)$), but that should not amount to very much.

We have seen that exposures are a very weak setting in
calculational terms, as they do not preserve equality. It is for
this reason that we are forced to externally impose equations,
such as those for cartesian products in \S\ref{sec:cartexp}.
However, we will shortly see that the idempotence of a comonadic
exposure is a particularly powerful tool that immediately allows
us to infer a lot about equality, especially intensional.

It follows, as in \S\ref{sec:intmodcat}, that following diagram
commutes for each $f : QA \rightarrow QB$: \[
  \begin{tikzcd}
    QA
      \arrow[r, "f"]
      \arrow[d, "\delta_A", swap]
    & QB
      \arrow[d, "\delta_B"] \\
    Q^2 A
      \arrow[r, "Qf", swap]
    & Q^2 B
  \end{tikzcd}
\] The proof is the same as before: nowhere in Theorem
\ref{thm:idem} did we use the `forbidden principle' \[
  f = g \quad\Longrightarrow Qf = Qg
\] Furthermore, the argument we produced just before that theorem
also still holds: each component of $\delta : Q \natexp{} Q^2$ is
a `reasonable quoting device,' in the sense that the diagram \[
  \begin{tikzcd}
  \mathbf{1}
    \arrow[r, "a"]
    \arrow[d, "m_0", swap]
  & Q A 
    \arrow[d, "\delta_A"] \\
  Q \mathbf{1}
    \arrow[r, "Qa", swap]
  & Q^2 A
  \end{tikzcd}
\] commutes: the only time we used the `forbidden principle,'
namely the demonstration that $Qm_0 \circ Q(m_0^{-1}) \sim
id_{Q\mathbf{1}}$, can be `simulated' with idempotence as follows:
we have \[
    Qm_0 \circ Q(m_0^{-1}) \circ \delta_\mathbf{1}
  \sim
    \delta_\mathbf{1} \circ m_0 \circ m_0^{-1}
  \sim
    \delta_\mathbf{1} \circ id_{Q\mathbf{1}}
  \sim
    Q(id_{Q\mathbf{1}}) \circ \delta_\mathbf{1}
\] by using idempotence twice, so that cancelling the iso
$\delta_{Q\mathbf{1}}$ then yields the result. We will later see
that this is due to a more general theorem.

So, even if this diagram commutes, where does the `degeneracy'
argument (Proposition \ref{prop:degeneracy}) break down? The key
lies precisely in the fact that $Q$ does not respect the PERs, and
hence $\textsf{in} : \mathfrak{C}(\mathbf{1}, A) \oper{}
\mathfrak{C}(\mathbf{1}, QA)$ is now only an operation, not a
P-function. Hence, there is no natural isomorphism
$\mathfrak{C}(\mathbf{1}, -) \cong \mathfrak{C}(\mathbf{1}, Q-)$.

Let us, however, take a closer look: the function $\textsf{in}$ is
defined by \[
  x : \mathbf{1} \rightarrow A
    \quad\longmapsto\quad
  Qf \circ m_0 : \mathbf{1} \rightarrow QA
\] That is, the only occurrence of $f$ is under $Q$, and the rest
is simply pre-composition with $m_0$. If $f \approx_Q f'$, i.e. if
$Qf \sim Qf'$, then we have that $\textsf{in}(f) \sim
\textsf{in}(f')$. Hence, $\textsf{in}$ changes $\approx_Q$ to
$\sim$, so it is actually more than an operation: it is a map \[
  \textsf{in} : 
    (\mathfrak{C}, \approx_Q)(\mathbf{1}, A)
      \rightarrow
    (\mathfrak{C}, \sim)(\mathbf{1}, QA)
\] Similarly, $\textsf{out}$ is defined by \[
  a : \mathbf{1} \rightarrow QA
    \quad\longmapsto\quad
  \epsilon_A \circ a : \mathbf{1} \rightarrow A
\] If $a \sim a'$, then $\delta_A \circ a \sim \delta_A \circ a'$,
so $Qa \circ m_0 \sim Qa' \circ m_0$. Cancelling the isomorphism
$m_0$ gives us $a \approx_Q a'$. Thus, if we have a reasonable
quoting device at $A$, $QA$ is point-univalent. For the time
being, notice that this means that $\textsf{out}$ takes $\sim$ to
$\approx_Q$, i.e. it is a map \[
  \textsf{out} : 
    (\mathfrak{C}, \sim)(\mathbf{1}, QA)
      \rightarrow
    (\mathfrak{C}, \approx_Q)(\mathbf{1}, A)
\] If we combine these facts with the previous calculations and
naturality of $Q$, we obtain a natural isomorphism
\begin{prop}
  $(\mathfrak{C}, \approx_Q)(\mathbf{1}, -)
    \cong
  (\mathfrak{C}, \sim)(\mathbf{1}, Q-)$
\end{prop} 

\noindent That is: the intensional structure of the points, now
visible in the x-ray category $(\mathfrak{C}, \approx_Q)$, is
represented by the points $\mathbf{1} \rightarrow QA$ under
extensional equality.

In the rest of this section let us fix a product-preserving
idempotent comonadic exposure $(Q, \epsilon, \delta)$ on
$\mathfrak{B}$.

\subsubsection*{Equal Intensional Transformations
are Intensionally Equal}

A central result is the generalisation of the argument we used to
define $\textsf{out}$, and it is the following. Think of arrows $f
: QA \rightarrow QB$ as \emph{intensional operations}: these
transform code of type $A$ to code of type $B$. It therefore
should transpire that, if $f \sim g : QA \rightarrow QB$, then $f$
and $g$ represent the same code transformation, and in fact should
be intensionally equal. If $Q$ is idempotent, then this is exactly
what happens.

\begin{thm}
  \label{thm:ideminteq}
  For any $f, g : QA \rightarrow QB$, \[
    f \sim g
      \quad\Longrightarrow\quad
    f \approx_Q g
  \]
\end{thm}
\begin{proof}
  We have \[
    Qf \circ \delta_A \sim \delta_B \circ f \sim \delta_B \circ g
    \sim Qg \circ \delta_A
  \] Pre-composing with the inverse of $\delta_A$ yields $Qf \sim
  Qg$, and hence $f \approx_Q g$.
\end{proof}

Idempotence also implies another crucial piece of information
regarding the product-preserving isomorphism $m_{A, B}: QA \times
QB \xrightarrow{\cong} Q(A \times B)$. Even though we know $m_{A,
B}$ is an isomorphism up to $\sim$, we ostensibly do not have any
information on its behaviour up to $\approx_Q$: the hom-operations
of $Q$ do not preserve the PERs. However, in the idempotent
setting it always iso, even up to $\approx_Q$.

\begin{lem}
  \label{lem:misoapprox}
  $m_{A, B}: QA \times QB \xrightarrow{\cong} Q(A \times B)$ is an
  isomorphism up to $\approx_Q$, i.e. it is an isomorphism in the
  x-ray category $(\mathfrak{B}, \approx_Q)$.
\end{lem}
\begin{proof}
  Calculate that
  \begin{derivation}
      Qm \circ Q\langle Q\pi_1, Q\pi_2 \rangle \circ \delta
    \since[\sim]{Proposition \ref{prop:prodpresangle}}
      Qm \circ m \circ \langle Q^2 \pi_1, Q^2 \pi_2 \rangle \circ \delta
    \since[\sim]{naturality of product bracket, $\delta$ natural}
      Qm \circ m \circ \langle \delta \circ Q\pi_1,
                               \delta \circ Q\pi_2 \rangle
    \since[\sim]{product equation}
      Qm \circ m \circ (\delta \times \delta)
        \circ \langle Q\pi_1, Q\pi_2 \rangle
    \since[\sim]{$\delta$ monoidal}
      \delta \circ m \circ \langle Q\pi_1, Q\pi_2 \rangle
    \since[\sim]{$m^{-1} \myeq \langle Q\pi_1, Q\pi_2 \rangle$}
      \delta
  \end{derivation} Since $\delta$ is an isomorphism, we can
  cancel it on both sides to yield $m \circ m^{-1} \approx_Q id$.
  The calculation is similar in the opposite direction, and relies
  on post-composing with the isomorphism $m \circ (\delta \times
  \delta)$, and then cancelling it.
\end{proof}

\noindent The same trick with $m \circ (\delta \times \delta)$
also shows that Proposition \ref{prop:prodpresproj} holds
intensionally.

\begin{lem}
  $Q\pi^{A, B}_i \circ m_{A, B} \approx_Q \pi^{QA, QB}_i$.
\end{lem}

\noindent These lemmas show that

\begin{cor}
  \label{cor:ideminteqprod}
  For any $f, g : \prod_{i=1}^n QA_i \rightarrow \prod_{j=1}^m
  QB_j$, \[
    f \sim g
      \quad\Longrightarrow\quad
    f \approx_Q g
  \]
\end{cor}
\begin{proof}
  Pre-and-post-compose with the appropriate isomorphisms $m^{(n)}$
  (see \S\ref{sec:scene}), use Theorem \ref{thm:ideminteq}, and
  then use Lemma \ref{lem:misoapprox} to cancel the $m^{(n)}$'s.
\end{proof}

\subsubsection*{Some more lemmas}

In this section we prove some more lemmas that hold in the
idempotent case. Firstly, we can show that the comonadic diagrams
commute intensionally.

\begin{lem}
  \label{lem:comonint}
  The comonadic diagrams \[
    \begin{tikzcd}
      QA
        \arrow[r, "\delta_A"]
        \arrow[d, "\delta_A", swap]
      & Q^2 A 
        \arrow[d, "\delta_{QA}"] \\
      Q^2 A 
        \arrow[r, "Q\delta_A", swap] 
      & Q^3 A
    \end{tikzcd} \quad
    \begin{tikzcd}
      QA
        \arrow[r, "\delta_A"]
        \arrow[d, "\delta_A", swap]
        \arrow[rd, "id_{QA}", swap]
      & Q^2 A 
        T\arrow[d, "\epsilon_{QA}"] \\
      Q^2 A
        \arrow[r, "Q\epsilon_A", swap]
      & QA
    \end{tikzcd}
  \] commute up to $\approx_Q$.
\end{lem}
\begin{proof}
  Simple calculations that mainly follow by pre-composing
  $\delta_A$ and then cancelling it. For example, \[
    Q^2 \epsilon_A \circ Q\delta_A \circ \delta_A
  \sim
    Q^2 \epsilon_A \circ \delta_{QA} \circ \delta_A
  \sim
    \delta_A \circ Q\epsilon_A \circ \delta_A
  \sim
    \delta_A
  \] by the equations and the naturality of $\delta$, which gives
  that $Q\epsilon_A \circ \delta_A \approx_Q id_{QA}$.
\end{proof}

\noindent Moreover, we have that

\begin{lem}
  \label{lem:epsilonepic}
  $\epsilon_A : QA \rightarrow A$ is epic up to $\approx_Q$.
\end{lem}
\begin{proof}
  If $f \circ \epsilon_A \approx_Q g \circ \epsilon_A : QA
  \rightarrow B$, then $Qf \circ Q\epsilon_A \sim Qg \circ
  Q\epsilon_A$. Pre-composing $\delta_A$ yields $f \approx_Q g$.
\end{proof}

\noindent The following lemma is also quite useful.

\begin{lem}[Quotation-Evaluation]
  \label{lem:qe}
  For any $f : QB \rightarrow QA$, the following diagram commutes
  up to $\sim$: \[
    \begin{tikzcd}
      QB
        \arrow[r, "{\delta_B}"]
        \arrow[d, "f", swap]
      & Q^2 B
        \arrow[d, "Qf"] \\
      Q A
      & Q^2 A
        \arrow[l, "{Q\epsilon_A}"]
    \end{tikzcd}
  \]
\end{lem}
\begin{proof}
  We may calculate \[
      Q\epsilon_A \circ Qf \circ \delta_B
    \sim
      Q\epsilon_A \circ \delta_A \circ f
    \sim
      f
  \] by idempotence and the comonadic equations.
\end{proof}

\noindent This has a simple corollary when it comes to quoted
points:

\begin{cor}
  \label{cor:qe}
  If $(Q, \delta, \epsilon)$ is a product-preserving idempotent
  comonadic exposure, then \[
    Q(\epsilon_A \circ a) \circ m_0 \sim a
  \] for any $a : \mathbf{1} \rightarrow QA$.
\end{cor}
\begin{proof}
  We may calculate \[
      Q\epsilon \circ Qa \circ m_0
    \sim
      Q\epsilon \circ \delta_A \circ a
    \sim
      a
  \] by $\delta$ being reasonable and $Q$ comonadic.
\end{proof}

\subsubsection*{Coalgebras, Idempotence, and Univalence}

Recall the definition of \emph{point-univalent objects}
(Definition \ref{def:univalent}): $U$ is point-univalent if $x
\sim y : \mathbf{1} \rightarrow U$ implies $x \approx_Q y :
\mathbf{1} \rightarrow U$. Intuitively, $U$ is point-univalent if
extensional and intensional equality coincide for its points, i.e.
if none of them contain intensional information. Now, the objects
$QA$ are supposed to contain intensions/codes corresponding to the
`elements' of $A$. It is no surprise then that we can prove that

\begin{lem}
  \label{lem:puni}
  $QA$ is point-univalent.
\end{lem}
\begin{proof}
  Suppose $x \sim y : \mathbf{1} \rightarrow QA$. Then $x \circ
  \epsilon_\mathbf{1} \sim y \circ \epsilon_\mathbf{1} :
  Q\mathbf{1} \rightarrow QA$. Invoking Theorem
  \ref{thm:ideminteq} yields $x \circ \epsilon_\mathbf{1}
  \approx_Q y \circ \epsilon_\mathbf{1}$.  If only we establish
  that $\epsilon_\mathbf{1} \circ m_0 \approx_Q id_\mathbf{1}$,
  then it would suffice to pre-compose $m_0$. But $Q$ is
  product-preserving, hence cartesian, with the result that \[
    \epsilon_\mathbf{1} \circ m_0
      \approx_Q
    {!}_\mathbf{1}
      \approx_Q
    id_\mathbf{1}
  \]
\end{proof}

\noindent In fact, this is a special case of a more general

\begin{thm}
  \label{thm:univalence}
  Let $\mathcal{Q} \myeq \setcomp{ QA }{ A \in \mathfrak{B}}$. Then
  $QA$ is $\mathcal{Q}$-univalent.
\end{thm}

\noindent This is just another way to state Theorem
\ref{thm:ideminteq}.

We would like to prove a kind of converse to this theorem, viz.
that, with idempotence, every $\mathcal{Q}$-univalent object $A$
is closely related to $QA$. Unfortunately, there is no systematic
way to obtain an arrow $A \rightarrow QA$ from the
$\mathcal{Q}$-univalence of $A$. But if we assume the existence of
such an arrow---with appropriate equations---then it is easy to
show that it is an isomorphism. This arrow is, of course, an old
friend:

\begin{defn}
  \label{def:Q-coalg}
  A \emph{$Q$-coalgebra} is an arrow \[
    \alpha : A \rightarrow QA
  \] such that the following diagrams commute up to $\sim$: \[
    \begin{tikzcd}
      A
        \arrow[r, "\alpha"]
        \arrow[rd, "id_A", swap]
      & QA
        \arrow[d, "\epsilon_A"] \\
      & A
    \end{tikzcd}
      \quad
    \begin{tikzcd}
      A
        \arrow[r, "\alpha"]
        \arrow[d, "\alpha", swap]
      & QA
        \arrow[d, "\delta_A"] \\
      QA
        \arrow[r, "Q\alpha", swap]
      & Q^2 A
    \end{tikzcd}
  \]
\end{defn}

In the modality-as-intension interpretation a $Q$-coalgebra has
an intuitive meaning: the equation $\epsilon_A \circ \alpha \sim
id$ states that $\alpha$ can `quote' the elements of $A$,
producing an element of $QA$ which---when evaluated---takes us
back to where we started. The second equation merely states that
$\alpha$ cooperates well with the quoter $\delta : Q \natexp{}
Q^2$. Thus, \emph{a $Q$-coalgebra exists when we can internally
quote the elements of an object}. 

\begin{lem}
  If $A$ is $\mathcal{Q}$-univalent, then a $Q$-coalgebra $\alpha
  : A \rightarrow QA$ is an isomorphism, with inverse $\epsilon_A
  : QA \rightarrow A$.
\end{lem}
\begin{proof}
  This is the classic proof that given idempotence all coalgebras
  are isomorphisms. However, a crucial step of that proof would
  rely on $Q$ preserving equality; in this case, this is where
  $\mathcal{Q}$-univalence steps in.

  We calculate that \[
    \delta_A \circ (\alpha \circ \epsilon_A)
      \sim
    Q(\alpha \circ \epsilon_A) \circ \delta_A
      \sim
    Q\alpha \circ Q\epsilon_A \circ \delta_A
      \sim
    Q\alpha
  \] by idempotence and the comonadic equations. If we post-compose
  $Q\epsilon_A$ to both sides of this equation, we obtain \[
    \alpha \circ \epsilon_A 
      \sim
    Q\epsilon_A \circ Q\alpha
      \sim
    Q(\epsilon_A \circ \alpha)
  \] So, if we show that $\epsilon_A \circ \alpha \approx_Q id_A$,
  then we would obtain $\alpha \circ \epsilon_A \sim id_{QA}$.  As
  we already know that $\alpha \circ \epsilon_A \sim id_{QA}$, we
  would conclude that $\alpha^{-1} \sim \epsilon_A$. Since
  $\epsilon_A \circ \alpha \sim id_A$, we obtain $\epsilon_A \circ
  \alpha \circ \epsilon_A \sim \epsilon_A : QA \rightarrow A$. But
  $A$ is $\mathcal{Q}$-univalent, so \[
    \epsilon_A \circ \alpha \circ \epsilon_A \approx_Q \epsilon_A
  \] But $\epsilon_A$ is epic up to $\approx_Q$ (Lemma
  \ref{lem:epsilonepic}), so $\epsilon_A \circ \alpha \approx_Q
  id_A$.
\end{proof}

So, when is $A$ univalent? Suppose that $\epsilon_A \circ \alpha
\approx_Q id_A$, i.e. quoting and then evaluating returns the same
intensional construction with which one began.  Then it must be
that $A$ has no real intensional structure, and conversely.

\begin{lem}
  \label{lem:intcoalgunivalent}
  Let $\alpha : A \rightarrow QA$ be a $Q$-coalgebra. Then $A$ is
  $\mathcal{Q}$-univalent if and only if $\epsilon_A \circ \alpha
  \approx_Q id_A$.
\end{lem}
\begin{proof}
  The `only if' part was shown in the preceding proof. As for the
  `if' part, let $f \sim g : QB \rightarrow A$. Then $\alpha \circ f \sim
  \alpha \circ g : QB \rightarrow QA$. By Theorem
  \ref{thm:univalence}/\ref{thm:ideminteq}, $\alpha \circ f
  \approx_Q \alpha \circ g$. Post-composing with $\epsilon_A$ and
  using the assumption gives $f \approx_Q g$.
\end{proof}

\noindent To sum, we can combine these facts to show the following

\begin{thm}
  If there is a $Q$-coalgebra $\alpha : A \rightarrow QA$ such
  that $\epsilon_A \circ \alpha \approx_Q id_A$, then $\alpha : A
  \xrightarrow{\cong} QA$ is an isomorphism.
\end{thm}

\noindent There is a partial converse, which is that

\begin{thm}
  If $Q$ is idempotent and there is an isomorphism $\alpha : A
  \xrightarrow{\cong} QA$ such that $\delta_A \circ \alpha \sim
  Q\alpha \circ \alpha$, then $\alpha : A \rightarrow QA$ is a
  $Q$-coalgebra, with $\epsilon_A \circ \alpha \approx_Q id_A$.
\end{thm}
\begin{proof}
  We compute that \[
    \alpha \sim Q\epsilon_A \circ \delta_A \circ \alpha
	   \sim Q\epsilon_A \circ Q\alpha \circ \alpha
  \] by the comonadic equations and the assumption. Cancelling
  $\alpha$ yields $\epsilon_A \circ \alpha \approx_Q id$, and
  hence $\alpha$ is a $Q$-coalgebra.
\end{proof}

\noindent Thus, either of the following data suffice to make $A$
$\mathcal{Q}$-univalent: \begin{enumerate}
  \item
    an isomorphism $A \xrightarrow{\cong} QA$  with the second
    $Q$-coalgebra equation; or
  \item
    a $Q$-coalgebra with the first equation holding intensionally.
\end{enumerate}

\subsection{Weakly Cartesian Closed Exposures}
  \label{sec:wcccexpo}

We close this section with a notion of cartesian closure for
exposures. Unlike the situation with products, the relevant notion
that will allow calculations under the exposure will be
\emph{weak}. We pick the weak notion because our motivating
example of an exposure (see \S\ref{sec:realexpo}) is a weakly
cartesian closed one. Whereas there is a good argument for an
exposure to distribute over products---i.e. products do not
contain any true intensional nature, they are simply
pairs---exposures truly reveal the internal structure of
morphisms: it makes sense that $\eta$ does not hold, cf. the
age-old discussion of the ordinary $\lambda$-theory $\lambda\beta$
and the extensional theory $\lambda\beta\eta$ in e.g.
\cite{Barendregt1984}.

\begin{defn}
  A product-preserving exposure $Q : \mathfrak{B} \rightarrow
  \mathfrak{C}$ from a P-ccc $\mathfrak{B}$ to a cartesian
  P-category $\mathfrak{C}$ is \emph{weakly cartesian closed} just
  if \[
    \textsf{ev} \circ (\lambda(f) \times id)
      \approx_{Q}
    f
  \] for all $f : C \times X \rightarrow Y$.
\end{defn}

\noindent These work as expected. For example, if for $f : A
\rightarrow B$ we let \[
  \ulcorner f \urcorner \myeq \lambda\left(
    \mathbf{1} \times A \xrightarrow{\pi_2}
    A \xrightarrow{f} B\right)
\] as before, then

\begin{lem}
  \label{lem:wcccexpopoint}
  For any $a : C \rightarrow A$, \[
    \textsf{ev} \circ \langle \ulcorner f \urcorner, a \rangle
      \approx_Q
    f \circ a
  \] 
\end{lem}
\begin{proof}
  All the necessary equations hold intensionally: \[
      \textsf{ev} \circ \langle \ulcorner f \urcorner, a \rangle
    \approx_Q
      \textsf{ev}
        \circ (\ulcorner f \urcorner \times id) 
        \circ \langle id_C, a \rangle
    \approx_Q
      f \circ \pi_2 \circ \langle id_C, a \rangle
    \approx_Q
      f \circ a
  \]
\end{proof}

\chapter[Three Examples of Exposures]
  {Three Examples of Exposures\footnote{A preliminary form of
  the results in this chapter was first published as
  \citep{Kavvos2017a}, which is available at Springer:
  \url{https://dx.doi.org/10.1007/978-3-662-54458-7_32}}}
  \label{chap:expoexamples}

  Having introduced the basics of P-categories and exposures in
the previous chapter, we now seek to prove that they are
indeed useful abstractions in the study of intensional phenomena.
Towards this goal, we shall present three examples of a P-category
and an endoexposure on it.

In \S\ref{sec:logicexpo} we will construct a P-category based on a
\emph{first-order classical theory}. In the particular case of
\emph{Peano Arithmetic (\textsf{PA})}, it will become apparent
that a well-behaved G\"odel numbering comprises an endoexposure.

Following that, in \S\ref{sec:realexpo} we turn to
\emph{realizability theory} in order to obtain a handle on
intensionality in settings related to higher-order computability
theory. The example of a comonadic exposure constructed therein is
particularly well-behaved, and is the motivating example for the
entire development of this thesis.

The third example is, we hope, somehow unfamiliar. Prompted by
\cite{Abramsky2014}, a natural question arises: is the phenomenon
of intensionality only relevant to logic and computation? We are
prepared to entertain the idea that it may be a more general
mathematical pattern, which has hitherto been either a nuisance
or---more often---invisible, and whose categorical formulation
will enable us to recognise it in more settings. In
\S\ref{sec:homoexpo}, we present a simple example of
intensionality found in \emph{homological algebra}, and submit it
to the reader for further discussion.

\section{Exposures as G\"odel Numbering}
  \label{sec:logicexpo}

Our first example of an exposure substantiates the claim that
exposures can be considered as abstract analogues of G\"odel
numberings. 

Following \cite{Lawvere1969, Lawvere2006}, we will construct a
P-category from a first-order theory \textsf{T}. Then, we will
sketch the proof that any sufficiently well-behaved G\"odel
numbering defines an exposure over this theory. We omit the
details, leaving them as a (probably rather complicated) exercise
in coding.

\subsection{The Lindenbaum P-category}

The construction in this section is called the \emph{Lindenbaum
P-category} of a first-order theory \textsf{T}, and it is simpler
than what one might imagine: we begin with three basic objects,
\begin{align*}
  \mathbf{1}, \qquad &\text{the terminal object} \\
  \mathbf{2}, \qquad &\text{the object of truth values} \\
  A,          \qquad &\text{the universe}
\end{align*} The objects of the category will be the \emph{formal
products} of these three objects; we write $\Gamma$ for a generic
product $A_1, \dots, A_n$ of these objects. The arrows of type \[
  \Gamma \rightarrow \mathbf{2}
\] will be construed as \emph{formulas}, with free variables in
$\Gamma$. Similarly, the arrows \[
  \Gamma \rightarrow A
\] will be construed as \emph{terms}, once again with free
variables in $\Gamma$. In this light, a formula $\phi(x, y, z)$ in
three free variables will be an arrow \[
  \phi(x, y, z) : A \times A \times A \rightarrow A
\] with the three copies of $A$ in the domain representing each
free variable in a fixed order---and similarly for terms.

However, this idea will be complicated by the presence of
$\mathbf{2}$ in the domain. This will represent a \emph{Boolean
hole} that may be filled by a formula. For example, consider the
expression \[
  \phi(x, P) \myeq \forall y.\ \exists z.\
    \left(f(x, y) = z \wedge P\right)
 \] is a formula with one free variable $x$, and one Boolean hole
$P$, for which a formula can be substituted. Hence, it will be an
arrow \[
  \phi(x, P) : A \times \mathbf{2} \rightarrow \mathbf{2}
\] This is a generalisation that Lawvere introduced, and of which
we shall be making use. The interesting thing about it is that the
Boolean connectives now appear as arrows, e.g.  $\wedge :
\mathbf{2} \times \mathbf{2} \rightarrow \mathbf{2}$, and $\lnot :
\mathbf{2} \rightarrow \mathbf{2}$. Of course, arrows $\Gamma
\rightarrow A$ with $\mathbf{2}$ occurring in $\Gamma$ do not make
particularly good sense: how could one have a Boolean hole in a
term?

Finally, we have the cases of $\mathbf{1}$ and products appearing
in the codomain. In the first case, there will be a unique arrow
${!}_\Gamma : \Gamma \rightarrow \mathbf{1}$. In the case of a
product $\Delta \myeq B_1 \times \dots \times B_n$, arrows $f :
\Gamma \rightarrow \Delta$ will be freely generated by brackets,
i.e.  they will be \[
  \langle f_1, \dots, f_n \rangle : \Gamma \rightarrow \Delta
\] where $f_i : \Gamma \rightarrow B_i$.
Almost everything in sight will act component-wise on these.

We shall say that \[
  f \sim g : \Gamma \rightarrow \mathbf{2}
    \quad\text{just if}\quad
  \textsf{T} \vdash f \leftrightarrow g
\] That is, two arrows that are predicates are extensionally equal
if and only if they can be proven equivalent in the theory. The
details are slightly more complicated than in most presentations
of first-order logic, since we also have to treat Boolean holes.
Similarly, two terms are extensionally equal if they can be proven
equal in the theory, i.e. \[
  t \sim s : \Gamma \rightarrow A
    \quad\text{just if}\quad
  \textsf{T} \vdash t = s
\] On brackets, $\sim$ acts component-wise.

It is not hard to see that we obtain the following theorem,
which---excluding the P-categorical twist---is essentially due to
Lawvere.

\begin{thm}
  \label{thm:fopcat}
  The above construction is a cartesian P-category, called the
  \emph{Lindenbaum P-category} $\mathfrak{Lind}(\mathsf{T})$ of
  the first-order theory $\mathsf{T}$.
\end{thm}

\subsection{Numbering as Exposure}

Let us now concentrate on the case of Peano Arithmetic, which we
denote by \textsf{PA}. A \emph{G\"odel numbering}
\citep{Boolos1994, Smullyan1992} is obtained when one assigns to
each formula $\phi(\vct{x})$ and each term $t(\vct{x})$ a
\emph{G\"odel number}, denoted by  \[
  \ulcorner \phi(\vct{x}) \urcorner,
    \quad
  \ulcorner t(\vct{x}) \urcorner
\] respectively. In the case of a first-order theory of
arithmetic, the G\"odel number of a term or formula is supposed to
represent the term or formula \emph{within the theory}. 

The notion of representation-within-the-theory is exactly what
exposures are meant to capture. Hence, we need to define the
action of an exposure, \[
  Q : \mathfrak{Lind}(\textsf{PA}) \expo{}
  \mathfrak{Lind}(\textsf{PA})
\] on the Lindenbaum P-category of \textsf{PA}. This is where we
need that the G\"odel numbering be well-behaved. Let us suppose
that we have a formula $\phi(x, y) : A \times A \rightarrow A$ in
two free variables. We would like to map this to a formula
$Q\phi(x, y)$, also in two free variables, that respects
substitution. Let us presume that we have functions \begin{align*}
  &sub_{\text{wff}, n}(x, z_1, \dots, z_n) \\
  &sub_{\text{t}, n}(x, z_1, \dots, z_n)
\end{align*} that are definable in \textsf{PA} by a term (denoted
by the same name), and moreover that these functions \emph{define
substitution for the G\"odel numbering}, in the sense that, for
example, if given---say---two closed terms $t, s$, we have \[
  \textsf{PA} \vdash
    sub_{\text{wff}, 2}\left(\ulcorner \phi(x, y) \urcorner,
    \ulcorner t \urcorner,
    \ulcorner s \urcorner\right)
    = \ulcorner \phi(t, s) \urcorner
\] We require more than most presentations of G\"odel numberings.
In particular, we require that these behave well under
substitution, e.g. we require that the terms \[
    sub_{\text{wff}, 2}\left(\ulcorner \phi(x, y) \urcorner,
    sub_{\text{t}, n}(\ulcorner t(\vct{z}) \urcorner, \vct{w}),
    sub_{\text{t}, n}(\ulcorner s(\vct{z}) \urcorner,
    \vct{w})\right)
\] and \[
    sub_{\text{wff}, n}(\ulcorner \phi\left(t(\vct{z}), s(\vct{z})\right)
    \urcorner, \vct{w})
\] be provably equal in $\textsf{PA}$. We can then define \[
  Q\phi(x, y) \myeq  sub_{\text{wff}, 2}(\ulcorner \phi(x, y) \urcorner,
  x, y)
\] Finally, in the case of a \emph{sentence} $\psi : \mathbf{1}
\rightarrow \mathbf{2}$, i.e. a closed formula, we shall define
$Q\psi : \mathbf{1} \rightarrow A$ to simply be the (numeral of
the) G\"odel number, i.e. \[
  Q\psi \myeq \ulcorner \psi \urcorner
\]

This certainly respects substitution! We have elided the concept
of Boolean holes, but we believe that to be a not so difficult
exercise.  Identities are a little stranger; the identity at $A$
is the term that is a single free variable, i.e. \[
  id_A \myeq x : A \rightarrow A
\] and this is mapped to $sub_{t, 1}(\ulcorner x \urcorner, x) : A
\rightarrow A$, which somehow needs to be provably equal to $x$
itself. This is a also strange requirement for a G\"odel
numbering, but we are mostly willing to believe that it is a
feasible desideratum. A similar situation occurs when we examine
the arrow \[
  \langle Q\pi_1, Q\pi_2 \rangle
  = \langle sub_{t, 2}(\ulcorner x \urcorner, x, y),
	    sub_{t, 2}(\ulcorner y \urcorner, x, y)
    \rangle
\] For the exposure $Q$ to be product-preserving, we would like
this to be an isomorphism, or---even better---the identity. And it
would indeed be, if we could prove that, in general, \[
  \textsf{PA} \vdash 
    sub_{t, n}(\ulcorner z_i \urcorner, \vct{z})
    = z_i
\]

More rigorously, we define \begin{align*}
  Q\mathbf{1} &\myeq \mathbf{1} \\
  Q(A) &\myeq A \\
  Q(\mathbf{2}) &\myeq A \\
  Q(B_1 \times \dots \times B_n) &\myeq QB_1 \times \dots \times
  QB_n
\end{align*} and, of course, $Q\langle f_1, \dots, f_n \rangle
\myeq \langle Qf_1, \dots, Qf_n \rangle$.

To complete the construction, we have to check the last axiom of
exposures, namely reflection of PERs. For that we need that \[
  \textsf{PA} \vdash
    sub_{\text{wff}, n}(\ulcorner \phi(\vct{x}) \urcorner, \vct{z})
    =
    sub_{\text{wff}, n}(\ulcorner \psi(\vct{x}) \urcorner, \vct{z})
\] implies \[
  \textsf{PA} \vdash \phi \leftrightarrow \psi
\] By substituting the G\"odel numbers of variables $\vct{y}$ in
the antecedent, we obtain that \[
  \textsf{PA} \vdash \ulcorner \phi(\vct{y}) \urcorner = \ulcorner
  \psi(\vct{y}) \urcorner
\] and so it suffices for the G\"odel numbering to be injective,
in the sense that \[
  \ulcorner \phi(\vct{y}) \urcorner = \ulcorner \psi(\vct{y})
  \urcorner
    \quad\Longrightarrow\quad
  \phi(\vct{y}) = \psi(\vct{y})
\] viz. that equality of G\"odel numbers implies syntactic
equality. In conclusion,

\begin{thm}
  If all of the above desiderata on G\"odel numberings are
  feasiable, then the construction is a product-preserving
  endoexposure, \[
    Q : \mathfrak{Lind}(\textsf{PA}) \expo{}
	\mathfrak{Lind}(\textsf{PA})
  \] on the Lindenbaum P-category of \textsf{PA}.
\end{thm}

\section{Exposures in Realizability}
  \label{sec:realexpo}

In this section we study our central example of an exposure,
which hails from realizability theory.

The basic objects in realizability are \emph{assemblies}, i.e.
sets of which every element is associated with a \emph{set of
realizers}. The elements of the set can be thought of as the
elements of an \emph{abstract datatype}, whereas the set of
realizers of each element contains its multiple
\emph{machine-level representations}. For example, realizers can
range over the natural numbers; then taking functions between
assemblies which can be `tracked' on the level of realizers by
partial recursive functions yields a category where `everything is
computable.'

In practice, the generalisation from natural numbers to an
arbitrary \emph{partial combinatory algebra (PCA)} is made. A PCA
is an untyped `universe' corresponding to some notion of
computability or realizability. There are easy tricks with which
one may encode various common first-order datatypes (such as
booleans, integers, etc.) in a PCA.\footnote{Most of these tricks
have their origins in untyped $\lambda$-calculus, and are hence
found in \cite{Barendregt1984}.} Moreover, it is easy to
show that, up to a simple representation of integers, one may
represent all partial recursive functions in a PCA. Before
proceeding with the construction, we recap in \S\ref{sec:pca} the
basics of PCAs. For the interested reader let us mention that
detailed discussions may be found in \citep{Beeson1985,
Longley1995, Longley1997, VanOosten2008, Longley2015}.

Once a PCA---and hence a notion of computability---is fixed,
defining a P-category $\mathfrak{Asm}(A)$ is reasonably
straightforward: objects are assemblies, and morphisms are
functions which are `computable' on the level of realizers. The
arrows are pairs $(f, r)$ where $f$ is a function on the
underlying set of each assembly, and $r$ is an element of the PCA
that tracks the function $f$. Two arrows are related if and only
if they define the same function on the underlying sets. Finally,
to define an endoexposure all we need to do is display the effect
of each tracking element $r$ on the realizers. It then transpires
that this exposure is a product-preserving idempotent comonadic
exposure on $\mathfrak{Asm}(A)$.

\subsection{Partial Combinatory Algebras}
  \label{sec:pca}

The definition of a PCA is deceptively simple:

\begin{defn}
  A \emph{partial combinatory algebra (PCA)} $(A, \cdot)$ consists
  of a carrier set $A$ and a partial binary operation $\cdot : A
  \times A \parfunc A$ such that there exist $\mb{K}, \mb{S} \in
  A$ with the properties that \[
    \mb{K} \cdot x\downarrow, 
      \quad
    \mb{K} \cdot x \cdot y \simeq y,
      \quad
    \mb{S} \cdot x \cdot y\downarrow,
      \quad
    \mb{S} \cdot x \cdot y \cdot z \simeq x \cdot z \cdot (y \cdot z)
  \] for all $x, y, z \in A$.
\end{defn}

\noindent The paradigmatic example comes from ordinary
computability theory, as presented by e.g. \cite{Cutland1980,
Odifreddi1992, Rogers1987}. It is not very difficult to use the
s-m-n theorem to cook up indices that behave like $\mb{S}$ and
$\mb{K}$, to the effect that

\begin{thm}[Kleene's First Model]
  The applicative structure $K_1 = (\mathbb{N},\ \cdot\ )$,
  where \[
    x \cdot y \simeq \phi_x(y)
  \] is a partial combinatory algebra.
\end{thm} 

\subsubsection*{Combinatory Completeness}

Even though the definition of a PCA is remarkably simple, there is
more to it than meets the eye. The structure of $\mb{S}$ and
$\mb{K}$ suffice to obtain a property known as \emph{combinatory
completeness}: every syntactic function on the PCA formed by
variables, applications and constants can be `internalised' as an
element of the PCA. This results originates from combinatory
logic, and is well-known in the study of untyped
$\lambda$-calculus---see \cite{Barendregt1984}.

Once combinatory completeness is obtained, standard tricks from
untyped $\lambda$-calculus can be used to represent first-order
data, but also greatly simplify calculations.  Nevertheless, let
us remind the reader that \emph{not all the common rules of the
untyped $\lambda$-calculus hold in a PCA}, so caution is
advised. The particular presentation in this section is due to
\cite{Longley1995}.

\begin{defn}
  Let $V$ be an infinite set of variables. The set $\mathcal{E}(A)$
  of \emph{terms} or \emph{formal expressions} over a PCA $(A,
  \cdot)$ is defined as the least set that satisfies the following
  conditions: \[
    A \subseteq \mathcal{E}(A), 
      \quad
    V \subseteq \mathcal{E}(A),
      \quad
    \begin{prooftree}
      s \in \mathcal{E}(A) 
        \quad
      t \in \mathcal{E}(A)
        \justifies
      (s \cdot t) \in \mathcal{E}(A)
    \end{prooftree}
  \]
\end{defn}

Conventionally, we will use $e, s, t, u, v, \dots$ as
metavariables ranging over $\mathcal{E}(A)$, and we will write
$s[t/x]$ for the formal expression obtained by substituting $t \in
\mathcal{E}(A)$ for every occurrence of the variable $x$ in the
formal expression $s \in \mathcal{E}(A)$.

A formal expression is \emph{closed} if it contains no variables.
Write $\fv{e}$ for the set of free variables of expression $e \in
\mathcal{E}(A)$. Also, write $e\downarrow$ (``$e$ denotes''),
where $e$ is a closed formal expression, to mean that that, if the
formal expression is interpreted as an actual algebraic expression
in the standard way, composition throughout is defined and it
denotes an element. This implies that all subexpressions also
denote: if $s \cdot t\downarrow$, then $s\downarrow$ and
$t\downarrow$. Otherwise, we write $e\uparrow$ to mean that
partiality kicks in, and the expression does not denote an element
in the PCA.

We also notationally distinguish two equalities: the strict
equality, $s = t$, where both $s\downarrow$ and $t\downarrow$ and
they denote the same element; and the Kleene equality, $s \simeq
t$, which holds when $s$ and $t$ are both undefined, or both
defined and denote the same element.

Finally, the above notions are straightforwardly extended to open
terms, by substituting for all variables: let the variables of $s,
t \in \mathcal{E}(A)$ be amongst $x_1, \dots, x_n$; then, for
example \begin{align*}
  &e\downarrow \quad &\text{just if} \quad
    &\forall a_1, \dots, a_n \in A. \;
    e[\vec{a}/\vec{x}]\downarrow \\
  &s \simeq t \quad &\text{just if} \quad
    &\forall a_1, \dots, a_n \in A. \;
    s[\vec{a}/\vec{x}] \simeq t[\vec{a}/\vec{x}]
\end{align*} for the obvious generalisation to simultaneous
substitution. We can now $\lambda$-abstract:

\begin{thm}[Combinatory Completeness]
  Let $(A, \cdot)$ be a PCA.
  For any $e \in \mathcal{E}(A)$, there exists a formal expression
  $\lambda^\ast x. e \in \mathcal{E}(A)$, where 
  $\fv{\lambda^\ast x. e} = \fv{e} - \{x\}$, such that \[
    \lambda^\ast x. e\downarrow
  \] and \[
    (\lambda^\ast x. e)a \simeq e[a/x]
  \] for all $a \in A$.
\end{thm}
\begin{proof}
  Define
  \begin{align*} 
    \lambda^\ast x.\; x 
      &\myeq \mb{S} \cdot \mb{K} \cdot \mb{K} & \\
    \lambda^\ast x.\; t 
      &\myeq \mb{K} \cdot t &\text{ if $t \in A \cup V$ and $t \not\equiv x$} \\
    \lambda^\ast x.\; s \cdot t
      &\myeq \mb{S} \cdot (\lambda^\ast x. s) \cdot (\lambda^\ast x. t) &
  \end{align*} Then $\lambda^\ast x.  e\downarrow$ by the
  definedness conditions for $\mb{S}$ and $\mb{K}$. The rest follows
  by induction.
\end{proof}

Let us be pedantic and reiterate the warnings: Kleene equality is
\emph{not} respected by the $\lambda^\ast x$ operation on terms,
and the obvious $\beta$-rules do not hold---observe that the
operand above has to be a constant!  Longley carefully develops
correct $\beta$-rules for this language of terms in \citep[\S
1.1.2]{Longley1995}.

\subsubsection*{Some common encodings}

We will need the following combinators:
\begin{align*}
  \mathbf{I} &\myeq \mathbf{S}\mathbf{K}\mathbf{K} \\
  \mathbf{B} &\myeq \lambda^\ast f. \lambda^\ast g. \lambda^\ast x. f (g x)
\end{align*}

\noindent It is easy to define selection and pairs.  Let
\begin{align*}
  \textsf{true}  &\myeq \lambda^\ast a b . a \\
  \textsf{false} &\myeq \lambda^\ast a b . b \\
  \textsf{if}    &\myeq \lambda^\ast xyz. xyz \\
  \textsf{pair}  &\myeq \lambda^\ast xyz. zxy \\
  \textsf{fst}   &\myeq \lambda^\ast p. p(\textsf{true})\\
  \textsf{snd}   &\myeq \lambda^\ast p. p(\textsf{false})
\end{align*} We always have $\textsf{if}\ x\ y \downarrow$,
and $\textsf{pair}\ x\ y \downarrow$. Furthermore, the following
equalities hold: \begin{align*}
  \textsf{fst}\ (\textsf{pair}\ x\ y) &= x \\
  \textsf{snd}\ (\textsf{pair}\ x\ y) &= y \\
  \textsf{if}\ \textsf{true}\ y\ z &= y \\
  \textsf{if}\ \textsf{false}\ y\ z &= z
\end{align*}

\noindent Encoding numbers is not more difficult, and---like
\citep{Longley1995} and \citep{Longley1997}---we use a trick due
to Curry. Let
\begin{align*}
  \overline{0} &= \mathbf{I} \\
  \overline{n+1} &= \textsf{pair}\ \textsf{false}\ \overline{n}
\end{align*}
Then we may let $\textsf{succ} \myeq \lambda^\ast x.\
\textsf{pair}\ \textsf{false}\ x$, so that $\textsf{succ}\
\overline{n} = \overline{n+1}$.  To check if a number is zero, use
$\textsf{iszero}\ \myeq \textsf{fst}$ so that \[
  \textsf{iszero}\ \overline{0} 
    =
  \mathbf{I}\ (\textsf{true})
    =
  \textsf{true}
\] whereas \[
  \textsf{iszero}\ \overline{n+1}
    =
  \textsf{fst}\ (\textsf{pair}\ \textsf{false}\ \overline{n})
    =
  \textsf{false}
\] Finally, we can define the predecessor function by \[
  \textsf{pred} \myeq \lambda^\ast x.\
    \textsf{if}\ (\textsf{iszero}\ x)\ 
      \overline{0}\
      (\textsf{snd}\ x)
\]

\subsection{Assemblies and Modest Sets}

\begin{defn}
  An \emph{assembly} $X$ on $A$ consists of a set $\left\lvert X
  \right\rvert$ and for each $x \in \left\lvert X \right\rvert$
  a non-empty subset $\left\|x\right\|_X \subseteq A$. If $a \in
  \left\|x\right\|_X$, we say that \emph{$a$ realizes $x$}.
\end{defn}

\begin{defn} For two assemblies $X$ and $Y$, a function $f :
\bars{X} \rightarrow \bars{Y}$ is said to be \emph{tracked by $r
\in A$} just if, for all $x \in \bars{X}$ and $a
\in \dbars{x}_X$, we have \[
  r \cdot a \downarrow
    \quad \text{and} \quad 
  r \cdot a \in \dbars{f(x)}_Y
\] \end{defn}

\begin{defn} An assembly $X$ on $A$ is a \emph{modest set} just if no
element of $A$ realizes two elements of $\bars{X}$. That is, \[
  x \neq x' \quad \Longrightarrow \quad \dbars{x}_X \cap
  \dbars{x'}_X = \emptyset
\] \end{defn}

\noindent It is not hard to see that for each PCA $A$ we can
define a category $\mathbf{Asm}(A)$, with objects all assemblies
$X$ on $A$, and morphisms $f : X \rightarrow Y$ being functions $f
: \bars{X} \rightarrow \bars{Y}$ that are tracked by some $r \in
A$. In fact,

\begin{thm} 
  Assemblies and trackable morphisms between them form a
  category $\mathbf{Asm}(A)$ that is cartesian closed, has
  coproducts, as well as a natural numbers object.
\end{thm}

\noindent It is not clear who originated the---admittedly very
intuitive---definition of $\mathbf{Asm}(A)$, and who first proved
the above theorem. The identification of assemblies as the
$\lnot\lnot$-separated objects of the effective topos can be found
in the work of \cite{Hyland1982}. \cite{Longo1991} refer to
$\mathbf{Asm}(A)$ as the category of $\omega$-sets, and so does
\cite{Jacobs1999}. For more details, see \citep{Longley1995} or
\citep{Longley1997}. 

A special subcategory of assemblies is of particular interest:

\begin{thm}
  The full subcategory of $\mathbf{Asm}(A)$ consisting only of
  objects which are modest sets, which we denote by
  $\mathbf{Mod}(A)$, inherits the cartesian closed, coproduct, and
  natural number object structure from the category of assemblies.
\end{thm}

\noindent The category of modest sets---or its equivalent
presentation in terms of PERs on the PCA $A$---seems to have
originated in unpublished work by Turing, and later used by
\cite{Gandy1956, Gandy1959}: see \citep{Hyland2016}. In semantics,
PERs were used independently by \cite{Scott1976} and
\cite{Girard1972}.  Accessible presentations may be found in
\cite{Mitchell1996} or \cite{Crole1993}.

\subsection{Passing to a P-category}
  \label{sec:passpcat}

The lack of intensionality in the category $\mathbf{Asm}(A)$ is
blatant: to elevate a function $f : \bars{X} \rightarrow \bars{Y}$
to a morphism $f : X \rightarrow Y$, we only require that
\emph{there exists} a $r \in A$ that tracks it: as soon as this is
witnessed, we throw away the witness. For all the reasons
discussed in \S\ref{sec:catint} we have to move to P-categories to
mend this.

The P-category of assemblies on $A$, denoted $\mathfrak{Asm}(A)$,
is defined as follows. Its objects are once more all assemblies
$X$ on $A$. Given assemblies $X$ and $Y$, the P-set
$\mathfrak{Asm}(X, Y)$ is defined by having underlying set \[
  \bars{\mathfrak{Asm}(X, Y)} \myeq
    \setcomp{ (f : \bars{X} \rightarrow \bars{Y}, r \in A) }
            { r \text{ tracks } f}
\] and \[
  (f, r) \sim_{\mathfrak{Asm}(X, Y)} (g, s)
    \quad\text{just if}\quad
  f = g
\] For $(f, r) : X \rightarrow Y$ and $(g, s) : Y \rightarrow Z$,
we define composition by \[
  (g, s) \circ (f, r) \myeq (g \circ f, \mb{B} \cdot s \cdot r)
\] Notice that for any $x \in \bars{X}$, $a \in \dbars{x}_X$
implies $p \cdot a \in \dbars{f(x)}_Y$, which implies $q \cdot (p
\cdot a) \in \dbars{g(f(x))}_Z$, and as $\mb{B} \cdot q \cdot p
\cdot a \simeq q \cdot (p \cdot a)$ it follows that $(g \circ f,
\mb{B} \cdot p \cdot q)$ is an arrow $X \rightarrow Z$. 
It is easy to see that composition is a P-function: $\sim$ only
refers to the underlying `extensional' functions. Composition of
set-theoretic functions is associative and the identity function
is its unit. That said, we define the identity $id_X : X
\rightarrow X$ to simply be $(id_{\bars{X}}, \mathbf{I})$.

\subsubsection*{Finite Products}

It is not hard to show that 

\begin{prop} 
  The category $\mathfrak{Asm}(A)$ has binary products.
\end{prop}

\begin{proof}
  The construction essentially follows the underlying
  structure of products in the category of sets, but augments it
  with tracking elements. For assemblies $X$ and $Y$, we define \[
    \bars{X \times Y} \myeq \bars{X} \times \bars{Y},
      \qquad
    \dbars{(x, y)}_{X \times Y} \myeq \setcomp{ \textsf{pair}\,a\ b }
    { a \in \dbars{x}_{X}, b \in \dbars{y}_Y }
  \] The projections are the following arrows:
  \begin{align*}
    \pi_1  &\myeq (\pi_1 : \bars{X} \times \bars{Y} \rightarrow \bars{X},
      \textsf{fst}) \\
    \pi_2 &\myeq (\pi_2 : \bars{X} \times \bars{Y} \rightarrow \bars{Y},
      \textsf{snd})
  \end{align*} Define the P-function $\langle -, -\rangle$ by \[
    \left\langle (f, r), (g, s) \right\rangle 
      \myeq \left(\langle f, g \rangle, \lambda^\ast c.\
    \textsf{pair}\ (r\;c)\ (s\;c)\right)
  \] It is easy to see that this is a P-function.
  We compute that \[
    \pi_1 \circ \langle (f, r), (g, s) \rangle
      \myeq
    (\pi_1 \circ f,
      \mathbf{B}
        \cdot \mathsf{fst} 
        \cdot (\lambda^\ast c.\, \mathsf{pair}\ (r\;c)\ (s\;c)))
      \sim
    (f, r)
  \] and similarly for the other two equations.
\end{proof}

\begin{prop}
  The P-category $\mathfrak{Asm}(A)$ has a terminal object.
\end{prop}

\begin{proof}
  Define $\mathbf{1} \in \mathfrak{Asm}(A)$ by \[
    \bars{\mathbf{1}} \myeq \{\ast\},
      \qquad
    \dbars{\ast}_\mathbf{1} \myeq \{ \overline{0} \}
  \] and, for $A \in \mathfrak{Asm}(A)$, let ${!}_A \myeq (a \mapsto
  \ast, \mathbf{K} \cdot \overline{0}) : A \rightarrow
  \mathbf{1}$, which is unique (up to $\sim$).
\end{proof}

\noindent Hence,

\begin{thm}
  $\mathfrak{Asm}(A)$ has finite products.
\end{thm}

\subsubsection*{Exponentials}

Given assemblies $X$ and $Y$, we define \[
  \bars{Y^X} \myeq \setcomp{ f }{ (f, r) : X \rightarrow Y }, \qquad
  \dbars{f}_{Y^X} \myeq \setcomp{ r }{ r \text{ tracks } f }
\] Let \[
  \textsf{ev}_{X, Y} \myeq
    \left(
      (f, x) \mapsto f(x),
      \lambda^\ast p.\, 
        (\textsf{fst } p)\,
        (\textsf{snd } p)
    \right) : Y^X \times X \rightarrow Y
\] Define a P-function \[
  \lambda_C : 
    \mathfrak{Asm}(A)(C \times X, Y)
      \rightarrow
    \mathfrak{Asm}(A)(C, Y^X)
\] by \[
  (f, r) \mapsto 
    \left(
      z \mapsto (x \mapsto f(z, x)),
      \lambda^\ast c.\
        \lambda^\ast a.\
          r (\textsf{pair}\ c\ a)
    \right)
\] It is again easy to see that this is a P-function, and one can
verify that this is the exponential. Therefore,

\begin{thm}
  $\mathfrak{Asm}(A)$ is cartesian closed.
\end{thm}

\subsubsection*{The Lifted Assembly for $K_1$}

We only mention one other indispensable construction, which
embodies partiality in the computable setting of assemblies on
$K_1$. Given an assembly $X \in \mathfrak{Asm}(K_1)$, the
\emph{lifted assembly} $X_\bot$ is defined to be \[
  \bars{X_\bot} \myeq \bars{X} + \{ \bot \},
    \qquad
  \dbars{x}_{X_\bot} \myeq 
    \begin{cases}
      \setcomp{ r }{ r \cdot \overline{0}\downarrow 
                      \text{ and }
                     r \cdot \overline{0} \in \dbars{x}_X }
        &\text{for } x \in \bars{X} \\
      \setcomp{ r }{r \cdot \overline{0}\uparrow}
        &\text{for } x = \bot
    \end{cases}
\] for some chosen element of the PCA $\overline{0}$.
Elements of $X_\bot$ are either elements of $X$, or the undefined
value $\bot$. Realizers of $x \in \bars{X}$ are `computations' $r
\in A$ which, when run (i.e. given the dummy value $\overline{0}$
as argument) return a realizer of $x$. A computation that does not
halt when run represents the undefined value.

Bear in mind that this definition of the lifted assembly is only
useful in $K_1$. In particular, it does not work at all if the PCA
is total. There are other, more involved ways of defining the
lifted assembly: see \cite{Longley1997} for a rather elegant and
uniform method.

\subsection{The Exposure}

\begin{thm}
  There is an exposure \[
    \Box : \mathfrak{Asm}(A) \looparrowright \mathfrak{Asm}(A)
  \] for any PCA $A$.
\end{thm}
\begin{proof}
  For an assembly $X \in \mathfrak{Asm}(A)$, let $\Box
  X$ be the assembly defined by \begin{align*}
    \bars{\Box X} &\myeq \setcomp{ (x, a) }
      { x \in \bars{X}, a \in \dbars{x}_X} \\
    \dbars{(x, a)}_{\Box X} &\myeq \{\ a\ \}
  \end{align*} Given $(f, r) : X \rightarrow Y$, we define $
    \Box(f, r) = (f_r, r) : \Box X \rightarrow \Box Y
  $ where \begin{align*}
    f_r : \bars{\Box X} &\longrightarrow \bars{\Box Y} \\
	     (x, a)     &\longmapsto (f(x), r \cdot a)
  \end{align*} Each element $(x, a) \in \bars{\Box X}$ has a unique
  realizer, $a$. As $a \in \dbars{x}_X$, and $f$ is tracked by $r$,
  we see that $r \cdot a \downarrow$ and $r \cdot a \in
  \dbars{f(x)}_Y$, so that $(f(x), r \cdot a) \in \bars{\Box Y}$.
  It follows that $r$ tracks $f_r$, and so $(f_r, r)$ is an arrow
  $\Box X \rightarrow \Box Y$.

  To prove that $\Box$ preserves composites, observe that for
  arrows $(f, r) : A \rightarrow B$ and $(g, s) : B \rightarrow C$,
  we have \[
    (x, a)
      \overset{f_r}{\longmapsto}
    (f(x), r \cdot a)
      \overset{g_s}{\longmapsto}
    \left(g(f(x)), s \cdot (r \cdot a)\right)
  \] and as $\mathbf{B} \cdot s \cdot r \cdot a \simeq s \cdot (r
  \cdot a)$, we have \[
    g_s \circ f_r = (g \circ f)_{\mathbf{B} \cdot s \cdot r}
  \] which is of course tracked by $\mathbf{B} \cdot s \cdot r$.
  Hence $\Box(g \circ f) \sim \Box g \circ \Box f$. Regarding
  identities, notice that if $id_X : X \rightarrow X$ is the
  identity arrow, then \[
    (x, a) 
      \xmapsto{\left(id_X\right)_{\mathbf{I}}}
    (x, \mathbf{I} \cdot a)
  \] and as $\mathbf{I} \cdot a \simeq a$, the latter is equal to
  $(x, a)$, so $\left(id_X\right)_\mathbf{I} = id_{\bars{\Box X}}$.
  Hence $\Box id_X \sim id_{\Box X}$.

  Finally, we need to show that intensional equality implies
  extensional equality. Suppose $\Box (f, r) \sim \Box (g, s)$.
  That is equivalent to $f_r = g_s$, which in turn gives us both
  $f = g$ and also $r \cdot a \simeq s \cdot a$ for all $a \in
  \dbars{x}_X$. The first implies $(f, r) \sim (g, s)$.
\end{proof}

\subsubsection*{Intensional Equality}

When showing that $\Box$ is an exposure, we inadvertently
characterised intensional equality up to $\Box$:

\begin{lem}
  $(f, r) \approx_\Box (g, s) : X \rightarrow Y$ precisely when $f_r =
  g_s$, i.e. \[
  f = g \quad \text{and} \quad 
    \forall x \in \bars{X}.\
    \forall a \in \dbars{x}_X.\
      r \cdot a \simeq s \cdot a
  \]
\end{lem}

\noindent This is indeed the meaning we expected of intensional
equality: $(f, r)$ and $(g, s)$ are not only equal extensionally,
but they also have \emph{the same effect on realizers}. Notice
that, unless we are in a highly extensional environment (which not
all PCAs are), this is very far from \emph{strict} equality.
Instead, it is something in between.

We can also use this characterisation of intensional equality to
speak about the realizer structure of assemblies, in particular by
characterising which objects of $\mathfrak{Asm}(A)$ are univalent
(recall Definition \ref{def:univalent}).

\begin{defn}
  An assembly $X$ has \emph{unique realizers} just if
  $\dbars{x}_X$ is a singleton, for each $x \in \bars{X}$.
\end{defn}

\begin{lem}
  The univalent objects of $\mathfrak{Asm}(A)$ are precisely those
  which have unique realizers.
\end{lem}
\begin{proof}
  Suppose $X$ is univalent. To each realizer $a \in
  \dbars{x}_X$, there corresponds an arrow \[
    \hat{x}_a \myeq (\ast \mapsto x, \lambda^\ast c.\ a)
      : \mathbf{1} \rightarrow X
  \] But if $a, b \in \dbars{x}_X$, then $\hat{x}_a \sim
  \hat{x}_b$, as they share the same function component $(\ast
  \mapsto x)$. As $X$ is univalent, it follows that $\hat{x}_a
  \approx_\Box \hat{x}_b$, so \[
    a \simeq (\lambda^\ast c.\ a) \cdot \overline{0} 
      \simeq (\lambda^\ast c.\ b) \cdot \overline{0}
      \simeq b
  \] Conversely, suppose $X$ has unique realizers. For any two
  extensionally equal arrows $(f, r), (f, s) : Y \rightarrow
  X$, it is not very hard to see that $f_r = f_s$: we have that
  $\pi_1(f_r(y, b)) = f(y) = \pi_1(f_s(y, b))$ for any $b \in
  \dbars{y}_Y$. Thus, as $f(y)$ is uniquely realized by only one
  $a$, their second components are equal too, and $(f, r)
  \approx_\Box (f, s)$.
\end{proof}

\subsubsection*{$\Box$ is cartesian}

It so happens that projections behave nicely under exposures.

\begin{thm}
  $\Box : \mathfrak{Asm}(A) \looparrowright \mathfrak{Asm}(A)$ is
  a cartesian exposure.
\end{thm}

\noindent Recall that $\pi_1 \circ \langle (f, r), (g, s) \rangle
= (f, d)$, where \[
  d \myeq \mathbf{B}
    \cdot \mathsf{fst}
    \cdot (\lambda^\ast c.\, \mathsf{pair}\ (r\;c) (s\;c))
\] so that the function component of $\Box (\pi_1 \circ \langle
(f, r), (g, s) \rangle)$ is $f_d(z, c) = (f(z), d \cdot c)$. We
compute that \[
    d \cdot c
  =
    \mathbf{B}
      \cdot \mathsf{fst}
      \cdot (\lambda^\ast c.\ \mathsf{pair}\ (r\;c) (s\;c))
      \cdot c
  =
    r \cdot c
\] which is to say that $f_d = f_r$, and hence $\Box(\pi_1 \circ
\langle (f, r), (g, s) \rangle) \sim \Box (f, r)$. The calculation
is similar for the other projection.

For the third equation, it is easy to calculate that, for any
arrow $h : C \rightarrow X \times Y$, the function component of
$\Box\langle \pi_1 \circ (h, r), \pi_2 \circ (h, r) \rangle$ is
$h_s$, where \[
  s = \lambda^\ast c.\
    \textsf{pair}\
      (\mathbf{B} \cdot \mathsf{fst} \cdot r \cdot c)\
      (\mathbf{B} \cdot \mathsf{snd} \cdot r \cdot c)
\] We proceed by calculating that \[
  s \cdot c
  = \textsf{pair}\ 
      (\mathbf{B} \cdot \mathsf{fst} \cdot r \cdot c)\
      (\mathbf{B} \cdot \mathsf{snd} \cdot r \cdot c)
  = \textsf{pair}\
      (\textsf{fst}\ (r \cdot c))\
      (\textsf{snd}\ (r \cdot c))
\] so that \[
  h_s(z, c) =
    (h(z),
      \textsf{pair}\ 
	(\textsf{fst}\ (r \cdot c))\ 
	(\textsf{snd}\ (r \cdot c)))
\] The argument would now be complete were our pairing surjective,
but this is not so in general PCAs. However, we know that $c \in
\dbars{z}_C$ for some $z \in \bars{C}$, so $r \cdot c \in
\dbars{(x, y)}_{X \times Y}$ with $h(z) = (x, y)$. Hence, \[
  r \cdot c = \textsf{pair}\ a\ b
\] for some $a \in \dbars{x}_X$ and $b \in \dbars{y}_Y$, and hence
\[
  s \cdot c = \textsf{pair}\ a\ b
\] as well. It follows that $h_s = h_r$, where $r$ was the
original tracking element of $(h, r)$. We have finally obtained
that \[
  \Box\langle \pi_1 \circ (h, r), \pi_2 \circ (h, r) \rangle 
    \sim \Box (h, r)
\] The final thing to check is that any arrow into the terminal
object $\mathbf{1}$ is intensionally equal to the canonical one;
this follows, as $\mathbf{1}$ has only one element with a unique
realizer. 

\subsubsection*{$\Box$ is weakly cartesian closed}

It is also the case that $\Box : \mathfrak{Asm}(A) \expo{}
\mathfrak{Asm}(A)$ is \emph{weakly cartesian closed}
(\S\ref{sec:wcccexpo}), in the sense that the equation \[
  \textsf{ev} \circ (\lambda(f, r) \times id_X)
    \approx_\Box
  (f, r)
\] holds for any $(f, r) : C \times X \rightarrow Y$. To prove
this one needs another tiring but easy calculation like the one
showing that $\Box$ is cartesian: it is of no interest, save
another use of the fact that it uses the trick that any $z \in
\dbars{(c, x)}_{C \times X}$ is always of the form $z =
\textsf{pair}\ i\ j$ for $i \in \dbars{c}_C$ and $j \in
\dbars{x}_X$.

\subsubsection*{Preservation of Products}

Define \[
  m_0 \myeq (\ast \mapsto (\ast, \overline{0}), \mathbf{I})
    : \mathbf{1} \rightarrow \Box \mathbf{1}
\] which maps the unique element of $\mathbf{1}$ to the pair of
itself and its unique realizer, and is realized by the identity
combinator. This is a P-isomorphism with inverse \[
  {!}_{\Box \mathbf{1}} \myeq
    ((\ast, \overline{0}) \mapsto \ast, \mathbf{I})
      : \Box \mathbf{1} \rightarrow \mathbf{1}
\] Also, define $m_{A, B} : \Box A \times \Box B \rightarrow \Box
(A \times B)$ by $m_{A, B} \myeq (w_{A, B}, \mathbf{I})$, where
\begin{align*}
  w_{A, B} : \bars{\Box A} \times \bars{\Box B} &\longrightarrow \bars{\Box (A \times B)}\\
             ((x, a), (y, b)) &\longmapsto ((x, y), \textsf{pair}\ a\ b)
\end{align*}

\noindent The only realizer for the pair $((x, a), (y, b))$ is
$\textsf{pair}\ a\ b$, so the identity combinator tracks $w_{A,
B}$. Then $\langle \Box\pi_1, \Box\pi_2 \rangle : \Box (A \times
B) \rightarrow \Box A \times \Box B$ is equal to
$(v_{A, B}, \mathbf{I})$, where \begin{align*}
  v_{A, B} : \bars{\Box (A \times B)} &\longrightarrow \bars{\Box A} \times \bars{\Box B} \\
	     ((x, y), r) &\longmapsto
	      ((x, \textsf{fst} \cdot r), (y, \textsf{snd} \cdot r))
\end{align*} and, as before, it is easy to see that this is an
inverse to $m_{A, B}$, as $r$ is necessarily of the form
$\textsf{pair}\ a\ b$. Hence,

\begin{thm}
  The exposure $\Box : \mathfrak{Asm}(A) \expo{}
  \mathfrak{Asm}(A)$ is product-preserving.
\end{thm}

\subsubsection*{Comonadicity and Idempotence}

\begin{prop}
  There exists a natural transformation of exposures, \[
    \epsilon : \Box \natexp \mathsf{Id}_{\mathfrak{Asm}(A)}
  \]
\end{prop}

\begin{proof}
  Define $\epsilon_X = (u_X, \mathbf{I})$, where \begin{align*}
    u_X : \bars{\Box X} &\rightarrow \bars{X} \\
	  (x, a)        &\longmapsto x
  \end{align*} If $b \in \dbars{(x, a)}_{\Box X} = \{ a \} $, then $b = a$ and
  $\mathbf{I} \cdot b \simeq b = a \in \dbars{x}_X$, so $u_X$ is
  indeed tracked by $\mathbf{I}$. To show naturality for $(f, r) :
  X \rightarrow Y$, we chase around the diagram: \[
    \begin{tikzcd}
      (x, a)
	\arrow[r, mapsto, "f_r"]
	\arrow[d, mapsto, swap, "u_X"]
      & (f(x), r \cdot a)
	\arrow[d, mapsto, "u_Y"] \\
      x
	\arrow[r, mapsto, swap, "f"]
      & f(x)
    \end{tikzcd}
  \] Hence $u_Y \circ f_r = f \circ u_X$, so the square commutes up
  to $\sim$.  Bear in mind that the tracking
  element along one composite is $\mathbf{B} \cdot r \cdot
  \mathbf{I}$, whereas along the other composite it is $\mathbf{B}
  \cdot \mathbf{I} \cdot r$.
\end{proof}

Let us now investigate the structure of $\Box^2 X$, for any
assembly $X$. For each $(x, a) \in \bars{\Box X}$, we have that
$\dbars{(x, a)}_{\Box X} = \{ a \}$. Thus, we can infer that \[
  \bars{\Box^2 X} = \setcomp{((x, a), a)}{a \in \dbars{x}_X}
\] and that, for any $(f, r) : X \rightarrow Y$, \begin{align*}
  \Box^2 f : \bars{\Box^2 X} &\longrightarrow \bars{\Box^2 Y} \\
        ((x, a), a)        &\longmapsto ((f(x), r \cdot a), r \cdot a)
\end{align*}

\begin{prop}
  There exists a natural transformation of exposures, \[
    \delta : \Box \natexp{} \Box^2
  \]
\end{prop}

\begin{proof}
  Define $\delta_X = (v_X, \mathbf{I})$, where \begin{align*}
    v_X : \bars{\Box X} &\longrightarrow \bars{\Box^2 X} \\
	  (x, a)        &\longmapsto ((x, a), a)
  \end{align*} If $a \in \dbars{x}_X$, then $\mathbf{I} \cdot a
  \simeq a \in \{a\} = \dbars{((x, a), a)}_{\Box^2 X}$, so
  $\mathbf{I}$ indeed tracks $v_X$. To show naturality for a given
  arrow $(f, r) : X \rightarrow Y$, we chase around the diagram: \[
    \begin{tikzcd}
      (x, a)
	\arrow[r, mapsto, "f_r"]
	\arrow[d, mapsto, swap, "v_X"]
      & (f(x), r \cdot a)
	\arrow[d, mapsto, "v_Y"] \\
      ((x, a), a) 
	\arrow[r, mapsto, swap, "\left(f_r\right)_r"]
      & \left((f(x), r \cdot a), r \cdot a\right)
    \end{tikzcd}
  \] and so the diagram commutes up to $\sim$.  Note that the
  tracking elements along the composites are respectively
  $\mathbf{B} \cdot \mathbf{I} \cdot r$ and $\mathbf{B} \cdot r
  \cdot \mathbf{I}$.
\end{proof}

\noindent It is not difficult to see that the components of 
$\delta : \Box \natexp{} \Box^2$ are actually isomorphisms.
Hence, putting everything together:

\begin{thm}
  $(\Box, \epsilon, \delta)$ is a product-preserving idempotent
  comonadic exposure.
\end{thm}

\begin{proof}
  It suffices to verify the coherence conditions.  Regarding the
  first one: \[
    \begin{tikzcd}
      (x, a)
        \arrow[r, mapsto, "v_X"]
        \arrow[d, mapsto, "v_X", swap]
      & ((x, a), a)
        \arrow[d, mapsto, "v_{\Box X}"] \\
      \left((x, a), a \right)
        \arrow[r, mapsto, "\left(v_X\right)_{\mathbf{I}}", swap]
      & (((x, a), a), \mathbf{I} \cdot a) = (((x, a), a), a)
    \end{tikzcd}
  \] which commutes, since $\mathbf{I} \cdot a \simeq
  a$. Both the tracking elements along each composite are $\mathbf{B}
  \cdot \mathbf{I} \cdot \mathbf{I}$, so the diagram actually
  commutes on the nose. Regarding the second one: \[
    \begin{tikzcd}
      (x, a)
        \arrow[r, mapsto, "v_X"]
        \arrow[d, mapsto, "v_X", swap]
      & ((x, a), a)
        \arrow[d, mapsto, "u_{\Box X}"] \\
          \left((x, a), a \right)
        \arrow[r, mapsto, "\left(u_X\right)_{\mathbf{I}}", swap]
      & (x, \mathbf{I} \cdot a) = (x, a)
    \end{tikzcd}
  \] which commutes, since $\mathbf{I} \cdot a \simeq
  a$. Once more, both tracking elements are $\mathbf{B} \cdot
  \mathbf{I} \cdot \mathbf{I}$, so commutation is again on the nose.
\end{proof}

\subsection{Weak Extensionality and Naturality}
  \label{sec:wextasm}

We have shown that nearly everything in sight behaves well, even
with respect to intensional equality $\approx_\Box$: Proposition
\ref{prop:presint} necessitates that, when we have a
product-preserving (and hence cartesian) exposure, the function \[
  \langle \cdot, \cdot \rangle :
    \mathfrak{Asm}(A)(C, X) \times \mathfrak{Asm}(C, Y)
      \rightarrow
    \mathfrak{Asm}(A)(C, X \times Y)
\] that is implicated in the definition of products respects
intensional equality $\approx_\Box$. The glaring exception, of
course, is the cartesian closed structure: it is not necessarily
that \[
  \lambda_C : 
    \mathfrak{Asm}(A)(C \times X, Y)
      \rightarrow
    \mathfrak{Asm}(A)(C, Y^X)
\] preserves intensional equality.

However, it is interesting to investigate when this might happen.
Let $(f, r) \approx_\Box (f, s) : C \times X \rightarrow Y$ be two
intensionally equal morphisms; we have \begin{equation}
  \label{eq:inteq}
  \forall d \in \dbars{(c, x)}_{C \times X}.\
    r \cdot d \simeq s \cdot d
\end{equation} We can calculate that \begin{align*}
  \lambda(f, r) \myeq (\lambda(f), 
    \lambda^\ast c\,a.\ r(\textsf{pair}\ c\ a)) \\
  \lambda(f, s) \myeq (\lambda(f), 
    \lambda^\ast c\,a.\ s(\textsf{pair}\ c\ a))
\end{align*} Thus, to prove that $\lambda(f, r) \approx_Q \lambda
(f, s)$, all we need to check is that \[
  \forall d \in \dbars{c}_C.\
    \lambda^\ast a.\ r(\textsf{pair}\ c\ a)
      \simeq
    \lambda^\ast a.\ s(\textsf{pair}\ c\ a)
\] By \eqref{eq:inteq}, it is indeed the case that
$r(\textsf{pair}\ c\ a) \simeq s(\textsf{pair}\ c\ a)$, because
the realizers $d \in \dbars{(c, x)}_{C \times X}$ are exactly of
the right form. Nevertheless, \emph{we are not allowed to use that
equation}  under an occurrence of $\lambda^\ast$! The situation
that allows this is the one where the PCA $A$ is \emph{weakly
extensional}.

\begin{defn}
  A PCA $(A, \cdot)$ is \emph{weakly extensional} if it satisfies
  the rule \[
    \begin{prooftree}
        M \simeq N
      \justifies
        \lambda^\ast x.\ M \simeq \lambda^\ast x.\ N
    \end{prooftree}
  \] for any two expressions $M, N$ and any variable $x$.
\end{defn}

\noindent Weak extensionality, also known as rule $(\xi)$, is a
notorious thorn in the study of the correspondence between
combinatory logic and untyped $\lambda$-calculus. \cite[\S
7.3.5(iii)]{Barendregt1984} sets out a finite set of equational
axioms (due to Curry) that suffice to ensure it.

The original plan for this thesis was that $\mathfrak{Asm}(K_1)$
would be a model of Intensional PCF, and L\"ob's rule would
directly correspond to Kleene's Second Recursion Theorem.
Unfortunately, $K_1$ is very likely \emph{not} weakly extensional.
In \S\ref{chap:intsem2} we will develop a slight restriction on
Intensional PCF, which will remove the requirement that
$\lambda(-)$ preserve intensional equality, which is otherwise
necessary.

Since we have come this far, let us also investigate the
\emph{naturality} of $\lambda(-)$, i.e. the equation \[
  \lambda\left(f \circ (g \times id)\right)
    \approx_\Box
  \lambda(f) \circ g
\] under intensional equality. Given $(f, r) :
C \times X \rightarrow Y$ and $(g, s) : C' \rightarrow C$, we
compute that \begin{align*}
  \lambda\left((f, r) \circ \left((g,s) \times id\right)\right)
    &\myeq (\lambda\left(f \circ (g \times id)\right),
            \lambda^\ast c'\,a.\
              (\mb{B} \cdot r \cdot h)
                \cdot
            (\textsf{pair}\ c'\ a)) \\
  \lambda(f, r) \circ (g, s)
    &\myeq (\lambda(f) \circ g,
            \mb{B}
	      \cdot
            (\lambda^\ast c'\;a.\
              r \cdot (\textsf{pair}\ c'\ a))
                \cdot
              s)
\end{align*} where $h \myeq \lambda^\ast d.\ \textsf{pair}\
(\mathbf{B} \cdot s \cdot \textsf{fst} \cdot d)\ (\mathbf{B} \cdot
\mathbf{I} \cdot \textsf{snd} \cdot d)$. It would suffice to prove
that these two realizers, when applied to anything, would return
the same $c'$, namely \[
  \lambda^\ast.\ r \cdot (\textsf{pair}\ (s \cdot c')\ a)
\] This is easy to check with weak extensionality, but it seems
impossible to ensure without it. Therefore, $\lambda(-)$ is
natural up to $\approx_\Box$ if $A$ is weakly extensional.

Finally, let us seize the opportunity to mention that if $A$ is
\emph{extensional}, in the sense that the $\eta$-rule \[
  \lambda^\ast x.\ e\;x \simeq e
\] holds for any expression $e \in \mathcal{E}(A)$, then \[
  \textsf{ev} \circ \left((f, r) \circ id\right)
    \approx_Q
  (f, r) : C \rightarrow Y^X
\] In that case we would say that $\Box : \mathfrak{Asm}(A)
\expo{} \mathfrak{Asm}(A)$ is \emph{cartesian closed}.

\section{Exposures in Homological Algebra}
  \label{sec:homoexpo}

This section aims to support the claim that, even though inspired
by logic and computability, exposures are to be found in other
contexts as well. This lends credibility to the idea that, even if
within logic exposures are a sort of abstract yet well-behaved
G\"odel numbering, the phenomenon of \emph{intensionality}, as
discussed in \S\ref{sec:intensionality} is more general, and can
be found in other areas of mathematics.

We will draw our example from \emph{homological algebra}.
Homological algebra begins once we have \emph{chain complexes}
$C(X)$, i.e. sequences of abelian groups $C_i$ with homomorphisms
\[
  \dots
    \xrightarrow{\partial_{n+1}}
  C_n
    \xrightarrow{\partial_{n}}
  C_{n-1}
    \xrightarrow{\partial_{n-1}}
  \dots
    \xrightarrow{\partial_1}
  C_1
    \xrightarrow{\partial_0}
  C_0
\] such that $\partial_d \circ \partial_{n+1} = 0$. One then forms
the groups of \emph{boundaries} and \emph{cycles}, namely \[
  B_n(X) \myeq im(\partial_{n+1})
    \qquad
  Z_n(X) \myeq ker(\partial_n)
\] The objects of study are then the \emph{homology groups} $H_n$,
defined by \[
  H_n(X) \myeq Z_n(X) / B_n(X)
\] 

\noindent A natural question arises: what if we make it so we
never \emph{actually} have to take quotients?

\subsection{The P-category $\mathfrak{Grp}$}

Instead of taking the quotient $G/N$ of a group $G$ by one of its
normal subgroups $N$, we will instead merely keep the tuple \[
  (G, N)
\] and work with it: we will consider this as $G/N$, even though
the two components are kept separately.

Often in homology one considers maps between homology groups,
which are group homomorphisms of type \[
  f_\ast : G_1/H_1 \rightarrow G_2/H_2
\] Nevertheless, rarely does one work out $G_1/H_1$ exactly before
defining such a $f_\ast$. More commonly, one picks out a
representative of the equivalence class, defines $f$ on that, and
then proves that the outcome is invariant under the choice of
representative. This amounts to defining some $f : G_1 \rightarrow
G_2$ such that \[
  f(H_1) \subseteq H_2
\] Thus, we pick the morphisms $f : (G_1, H_1) \rightarrow (G_2,
H_2)$ to be exactly those homomorphisms. Each one of them induces
a homomorphism $f_\ast : G_1/H_1 \rightarrow G_2/H_2$ as is
customary.

To prove that two such maps $f_\ast, g_\ast : G_1/H_1 \rightarrow
G_2/H_2$ are equal, it suffices to prove that they are pointwise
\emph{homologous}, i.e.  that that $f - g$ takes values only in
$H_2$. This will be exactly our definition of extensional
equality: \[
  f \sim g : (G_1, H_1) \rightarrow (G_2, H_2)
    \quad \text{ just if } \quad
  im(f - g) \subseteq H_2
\] A classic result of basic group theory, viz. \[
  G_1 / N_1 \times G_2 / N_2
    \cong
  (G_1 \times G_2) / (N_1 \times N_2)
\] also implies that we can define \[
  (G_1, N_1) \times (G_2, N_2)
    \myeq (G_1 \times G_2, N_1 \times N_2)
\] and use it to prove that

\begin{thm}
  The P-category $\mathfrak{Grp}$ is cartesian.
\end{thm}

We can now think of the $n$-th homology functor as taking values
in this P-category, or---even better---its subcategory
$\mathfrak{Ab}$ of \emph{abelian groups}, \begin{align*}
  H_n : \textbf{Top}         &\longrightarrow \mathfrak{Ab} \\
        X                    &\longmapsto (Z_n(X), B_n(X)) \\
        f : X \rightarrow Y  &\longmapsto f_\# : (Z_n(X), B_n(X))
        \rightarrow (Z_n(Y), B_n(Y))
\end{align*}

\subsection{Intensionality and Homomorphisms}

The point of not forcing $f$ to be $f_\ast$ is that the action of
$f : (G_1, H_1) \rightarrow (G_2, H_2)$ on each cycle of $G_1$ is
`visible,' \emph{even if that cycle is a boundary}. This is
because $f : G_1 \rightarrow G_2$ is still an actual homomorphism,
which only happens to `respect' a normal subgroup. The way we will
define an exposure on this category is precisely by `exposing' the
action of $f$ on individual cycles, \emph{even if they are
boundaries}.

We shall then define an endoexposure, \[
  C : \mathfrak{Grp} \expo{} \mathfrak{Grp}
\] by \begin{align*}
  C(G, H) &\myeq (G, \{e_G\}) \\
  Cf &\myeq f : (G_1, \{e_{G_1}\}) \rightarrow (G_2, \{e_{G_2}\})
\end{align*}

It is not at all difficult to prove that this is indeed an
exposure: it preserves composition and identities, and indeed if
$Cf \sim Cg$ then $f$ and $g$ are equal, so $im(f - g)$ is just
the identity element.

In fact, this exposure has comonadic structure, which comes for
free. For evaluators, we notice that $\epsilon_{(G, H)} : C(G, H)
\rightarrow (G, H)$ is actually of type $(G, \{e_G\}) \rightarrow
(G, H)$, so it suffices to take the identity, which---in this
context---is a kind of quotient map. Similarly, $\delta_{(G, H)} :
(G, \{e_G\}) \rightarrow (G, \{e_G\})$, so again it suffices to
take the identity. It is trivial that every single diagram in the
definition of evaluator and quoter, as well as that of comonadic
exposure, commutes: all the arrows are identities.

Furthermore, it is not hard to see that $C$ preserves products:
the candidate isomorphisms \[
  m : Q(G_1, H_1) \times Q(G_2, H_2) 
        \rightarrow 
      Q(G_1 \times G_2, H_1 \times H_2)
\] have the same source and target, namely $(G_1 \times G_2,
\{(e_{G_1}, e_{G_2})\})$, so it suffices to take the identity.

\chapter[Intensional Recursion in P-Categories]
  {Intensional Recursion in P-Categories\footnote{A preliminary
  form of the results in this chapter was first published as
  \citep{Kavvos2017a}, which is available at Springer:
  \url{https://dx.doi.org/10.1007/978-3-662-54458-7_32}}}

  \label{chap:irec}

  Armed with the framework of exposures, we can now speak of both
extensional and intensional recursion in categorical terms.

The case of \emph{extensional fixed points (EFPs)} was first
treated by \cite{Lawvere1969, Lawvere2006} in the late 1960s.
However, we will argue that his notion of fixed point is far too
coarse for most applications in logic and computer science.

Instead, we will use exposures to replace that definition with one
that captures intensional recursion, namely that of
\emph{intensional fixed points (IFPs)} (\S\ref{sec:extintfp}). We
begin our investigation by showing that our framework allows for
clear and concise formulations of the classic theorems of
G\"odel, Tarski, and Rice.  The relevant arguments are entirely
algebraic, and it is very clear what logical devices or
assumptions each one requires. The conclusion to be drawn is that,
despite their common use of IFPs, these three arguments have a
fundamentally different flavour. In (\S\ref{sec:lob}) we discuss
the relationship between IFPs and L\"ob's rule in provability
logic.

Then, in \S\ref{sec:lawvere}, we then ask the natural question:
where do IFPs come from? We recall in detail a theorem of Lawvere
which guarantees the existence of EFPs under certain assumptions.
We use exposures to prove the Intensional Recursion Theorem, a
similar theorem that pertains to IFPs instead.

Finally, we examine the nature of both EFPs and IFPs in the three
examples that we presented at length in \S\ref{chap:expoexamples}.
In particular, when viewed through the lens of the exposure on
assemblies (\S\ref{sec:realexpo}), Lawvere's theorem and our
Intensional Recursion Theorem are revealed to be categorical
versions of the First and Second Recursion Theorems of Kleene
respectively, as discussed in \S\ref{chap:srtht}.

\section{Extensional and Intensional Fixed Points}
  \label{sec:extintfp}

\cite{Lawvere1969, Lawvere2006} famously proved a theorem which
guarantees that, under certain assumptions, which we discuss in
\S\ref{sec:lawvere}, there exist fixed points of the following
sort.

\begin{defn}
  An \emph{extensional fixed point (EFP)} of an arrow $t : Y
  \rightarrow Y$ is a point $y : \mathbf{1} \rightarrow Y$ such
  that \[
    t \circ y = y
  \] If every arrow $t : Y \rightarrow Y$ has a EFP, then we say
  that \emph{$Y$ has EFPs}.
\end{defn}

In Lawvere's paper EFPs are a kind of fixed point that, for
logical purposes, \emph{oughtn't} exist. After constructing a
category based on a logical theory (e.g. \textsf{PA}), in a manner
that we have quite closely followed in \S\ref{sec:logicexpo}, he
argues that there can be no \[
  \text{sat} : A \times A \rightarrow \mathbf{2}
\] such that for every formula $\phi : A \rightarrow \mathbf{2}$
there is a point $c_\phi : \mathbf{1} \rightarrow A$ such that \[
  \begin{tikzcd}
    \mathbf{1}
      \arrow[r, "a"]
      \arrow[d, "\langle a{,} c_\phi \rangle", swap]
    & A
      \arrow[d, "\phi"] \\
    A \times A
      \arrow[r, "\text{sat}", swap]
    & \mathbf{2}
  \end{tikzcd}
\] for every point $a : \mathbf{1} \rightarrow A$. In logical
terms, this would amount to the existence of a two-variable
predicate $\text{sat}(-,-)$, and a G\"odel number $\ulcorner \phi
\urcorner$ for each unary predicate $\phi(x)$, such
that \[
  \mathsf{T} \vdash \text{sat}(\ulcorner \phi(x) \urcorner, n)
    \leftrightarrow
  \phi(n)
\] for each $n$. If such a predicate existed, then
\emph{`satisfaction would be definable,'} and we would obtain a
EFP of the arrow $\mathbf{2} \rightarrow \mathbf{2}$ encoding the
logical `not' function. In logical terms, we could obtain a closed
formula $\psi$ such that $\mathsf{T} \vdash \psi \leftrightarrow
\lnot \psi$. This leads to a categorical version of Tarski's
\emph{Undefinability Theorem}: if `truth were definable,' then
substitution would be too, and $\lnot$ would have a fixed point.

Finally, Lawvere obtains a version of G\"odel's \emph{First
Incompleteness Theorem} as follows: given an (external) relation
between closed formulas and points, relating closed formulas to
their `G\"odel number,' if we assume that provability is
`internally decidable' on these G\"odel numbers, and if we assume
that all `truth values' are either true or false, then truth would
be definable, which---by the categorical version of Tarski's
undefinability theorem---it is not.

We can already see that Lawvere's notion of EFPs do not encompass
fixed points that \emph{ought} to exist. For example, the
\emph{diagonal lemma} for \emph{Peano Arithmetic} (henceforth
\textsf{PA}) manufactures a closed formula $\mathbf{fix}(\phi)$
for every formula $\phi(x)$, such that \[
  \textsf{PA} \vdash
    \mathbf{fix}(\phi) 
      \leftrightarrow
      \phi(\ulcorner \mathbf{fix}(\phi) \urcorner)
\] The formula $\mathbf{fix}(\phi)$ occurs asymmetrically: on the
left hand side of the bi-implication it appears as a truth value,
but on the right hand side it appears under a \emph{G\"odel
numbering}, i.e. an assignment $\ulcorner \cdot \urcorner$ of a
numeral to each term and formula of \textsf{PA}. Taking our cue
from the exposure on Peano arithmetic (\S\ref{sec:logicexpo}), we
can generalise this idea to the following, which encompasses this
kind of `asymmetric' fixed point.

\begin{defn}
  Let $Q : \mathfrak{B} \expo{} \mathfrak{B}$ be a
  product-preserving endoexposure.  An \emph{intensional fixed
  point (IFP)} (w.r.t. to $Q$) of an arrow $t : QA \rightarrow A$
  is a point $a : \mathbf{1} \rightarrow A$ such that the
  following diagram commutes up to $\sim$: \[
    \begin{tikzcd}
      \mathbf{1}
	\arrow[r, "a"]
	\arrow[d, "m_0", swap]
      & A \\
      Q\mathbf{1}
	\arrow[r, "Qa", swap]
      & QA
	\arrow[u, "t", swap]
    \end{tikzcd}
  \] An object $A$ \emph{has IFPs (w.r.t. $Q$)} if every arrow $t
  : QA \rightarrow A$ has a IFP.
\end{defn}

\noindent This makes intuitive sense: $a : \mathbf{1} \rightarrow
A$ is extensionally equal to $t$ `evaluated' at the point $Qa
\circ m_0 : \mathbf{1} \rightarrow Qa$, which is the `quoted'
version of $a$. 

\subsection{Consistency, Truth and Provability: G\"odel and Tarski}
  \label{sec:consistency}

We are now in a position to argue that the two well-known theorems
that were discussed by Lawvere can be reduced to very simple
algebraic arguments involving exposures. In fact, the gist of both
arguments relies on the existence of IFPs for an `object of truth
values' in a P-category. For background in G\"odel's First
Incompleteness Theorem and Tarski's \emph{Undefinability Theorem},
see \cite{Smullyan1992} and \cite{Boolos1994}.
 
Suppose that we have some sort of object $\mathbf{2}$ of `truth
values.' This need not be fancy: we require that it has two
points, \begin{align*}
  &\top : \mathbf{1} \rightarrow \mathbf{2} \\
  &\bot : \mathbf{1} \rightarrow \mathbf{2}
\end{align*} standing for truth and falsehood respectively. We
also require an arrow $\lnot : \mathbf{2} \rightarrow \mathbf{2}$
that encodes the logical negation, satisfying \begin{align*}
  \lnot \circ \top &\sim \bot \\
  \lnot \circ \bot &\sim \top
\end{align*} and \[
  \lnot \circ f \sim \bot \quad\Longrightarrow\quad f \sim \top
\]

\noindent A simplified version of G\"odel's First Incompleteness
theorem for \textsf{PA} is this:

\begin{thm}[G\"odel]
  \label{thm:GodelPA}
  If \textsf{PA} is consistent, then there are sentences $\phi$ of
  $\textsf{PA}$ such that neither $\textsf{PA} \vdash \phi$ nor
  $\textsf{PA} \vdash \lnot \phi$.
\end{thm}

\noindent The proof relies on two constructions: the diagonal
lemma, and the fact that provability is definable within the
system. The definability of provability amounts to the fact that
there is a formula $\text{Prov}(x)$ with one free variable $x$
such that \[
  \textsf{PA} \vdash \phi
    \quad \text{if and only if} \quad
  \textsf{PA} \vdash \text{Prov}(\ulcorner \phi \urcorner)
\] That is: modulo G\"odel numbering, the system can internally
`talk' about its own provability. It is not then hard to sketch
the proof to G\"odel's theorem.

\begin{proof}[Proof of Theorem \ref{thm:GodelPA}]
  Use the diagonal lemma to construct $\psi$ such that \[
    \textsf{PA} \vdash 
      \psi
	\leftrightarrow
      \lnot \text{Prov}(\ulcorner \psi \urcorner)
  \] Then $\psi$ is provable if and only if it is not, so if
  either $\textsf{PA} \vdash \psi$ or $\textsf{PA} \vdash
  \lnot\psi$ we would observe inconsistency. Thus, if
  $\textsf{PA}$ is consistent, neither $\psi$ nor $\lnot\psi$ are
  provable. 
\end{proof}

\noindent It follows that $\psi$ is not equivalent to either truth
value. In a way, $\psi$ has some other eerie truth value, which is
neither $\top$ nor $\bot$. Classical logicians would say that it
is \emph{undecidable}.

Let us represent the provability predicate as an arrow $p :
Q\mathbf{2} \rightarrow \mathbf{2}$ such that $y \sim \top$ if and
only if $p \circ Qy \circ m_0 \sim \top$. Consistency is
captured by the following definition:
\begin{defn}
  An object of truth values $\mathbf{2}$ as above is \emph{simply
  consistent} just if \[
    \top \not\sim \bot
  \]
\end{defn}

\noindent Armed with this machinery, we can now transport the
argument underlying G\"odel's proof to our more abstract setting.

\begin{thm}
  If a $p : Q\mathbf{2} \rightarrow \mathbf{2}$ is as
  above, and $\mathbf{2}$ has IFPs, then one of the following
  things is true:
  \begin{itemize}
    \item
      either there are points of $\mathbf{2}$ other than $\top :
      \mathbf{1} \rightarrow \mathbf{2}$ and $\bot : \mathbf{1}
      \rightarrow \mathbf{2}$, or
    \item
      $\mathbf{2}$ is \emph{not} simply consistent, i.e. $\top
      \sim \bot$.
  \end{itemize}
\end{thm}

\begin{proof}
  As $\mathbf{2}$ has IFPs, take $y : \mathbf{1} \rightarrow
  \mathbf{2}$ such that \[
    y \sim \lnot \circ p \circ Qy \circ m_0
  \] Now, if $y \sim \top$, then by the property of $p$ above, $p
  \circ Qy \circ m_0 \sim \top$, hence $\lnot \circ p \circ Qy
  \circ m_0 \sim \bot$, hence $y \sim \bot$. So either $y \not\sim
  \top$ or $\mathbf{2}$ is not simply consistent.  Similarly,
  either $y \not\sim \bot$ or $\mathbf{2}$ is not simply
  consistent.
\end{proof}

Tarski's \emph{Undefinability Theorem}, on the other hand is the
result that \emph{truth cannot be defined in arithmetic}
\citep{Smullyan1992}.

\begin{thm}[Tarski]
  If \textsf{PA} is consistent, then there is no predicate
  $\text{True}(x)$ such that \[
    \textsf{PA} \vdash
      \phi
	\leftrightarrow
      \text{True}(\ulcorner \phi \urcorner)
  \] for all sentences $\phi$.
\end{thm}

\begin{proof}
  Use the diagonal lemma to obtain a closed
  $\psi$ such that \[
    \textsf{PA} \vdash
      \psi
	\leftrightarrow
      \lnot \text{True}(\ulcorner \psi \urcorner)
  \] Then $\textsf{PA} \vdash \psi \leftrightarrow \lnot \psi$,
  which leads to inconsistency.
\end{proof}

\noindent A truth predicate would constitute an evaluator
$\epsilon : Q \natexp{} \textsf{Id}_\mathfrak{B}$. If we had one,
we would have that \[
  \epsilon_\mathbf{2} \circ Q(y) \circ m_0
    \sim y \circ \epsilon_\mathbf{1} \circ m_0
    \sim y
\] where the last equality is because $\mathbf{1}$ is terminal.
This is actually a more general
\begin{lem}
  Let $Q : \mathfrak{B} \expo{} \mathfrak{B}$ be an endoexposure,
  and let $\epsilon : Q \natexp{} \textsf{Id}_\mathfrak{B}$ be an
  evaluator. Then, if $A$ has IFPs then it also has EFPs.
\end{lem}

\begin{proof}
  Given $t : A \rightarrow A$, consider $t \circ \epsilon_A : QA
  \rightarrow A$. A IFP for this arrow is a point $y : \mathbf{1}
  \rightarrow A$ such that $y \sim t \circ \epsilon_A \circ Qy
  \circ m_0 \sim t \circ y$.
\end{proof}

In proving Tarski's theorem, we constructed a sentence $\psi$ such
that $\textsf{PA} \vdash \psi \leftrightarrow \lnot \psi$. This
can be captured abstractly by the following definition.
\begin{defn}
  An object $\mathbf{2}$ as above is \emph{fix-consistent} just if
  the arrow $\lnot : \mathbf{2} \rightarrow \mathbf{2}$ has no
  EFP: that is, there is no $y : \mathbf{1} \rightarrow
  \mathbf{2}$ such that $\lnot \circ y \sim y$. 
\end{defn}

Putting these together, we get

\begin{thm}
  If $\mathbf{2}$ has IFPs in the presence of an evaluator, then
  it is \emph{not} fix-consistent.
\end{thm}

\subsection{Rice's theorem}
  \label{sec:rice}

To further illustrate the applicability of the language of
exposures, we state and prove an abstract version of \emph{Rice's
theorem}. Rice's theorem is a result in computability which states
that no computer can decide any non-trivial property of a program
by looking at its code. A short proof relies on the SRT.

\begin{thm}[Rice]
  Let $\mathcal{F}$ be a non-trivial set of partial recursive
  functions, and let $A_\mathcal{F} \myeq \setcomp{ e \in
  \mathbb{N} }{ \phi_e \in \mathcal{F}}$ be the set of indices of
  functions in that set. Then $A_\mathcal{F}$ is undecidable.
\end{thm}
\begin{proof}
  Suppose $A_\mathcal{F}$ is decidable. The fact $\mathcal{F}$ is
  non-trivial means that there is some $a \in \mathbb{N}$ such
  that $\phi_a \in \mathcal{F}$ and some $b \in \mathbb{N}$ such
  that $\phi_b \not\in\mathcal{F}$. Consequently, $a \in
  A_\mathcal{F}$ and $b \not\in A_\mathcal{F}$.

  Define $f(e, x) \simeq \textbf{if } e \in A_\mathcal{F} \textbf{
  then } \phi_b(x) \textbf{ else } \phi_a(x)$.  By Church's
  thesis, $f : \mathbb{N} \times \mathbb{N} \parfunc{} \mathbb{N}$
  is partial recursive. Use the SRT to obtain $e \in \mathbb{N}$
  such that $\phi_e(x) \simeq f(e, x)$.  Now, either $e \in
  A_\mathcal{F}$ or not. If it is, $\phi_e(x) \simeq f(e, x)
  \simeq \phi_b(x)$, so that $\phi_e \not\in \mathcal{F}$, a
  contradiction.  Similarly if $e \not\in A_\mathcal{F}$.
\end{proof}

Constructing the function $f$ in the proof required three basic
elements: (a) the ability to evaluate either $\phi_a$ or $\phi_b$
given $a$ and $b$; (b) the ability to decide which one to use
depending on the input; and (c) intensional recursion. For (a), we
shall need evaluators, for (b) we shall need that the truth object
$\mathbf{2}$ is a \emph{weak coproduct} of two copies of
$\mathbf{1}$, and for (c) we shall require IFPs.

\begin{thm}
  Let $\mathbf{2}$ be a simply consistent truth object which also
  happens to be a a weak coproduct of two copies of $\mathbf{1}$,
  with injections \begin{align*}
    &\top : \mathbf{1} \rightarrow \mathbf{2} \\
    &\bot : \mathbf{1} \rightarrow \mathbf{2}
  \end{align*} Suppose that $A$ has EFPs. If $f : A \rightarrow
  \mathbf{2}$ is such that for all $x : \mathbf{1} \rightarrow A$,
  either $f \circ x \sim \top$ or $f \circ x \sim \bot$. Then $f$
  is trivial, in the sense that either \[
    \forall x : \mathbf{1} \rightarrow A.\
      f \circ x \sim \top
      \qquad\text{or}\qquad
    \forall x : \mathbf{1} \rightarrow A.\
      f \circ x \sim \bot
  \]
\end{thm}
\begin{proof}
  Suppose that there are two such distinct $a, b : \mathbf{1}
  \rightarrow A$ such that $f \circ a \sim \top$ and $f \circ b
  \sim \bot$. Let \[
    g \myeq [b, a] \circ f : A \rightarrow A
  \] and let $y : \mathbf{1} \rightarrow A$ be its EFP. Now,
  either $f \circ y \sim \top$ or $f \circ y \sim \bot$. In the
  first case, we can calculate that \[
    \top \sim\ f \circ y \sim f \circ g \circ y \sim\ f \circ [b,
    a] \circ f \circ y \sim\ f \circ [b, a] \circ \top \sim\ f
    \circ b \sim\ \bot
  \] so that $\mathbf{2}$ is \emph{not simply consistent}. A
  similar situation occurs if $f \circ y \sim \bot$.
\end{proof}

\noindent Needless to say that the premises of this theorem are
easily satisfied in our exposure on assemblies from
\S\ref{sec:realexpo} if we take $A = {\mathbb{N}_\bot}^\mathbb{N}$
and $\mathbf{2}$ to be the lifted coproduct $(\mathbf{1} +
\mathbf{1})_\bot$.

\section{The relationship to L\"ob's rule}
  \label{sec:lob}

In this chapter we have considered two sorts of fixed points,
\emph{extensional} and \emph{intensional}. Figure \ref{fig:fpsumm}
summarises the definition of these two types of fixed points, in
1-category theory and P-category theory respectively.

\begin{figure}[h]
  \caption{Types of Fixed Points (without parameters)}
  \label{fig:fpsumm}
  \centering
  \renewcommand{\arraystretch}{2}
  \begin{tabular}{c c p{3cm}}
    Type & Morphism  & Fixed Point \\
    % ----------------------------------------------------
    \hline
    Extensional &
    $t : A \rightarrow A$ &
      $\begin{aligned}
	&a : \mathbf{1} \rightarrow A \\
	&t \circ a = a
      \end{aligned}$ \\
    Intensional &
    $t : QA \rightarrow A$ &
      $\begin{aligned}
	&a : \mathbf{1} \rightarrow A \\
	&t \circ Qa \circ m_0 \sim a
      \end{aligned}$
  \end{tabular}
\end{figure}

Viewed through the lens of the Curry-Howard isomorphism, the
existence of extensional fixed points at $A$ can be written as the
logical inference rule \[
  \begin{prooftree}
      A \rightarrow A
    \justifies
      A
  \end{prooftree}
\] which is exactly the type of the $Y$ combinator of PCF. In
fact, this exact inference rule corresponds to an equivalent
formulation of PCF that proceeds through the binding construct
$\mu x{:}A. M$ with equation \[
  \mu x{:}A. M = M[\mu x{:}A. M]
\] For more details, see \cite{Gunter1992}.

This rule is obviously logically catastrophic, as it produces a
closed term $\mu x{:}A.\;x$ at each type $A$. Consequently, there
is no honest Curry-Howard isomorphism for PCF: every type is
inhabited, and thus every `formula' is provable, leading to the
trivial logic. However, \emph{the terms do matter, because they
have computational behaviour}: it might be that the types do not
correspond to logical formulae, but they are there to stop basic
programming errors. And, in the end, the purpose of PCF is simply
typed general recursive programming.

Viewing the notion of intensional fixed points through the lens of
Curry-Howard, the result is much more impressive: IFPs correspond
to \emph{L\"ob's rule}, namely \[
  \begin{prooftree}
      \Box A \rightarrow A
    \justifies
      A
  \end{prooftree}
\] To see this, it suffices to read $\Box \myeq Q$ and to look at
the definition of intensional fixed points as an inference rule,
i.e. \[
  \begin{prooftree}
      f : QA \rightarrow A
    \justifies
      f^\circ : \mathbf{1} \rightarrow A
  \end{prooftree}
\] such that \[
  \begin{tikzcd}[column sep=large]
    \mathbf{1}
      \arrow[r, "Qf^\circ \circ m_0"]
      \arrow[d, "{f^\circ}", swap]
    & QA
      \arrow[dl, "f"] \\
    A
    &
  \end{tikzcd}
\] commutes up to $\sim$.

Unfortunately, we have seen in this chapter that a key ingredient
in the theory of exposures are the so-called evaluators, which are
natural transformations $\epsilon : Q \natexp{} \mathsf{Id}$.
These encapsulate the modal axiom \textsf{T}, namely \[
  \Box A \rightarrow A
\] Coupled with the above inference rule, this will also have the
catastrophic consequence that every type is inhabited. We shall
not be alarmed by this fact, for we will want general recursion,
and there seems no way around partiality in that case, as we
discussed in \S\ref{sec:typethrec}. We shall still use the
Curry-Howard isomorphism, but only \emph{heuristically}.

However, let us for the moment revert to the mindset of a purely
categorical logician. In our parallel work \citep{Kavvos2017b,
Kavvos2017c} we have investigated an extension of the Curry-Howard
isomorphism to box modalities. Our methodology consisted of
mimicking the rules of sequent calculus in natural deduction. A
quick perusal suffices to bring to light the fact that in that
work we used a \emph{stronger form of L\"ob's rule}, namely \[
  \begin{prooftree}
      \Box A \rightarrow A
    \justifies
      \Box A
  \end{prooftree}
\] Historically speaking, this variant of L\"ob's rule is
proof-theoretically well-behaved, and yields cut-free sequent
calculi. It was introduced in the context of intuitionistic modal
logic by \cite{Ursini1979}. If we were to use it to formulate
iPCF,\footnote{In fact, the first versions of this thesis and
\citep{Kavvos2017d} both used this version of iPCF.} we would
obtain something akin to \[
  \begin{prooftree}
    \ctxt{\Delta}{z : \Box A} \vdash M : A
      \justifies
    \ctxt{\Delta}{\Gamma} \vdash \fixbox{z}{M} : \Box A
  \end{prooftree} 
\] with \[
  \fixbox{z}{M} \red{} \ibox{M[\fixbox{z}{M}/z]}
\] However, this does not have a clear intensional interpretation.
If, as in \S\ref{sec:intrec}, we try to devise a `reading' of this
in the untyped $\lambda$-calculus, this rule would require that
for each term $f$ there exists a $u$ such that \[
  u =_\beta \ulcorner f\ulcorner u \urcorner \urcorner
\] which is obviously nonsense. This is why in this thesis we have
\emph{weakened} our formulation to L\"ob's original rule.

Nevertheless, we can still `transport' this version of L\"ob's
rule over to P-categories: it amounts to the inference rule \[
  \begin{prooftree}
      f : QA \rightarrow A
    \justifies
      f^\dagger : \mathbf{1} \rightarrow QA
  \end{prooftree}
\] such that \[
  \begin{tikzcd}[column sep=large]
    \mathbf{1}
      \arrow[r, "Qf^\dagger \circ m_0"]
      \arrow[d, "{f^\circ}", swap]
    & Q^2 A
      \arrow[dl, "Qf"] \\
    QA
    &
  \end{tikzcd}
\] commutes up to $\sim$.

Now: can we use the framework of comonadic exposures to
\emph{prove} that this version of L\"ob's rule is stronger? What
is the exact relationship between this and the original form?

Surprisingly, the answer is positive. But before we show that, let
us give a name to these two kinds of IFPs so we can talk about
them efficiently.

\begin{defn}
  \label{def:ifps}
  Let $Q : \mathfrak{B} \expo{} \mathfrak{B}$ be a
  product-preserving endoexposure.
  \begin{enumerate}
  \item
    A \emph{meek intensional fixed point (meek IFP)} of an arrow $f :
    QA \rightarrow A$ is a point $a : \mathbf{1} \rightarrow a$ such
    that the following diagram commutes up to $\sim$: \[
      \begin{tikzcd}
        \mathbf{1}
          \arrow[r, "a"]
          \arrow[d, "m_0", swap]
        & A \\
          Q\mathbf{1}
          \arrow[r, "Qa", swap]
        & QA
          \arrow[u, "f", swap]
      \end{tikzcd}
    \] i.e. $a \sim f \circ Qa \circ m_0$.
  \item
    A \emph{vehement intensional fixed point (vehement IFP)} of an
    arrow $f : QA \rightarrow A$ is a point $a : \mathbf{1}
    \rightarrow QA$ such that the following diagram commutes up to
    $\sim$: \[
      \begin{tikzcd}
        \mathbf{1}
          \arrow[r, "a"]
          \arrow[d, "m_0", swap]
        & QA \\
          Q\mathbf{1}
          \arrow[r, "Qa", swap]
        & Q^2 A
          \arrow[u, "Qf", swap]
      \end{tikzcd}
    \] i.e. $a \sim Qf \circ Qa \circ m_0$.
  \end{enumerate}
\end{defn}

\noindent Thus the intensional fixed points we have been working
with up to this point are \emph{meek}, whereas the proof theory in
\citep{Kavvos2017b, Kavvos2017c} use a pattern closer to
\emph{vehement} ones.

If we have a meek IFP whose defining equation holds up to
intensional equality ($\approx_Q$), then that is also a vehement
IFP. Conversely, if we have a vehement IFP, and our comonadic
exposure is idempotent, then we obtain a meek IFP.

\begin{thm}
  \label{thm:relintfp}
  Let $f : QA \rightarrow A$.
  \begin{enumerate}
    \item
      If we have a meek IFP of $f : QA \rightarrow
      A$, which moreover is so intensionally, i.e. \[
        f^\circ \approx_Q f \circ Qf^\circ \circ m_0
      \] then we can obtain a vehement IFP of $f$, defined by \[
        f^\dagger \myeq Qf^\circ \circ m_0
      \]
    \item
      If we have a vehement IFP $f : QA \rightarrow
      A$, and moreover the comonadic exposure $(Q, \epsilon,
      \delta)$ is idempotent, then we can obtain a meek IFP of of
      $f$, defined by \[
        f^\circ \myeq \epsilon_A \circ f^\dagger
      \]
  \end{enumerate}
\end{thm}
\begin{proof} \hfill
  \begin{enumerate}
    \item
      We calculate:
        \begin{derivation}
          f^\dagger
        \since[\sim]{definition}
          Qf^\circ \circ m_0
        \since[\sim]{assumption}
          Q(f \circ Qf^\circ \circ m_0) \circ m_0
        \since[\sim]{exposure}
          Qf \circ Q(Qf^\circ \circ m_0) \circ m_0
        \since[\sim]{definition}
          Qf \circ Qf^\dagger \circ m_0
      \end{derivation} so that $f^\dagger$ is a vehement IFP.
    \item
      We calculate \begin{derivation}
          f^\circ
        \since[\sim]{definition}
          \epsilon_A \circ f^\dagger
        \since[\sim]{definition}
          \epsilon_A 
            \circ Qf \circ Qf^\dagger \circ m_0
        \since[\sim]{$\epsilon$ natural}
          f
            \circ \epsilon_{QA} \circ Qf^\dagger \circ m_0
        \since[\sim]{$\epsilon$ natural}
          f
            \circ f^\dagger \circ \epsilon_{\mathbf{1}} \circ m_0
        \since[\sim]{$\epsilon$ monoidal}
          f
            \circ f^\dagger
      \end{derivation} But, by the corollary of the
      Quotation-Evaluation lemma (Lemma \ref{cor:qe}), \[
        f^\dagger 
          \sim
        Q(\epsilon_A \circ f^\dagger) \circ m_0
          \sim
        Qf^\circ \circ m_0
      \] so $f^\circ \sim f \circ Qf^\circ \circ m_0$ is a meek
      IFP.
  \end{enumerate}
\end{proof}

\section{Whence fixed points?}
  \label{sec:lawvere}

\subsection{Lawvere's Theorem}

\cite{Lawvere1969, Lawvere2006} proved a fixed point theorem that
generalises a number of `diagonal' constructions, including Cantor's
theorem, G\"odel's First Incompleteness Theorem, and the Tarski
undefinability theorem. In order to state it, we will first need
some notation for cartesian closed categories (CCCs).  A cartesian
closed category \citep{Eilenberg1966} is a great place: it is a
mathematical universe where morphisms of the category correspond
exactly to points of certain objects, the \emph{exponentials}.
Indeed, to every morphism $f : A \rightarrow Y$ there corresponds
a point of the exponential object $Y^A$, namely \[
  \ulcorner f \urcorner : 
    \mathbf{1}
      \rightarrow
    Y^A
\] which is defined by \[
  \ulcorner f \urcorner \myeq 
    \lambda\left(
      \mathbf{1} \times A
	\xrightarrow{\cong}
	A \xrightarrow{f} Y
    \right)
\] and to each point $y : \mathbf{1} \rightarrow Y^A$ there
corresponds a morphism $y^{\wr} : A \rightarrow Y$, defined by \[
  y^{\wr} \myeq
    A
      \xrightarrow{\langle y \circ {!}_A, id_A \rangle}
    Y^A \times A
      \xrightarrow{\mathsf{ev}}
    Y
\] The above operations are mutually inverse: \[
  {\ulcorner f \urcorner}^{\wr} = f,
    \quad\quad\quad
  \left\ulcorner{y^{\wr}}\right\urcorner = y
\] For a classic exposition, see \cite{Lambek1988}.

The main idea in Lawvere's paper is this: a morphism \[
  X \xrightarrow{r} Y^A
\] from an object to an exponential may be thought as `indexing'
morphisms of type $A \rightarrow Y$; for, given $x : \mathbf{1}
\rightarrow X$, we have $(r \circ x)^{\wr} : A \rightarrow Y$.
Such an indexing may be considered an \emph{enumeration} if it is,
in some sense, \emph{surjective}. There are many ways in which
an arrow $r : X \rightarrow A$ can be surjective. Here are four:
\begin{itemize}
  \item
    It could a \emph{retraction}; that is, there could exist $s :
    A \rightarrow X$ such that \[
      \begin{tikzcd}
        A
          \arrow[rd, "id_A"]
          \arrow[d, "s", swap]
        & \\
        X
          \arrow[r, "r", swap]
        & A
      \end{tikzcd}
    \] The arrow $s$ may be thought of as `choosing a preimage' of
    elements of $A$ with respect to $r$. In the category of sets,
    surjective functions are always retractions (if one assumes
    the axiom of choice).

  \item 
    It could be \emph{point-surjective}: for each point $a :
    \mathbf{1} \rightarrow A$ there could be a point $x :
    \mathbf{1} \rightarrow X$ such that \[
      \begin{tikzcd}
        \mathbf{1}	
          \arrow[rd, "a"]
          \arrow[d, dashed, "x", swap]
        & \\
        X
          \arrow[r, "r", swap]
        & A
      \end{tikzcd}
    \]

  \item 
    It could be \emph{$N$-path-surjective}: it could be that, for
    each `$N$-path' $q : N \rightarrow A$, there is a $N$-path $p
    : N \rightarrow X$ such that \[
      \begin{tikzcd}
        N
          \arrow[rd, "p"]
          \arrow[d, dashed, "q", swap]
        & \\
        X
          \arrow[r, "r", swap]
        & A
      \end{tikzcd}
    \] The name of the object $N$ has been chosen to suggest the
    natural numbers, so that $p : N \rightarrow A$ can be thought
    of as tracing out a \emph{discrete path} of points in $A$. A
    point-surjective arrow is, of course, a
    $\mathbf{1}$-path-surjective arrow.

  \item
    It could be \emph{weakly point-surjective} (only if the
    codomain is an exponential): if $r : X \rightarrow Y^A$, then,
    for each $x : \mathbf{1} \rightarrow X$, we obtain $r \circ x
    : \mathbf{1} \rightarrow Y^A$, which corresponds to a morphism \[
      (r \circ x)^\wr : A \rightarrow Y
    \] It could then be that \emph{every} morphism $A \rightarrow Y$
    is `pointwise emulated' by $(r \circ x)^\wr$ for some $x$.
    That is, for each $f : A \rightarrow Y$, there exists $x_f :
    \mathbf{1} \rightarrow X$ such that \[
      \forall a : \mathbf{1} \rightarrow A.\
        (r \circ x_f)^\wr \circ a = f \circ a
    \] So a weak point-surjection is a bit like `pointwise cartesian
    closure.'
\end{itemize} Evidently, \begin{align*}
  r \text{ is a retraction}
    \Longrightarrow
  r &\text{ is $N$-path-surjective} \\
    \xRightarrow{\text{if } N \text{ non-empty}}
  r &\text{ is point-surjective} \\
    \Longrightarrow
  r &\text{ is weakly point-surjective}
\end{align*} where the third implication only makes sense if codomain of
$r$ is an exponential, but may be skipped otherwise. For the
second implication, notice that $N \xrightarrow{{!}} \mathbf{1}
\xrightarrow{x} A$ is a $N$-path, factorise it through $X$, and
pre-compose with any point $n : \mathbf{1} \rightarrow N$.

Lawvere then observed that, if the codomain of the weak
point-surjection is an exponential $Y^A$, and the `indexing
object' $X$ coincides with $A$, a curious phenomenon occurs.

\begin{thm}[Lawvere]
  If $r : A \rightarrow Y^A$ is a weak point-surjection, then then
  every arrow $t : Y \rightarrow Y$ has a fixed point.
\end{thm}
\begin{proof}
  Let \[
    f \myeq A \xrightarrow{\langle r, id_A \rangle} Y^A
    \times A \xrightarrow{\mathsf{ev}} Y \xrightarrow{t} Y
  \] As $r$ is a weak point-surjection, there exists a $x_f :
  \mathbf{1} \rightarrow A$ such that, for all $a : \mathbf{1}
  \rightarrow A$, we have \begin{derivation}
      (r \circ x_f)^\wr \circ a
    \since{$r$ is a weak point-surjection}
      f \circ a
    \since{definition of $f$}
      t \circ \mathsf{ev} \circ \langle r, id_A \rangle \circ a
    \since{naturality of product}
      t \circ \mathsf{ev} \circ \langle r \circ a, a \rangle
    \since{terminal object: $a \circ {!}_A = id_\mathbf{1}$ }
      t \circ \mathsf{ev} \circ \langle r \circ a \circ {!}_A \circ a, a \rangle
    \since{naturality of product}
      t \circ \mathsf{ev} \circ \langle r \circ a \circ {!}_A, id_{A} \rangle \circ a
    \since{definition of $(-)^\wr$}
      t \circ (r \circ a)^\wr \circ a
  \end{derivation} Taking $a \myeq x_f$ produces a fixed point.
\end{proof}

\noindent Lawvere also hinted at a `cartesian' version of the
above result that does not require exponentials. In this version,
the diagonal nature of the argument is even more evident. To prove
it, we need to introduce the following definition:

\begin{defn}
  \label{def:wps}
  An arrow $r : X \times A \rightarrow Y$ is a \emph{(cartesian)
  weak point-surjection} if for every $f : A \rightarrow Y$
  there exists a $x_f : \mathbf{1} \rightarrow X$ such that \[
    \forall a : \mathbf{1} \rightarrow A.\
      r \circ \langle x_f, a \rangle = f \circ a
  \]
\end{defn} 

\noindent We will not bother to qualify weak point-surjections as
ordinary or cartesian, as it will be clear by the context. We can
now prove the

\begin{thm}[Lawvere]
  \label{thm:lawvere}
  If $r : A \times A \rightarrow Y$ is a weak point-surjection,
  then every arrow $t : Y \rightarrow Y$ has a fixed point.
\end{thm}
\begin{proof}
  Let \[
    f \myeq t \circ r \circ \langle id_A, id_A \rangle
  \] Then there exists a $x_f : \mathbf{1} \rightarrow A$ such
  that \[
    r \circ \langle x_f, a \rangle = f \circ a
  \] for all $a : \mathbf{1} \rightarrow A$. We compute that \[
       	 r \circ \langle x_f, x_f \rangle
    = t \circ r \circ \langle id_A, id_A \rangle \circ x_f
    = t \circ r \circ \langle x_f, x_f \rangle
  \] so that $r \circ \langle x_f, x_f \rangle$ is a
  fixed point of of $t$.
\end{proof}

We have seen in \S\ref{sec:extintfp} that the extensional kind of
fixed points produced by this theorem are of a sort that
\emph{oughtn't} exist. We believe that this is one of the reasons
that Lawvere's result has not found wider applications.
Nevertheless, we will also see in \S\ref{sec:catsrt} that---in a
certain setting---this theorem corresponds to a very weak form of
Kleene's \emph{First Recursion Theorem}, which has been a central
theorem in the semantics of programming languages (see
\S\ref{chap:srtht}).

\subsection{An intensional Lawvere theorem}

Can we adapt Lawvere's result to IFPs? The answer is positive: all
we need is a cartesian P-category $\mathfrak{B}$, a
product-preserving exposure $Q : \mathfrak{B} \expo{}
\mathfrak{B}$, and a reasonable quoting device. What remains is to
`embellish' Lawvere's argument with appropriate instances of $Q$.

\begin{thm}[Intensional Recursion]
  \label{thm:intrec}
  Let $Q : \mathfrak{B} \expo{} \mathfrak{B}$ be a
  product-preserving exposure, and let $\delta_A : QA \rightarrow
  Q^2 A$ be a reasonable quoting device. If $r : QA \times QA
  \rightarrow Y$ is a weak point-surjection, then every arrow \[
    t : QY \rightarrow Y
  \] has an intensional fixed point.
\end{thm}
\begin{proof}
  Let \[
    f \myeq
      QA
	\xrightarrow{\langle \delta, \delta \rangle}
      Q^2 A \times Q^2 A
	\xrightarrow{m}
      Q(QA \times QA)
	\xrightarrow{Qr}
      QY
	\xrightarrow{t}
      Y
  \] Then, there exists a $x_f : \mathbf{1} \rightarrow QA$ such
  that \[
    r \circ \langle x_f, a \rangle \sim f \circ a
  \] for all $a : \mathbf{1} \rightarrow QA$. We compute that
  \begin{derivation}
      r \circ \langle x_f, x_f \rangle
    \since[\sim]{ definition }
      t \circ Qr \circ m \circ \langle \delta_A, \delta_A \rangle \circ x_f
    \since[\sim]{ naturality }
      t \circ Qr \circ m \circ \langle \delta_A \circ x_f, \delta_A \circ x_f \rangle
    \since[\sim]{ $\delta_A$ is a reasonable quoting device }
      t \circ Qr \circ m \circ \langle Qx_f \circ m_0, Qx_f \circ m_0 \rangle
    \since[\sim]{ naturality }
      t \circ Qr \circ m \circ \langle Qx_f, Qx_f \rangle \circ m_0
    \since[\sim]{ Proposition \ref{prop:prodpresangle} }
      t \circ Qr \circ Q\langle x_f, x_f \rangle \circ m_0
    \since[\sim]{ exposures preserve composition }
      t \circ Q(r \circ \langle x_f, x_f \rangle) \circ m_0
  \end{derivation} so that $r \circ \langle x_f, x_f \rangle$ is a IFP
  of $t$.
\end{proof}

\section{Examples of Fixed Points}
  \label{sec:fpexamples}

In this final section we shall briefly examine what extensional
and intensional fixed points mean in the first two examples of
exposures presented in \S\ref{chap:expoexamples}, namely the
Lindenbaum P-category and the P-category of assemblies.

Unfortunately, since the trivial group is a \emph{zero object} in
the P-category of groups, points do not carry any interesting
structure, and thus neither notion of fixed point is interesting
in the third example.

\subsection{Fixed Points in G\"odel numbering: the Diagonal Lemma}

We will now carefully consider what we have hinted at throughout
the development of the abstract analogues of G\"odel and Tarski's
results in \S\ref{sec:consistency}, viz. the diagonal lemma of
Peano arithmetic is precisely the existence of IFPs in the case of
the exposure on arithmetic presented in \S\ref{sec:logicexpo}.

Recall once more the diagonal lemma: for every formula $\phi(x)$
there exists a closed formula $\mathbf{fix}(\phi)$ such that \[ 
  \textsf{PA} \vdash
    \mathbf{fix}(\phi) \leftrightarrow \phi(\ulcorner
  \mathbf{fix}(\phi) \urcorner)
\] Take the formula $\phi(x)$. In
$\mathfrak{Lind}(\textsf{PA})$, it is an arrow \[
  \phi(x) : A \rightarrow \mathbf{2}
\] Let there be a sentence $\psi : \mathbf{1} \rightarrow
\mathbf{2}$. When the exposure acts on it, it produces (the
numeral of) the G\"odel number of $\psi$, which is a closed term.
Suppose $\psi$ is an IFP of $\phi$, i.e \[
  \psi \sim \phi \circ Q\psi \circ m_0
\] We can then see that the RHS simplifies by substitution to
$\phi(\ulcorner \psi \urcorner)$, so this is precisely the
conclusion of the diagonal lemma.

\subsection{Fixed Points in Assemblies: Kleene's Recursion Theorems}
  \label{sec:catsrt}

We now turn to the consideration of fixed points in the P-category
of assemblies $\mathfrak{Asm}(A)$ that we considered in
\S\ref{sec:realexpo}, along with the paradigmatic exposure \[
  \Box : \mathfrak{Asm}(A) \expo{} \mathfrak{Asm}(A)
\] 

EFPs are exactly what one would expect: given an arrow $(f, r) : X
\rightarrow X$, a EFP of this arrow is a $x \in \bars{X}$ such
that \[
  f(x) = x
\] We will shortly argue that EFPs in this setting are strongly
reminiscent of \emph{Kleene's First Recursion Theorem}, which we
discussed at length in \S\ref{chap:srtht}.

On the contrary, a IFP of an arrow \[
  (f, r) : \Box X \rightarrow X
\] is more than what we had before: it is an element $x \in
\bars{X}$ along with a realizer $a \in \dbars{x}_X$ of it, such
that \[
  f(x, a) = x
\] That is, it is a kind of fixed point of $f$, but the
computation of $f$ also depends on the chosen realizer $a$ rather
than simply $x$.

It is worth pausing for a moment to ask what a \emph{vehement}
IFP (as discussed in \S\ref{sec:lob}) is in this case. It is not
hard to compute that it consists once again of both an element $x
\in \bars{X}$ and a realizer $a \in \dbars{x}_X$ for it, but the
pair now needs to satisfy not only $f(x, a) = x$, but also \[
  r \cdot a \simeq a
\] That is, the two sides need to have the same realizer. The
above theorem is not too surprising in light of Theorem
\ref{thm:ideminteq} and Corollary \ref{cor:ideminteqprod}: in the
idempotent case, the fact that two arrows $\mathbf{1} \rightarrow
QA$ are equal immediately yields that they are intensionally
equal. It therefore follows that vehement IFPs are too strong a
notion for what we consider to be intensional recursion.

\subsubsection*{Kleene's Recursion Theorems}

We will now argue that, in the case of $\mathfrak{Asm}(K_1)$, the
notions of EFPs and IFPs indeed match the conclusions of the two
main theorems that are used to make recursive definitions in
computability theory, namely the First and Second Recursion
Theorems of Kleene. In fact, we will argue that Lawvere's fixed
point theorem (Theorem \ref{thm:lawvere}) and our own Intensional
Recursion theorem (Theorem \ref{thm:intrec}) each correspond to
abstract versions of them. We have discussed the relationship
between the two theorems at length in \S\ref{chap:srtht},
but---for the benefit of the reader---we recapitulate some basic
points of this discussion here, in a form tailored to our needs in
this chapter.

Let us fix some notation.  We write $\simeq$ for \emph{Kleene
equality}: we write $e \simeq e'$ to mean either that both
expressions $e$ and $e'$ are undefined, if either both are
undefined, or both are defined and of equal value. Let $\phi_0$,
$\phi_1$, $\dots$ be an enumeration of the partial recursive
functions. We will also require the \emph{s-m-n theorem} from
computability theory. Full definitions and statements may be found
in the book by \cite{Cutland1980}.

\begin{thm}[First Recursion Theorem]
  \label{thm:frtexpo}
  Let $\mathcal{PR}$ be the set of unary partial recursive
  functions, and let $F : \mathcal{PR} \rightarrow \mathcal{PR}$
  be an \emph{effective operation}. Then $F : \mathcal{PR}
  \rightarrow \mathcal{PR}$ has a fixed point.
\end{thm}

\begin{proof}
  That $F : \mathcal{PR} \rightarrow \mathcal{PR}$ is an effective
  operation means that there is a partial recursive $f :
  \mathbb{N} \times \mathbb{N} \parfunc{} \mathbb{N}$ such that
  $f(e, x) \simeq F(\phi_e)(x)$. Let $d \in \mathbb{N}$ a code for
  the partial recursive function $\phi_d(y, x) \myeq f(S(y, y),
  x)$, where $S : \mathbb{N} \times \mathbb{N} \parfunc{}
  \mathbb{N}$ is the s-1-1 function of the s-m-n theorem. Then, by
  the s-m-n theorem, and the definitions of $d \in \mathbb{N}$ and
  $f$, \[
    \phi_{S(d, d)}(x)
      \simeq \phi_d(d, x)
      \simeq f(S(d, d), x)
      \simeq F(\phi_{S(d,d)})(x)
  \] so that $\phi_{S(d, d)}$ is a fixed point of $F :
  \mathcal{PR} \rightarrow \mathcal{PR}$.
\end{proof}

\noindent Lawvere's theorem is virtually identical to a point-free
version of this proof. Indeed, if we let \begin{align*}
  \mathbb{S} : 
    \mathbb{N} \times \mathbb{N} &\longrightarrow \mathcal{PR} \\
                      (a, b)     &\longmapsto     \phi_{S(a, b)}
\end{align*} We can then show that this is a weak
point-surjection: any `computable' $f : \mathbb{N} \rightarrow
\mathcal{PR}$ is actually just an index $d \in \mathbb{N}$ such
that \[
  \phi_d(x, y) \simeq f(x)(y)
\] for all $x, y \in \mathbb{N}$. We can then show that \[
  \mathbb{S}(d, x)(y)
    \simeq
  \phi_{S(d, x)}(y)
    \simeq
  \phi_d(x, y)
    \simeq
  f(x)(y)
\] so that $\mathbb{S}(d, x) = f(x) \in \mathcal{PR}$. Thus the
resemblance becomes formal, and the argument applies to yield the
FRT.

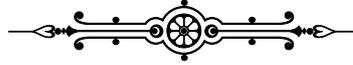
\begin{center}
  \pgfornament[scale=0.45]{84}
\end{center}

Yet, one cannot avoid noticing that we have proved
more than that for which we bargained. The $f : \mathbb{N} \times
\mathbb{N} \parfunc{} \mathbb{N}$ in the proof implemented a
certain effective operation $F: \mathcal{PR} \rightarrow
\mathcal{PR}$. It follows that $f$ has a special property: it is
\emph{extensional}, in the sense that \[
  \phi_e = \phi_{e'} \quad\Longrightarrow\quad
    \forall x \in \mathbb{N}.\ f(e, x) \simeq f(e', x)
\] Notice, however, that the main step that yields the fixed point
in this proof also holds for any such $f$, not just the
extensional ones. This fact predates the FRT, and was shown by
\cite{Kleene1938}.

\begin{thm}[Second Recursion Theorem]
  For any partial recursive $f: \mathbb{N} \times \mathbb{N}
  \parfunc \mathbb{N}$, there exists $e \in \mathbb{N}$ such that
  $\phi_e(y) \simeq f(e, y)$ for all $y \in \mathbb{N}$.
\end{thm} 

This is significantly more powerful than the FRT, as $f(e, y)$ can
make arbitrary decisions depending on the source code $e$,
irrespective of the function $\phi_e$ of which it is the source
code. Moreover, it is evident that the function $\phi_e$ has
\emph{access to its own code}, allowing for a certain degree of
reflection. Even if $f$ is extensional, hence defining an
effective operation, the SRT grants us more power than the FRT:
for example, before recursively calling $e$ on some points, $f(e,
y)$ could `optimise' $e$ depending on what $y$ is, hence ensuring
that the recursive call will run faster than $e$ itself would.

Lawvere's argument, as presented above, cannot account for the
Second Recursion Theorem: $\mathbb{S} : \mathbb{N} \times \mathbb{N}
\rightarrow \mathcal{PR}$ had $\mathcal{PR}$ in the codomain, and
thus this argument only works to yield fixed points of effective
operations $\mathcal{PR} \rightarrow \mathcal{PR}$. We do not see
a meaningful way of replacing this with, say, $\mathcal{N}$.
We could certainly replace it with an object $\mathbb{N}_\bot$
that accounts for non-termination, but that is not what we want.
we wish to take here.

What we really want is to define a computable \emph{operation}
$\mathcal{PR} \oper{} \mathcal{PR}$. In order to explain the
meaning of this, a shift in perspective is required. To say that
$F : \mathcal{PR} \rightarrow \mathcal{PR}$ is an effective
operation, we need an extensional, total recursive function $f$,
implemented by some index $d \in \mathbb{N}$ that tracks it on
codes, along with a proof that $f$ is extensional. But what about
indices $d \in \mathbb{N}$ that describe a total $\phi_d$, yet are
\emph{not} extensional? These still map every $e \in \mathbb{N}$,
which is an index for $\phi_e$, to $\phi_d(e)$, which is an index
for $\phi_{\phi_d(e)}$. Hence, while $f$ may map codes of partial
recursive functions to codes of partial recursive functions, it
may do so \emph{without being extensional}. In that case, it
defines a \emph{non-functional operation} $G : \mathcal{PR}
\oper{} \mathcal{PR}$, which is exactly the case where
IFPs will apply.

We can see this far more clearly in the setting of
$\mathfrak{Asm}(K_1)$. Arrows \[
  \mathbb{N} \rightarrow \mathbb{N}_\bot
\] are easily seen to correspond to partial recursive functions.
The weak point-surjection $\mathbb{S} : \mathbb{N} \times
\mathbb{N} \rightarrow \mathcal{PR}$ we produced above can now be
seen as an arrow $r : \mathbb{N} \times \mathbb{N} \rightarrow
\mathbb{N}_\bot^{\mathbb{N}}$ in $\mathfrak{Asm}(K_1)$, and
invoking Lawvere's theorem indeed shows that every arrow\[
  \mathbb{N}_\bot^{\mathbb{N}} \rightarrow \mathbb{N}_\bot^{\mathbb{N}}
\] has an extensional fixed point. Now, by Longley's generalised
Myhill-Shepherdson theorem \citep{Longley1995,Longley2015}, these
arrows exactly correspond to effective operations. Hence, in this
context Lawvere's theorem states that each effective operation has
a fixed point, and indeed corresponds to the simple diagonal
argument above.\footnote{But note that this is not the
complete story, as there is no guarantee that the fixed point
obtained in \emph{least}, which is what Kleene's original proof in
\cite{Kleene1952} gives. See also \S\ref{chap:srtht}.}

However, let us look now look at arrows of type \[
  \Box \mathbb{N}_\bot^{\mathbb{N}}
    \rightarrow \mathbb{N}_\bot^{\mathbb{N}}
\] These correspond to `non-functional' transformations, mapping
functions to functions, but without respecting extensionality. As
every natural number indexes a partial recursive function, these
arrows really correspond to all partial recursive functions.
It is not hard to see that $\Box \mathbb{N}$ is P-isomorphic to
$\mathbb{N}$: this fact can be used with $r$ to immediately
produce a weak point-surjection $q: \Box \mathbb{N} \times \Box
\mathbb{N} \rightarrow \mathbb{N}_\bot^{\mathbb{N}}$, so that, by
our Intensional Recursion theorem (Theorem \ref{thm:intrec}),
every arrow of type \[ 
  \Box \mathbb{N}_\bot^{\mathbb{N}}
    \rightarrow \mathbb{N}_\bot^{\mathbb{N}}
\] has a IFP. It is thus evident that there is considerable formal
similarity between this application of Theorem \ref{thm:intrec}
and Kleene's Second Recursion Theorem!

\chapter{Intensional Semantics of iPCF I}

  \label{chap:intsem1}

  This is the first of the two chapters that address the issue of
finding a truly intensional semantics for the language iPCF which
we introduced in \S\ref{chap:ipcf}. Both of these chapters can be
seen as evidence towards the claim that iPCF is a truly
intensional language.

To make progress towards our goal, we will use the theory of
exposures, as developed in \S\ref{chap:expo} and \ref{chap:irec}.
The desired outcome is that the category of assemblies
$\mathfrak{Asm}(K_1)$ over the PCA $K_1$, which---as we showed in
\S\ref{sec:realexpo}---is a cartesian closed P-category equipped
with an idempotent comonadic exposure, is a model of iPCF.

However, before we delve into the details, we need to develop some
algebraic machinery for interpreting the rules of iPCF.  Because
of the delicate behaviour of our two notions of
equality---extensional ($\sim$) and intensional
($\approx_Q$)---this will be more involved than it sounds.
Nevertheless, a lot of the ground work has been done before, and
is still applicable to this setting. 

We have discussed the categorical semantics of modal
$\lambda$-calculi at length in previous work---see
\citep{Kavvos2017b, Kavvos2017c}: therein we found that
categorical semantics of a \textsf{S4} modality comprise a
\emph{Bierman-de Paiva category} \citep{Bierman2000}, viz. a
cartesian closed category $(\mathcal{C}, \times, \mathbf{1})$
along with a product-preserving comonad $(Q, \epsilon, \delta)$.
Adapting this to the intensional setting involves exchanging $Q :
\mathcal{C} \longrightarrow \mathcal{C}$ for an exposure $Q :
(\mathfrak{C}, \sim) \expo{} (\mathfrak{C}, \sim)$ that is
comonadic.  The crucial factor that one should be aware of is that
no calculation in our previous work depended on the `forbidden
principle' $f = g \Longrightarrow Qf = Qg$. Hence, a lot of the
groundwork may be immediately transferred to the intensional
setting.

Nevertheless, there are some differences, and we shall deal with
them in this chapter. One of the main ones is our emphasis on
\emph{idempotence}. In the ensuing development it will become
clear that the \emph{idempotent} case is particularly elegant and
useful. What is more, the argument made in \S\ref{sec:idempotent},
viz. that idempotence is the right notion for intensionality, will
be corroborated in this chapter.

\section{Setting the scene}
  \label{sec:scene}

First, we shall define certain basic ingredients that we will use
for our semantics. These amount to algebraic short-hands that will
prove incredibly useful for calculations. 

We shall define the following arrows by induction (and be lax
about the subscripts): \begin{align*}
  m^{(0)} &\myeq \mathbf{1} \xrightarrow{m_0} Q\mathbf{1} \\
  m^{(n+1)} &\myeq
    \prod_{i = 1}^{n+1} QA_i
      \xrightarrow{m^{(n)} \times id }
    Q\left(\prod_{i = 1}^{n} A_i\right) \times QA_{n+1}
      \xrightarrow{m}
    Q\left(\prod_{i = 1}^{n+1} A_i\right)
\end{align*}

\noindent Then, the $m^{(n)}$'s are natural, in the sense that \[
  m^{(n)} \circ \prod_{i=1}^n Qf_i
   \sim
  Q\left(\prod_{i=1}^n f_i\right) \circ m^{(n)}
\]

\noindent The main contraption in the semantics of \textsf{S4} was
a hom-set map that generalised the notion of \emph{co-Kleisli
lifting}. In our setting, this will be an operation: \[
  (-)^\ast :
    \mathfrak{C}\left(\prod_{i = 1}^n QA_i , B\right)
      \oper{}
    \mathfrak{C}\left(\prod_{i = 1}^n QA_i, QB\right)
\] which we define as follows: \[
  \begin{prooftree}
    f : 
      \prod_{i = 1}^n QA_i
        \rightarrow
      B
      \justifies
    f^\ast \myeq
      \prod_{i = 1}^n QA_i
        \xrightarrow{\prod_{i=1}^n \delta_{A_i}}
      \prod_{i = 1}^n Q^2 A_i
        \xrightarrow{m^{(n)}}
      Q\left(\prod_{i = 1}^n QA_i\right)
        \xrightarrow{Qf}
      QB
  \end{prooftree}
\]

\noindent This operation is the categorical counterpart of an
admissible rule of \textsf{S4}, which from $\Box\Gamma \vdash A$
allows one to infer $\Box\Gamma \vdash \Box A$. In the weaker
setting of \textsf{K}, we only have \emph{Scott's rule}: from
$\Gamma \vdash A$ infer $\Box \Gamma \vdash \Box A$. This is also
categorically also an operation: \[
  (-)^\bullet : 
    \mathfrak{C}\left(\prod_{i = 1}^n A_i, B\right)
      \oper{}
    \mathfrak{C}\left(\prod_{i = 1}^n QA_i, QB\right)
\] which is defined as follows: \[
  \begin{prooftree}
    f : \prod_{i = 1}^n A_i \rightarrow B
      \justifies
    f^\bullet \myeq
      \prod_{i = 1}^n QA_i
	\xrightarrow{m^{(n)}}
      Q\left(\prod_{i = 1}^n A_i\right)
	\xrightarrow{Qf}
      QB
  \end{prooftree}
\]

It might, of course, seem that both $(-)^\bullet$ and $(-)^\ast$
are operations, but what they have in common is that they both act
by applying the exposure $Q$ to their argument. It follows then
that they are \emph{functions} with respect to intensional
equality, and hence well-defined on the x-ray category of
$\mathfrak{C}$. We write \begin{align*}
  (-)^\ast &:
    (\mathfrak{C}, \approx_Q)\left(\prod_{i = 1}^n QA_i , B\right)
      \rightarrow
    \mathfrak{C}\left(\prod_{i = 1}^n QA_i, QB\right) \\
  (-)^\bullet &: 
    (\mathfrak{C}, \approx_Q)\left(\prod_{i = 1}^n A_i, B\right)
      \rightarrow
    \mathfrak{C}\left(\prod_{i = 1}^n QA_i, QB\right)
\end{align*}

\noindent However, if $(Q, \epsilon, \delta)$ is idempotent, we
can show that $(-)^\ast$ actually preserves intensional equality.
For this, we will need the following proposition.

\begin{prop}
  \label{prop:deltaQast}
  If $(Q, \epsilon, \delta)$ is idempotent, then for $f :
  \prod_{i=1}^n QA_i \rightarrow B$, \[
    Qf^\ast \sim \delta_B \circ Qf
  \]
\end{prop}
\begin{proof}
    \begin{derivation}
        Qf^\ast
      \since[\sim]{definition, $Q$ exposure}
        Q^2 f \circ Qm^{(n)}
          \circ Q\left(\prod_{i=1}^n \delta_{A_i} \right)
      \since[\sim]{Proposition \ref{prop:prodpresangle}}
        Q^2 f 
          \circ Qm^{(n)}
          \circ m^{(n)} 
          \circ \langle \vct{Q\delta_{A_i} \circ Q\pi_{A_i}} \rangle
      \since[\sim]{$Q$ idempotent, so $Q\delta_{A_i} \sim \delta_{QA_i}$}
        Q^2 f \circ Qm^{(n)} \circ m^{(n)} 
          \circ \langle \vct{\delta_{QA_i} \circ Q\pi_i} \rangle
      \since[\sim]{product equation, $\delta$ monoidal}
        Q^2 f \circ \delta \circ m^{(n)} 
          \circ \langle \vct{Q\pi_i}\rangle
      \since[\sim]{$\delta$ natural, inverse of $m^{(n)}$ is
 		   $\langle \vct{Q\pi_i}\rangle$}
        \delta \circ Qf
    \end{derivation}
\end{proof}

\noindent This allows us to infer that
\begin{cor}
  \label{cor:idemastpresint}
  If $Q$ is idempotent then $(-)^\ast$ preserves $\approx_Q$: it
  is a map \[
    (-)^\ast :
      (\mathfrak{C}, \approx_Q)\left(\prod_{i = 1}^n QA_i , B\right)
        \rightarrow
      (\mathfrak{C}, \approx_Q)\left(\prod_{i = 1}^n QA_i, QB\right) \\
  \]
\end{cor}
\begin{proof}
  If $f \approx_Q g$, then $Qf^\ast \sim \delta \circ Qf \sim
  \delta \circ Qg \sim Qg^\ast$.
\end{proof}

\section{Distribution and naturality laws}

We now look at the relationship between the two operations
$(-)^\bullet$ and $(-)^\ast$, and we state and prove certain
`distribution' laws that apply to them, and describe their
interaction.  First, we note that it is definitionally the case
that \[
  f^\ast \myeq f^\bullet \circ \prod_{i=1}^n \delta_{A_i}
\] for $f : \prod_{i = 1}^n QA_i \rightarrow B$. 

\begin{prop}
  Given a comonadic exposure $(Q, \epsilon, \delta)$, the
  following equations hold up to $\sim$. Furthermore, if $Q$ is
  idempotent, they hold up to $\approx_Q$.
  \label{prop:astdist}
  \begin{enumerate}[label=(\roman{*})]
    \item \label{itm:idast}
      $id_{QA}^\ast \sim \delta_{QA}$
    \item \label{itm:epsast}
      $\epsilon_A^\ast \sim id_{QA}$
    \item \label{itm:astastdist}
      For $k : \prod_{i=1}^n QA_i \rightarrow B$ and $l : QB
      \rightarrow C$, \[
        (l \circ k^\ast)^\ast \sim l^\ast \circ k^\ast
      \]
    \item \label{itm:astbulletdist}
      For $k : \prod_{i=1}^n A_i \rightarrow B$ and $l : QB \rightarrow C$, \[
        (l \circ k^\bullet)^\ast \sim l^\ast \circ k^\bullet
      \]
    \item \label{itm:astdist}
      Let $f : \prod_{i = 1}^n B_i \rightarrow C$ and $g_i : \prod_{j
      = 1}^k QA_j \rightarrow B_i$ for $i = 1, \dots, n$.  Then \[
        \left(f \circ \langle \vct{g_i} \rangle\right)^\ast
         \sim
        f^\bullet \circ \left\langle \vct{g^\ast_i} \right\rangle
      \] 
    \item \label{itm:astprojdist}
      For $f : \prod_{i=1}^n QA_i \rightarrow B$ and $\langle
      \vct{\pi_j} \rangle : \prod_{i=1}^n QA_i \rightarrow
      \prod_{j \in J} QA_j$ for $J$ a list with elements from
      $\{1, \dots, n\}$, \[
        \left(f \circ \langle \vct{\pi_j} \rangle\right)^\ast
          \sim
        f^\ast \circ \langle \vct{\pi_j} \rangle
      \]
%    \item
%      For $f : \prod_{i=1}^n QA_i \rightarrow B$ and $g_i : D_i
%      \rightarrow A_i$, \[
%        \left(f \circ \prod_{i=1}^n Qg_i\right)^\ast
%          \sim
%        f^\ast \circ \prod_{i=1}^n Qg_i
%      \]
    \item \label{itm:astdistidem}
      If $(Q, \epsilon, \delta)$ is idempotent then for $f :
      \prod_{i=1}^n QA_i \rightarrow B$ and $k : \prod_{j=1}^m
      QD_j \rightarrow \prod_{i=1}^n QA_i$ we have \[
        (f \circ k)^\ast \approx_Q f^\ast \circ k
      \] and hence $(f \circ k)^\ast \sim f^\ast \circ k$. 
%      As a
%      corollary, for $f : \prod_{i=1}^n QA_i \rightarrow B$ and
%      $g_i : QD_i \rightarrow QA_i$, \[
%        \left(f \circ \prod_{i=1}^n g_i\right)^\ast
%          \sim
%        f^\ast \circ \prod_{i=1}^n g_i
%      \]
  \end{enumerate}
\end{prop}
\begin{proof}
  Straightforward calculations involving the comonadic equations.
  \ref{itm:idast} and \ref{itm:epsast} are standard from the
  theory of comonads and functional programming. In the case of
  idempotence we can use them alongside Proposition
  \ref{prop:deltaQast} to prove the intensional
  equation, e.g. \[
    Q(id_A^\ast) 
      \sim
    \delta_{QA} \circ Qid_A
      \sim
    \delta_{QA} 
      \sim
    Q\delta_A
  \] and hence $id_{QA}^\ast \approx_Q \delta_{QA}$. For
  \ref{itm:epsast} use $\delta_A \circ Q\epsilon_A \sim id_{Q^2
  A}$. \ref{itm:astastdist} and \ref{itm:astbulletdist} are easy
  calculations; e.g. for \ref{itm:astbulletdist}:
  \begin{derivation}
      (l \circ k^\bullet)^\ast
    \since[\sim]{definitions}
      Ql \circ Q^2 k 
        \circ Qm^{(n)} \circ m^{(n)} \circ \prod_{i=1}^n \delta_{A_i}
    \since[\sim]{$\delta$ monoidal}
      Ql \circ Q^2 k \circ \delta \circ m^{(n)}
    \since[\sim]{$\delta$ natural}
      Ql \circ \delta \circ Qk \circ m^{(n)}
  \end{derivation} which by definition is $l^\ast \circ
  k^\bullet$. Given idempotence it is simple to use Proposition
  \ref{prop:deltaQast} to prove \ref{itm:astastdist} and
  \ref{itm:astbulletdist} up to $\approx_Q$, without even using
  the non-idempotent result; e.g. for \ref{itm:astbulletdist}: \[
    Q\left(l \circ k^\bullet\right)^\ast
      \sim
    \delta_C \circ Ql \circ Qk^\bullet
      \sim
    Ql^\ast \circ Qk^\bullet
  \] \ref{itm:astdist} is a simple but lengthy calculation. With
  idempotence we have \begin{derivation}
      Q\left(f \circ \langle \vct{\pi_j} \rangle\right)^\ast
    \since[\sim]{Proposition \ref{prop:deltaQast}}
      \delta
        \circ Qf
        \circ Q\langle \vct{g_i} \rangle
    \since[\sim]{$\delta$ natural, Proposition \ref{prop:prodpresangle}}
      Q^2 f
        \circ \delta
        \circ m^{(n)}
        \circ \langle \vct{Qg_i} \rangle
    \since[\sim]{$\delta$ monoidal}
      Q^2 f
        \circ Qm^{(n)}
        \circ m^{(n)}
        \circ \prod_{i=1}^n \delta
        \circ \langle \vct{Qg_i} \rangle
    \since[\sim]{product equation, Proposition \ref{prop:deltaQast}}
      Q^2 f
        \circ Qm^{(n)}
        \circ m^{(n)}
        \circ \langle \vct{Q(g_i^\ast)} \rangle
    \since[\sim]{Proposition \ref{prop:prodpresangle}}
      Q^2 f
        \circ Qm^{(n)}
        \circ Q\langle \vct{(g_i^\ast)} \rangle
  \end{derivation} and hence $(f \circ \langle \vct{g_i})^\ast
  \approx_Q f^\bullet \circ \langle \vct{g_i^\ast} \rangle$.
  \ref{itm:astprojdist} is a corollary of \ref{itm:astdist}, once we
  notice that $\pi_i^\ast \sim \delta_{A_i} \circ \pi_i$, and use
  the definition of $f^\ast \myeq f^\bullet \circ \prod \delta$.
  For the idempotent case, notice that $\pi_i^\ast \approx_Q
  \delta_{A_i} \circ \pi_i$, or derive it as a corollary of
  \ref{itm:astdistidem}.

  \ref{itm:astdist} is an easy calculation: \[
    Q(f \circ k)^\ast 
      \sim
    \delta \circ Qf \circ Qk
      \sim 
    Qf^\ast \circ Qk
  \] which yields $(f \circ k)^\ast \approx_Q f^\ast \circ k$, and
  hence $(f \circ k)^\ast \sim f^\ast \circ k$. It is worth
  noting that we know of no direct calculation that proves (7) up
  to $\sim$ without going through $\approx_Q$.
\end{proof}

\noindent In other news, $(-)^\ast$ interacts predictably with
$\delta$ and $\epsilon$.

\begin{prop} \hfill
  \label{prop:deltaepsilonast}
  \begin{enumerate}[label=(\roman{*})]
    \item \label{itm:deltaafterast}
      Let $f : \prod_{i = 1}^n QA_i \rightarrow B$. Then 
      $\delta_B \circ f^\ast \sim \left(f^\ast\right)^\ast$.
      Furthermore, if $Q$ is idempotent then this equation holds
      intensionally.
    \item \label{itm:epsilonafterast}
      Let $f : \prod_{i=1}^n QA_i \rightarrow B$. Then
      $\epsilon_B \circ f^\ast \sim f$.  Furthermore, if $Q$ is
      idempotent then this equation holds intensionally.
  \end{enumerate}
\end{prop}

\begin{proof} \hfill
  \begin{enumerate}
    \item
      Let $E \myeq  \prod_{i = 1}^n QA_i$.  Then 
      \begin{derivation}
          \delta_B \circ f^\ast
        \since[\sim]{definition}
          \delta_B 
            \circ Qf
            \circ m^{(n)}
            \circ \prod_{i=1}^n \delta_{A_i}
        \since[\sim]{$\delta$ natural}
          Q^2 f
            \circ \delta_E 
            \circ m^{(n)}
            \circ \prod_{i=1}^n \delta_{A_i}
        \since[\sim]{$\delta$ monoidal}
          Q^2 f
            \circ Qm^{(n)}
            \circ m^{(n)}
            \circ \prod_{i=1}^n \delta_{QA_i}
            \circ \prod_{i=1}^n \delta_{A_i}
        \since[\sim]{product is functorial, comonadic equation}
          Q^2 f
            \circ Qm^{(n)}
            \circ m^{(n)}
            \circ \prod_{i=1}^n Q\delta_{A_i}
	    \circ \prod_{i=1}^n \delta_{A_i}
        \since[\sim]{$Q$ product-preserving}
          Q^2 f 
            \circ Qm^{(n)}
            \circ Q\left(\prod_{i=1}^n \delta_{A_i}\right)
            \circ m^{(n)}
            \circ \prod_{i=1}^n \delta_{A_i}
        \since[\sim]{definitions}
          \left(f^\ast\right)^\ast
      \end{derivation} If $Q$ is idempotent, we can also compute that
      \begin{derivation}
          Q\delta \circ Qf^\ast
        \since[\sim]{Proposition \ref{prop:deltaQast}}
          Q\delta \circ \delta \circ Qf
        \since[\sim]{comonadic equation}
          \delta \circ \delta \circ Qf
        \since[\sim]{Proposition \ref{prop:deltaQast}}
          \delta \circ Qf^\ast
        \since[\sim]{Proposition \ref{prop:deltaQast}}
          Q\left(\left(f^\ast\right)^\ast\right)
      \end{derivation} and hence $\delta \circ f^\ast \approx_Q
      \left(f^\ast\right)^\ast$.
    \item
      Straightforward calculation involving---amongst other
      things---the naturality and monoidality of $\epsilon$. If $Q$ is
      idempotent, then we calculate that
      \begin{derivation}
          Q\epsilon_B \circ Qf^\ast
        \since[\sim]{Proposition \ref{prop:deltaQast}}
          Q\epsilon_B \circ \delta_B \circ Qf
        \since[\sim]{comonadic equation}
          Qf
      \end{derivation} and hence $\epsilon_B \circ f^\ast \approx_Q
      f$.
  \end{enumerate}
\end{proof}

\noindent To conclude this section, we note that a special case of
Proposition \ref{prop:astdist}(7) generalises the
Quotation-Evaluation lemma (Lemma \ref{lem:qe}).

\begin{cor}
  \label{cor:qeparam}
    If $(Q, \epsilon, \delta)$ is idempotent then for $k :
    \prod_{i=1}^m QB_i \rightarrow QA$ we have \[
      (\epsilon_A \circ k)^\ast \approx_Q k
    \] and hence $(f \circ k)^\ast \sim f^\ast \circ k$. 
\end{cor}
\begin{proof}
  By Proposition \ref{prop:astdist} \ref{itm:astdistidem} \&
  \ref{itm:epsast},
  $
    (\epsilon_A \circ k)^\ast
  \approx_Q
    \epsilon_A^\ast \circ k
  \approx_Q
    k
  $ 
\end{proof}

\section{Fixed Points with Parameters}
  \label{sec:fpparam}

In \S\ref{chap:irec} we discussed two types of fixed points, EFPs
and IFPs. The first type concerned arrows $f : A \rightarrow A$,
whereas the second pertained to arrows $f : QA \rightarrow A$.
However, if we are to formulate a categorical semantics of PCF and
iPCF, we are going to need a slightly more general notion of each
fixed point, namely a \emph{fixed point with parameters}. The
parameters correspond to the \emph{context of free variables} in
the presence of which we are taking the relevant fixed point.

In the case of extensional fixed points, the context appears as a
cartesian product in the domain of the morphism, which is of type
$t : B \times Y \rightarrow Y$. $B$ is usually of the form
$\prod_{i=1}^n B_i$. Indeed, this is what happens in the
categorical semantics of PCF, for which see \cite{Hyland2000},
\cite{Poigne1992}, or \cite{Longley1995}. It is not at all
difficult to generalise Lawvere's theorem to produce this kind of
fixed point.

We then move on to intensional fixed points. The situation in this
case is slightly more nuanced, for---as we saw in
\S\ref{sec:lob}---we essentially need to model our
fixed points after \emph{L\"ob's rule}, viz. \[
  \begin{prooftree}
      \Box A \rightarrow A
    \justifies
      A
  \end{prooftree}
\] Adapting this rule to a parametric version is not a trivial
task, as it is almost equivalent to developing proof theory for
it. However, we can look to our previous work on the proof theory
of \textsf{GL} \citep{Kavvos2017b, Kavvos2017c} to find a good
pattern. We briefly discussed that work in \S\ref{sec:lob}.
The right form of the generalised rule is \[
  \begin{prooftree}
      \Box \Gamma, \Gamma \vdash \Box A \rightarrow A
    \justifies
      \Box \Gamma \vdash \Box A
  \end{prooftree}
\] However, since iPCF is based on an \textsf{S4}-like setting,
the \textsf{4} axiom is available. Thus, it suffices to consider a
rule of the following shape: \[
  \begin{prooftree}
      \Box \Gamma \vdash \Box A \rightarrow A
    \justifies
      \Box \Gamma \vdash \Box A
  \end{prooftree}
\] Indeed, this is very close to the rule we used for iPCF in
\S\ref{chap:ipcf}: the only remaining step is the weaken the fixed
point by dropping the box in the conclusion, thereby mimicking the
form of L\"ob's rule more commonly found in the literature.
Therefore, IFPs with parameters will pertain to arrows of type \[
  f : \prod_{i=1}^n QB_i \times QA \rightarrow A
\] and yield \[
  f^\dagger : \prod_{i=1}^n QB_i \rightarrow A
\] We thus arrive at a kind of context for IFPs that consists of a
handful of intensional assumptions $QB_i$. We elaborate on that in
\S\ref{sec:intfpparam}.

But, before that, let us recall the extensional case. 

\subsection{Extensional Fixed Points with Parameters}
  \label{sec:extfpparam}

Suppose we have
a cartesian 1-category $\mathcal{C}$. An arrow \[ 
  f : B \times A \rightarrow A
\] can be considered as a sort of endomorphism of $A$ that is
`parameterised' by $B$. Type-theoretically, we will consider $B$
to be the context, and we can take the EFP `at $A$.'

\begin{defn}
  Let $\mathcal{C}$ be a cartesian 1-category. A \emph{parametric
  extensional fixed point} (parametric EFP) of $f : B \times A
  \rightarrow A$ is an arrow $f^\dagger : B \rightarrow A$ such
  that the following diagram commutes: \[
    \begin{tikzcd}
      B
	\arrow[r, "{\langle id_B, f^\dagger \rangle}"]
	\arrow[d, "{f^\dagger}", swap]
      & B \times A
	\arrow[dl, "f"] \\
      A
      &
    \end{tikzcd}
  \]
\end{defn}

\begin{defn}
  [Extensional Fixed Points with Parameters]
  A cartesian category $\mathcal{C}$ has \emph{parametric
  extensional fixed points} (parametric EFPs) at $A \in
  \mathcal{C}$ just if for all $B \in \mathcal{C}$ there exists a
  map \[
    (-)_B^\dagger : \mathcal{C}(B \times A, A)
       	\longrightarrow
      \mathcal{C}(B, A)
  \] such that for each $f : B \times A \rightarrow A$, the
  morphism $f^\dagger : B \rightarrow A$ is a EFP of $f$.
\end{defn}

\noindent This, of course, is an `external view' of what it means
to have parametric EFPs in a category. However, in typed
$\lambda$-calculi we often include a fixed point combinator $Y_A :
(A \rightarrow A) \rightarrow A$ in our calculus, which is rather
more `internal.' This fixed point combinator can be either
\emph{weak} or \emph{strong}, and always occurs in the context of
a CCC.

\begin{defn}[Fixed Point Combinators]
  Let $\mathcal{C}$ be a cartesian closed category.
  
  \begin{enumerate}
    \item
      A \emph{strong fixed point combinator at $A$} is an arrow $Y
      : A^A \rightarrow A$ such that \[
	\begin{tikzcd}
	  A^A 
	    \arrow[r, "{\langle id, Y \rangle}"]
	    \arrow[d, "{Y}", swap]
	  & A^A \times A
	    \arrow[dl, "\textsf{ev}"] \\
	  A
	  &
	\end{tikzcd}
      \]
    \item
      A \emph{weak fixed point combinator at $A$} is an arrow $Y :
      A^A \rightarrow A$ such that, for each $f : B \times A
      \rightarrow A$, the arrow \[ 
	Y \circ \lambda(f) : B \rightarrow A
      \] is a EFP of $f$.
  \end{enumerate}
\end{defn}

\noindent Surprisingly, we have the following theorem.

\begin{thm}
  In a cartesian closed category $\mathcal{C}$, the following are
  equivalent: \begin{enumerate}
    \item A strong fixed point combinator at $A$.
    \item A weak fixed point combinator at $A$.
    \item Extensional fixed points at $A$.
  \end{enumerate}
\end{thm}
\begin{proof}
  We will prove a circular chain of implications.
  \begin{indproof}
    \case{$1 \Rightarrow 2$}
      If $Y : A^A \rightarrow A$ is a strong FPC, then \[
          Y \circ \lambda(f)
        = 
          \textsf{ev}
            \circ \langle id, Y \rangle
            \circ \lambda(f)
        =
          \textsf{ev}
            \circ \langle
                    \lambda(f),
                    Y \circ \lambda(f)
                  \rangle
        =
          f
            \circ 
              \langle
                id,
                Y \circ \lambda(f)
              \rangle
      \] so $Y \circ \lambda(f)$ is a fixed point of $f$.
    \case{$2 \Rightarrow 3$}
      Trivial: define $(-)^\dagger$ by $\lambda$-abstraction and
      composition with $Y$.
    \case{$3 \Rightarrow 1$}
      A strong FPC $Y$ at $A$ is a fixed point of $\textsf{ev} :
      A^A \times A \rightarrow A$. Hence, if we let \[
	Y \myeq A^A \xrightarrow{\textsf{ev}^\dagger} A
      \] then $Y = \textsf{ev} \circ \langle id, Y \rangle$.
  \end{indproof}
\end{proof}

\noindent Finally, we note the following naturality property,
which we learned from \cite{Simpson2000}.

\begin{lem}
  EFPs induced by a (strong or weak) fixed point combinator
  satisfy the following equation: for any $f : B \times A
  \rightarrow A$ and $g : C \rightarrow B$, \[
    (f \circ (g \times id_A))^\dagger = f^\dagger \circ g
  \]
\end{lem}
\begin{proof}
  We have \[
    (f \circ (g \times id_A))^\dagger = Y \circ \lambda\left(f
    \circ (g \times id_A)\right) = Y \circ \lambda(f) \circ g =
    f^\dagger \circ g
  \] by naturality of the $\lambda(-)$ operation.
\end{proof}

It is easy to extend the cartesian version of Lawvere's theorem to
one that also yields fixed points with parameters: we redefines
weak point-surjections to include a parameter.

\begin{defn}[Parametric weak point-surjection]
  \label{def:parametricwps}
  An arrow $r : X \times A \rightarrow Y$ is a \emph{parametric
  weak point-surjection} if for every $f : B \times A \rightarrow
  Y$ there exists a $x_f : B \rightarrow X$ such that \[
    \forall a : B \rightarrow A.\
      r \circ \langle x_f, a \rangle
	= f \circ \langle id_B,  a \rangle
  \]
\end{defn} 

\noindent The only change in the main theorem is that the fixed
points of an arrow of type $B \times Y \rightarrow Y$ now have
type $B \rightarrow Y$ instead of being points of $Y$.

\begin{thm}[Parametric Recursion]
  \label{thm:paramrec}
  If $r : A \times A \rightarrow Y$ is a parametric weak-point
  surjection, then every arrow \[
    t : B \times Y \rightarrow Y
  \] has a EFP.
\end{thm}
\begin{proof}
  Let \[
    f \myeq 
      B \times A
	\xrightarrow{id_B \times \langle id_A, id_A \rangle}
      B \times (A \times A)
	\xrightarrow{id_B \times r}
      B \times A
	\xrightarrow{t}
      A
  \] Then there exists a $x_f : B \rightarrow A$ such that \[
    r \circ \langle x_f, a \rangle = f \circ a
  \] for all $a : B \rightarrow A$. Then
  \begin{derivation}
      r \circ \langle x_f, x_f \rangle
    \since{definition of parametric weak point-surjection}
      t 
	\circ (id_B \times r)
	\circ (id_B \times \langle id_A, id_A \rangle)
	\circ \langle id_B, x_f \rangle
    \since{various product equations}
      t
	\circ \langle
		id_B,
		r \circ \langle x_f, x_f \rangle
	      \rangle
  \end{derivation} so that $r \circ \langle x_f, x_f \rangle$ is a
  EFP of $t$.
\end{proof}

\subsection{Intensional Fixed Points with Parameters}
  \label{sec:intfpparam}

Without further ado, we generalise IFPs to include parameters.

\begin{defn}
  \label{def:intfp}
  Let $(Q, \epsilon, \delta)$ be a product-preserving comonadic
  exposure. A \emph{parametric intensional fixed point}
  (parametric IFP) of $f : \prod_{i=1}^n QB_i \times QA
  \rightarrow A$ (w.r.t. $Q$) is an arrow \[
    f^\dagger : \prod_{i=1}^n QB_i \rightarrow A
  \] such that the following diagram commutes up to $\sim$: \[
    \begin{tikzcd}[column sep=large]
      \prod_{i=1}^n QB_i
	\arrow[r, "\langle id{,} \left(f^\dagger\right)^\ast \rangle"]
	\arrow[d, "f^\dagger", swap]
      & \prod_{i=1}^n QB_i \times QA
	\arrow[dl, "f"] \\
      A
      &
    \end{tikzcd}
  \]
\end{defn}

\noindent If $f : \prod_{i=1}^n QB_i \times QA \rightarrow
A$, then $f^\dagger : \prod_{i=1}^n QB_i \rightarrow A$, so
$\left(f^\dagger\right)^\ast : \prod_{i=1}^n QB_i \rightarrow QA$,
so this diagram has the right types.

This definition is a generalisation analogous to the one for EFPs.
The context is intensional ($\prod_{i=1}^n QB_i$) instead of
extensional ($B = \prod_{i=1}^n B_i$), and $f^\dagger$ appears
under the co-Kleisli lifting, so essentially under an occurrence of
$Q$. Notice that we have used both product-preservation and the
quoter $\delta : Q \natexp{} Q^2$ for this definition; this
probably corresponds to the fact that that the \textsf{4} axiom
($\Box A \rightarrow \Box\Box A$) is a theorem of provability
logic: see \citep{Boolos1994, Kavvos2017b, Kavvos2017c}. Since we
are in a categorical logic setting, an appropriate gadget standing
for the \textsf{4} axiom must be given as a primitive, alongside
appropriate coherence conditions (e.g. the first comonadic
equation).

\begin{defn}
  [Intensional Fixed Points with Parameters]
  \label{def:intfpparam}
  Let $(Q, \epsilon, \delta)$ be a product-preserving comonadic
  exposure on $\mathfrak{B}$. We say that $\mathfrak{B}$ has
  \emph{parametric intensional fixed points at $A$} (parametric
  IFPs) just if for any objects $B_i \in \mathcal{C}$ there exists
  an operation \[
    (-)_{\vct{B_i}}^\dagger :
      \mathfrak{B}\left(\prod_{i=1}^n QB_i \times QA, A\right)
       	\oper{}
      \mathfrak{B}\left(\prod_{i=1}^n QB_i, A\right)
  \] such that for each arrow $f : \prod_{i=1}^n QB_i \times QA
  \rightarrow A$, the arrow $f^\dagger : \prod_{i=1}^n QB_i
  \rightarrow A$ is an intensional fixed point of $f$.
\end{defn}

\noindent We often write $f^\dagger$ for succinctness, without
specifying the context $\vct{B_i}$.

In analogy to EFPs, there is a similar `internal' view of this
definition, related to the \emph{G\"odel-L\"ob axiom}. However, in
contrast to what we had before, it comes with certain caveats. But
first, some more definitions:

\begin{defn}[Intensional Fixed Point Combinators]
  Let there be a cartesian closed P-category $\mathfrak{B}$, along
  with a product-preserving comonadic exposure $(Q, \epsilon,
  \delta)$ on it.
  \begin{enumerate}
    \item
      A \emph{strong intensional fixed point combinator at
      $A$} (w.r.t. $Q$) is an arrow \[
        Y : Q(A^{QA}) \rightarrow QA
      \] such that the following diagram commutes up to $\sim$: \[
        \begin{tikzcd}[column sep=large]
          Q(A^{QA}) 
            \arrow[r, "{\langle \epsilon, Y\rangle}"]
            \arrow[d, "{Y}", swap]
          & A^{QA} \times QA
            \arrow[d, "\textsf{ev}"] \\
          QA
            \arrow[r, "\epsilon", swap]
          & A
        \end{tikzcd}
      \]
    \item
      A \emph{weak intensional fixed point combinator} (w.r.t.
      $Q$) is an arrow \[
        Y : Q(A^{QA}) \rightarrow QA
      \] such that, for each $f : \prod_{i=1}^n QB_i \times QA
      \rightarrow A$, the arrow \[ 
        \prod_{i=1}^n QB_i
          \xrightarrow{\left(\lambda\left(f\right)\right)^\ast}
        Q\left(A^{QA}\right)
          \xrightarrow{Y}
        QA
          \xrightarrow{\epsilon_A}
        A
      \] is a IFP of $f$.
  \end{enumerate}
\end{defn}

\begin{thm}
  \label{thm:ifpcs}
  Let $\mathfrak{B}$ be a cartesian closed P-category, along with
  a product-preserving comonadic exposure $(Q, \epsilon, \delta)$
  on it.
  \begin{enumerate}
    \item 
      If $(Q, \epsilon, \delta)$ is idempotent then a strong 
      fixed point combinator is also a weak fixed point
      combinator.
    \item
      A weak fixed point combinator $Y : Q(A^{QA}) \rightarrow QA$
      implies the existence of intensional fixed points at $A$.
    \item
      The existence of intensional fixed points at $A$ implies the
      existence of a strong intensional fixed point combinator at
      $A$.
  \end{enumerate}
\end{thm}

\begin{proof} \hfill
  \begin{enumerate}
    \item
      Let $Y : Q(A^{QA}) \rightarrow QA$ be a strong FPC.  Then,
      given $f : \prod_{i=1}^n QB_i \times QA \rightarrow
      A$, we may calculate that \begin{derivation}
          \epsilon \circ Y \circ (\lambda f)^\ast
        \since[\sim]{definition of strong FPC}
	  \textsf{ev}
            \circ \langle \epsilon, Y \rangle
            \circ (\lambda f)^\ast
        \since[\sim]{naturality of product, Proposition
		     \ref{prop:deltaepsilonast}
		     \ref{itm:epsilonafterast}}
	  \textsf{ev}
            \circ \langle \lambda f, Y \circ (\lambda f)^\ast \rangle
        \since[\sim]{cartesian closure}
          f \circ \langle id, Y \circ (\lambda f)^\ast \rangle
      \end{derivation} But as $Q$ is idempotent and $Y \circ
      (\lambda f)^\ast : \prod_{i=1}^n QB_i \rightarrow QA$, we
      can use quotation-evaluation (Corollary \ref{cor:qeparam})
      to conclude that $(\epsilon \circ Y \circ (\lambda
      f)^\ast)^\ast \approx_Q Y \circ (\lambda f)^\ast$, and hence
      that we have a IFP.
    \item
      Trivial.
    \item
      Let \[
        g \myeq
          Q(A^{QA}) \times QA
            \xrightarrow{\epsilon \times id}
          A^{QA} \times QA
            \xrightarrow{\textsf{ev}}
          A
      \] Then we can show that $(g^\dagger)^\ast : Q(A^{QA})
      \rightarrow QA$ is a strong FPC. It is easy to calculate
      that $
        g^\dagger \sim 
          \textsf{ev} 
            \circ \langle 
                    \epsilon,
                    \left(g^\dagger\right)^\ast
                  \rangle
      $ and hence that \[
	\epsilon \circ \left(g^\dagger\right)^\ast
	  \sim
	g^\dagger
	  \sim
	\textsf{ev} 
	  \circ \langle 
		  \epsilon,
		  \left(g^\dagger\right)^\ast
		\rangle
      \] by using Proposition \ref{prop:deltaepsilonast}
      \ref{itm:epsilonafterast}.
      \end{enumerate}
\end{proof}

It is interesting to examine if and when the map $(-)^\dagger$
might turn $\approx_Q$ to $\sim$, or even preserve it. In fact, it
is easy to see that if $\lambda(-)$ preserves $\approx_Q$, then we
can build a $(-)^\dagger$ that turns it into $\sim$: a weak FPC
suffices. If $Q$ is idempotent then $(-)^\dagger$ also preserves
$\approx_Q$.

\begin{lem}
  \label{lem:circpresint}
  If $Y : Q(A^{QA}) \rightarrow QA$ is a weak FPC, and
  the map \[
    \lambda_C(-) :
      \mathfrak{C}(C \times A, B)
        \rightarrow
      \mathfrak{C}(C, B^A)
  \] preserves the intensional equality $\approx_Q$, i.e. is a
  map \[
    \lambda_C(-) :
      (\mathfrak{C}, \approx_Q)(C \times A, B)
        \rightarrow
      (\mathfrak{C}, \approx_Q)(C, B^A)
  \] then the IFPs induced by $f^\dagger \myeq \epsilon_A \circ Y
  \circ \left(\lambda(f)\right)^\ast$ turn $\approx_Q$ into
  $\sim$, i.e.  $(-)^\dagger$ is a map \[
    (-)^\dagger :
      (\mathfrak{B}, \approx_Q)\left(\prod_{i=1}^n QB_i \times QA, A\right)
       	\rightarrow
      \mathfrak{B}\left(\prod_{i=1}^n QB_i, A\right)
  \] Moreover, if $Q$ is idempotent, then $(-)^\dagger$ preserves
  $\approx_Q$, and hence is a map \[
    (-)^\dagger :
      (\mathfrak{B}, \approx_Q)\left(\prod_{i=1}^n QB_i \times QA, A\right)
       	\rightarrow
      (\mathfrak{B}, \approx_Q)\left(\prod_{i=1}^n QB_i, A\right)
  \] 
\end{lem}
\begin{proof}
  Since $\lambda(-)$ preserves $\approx_Q$ (by assumption), and
  $(-)^\ast$ turns that into $\sim$ (\S\ref{sec:scene}) the result
  follows. In the case of idempotence observe that $(-)^\ast$ also
  preserves $\approx_Q$ (Corollary \ref{cor:idemastpresint}), and
  so does post-composition with $\epsilon \circ Y$.
\end{proof}

\noindent A similar argument will yield that, if $\lambda(-)$ is
natural up to $\approx_Q$, then so are the fixed points in a
certain sense.

\begin{lem}
  \label{lem:fixnat}
  If $(Q, \epsilon, \delta)$ is idempotent, $Y : Q(A^{QA})
  \rightarrow QA$ is a weak FPC, and \[
    \lambda_C(-) :
      \mathfrak{C}(C \times A, B)
        \rightarrow
      \mathfrak{C}(C, B^A)
  \] is natural up to $\approx_Q$, i.e. \[
    \lambda\left(f \circ (g \times id)\right) 
      \approx_Q
    \lambda(f) \circ g
  \] then the IFPs induced by $f^\dagger \myeq \epsilon_A \circ Y
  \circ \left(\lambda(f)\right)^\ast$ are also natural,
  i.e. for $f : \prod_{i=1}^n QB_i \times QA \rightarrow A$ and $k
  : \prod_{j=1}^k QC_j \rightarrow \prod_{i=1}^n QB_i$, we have \[
    \left(f \circ (k \times id)\right)^\dagger
      \approx_Q
    f^\dagger \circ k
  \]
\end{lem}
\begin{proof}
  Use naturality for $\lambda(-)$ up to $\approx_Q$, Corollary
  \ref{cor:idemastpresint} for the preservation of $\approx_Q$ by
  $(-)^\ast$, and finally Proposition \ref{prop:astdist}
  \ref{itm:astdistidem} to show that $(\lambda(f) \circ k)^\ast
  \approx_Q \left(\lambda(f)\right)^\ast \circ k$.
\end{proof}

\noindent There is also a weaker version of this lemma that does
not require idempotence: it states that if $f : \prod_{i=1}^n QB_i
\times QA \rightarrow A$ and $g_i : C_i \rightarrow B_i$, then $
  \left(f \circ 
      \left(\prod_{i=1}^n Qg_i \times id\right)\right)^\dagger
    \sim 
  f^\dagger \circ \prod_{i=1}^n g_i
$. However, we do not find any use for it in the sequel.

%  The proof for this version is the following:
%
%  \begin{derivation}
%      \epsilon \circ Y 
%        \circ \left(
%                \lambda\left(
%                        f
%                          \circ \left(\prod_{i=1}^n Qg_i \times id\right)
%                       \right)
%              \right)^\ast
%    \since[\sim]{$\lambda(-)$ natural up to $\approx_Q$, 
%                  which $(-)^\ast$ then turns into $\sim$}
%      \epsilon \circ Y 
%        \circ \left(
%                \lambda(f) \circ \prod_{i=1}^n Qg_i
%              \right)^\ast
%    \since[\sim]{Proposition \ref{prop:astdist}}
%      \epsilon \circ Y 
%        \circ \left(
%                \lambda(f) 
%              \right)^\ast \circ \prod_{i=1}^n g_i
%  \end{derivation}

We can summarise the above results by saying that if we have
\begin{itemize}
  \item 
    an idempotent comonadic exposure $(Q, \epsilon, \delta)$; 
  \item
    any of our three flavours of IFPs (strong, weak,
    $(-)^\dagger$);
  \item
    a $\lambda(-)$ that preserves $\approx_Q$ and is natural up to
    it
\end{itemize} then we obtain IFPs at $A$, which preserve
$\approx_Q$ and are natural up to $\approx_Q$. While this is nice
and useful, we will see in \S\ref{sec:asmipcf} that in the most
intensional of P-categories it will emphatically \emph{not} be the
case that $\lambda(-)$ preserves $\approx_Q$.

\section{A Parametric Intensional Lawvere Theorem}
  \label{sec:paramlawvere}

In the final section of this chapter we shall generalise the
Intensional Recursion Theorem (Theorem \ref{thm:intrec}) to a
version that admits parameters. The theorem will now guarantee the
existence of parametric IFPs.

Accordingly, we will have to also change the antecedent: neither
weak point-surjections (Def. \ref{def:wps}) nor parametric weak
point-surjections (Def.  \ref{def:parametricwps}) are adequate
anymore. We shall therefore introduce a variant, which---due to
lack of imagination on the part of the author---we shall call an
\emph{intensional parametric weak point-surjection}. The only
essential difference is the restriction of the arbitrary context
$B$ to an intensional one, viz. of the form $\prod_{i=1}^n QB_i$.

\begin{defn}[IPWPS]
  \label{def:parametriciwps}
  An arrow $r : X \times A \rightarrow Y$ is a \emph{intensional
  parametric weak point-surjection (IPWPS)} if, for every $f :
  \prod_{i=1}^n QB_i \times A \rightarrow Y$, there exists a $x_f
  : \prod_{i=1}^n QB_i \rightarrow X$ such that \[
    \forall a : \prod_{i=1}^n QB_i \rightarrow A.\
      r \circ \langle x_f, a \rangle
	= f \circ \langle id,  a \rangle
  \]
\end{defn} 

\noindent Armed with this, we can now prove a theorem analogous to
the Parametric Recursion Theorem (Theorem \ref{thm:paramrec}), but
yielding IFPs instead. Recall that in the Intensional Recursion Theorem
(Theorem \ref{thm:intrec}) we only needed a particular component
of $\delta$ to be `reasonable,' but in this case we will require
full idempotence.

\begin{thm}[Parametric Intensional Recursion]
  \label{thm:paramintrec}
  Let $(Q, \epsilon, \delta)$ be a product-preserving idempotent
  comonadic exposure. If $r : QA \times QA \rightarrow Y$ is a
  IPWPS, then every arrow \[
    t : \prod_{i=1}^n QB_i \times QY \rightarrow Y
  \] has an intensional fixed point.
\end{thm}
\begin{proof}
  Let \begin{align*}
    f \myeq
      \prod_{i=1}^n QB_i \times QA
        &\xrightarrow{id \times \langle id, id \rangle}
          \prod_{i=1}^n QB_i \times (Q^2 A \times Q^2 A) \\
        &\xrightarrow{id \times r^\ast}
          \prod_{i=1}^n QB_i \times QY \\
        &\xrightarrow{t}
          Y
  \end{align*} Then, there exists a $x_f : \prod_{i=1}^n QB_i
  \rightarrow QA$ such that \[
      r \circ \langle x_f, a \rangle \sim f \circ \langle id, a \rangle
  \] for all $a : \prod_{i=1}^n QB_i \rightarrow QA$. We compute that
  \begin{derivation}
      r \circ \langle x_f, x_f \rangle
    \since[\sim]{definition of IPWPS}
      t 
        \circ (id \times r^\ast)
	\circ (id \times \langle id, id \rangle)
        \circ \langle id, x_f \rangle
    \since[\approx_Q]{product equation, naturality of brackets}
      t 
        \circ \left\langle
                      id,
                      r^\ast \circ \langle x_f, x_f \rangle
              \right\rangle
    \since[\approx_Q]{idempotence: Prop.
			\ref{prop:astdist}\ref{itm:astdistidem}}
      t 
        \circ \left\langle
                      id,
                      \left(r 
                              \circ \langle x_f, x_f \rangle \right)^\ast
              \right\rangle
  \end{derivation} so that $r \circ \langle x_f, x_f \rangle$ is
  a IFP of $t$.
\end{proof}

\noindent Notice that we are allowed to use Proposition
\ref{prop:astdist}\ref{itm:astdistidem} only because $\langle x_f,
x_f \rangle : \prod_{i=1}^n QB_i \rightarrow QA \times QA$ is of
the right type, i.e. there are occurrences of $Q$ `guarding' all
the types.

\subsection*{Naturality}

As we saw in \S\ref{sec:intfpparam}, the main reason for
introducing parametric IFPs is to have a context $\prod_{i=1}^n
QB_i$. In turn, a context is useful because one can
\emph{substitute} for any objects in it, as they represent free
variables. In this light, naturality is a key property: it states
that substitution (composition) commutes with the
(type-theoretic/categorical) construct in question; we will
discuss that more in \S\ref{chap:intsem2}.

We showed in Lemmata \ref{lem:circpresint} and \ref{lem:fixnat}
that, in the idempotent setting, if IFPs are induced by a weak
FPC, and $\lambda(-)$ preserves $\approx_Q$, then so does
$(-)^\dagger$, and moreover it is natural. But what about the IFPs
produced by the Parametric Intensional Recursion Theorem?

A IPWPS $r : QA \times QA \rightarrow Y$ induces a map \[
  f : \prod_{i=1}^n QB_i \times QA \rightarrow Y
    \longmapsto
  x_f : \prod_{i=1}^n QB_i \rightarrow QA
\] and then we can define $(-)^\dagger : 
      \mathfrak{B}\left(\prod_{i=1}^n QB_i \times QY, Y\right)
       	\oper{}
      \mathfrak{B}\left(\prod_{i=1}^n QB_i, Y\right)
$ to be \[
  t
    \longmapsto
  f_t
    \longmapsto
  x_{f_t}
    \longmapsto
  r \circ \langle x_{f_t}, x_{f_t} \rangle :
\] where $f_t \myeq t \circ (id \times r^\ast) \circ (id \times
\langle id, id \rangle) : \prod_{i=1}^n QB_i \times QA \rightarrow
Y$.

The final map $x_{f_t} \mapsto r \circ \langle x_{f_t}, x_{f_t}
\rangle$ is clearly natural and preserves both $\sim$ and
$\approx_Q$, as $\langle \cdot, \cdot \rangle$ does. But what
about the rest? Suppose that we pre-compose $t$ with $g \times id$,
where $g : \prod_{j=1}^m QD_j \rightarrow \prod_{i=1}^n QB_i$. The
map $t \mapsto f_t$ clearly preserves both $\sim$ and $\approx_Q$,
and moreover it is natural, in the sense that \[
  f_{t \circ (g \times id)}
    \approx_Q
  f_t \circ (g \times id)
\] because of the interchange law for cartesian products, which
holds up to $\approx_Q$.

This leaves only the difficult case of $f \mapsto x_f$. For
naturality, we would need\footnote{Note that because of
idempotence and the type of $x_f$, $\sim$ suffices.} \[
  x_{f_{t \circ (g \times id)}}
    \approx_Q
  x_{f_t} \circ g 
\] which easily follows if $f \mapsto x_f$ preserves $\approx_Q$,
and\footnote{Ditto.} \[
  x_{h \circ (g \times id)}
    \approx_Q
  x_h \circ g
\] In that case $x_{f_{t \circ (g \times id)}} \approx_Q x_{f_t
\circ (g \times id)} \approx_Q x_{f_t} \circ g$, and hence \[
  \left(t \circ (g \times id)\right)^\dagger
    \approx_Q
  r \circ \langle x_{f_{t \circ (g \times id)}} ,
                  x_{f_{t \circ (g \times id)}}
          \rangle
    \approx_Q
  r \circ \langle x_{f_t} \circ g, x_{f_t} \circ g \rangle
    \approx_Q
  r \circ \langle x_{f_t}, x_{f_t} \rangle \circ g
    \approx_Q
  t^\dagger \circ g
\] Hence, to obtain naturality we would need that the IPWPS be
`natural,' and that $f \mapsto x_f$ turn $\approx_Q$ to $\sim$.

\begin{cor}
  If the IPWPS $r : QA \times QA \rightarrow Y$ of Theorem
  \ref{thm:paramintrec} is such that $f \mapsto x_f$ preserves 
  $\approx_Q$ (or, equivalently, turns $\approx_Q$ into $\sim$),
  and is `natural' insofar as \[ 
    x_{f \circ (g \times id)} 
      \approx_Q
    x_f \circ g
  \] (equivalently, with $\sim$) then the induced $(-)^\dagger$ is
  a map \[
    (-)^\dagger : 
      (\mathfrak{B}, \approx_Q)\left(\prod_{i=1}^n QB_i \times QY, Y\right)
       	\rightarrow
      (\mathfrak{B}, \approx_Q)\left(\prod_{i=1}^n QB_i, Y\right)
  \] which is natural up to $\approx_Q$, i.e. \[
    \left(f \circ (g \times id)\right)^\dagger
      \approx_Q
    f^\dagger \circ g
  \]
\end{cor}

\chapter{Intensional Semantics of iPCF II}

  \label{chap:intsem2}

  In this final chapter, we shall use all our results up to this
point to attempt to produce an intensional semantics for iPCF.
The intended---but, as it transpires, unachievable---result is
that the category of assemblies $\mathfrak{Asm}(K_1)$ is a model
of iPCF.

We have shown in previous chapters that a \emph{product-preserving
comonadic exposure} satisfies most of the standard equations of a
product-preserving comonad. It should therefore be the case that,
given such an exposure, one can use it to almost directly
interpret the Davies-Pfenning fragment of iPCF. However, this is
not so straightforward. The main ingredient of our soundness
result is a \emph{substituion lemma}, which relates the
interpretation with substition. Since we are allowed to substitute
terms under $\ibox{(-)}$ constructs---which we model by
$(-)^\ast$---we need the substitution lemma to hold up to
$\approx_Q$ at that location. Furthermore, since we may have
multiple nested occurrences of boxes, we will need that $(-)^\ast$
preserve $\approx_Q$, which is the case if the comonadic exposure
is \emph{idempotent} (Cor.  \ref{cor:idemastpresint}). Thus
idempotence is essential.

Once this is decided, we run into a second issue. Suppose that
there is a $\lambda$ in a term, which we model by the function
$\lambda(-)$ of a cartesian closed category. Since this $\lambda$
might occur in the scope of a $\ibox{(-)}$, and a (modal) variable
for which we want to substitute might occur under that $\lambda$,
the above soundness result requires that $\lambda(-)$ be natural
up to $\approx_Q$. Whereas in the case of $\mathfrak{Asm}(K_1)$,
the exposure is idempotent, we have no guarantees at all about the
naturality of $\lambda(-)$---see \S\ref{sec:wextasm}. We will
therefore need to \emph{forbid $\lambda$-abstractions with free
variables under boxes}.

The above suffices to interpret the Davies-Pfenning fragment, in
an intensional sense. The remaining piece of the puzzle pertains
to the construction of IFPs. This is easy for booleans and
naturals, but at higher types we are only able to use an inductive
construction that only builds IFPs for a certain kind of
\emph{intensional exponential ideal}: if $X$ is any object and $Y$
is in the ideal, then $Y^{QX}$ is in the ideal as well. We also
face certain difficulties in showing that the IFPs are natural, as
per \S\ref{sec:paramlawvere}, and lack of naturality entails lack
of substitution. It follows that \emph{we must constrain the
taking of IFPs to closed terms, at only at certain types}.

We are thus led to reformulate iPCF, or---more specifically---to
admit only a subset of it. To reach this subset, we first
\emph{layer} it, so as to separate the \emph{intensional
layer}---where everything will hold up to $\approx_Q$---and the
\emph{extensional layer}, where things will hold up to $\sim$.
Since the modality already implicitly enables this kind of
layering, this will be a minor change. Nevertheless, it decisively
quells the problem, as we will never need to substitute in a
$\lambda$-abstraction under the box, so the desired lemma holds up
to $\approx_Q$ at that location. In the case of IFPs, our only
option is to alter the fixed point rule.

The resulting language is called \emph{iPCF v2.0}. On the one
hand, the new language has $\mathfrak{Asm}(K_1)$ as a model; on
the other, a lot of expressiveness has been lost. We discuss what
has been lost, and when one can regain it. In particular, we can
model iPCF itself when given a \emph{natural} iPCF v2.0 model.
Furthermore, a \emph{weakly extensional} model of iPCF v2.0 is
already natural; the terminology comes from PCAs: if $A$ is a
weakly extensional PCA, then $\mathfrak{Asm}(A)$ is a weakly
extensional model, and hence a model of iPCF; see also
\S\ref{sec:wextasm}.

\section{iPCF v2.0}
  \label{sec:ipcfv2.0}

We promptly introduce iPCF v2.0. We shall not prove any theorems
in this section, because we have formally proven all of them in
\textsc{Agda}: see Appendix \ref{app:agda} for the proofs.

Each typing judgement of iPCF v2.0 will be \emph{annotated} by a
$\mathcal{J}$, like so: \[ \ctxt{\Delta}{\Gamma} \vj{} M : A \]
The possible options for $\mathcal{J}$ will be ``int.'' for
intensional, and ``ext.'' for extensional. 

\begin{figure}
  \caption{Syntax and Typing Rules for Intensional PCF v2.0}
  \label{fig:ipcf2}
  \begin{align*}
  \textbf{Ground Types} \quad &
    G & ::=\quad &\textsf{Nat} \;|\; \textsf{Bool}
   \\ \\
  \textbf{Types} \quad &
    A, B & ::=\quad &G \;|\; A \rightarrow B \;|\; \Box A
   \\ \\
  \textbf{Fixable Types} \quad &
    A_\textsf{fix} & ::=\quad &G \;|\; \Box A \rightarrow A_\textsf{fix}
   \\ \\
  \textbf{Terms} \quad &
    M, N & ::=\quad &x
	    \;|\; \lambda x{:}A.\ M
	    \;|\; M N
	    \;|\; \ibox{M}
	    \;|\; \letbox{u}{M}{N} \;| \\
   &      &    &\widehat{n}
	    \;|\; \textsf{true}
	    \;|\; \textsf{false}
	    \;|\; \textsf{succ}
	    \;|\; \textsf{pred}
	    \;|\; \textsf{zero?}
	    \;|\; \supset_G
	    \;|\; \fixlob{z}{M}
    \\ \\
%  \textbf{Canonical Forms} \quad &
%    V & ::= \quad &\widehat{n}
%	    \;|\; \textsf{true}
%	    \;|\; \textsf{false}
%	    \;|\; \lambda x{:}A.\ M
%	    \;|\; \textsf{box } M
%   \\ \\
  \textbf{Contexts} \quad &
    \Gamma, \Delta & ::=\quad &\cdot \;|\; \Gamma, x: A
\end{align*}

\vfill

\renewcommand{\arraystretch}{3}

% TABLE FOR CONSTANTS
\begin{tabular}{c c}
  %% CONSTANT RULES
  $
    \begin{prooftree}
      \justifies
        \ctxt{\Delta}{\Gamma} \vj \widehat{n} : \textsf{Nat}
    \end{prooftree}
  $
  
  &

  $	  
    \begin{prooftree}
      \justifies
        \ctxt{\Delta}{\Gamma} \vj b : \textsf{Bool}
      \using
        (b \in \{\textsf{true}, \textsf{false}\})
    \end{prooftree}
  $
  
  \\

  $	
    \begin{prooftree}
      \justifies
        \ctxt{\Delta}{\Gamma} 
          \vj \textsf{zero?} : \textsf{Nat} \rightarrow \textsf{Bool}
    \end{prooftree}
  $
  
  &

  $	
    \begin{prooftree}
      \justifies
        \ctxt{\Delta}{\Gamma}
            \vj f : \textsf{Nat} \rightarrow \textsf{Nat}
      \using
        (f \in \{\textsf{succ}, \textsf{pred}\})
    \end{prooftree}
  $
  
  \\

  \multicolumn{2}{c}{
  $
    \begin{prooftree}
      \justifies
        \ctxt{\Delta}{\Gamma}
            \vj {\supset_G}  : 
              \textsf{Bool} \rightarrow G \rightarrow G \rightarrow G
    \end{prooftree}
  $
  }

  \\

  %% VARIABLE RULES

  $
    \begin{prooftree}
      \justifies
        \ctxt{\Delta}{\Gamma, x{:}A, \Gamma'} \vj x:A
      \using
        {(\textsf{var})}
    \end{prooftree}
  $

  &

  $
    \begin{prooftree}
      \justifies
        \ctxt{\Delta, u{:} A, \Delta'}{\Gamma} \vj u:A
      \using
        {(\Box\textsf{var})}
    \end{prooftree}
  $

  \\

  %% IMPLICATION RULES

  $
    \begin{prooftree}
        \ctxt{\Delta}{\Gamma}, x{:}A \vext M : B
      \justifies
        \ctxt{\Delta}{\Gamma} \vext \lambda x{:}A. \; M : A \rightarrow B
      \using
        {(\rightarrow\mathcal{I})}
    \end{prooftree}
  $

  &

  $
    \begin{prooftree}
        \ctxt{\Delta}{\Gamma} \vj M : A \rightarrow B
      \quad
        \ctxt{\Delta}{\Gamma} \vj N : A
      \justifies
        \ctxt{\Delta}{\Gamma} \vj M N : B
      \using
        {(\rightarrow\mathcal{E})}
    \end{prooftree}
  $

  \\

  %% BOX RULES

  $
    \begin{prooftree}
        \ctxt{\Delta}{\cdot} \vint M : A
      \justifies
        \ctxt{\Delta}{\Gamma} \vj \ibox{M} : \Box A
      \using
        {(\Box\mathcal{I})}
    \end{prooftree}
  $

  &

  $
    \begin{prooftree}
        \ctxt{\Delta}{\Gamma} \vj M : \Box A
      \quad\quad
        \ctxt{\Delta, u{:}A}{\Gamma} \vj N : C
      \justifies
        \ctxt{\Delta}{\Gamma} \vj \letbox{u}{M}{N} : C
      \using
        {(\Box\mathcal{E})}
    \end{prooftree}
  $

  \\

  \multicolumn{2}{c}{
    $
      \begin{prooftree}
        \ctxt{\cdot}{z : \Box A_\textsf{fix}} \vint M : A_\textsf{fix}
          \justifies
        \ctxt{\Delta}{\Gamma} \vj \fixlob{z}{M} : A_\textsf{fix}
          \using{(\Box\textsf{fix})}
      \end{prooftree}
    $
  }

  \\

  \multicolumn{2}{c}{
  $
    \begin{prooftree}
      \ctxt{\cdot}{x{:}A} \vj M : B
        \justifies
      \ctxt{\Delta}{\Gamma} \vint \lambda x{:}A.\ M : A \rightarrow B
        \using{(\rightarrow\mathcal{I}_\text{int.})}
    \end{prooftree}
  $
  }
\end{tabular}

\end{figure}

The revised system appears in Figure \ref{fig:ipcf2}. The
occurrence of a generic $\mathcal{J}$ in these rules is universally
quantified. There is little else to say apart from the fact that
these rules enforce the prohibition of free variables under a
$\lambda$ in the intensional judgements. In programming language
terms this could be transliterated as the prohibition of the
creation of \emph{closures} (= $\lambda$-abstraction + environment
for free variables). Of course, a term can then only be placed
under a box if it is intensional, but the boxed term itself can be
either intensional or extensional: \[
  \begin{prooftree}
    \ctxt{\Delta}{\cdot} \vint{} M : A
      \justifies
    \ctxt{\Delta}{\Gamma} \vj{} \ibox{M} : \Box A
  \end{prooftree}
\] We can only $\lambda$-abstract in an extensional term, yielding
another extensional term: \[
  \begin{prooftree}
    \ctxt{\Delta}{\Gamma, x{:}A } \vext{} M : B
      \justifies
    \ctxt{\Delta}{\Gamma} \vext{} \lambda x{:}A.\ M : A \rightarrow B
  \end{prooftree}
\] But if this term has no other free variables, then the result
can be an intensional term. Of course, we must not forget to
include the `opportunity' to weaken the context: \[
  \begin{prooftree}
    \ctxt{\cdot}{x{:}A} \vj{} M : B
      \justifies
    \ctxt{\Delta}{\Gamma} \vint{} \lambda x{:}A.\ M : A \rightarrow B
  \end{prooftree}
\] We shall also change the rule for intensional fixed points,
which now reads \[
  \begin{prooftree}
    \ctxt{\cdot}{z : \Box A_\textsf{fix}} \vint M : A_\textsf{fix}
      \justifies
    \ctxt{\Delta}{\Gamma} \vj 
  \fixlob{z}{M} = M[\ibox{(\fixlob{z}{M})}/z] : A_\textsf{fix}
      \using
      {(\Box\textsf{fix})}
  \end{prooftree}
\] So $\fixlob{z}{M}$ recurses, but it can only do so when $M$ is
closed to everything else save the `diagonal' variable $z$.
What is more, we can only invoke this rule for types
$A_\textsf{fix}$ generated by the following grammar, where $A$ is
any type at all: \[
  A_\textsf{fix} \quad ::= \quad
    \textsf{Nat} 
    \quad|\quad \textsf{Bool} 
    \quad|\quad \Box A \rightarrow A_\textsf{fix}
\] 

\begin{thm}
  The following rules are admissible in iPCF v2.0: \[
    \begin{prooftree}
      \ctxt{\Delta}{\Gamma} \vj{} M : A
  \justifies
      \ctxt{\Delta}{\Gamma} \vext{} M : A
    \end{prooftree}
      \qquad\qquad
    \begin{prooftree}
      \ctxt{\Delta}{\Gamma} \vint{} M : A
  \justifies
      \ctxt{\Delta}{\Gamma} \vj{} M : A
    \end{prooftree}
  \]
\end{thm}

\noindent The standard admissibility results for iPCF are also
valid in iPCF v2.0, but they are now parametric in $\mathcal{J}$.

\begin{thm}[Structural]
  \label{thm:struct2}
    The standard structural rules of iPCF (as stated in Theorem
    \ref{thm:scut}) are admissible in iPCF v2.0, parametrically up
    to $\mathcal{J}$.
\end{thm}

\noindent The situation with the cut rule, however, is slightly
more complicated, and this has to do with the nature of the term
being substituted. If we substitute an intensional term for a
variable, the resulting term will still retain its original
disposition (int. or ext.).  However, if we substitute an
extensional term, we force the resulting term to be extensional.
Similarly, and because of $(\Box\mathcal{I})$, we may only
substitute an intensional term for a modal variable, and that
leaves the disposition of the term invariant.

\begin{thm}[Cut for iPCF v2.0]
  \label{thm:cut2} 
  The following rules are admissible in iPCF v2.0.
  \begin{enumerate}
    \item (Cut-Ext) \[
      \begin{prooftree}
        \ctxt{\Delta}{\Gamma} \vext{} N : A
          \qquad
        \ctxt{\Delta}{\Gamma, x{:}A, \Gamma'} \vj{} M : A
          \justifies
        \ctxt{\Delta}{\Gamma, \Gamma'} \vext{} M[N/x] : A
      \end{prooftree}
    \]
    \item (Cut-Int) \[
      \begin{prooftree}
        \ctxt{\Delta}{\Gamma} \vint{} N : A
          \qquad
        \ctxt{\Delta}{\Gamma, x{:}A, \Gamma'} \vj{} M : A
          \justifies
        \ctxt{\Delta}{\Gamma, \Gamma'} \vj{} M[N/x] : A
      \end{prooftree}
    \]
    \item (Modal Cut) \[
      \begin{prooftree}
        \ctxt{\Delta}{\cdot} \vint{} N :  A
          \quad
        \ctxt{\Delta, u{:}A, \Delta'}{\Gamma} \vj{} M : C
          \justifies
        \ctxt{\Delta, \Delta'}{\Gamma} \vj{} M[N/u] : C
      \end{prooftree}
    \]
  \end{enumerate}
\end{thm}

\begin{figure}
  \caption{Equational Theory for Intensional PCF v2.0}
  \label{fig:ipcf2eq}
  \renewcommand{\arraystretch}{4}

\begin{tabular}{c c}

%  \textbf{Equivalence Relation} & \\ \\
%
%  \multicolumn{2}{c}{
%    $ \begin{prooftree}
%        \ctxt{\Delta}{\Gamma} \vdash M : A
%          \justifies
%        \ctxt{\Delta}{\Gamma} \vdash M = M : A
%      \end{prooftree} $
%  } \\ \\
%
%  $ \begin{prooftree}
%      \ctxt{\Delta}{\Gamma} \vdash M = N : A
%        \justifies
%      \ctxt{\Delta}{\Gamma} \vdash N = M : A
%    \end{prooftree} $
%  &
%  $ \begin{prooftree}
%      \ctxt{\Delta}{\Gamma} \vdash P = Q : A
%        \qquad
%      \ctxt{\Delta}{\Gamma} \vdash Q = R : A
%        \justifies
%      \ctxt{\Delta}{\Gamma} \vdash P = R : A
%    \end{prooftree} $
%

  \textbf{Function Spaces} & \\

  \multicolumn{2}{c}{
    $
      \begin{prooftree}
        \ctxt{\Delta}{\Gamma} \vext N : A
          \qquad
        \ctxt{\Delta}{\Gamma, x{:}A, \Gamma'} \vj M : B
          \justifies
        \ctxt{\Delta}{\Gamma, \Gamma'}
            \vext (\lambda x{:}A. M)\,N = M[N/x] : B
          \using
        {(\rightarrow\beta)}
      \end{prooftree}
    $
  }

  \\

%  \multicolumn{2}{c}{
%    $
%      \begin{prooftree}
%        \ctxt{\Delta}{\Gamma} \vint N : A
%          \qquad
%        \ctxt{\Delta}{\Gamma, x{:}A, \Gamma'} \vj M : B
%          \justifies
%        \ctxt{\Delta}{\Gamma, \Gamma'}
%            \vj (\lambda x{:}A. M)\,N = M[N/x] : B
%          \using
%        {(\rightarrow\beta_\text{int.})}
%      \end{prooftree}
%    $
%  }

  \multicolumn{2}{c}{
    $
      \begin{prooftree}
        \ctxt{\Delta}{\Gamma} \vj M : A \rightarrow B
          \qquad
        x \not\in \text{fv}(M)
          \justifies
        \ctxt{\Delta}{\Gamma} \vext M = \lambda x{:}A. Mx : A \rightarrow B
          \using
        {(\rightarrow\eta)}
      \end{prooftree}
    $
  }

  \\ 

  \textbf{Modality} & \\

  \multicolumn{2}{c}{
    $
      \begin{prooftree}
        \ctxt{\Delta}{\cdot} \vint M : A
          \qquad
        \ctxt{\Delta, u : A}{\Gamma} \vj N : C
          \justifies
        \ctxt{\Delta}{\Gamma} \vj \letbox{u}{\ibox{M}}{N} = N[M/x] : C
          \using
        {(\Box\beta)}
      \end{prooftree}
    $
  }
    
  \\

  \multicolumn{2}{c}{
  $ \begin{prooftree}
      \ctxt{\Delta}{\Gamma} \vint M : \Box A
        \justifies
      \ctxt{\Delta}{\Gamma} \vj \letbox{u}{M}{\ibox{u}} = M : \Box A
        \using
      {(\Box\eta)}
    \end{prooftree} $
  }

  \\ 

  \multicolumn{2}{c}{
    $
      \begin{prooftree}
        \ctxt{\Delta}{\cdot} \vint M = N : A
          \justifies
        \ctxt{\Delta}{\Gamma}
	    \vj \ibox{M} = \ibox{N} : \Box A
          \using
        {(\Box\textsf{cong})}
      \end{prooftree}
    $
  }

  \\

  \multicolumn{2}{c}{
    $
      \begin{prooftree}
        \ctxt{\cdot}{z : \Box A} \vint M : A
          \justifies
        \ctxt{\Delta}{\Gamma} \vj 
            \fixlob{z}{M} = M[\ibox{(\fixlob{z}{M})}/z] : A
          \using
          {(\Box\textsf{fix})}
      \end{prooftree}
    $
  }

  \\

%  \multicolumn{2}{c}{
%    $
%      \begin{prooftree}
%        \ctxt{\cdot}{\cdot} \vdash M : A
%          \quad
%        f \in \mathcal{F}(A, B)
%          \justifies
%        \ctxt{\Delta}{\Gamma} \vext
%            \tilde f (\ibox{M}) = \ibox{f(M)} : \Box B
%          \using
%        {(\Box\textsf{int})}
%      \end{prooftree}
%    $
%  }
%
%  \\

  \multicolumn{2}{c}{
    $
      \begin{prooftree}
        \ctxt{\Delta}{\Gamma} \vj M = N : \Box A
          \qquad
        \ctxt{\Delta}{\Gamma} \vj P = Q : C
          \justifies
        \ctxt{\Delta}{\Gamma} \vj
            \letbox{u}{M}{P} = \letbox{u}{N}{Q} : B
          \using
        {(\Box\textsf{let-cong})}
      \end{prooftree}
    $
  }

  \\

  \multicolumn{2}{c}{
    \begin{minipage}{\textwidth}
      \textbf{Remark}. In addition to the above, one should also
      include (a) rules that ensure that equality is an equivalence
      relation, (b) congruence rules for $\lambda$-abstraction and
      application, and (c) rules corresponding to the behaviour of
      constants, as e.g. in Figure \ref{fig:ipcfbeta}. 
    \end{minipage}
  }

\end{tabular}

\end{figure}

In light of those theorems we reformulate the equational theory of
iPCF (Figure \ref{fig:ipcfeq}) for iPCF v2.0. The resulting theory
can be found in Figure \ref{fig:ipcf2eq}. Curiously, we see that
\emph{a form of the congruence rule for $\Box$ reappears}, even
though the considerations of Davies and Pfenning led us to banish
such rules in \S\ref{chap:ipcf}. When proving soundness, we will
see that this rule reflects the fact shown in Corollary
\ref{cor:idemastpresint}, viz. that $(-)^\ast$ preserves
intensional equality $\approx_Q$.

\subsection*{Expressitivity of iPCF v2.0}

It is easy to see that every typing judgment of iPCF v2.0 is also
a typing judgment of iPCF: each rule of v2.0 is a special case of
the rule for iPCF. Hence, iPCF v2.0 is in some sense a proper
subset of iPCF.

We can thus conclude that, by moving to v2.0, we have lost some
expressivity. This is centred around three limitations:
\begin{enumerate}
  \item no free variables under $\lambda$-abstractions in the
  modal/intensional fragment, i.e. under a $\ibox{(-)}$;
  \item IFPs can only be taken when there is exactly one free
  variable, the \emph{diagonal} variable, and that must be of
  modal type; and
  \item IFPs can only be taken at certain types
\end{enumerate}

The first limitation is, in a way, double-edged. On the one hand,
it is reasonable and familiar from someone coming from the
computability theory, especially from the perspective of
\cite{Jones1997}. Indices do not have ``free variables'';
sometimes they are meant to have more than one argument, and in
those cases we use the s-m-n theorem to substitute for one of
those; this can be simulated here using $\lambda$-abstraction.  On
the other hand, this limitation invalidates \emph{every single one
of the original examples of \textsf{S4}-typed staged
metaprogramming of \cite{Davies2001a}} ($power$, $acker$, $ip$,
etc.---see \S7 of that paper): in almost all cases, a $\ibox{(-)}$
containing a $\lambda$-abstraction with free variables is
implicated in the result.

The second limitation seems reasonably innocuous, but it is
actually more severe than it looks. As an example, it renders the
\emph{intensional fixed-point combinator}---whose type is the
G\"odel-L\"ob axiom, see \S\ref{sec:ipcfterms}---untypable:
\begin{align*}
  &\mathbb{Y}_A \myeq
    \lambda x : \Box (\Box A \rightarrow A). \;
      \letbox{f}{x}{
          \ibox{\left(\fixlob{z}{f\,z}\right)}
      } \\
 \not\vdash\ &\mathbb{Y}_A : 
  \Box(\Box A \rightarrow A) \rightarrow \Box A
\end{align*} Notice the characteristic occurrence of $f$ within
the $\fixlob{z}{(-)}$ construct. This particular limitation is
severe, and increases the distance between iPCF and provability
logic: L\"ob's rule does not even imply the G\"odel-L\"ob
axiom. This has to do with the `loss of naturality' that occurs
when we constrain the fixed point rule: the $\Box$ in the
antecedent $\Box(\Box A \rightarrow A)$ means that `only code
variables can be found in the input data,' and this is too weak an
assumption to use L\"ob's rule, which requires exactly one
free variable, the \emph{diagonal} one. This could be fixed by
redefining $\ibox{(-)}$ to only enclose \emph{completely closed}
terms, but then we completely obliterate the expressive content of
iPCF. If we did that, modal types would not be useful for anything
at all, leading us to add more and more \emph{combinators} that do
specific things, e.g. something like $\textbf{app} : \Box (A
\rightarrow B) \rightarrow \Box A \rightarrow \Box B$ with a
$\delta$-reduction of the form $\textbf{app}\ (\ibox{F})(\ibox{M})
\rightarrow \ibox{FM}$. We are not sure that this approach is
sustainable.

The relationship of this second limitation with computability
theory is more subtle.  Using the SRT on a program with a `free
variable' made no sense at all, as using it on a program with `two
arguments' was the norm; thus, not much seems to be lost. However,
we remind the reader that \emph{`constructive'} versions of the
SRT are available in computability theory, e.g. there is a a
partial recursive function $n(-)$ that, given an index $e$,
returns $n(e)$ which would be a sort of IFP of $e$: see
\S\ref{sec:compsrt}. These, one would expect, are
\emph{intensional fixed point combinators}. The fact that we
cannot define an intensional fixed-point combinator with the
G\"odel-L\"ob axiom as type means that this is not possible to do
\emph{within} iPCF v2.0, not unless we redefine $\Box$ to
mean \emph{completely closed}.  This is a serious limitation
towards the goal of making iPCF v2.0 a typed `language of indices'
that works more or less in the style of \cite{Jones1997}. The only
way out of this impasse would be to show that our intended model
is \emph{natural}, so that we could model iPCF itself. We discuss
this further in \S\ref{sec:natwextmod}.

\section{Interpreting iPCF v2.0}

We have finally made it to the categorical interpretation of iPCF
v2.0, which we will use the algebraic machinery developed
in \S\ref{chap:intsem1} to define. We assume that the reader
has some background on the categorical semantics of simply-typed
$\lambda$-calculus. Useful expositions include the classics by
\cite{Lambek1988} and \cite{Crole1993}, as well as the detailed
presentation of \cite{Abramsky2011a}. For the details of the
categorical semantics of modal $\lambda$-calculi see
\citep{Kavvos2017b, Kavvos2017c}.

First, we define the notion of a iPCF v2.0 model.

\begin{defn}[iPCF v2.0 model]
  \label{def:ipcfmodel}
  An iPCF v2.0 model consists of
  \begin{enumerate}[label=(\roman{*})]
    \item 
      a cartesian closed P-category $(\mathfrak{C}, \sim, \times,
      \mathbf{1})$;
    \item 
      a product-preserving idempotent comonadic exposure $(Q,
      \epsilon, \delta)$;
    \item
      a choice of objects $\mathbb{N}$ and $\mathbb{B}$, suitable
      for interpreting the constants (see \cite{Hyland2000}); and
    \item
      maps $(-)^\dagger$ yielding IFPs at all the objects
      generated by \[
        I \quad ::= \quad         \mathbb{B}
                    \quad | \quad \mathbb{N}
                    \quad | \quad I^{QZ}
      \] for all $Z \in \mathfrak{C}$, as per Definition
      \ref{def:intfpparam}.
  \end{enumerate}
\end{defn}

\noindent Given an iPCF v2.0 model we define an object $\sem{A}{}
\in \mathfrak{C}$ for every type $A$ of iPCF, by induction:
\begin{align*}
  \sem{\textsf{Nat}}{} &\myeq \mathbb{N} \\
  \sem{\textsf{Bool}}{} &\myeq \mathbb{B} \\
  \sem{A \rightarrow B}{} &\myeq \sem{B}{\sem{A}{}} \\
  \sem{\Box A}{} &\myeq Q\sem{A}{}
\end{align*} Then, given a well-defined context
$\ctxt{\Delta}{\Gamma}$ where $\Delta = u_1{:}B_1, \dots
u_n{:}B_n$ and $\Gamma = x_1{:}A_1, \dots, x_m{:}A_m$, we let \[
  \sem{\ctxt{\Delta}{\Gamma}}{}
    \myeq 
      QB_1 \times \dots \times QB_n 
  \times
      A_1 \times \dots \times A_m
\] where the product is, as ever, left-associating.

\begin{figure}
  \caption{Categorical Semantics for Intensional PCF v2.0}
  \label{fig:catsemdef}
  \input{ipcf/catsemdef.tex}
\end{figure}

We then extend the semantic map $\sem{-}{}$ to one that associates
an arrow \[
  \sem{\ctxt{\Delta}{\Gamma} \vdash M : A}{} :
  \sem{\ctxt{\Delta}{\Gamma}}{} \rightarrow \sem{A}{}
\] of the P-category $\mathfrak{C}$ to each derivation
$\ctxt{\Delta}{\Gamma} \vdash M : A$. The full definition is
given in Figure \ref{fig:catsemdef}. The map \[
  \pi_\Delta^{\ctxt{\Delta}{\Gamma}} :
    \sem{\ctxt{\Delta}{\Gamma}}{} \rightarrow
    \sem{\ctxt{\Delta}{\cdot}}{}
\] is the obvious projection. Moreover, the notation $\langle
\vct{\pi_\Delta}, f, \vct{\pi_\Gamma} \rangle$ stands for \[
  \langle \vct{\pi_\Delta}, f, \vct{\pi_\Gamma} \rangle
    \myeq
      \langle
        \pi_1, \dots, \pi_n, 
        f,
        \pi_{n+1}, \dots, \pi_{n+m}
      \rangle
\]

The first thing we need to observe is that there is no difference
in the interpretation if the term is intensional or extensional:
if a term can be both, it has the same interpretation.

\begin{lem}
  \label{lem:intextsame}
  If $\ctxt{\Delta}{\Gamma} \vint{} M : A$, then \[
    \sem{\ctxt{\Delta}{\Gamma} \vint{} M : A}{}
      =
    \sem{\ctxt{\Delta}{\Gamma} \vext{} M : A}{}
  \] where $=$ stands for strict equality.
\end{lem}

\section{Soundness}
  \label{sec:ipcfsound}

The main tools in proving soundness of our interpretation are (a)
lemmas giving the categorical interpretation of various admissible
rules, and (b) a fundamental lemma relating substitution of terms
to composition in the category. In the sequel we often use
informal vector notation for contexts: for example, we write
$\vec{u} : \vec{B}$ for the context $u_1 : B_1, \dots, u_n : B_m$.
We also write $[\vec{N}/\vec{u}]$ for the simultaneous,
capture-avoiding substitution $[N_1/u_1, \dots, N_m/u_n]$.

First, we interpret weakening and exchange.

\begin{lem}[Semantics of Weakening]
  \label{lem:semweak} \hfill
  \begin{enumerate}
    \item
      Let $\ctxt{\Delta}{\Gamma, x{:}C, \Gamma'} \vint{} M : A$ with $x
      \not\in \fv{M}$. Then \[
        \sem{\ctxt{\Delta}{\Gamma, x{:}C, \Gamma'} \vint{} M : A}{}
          \approx_Q
        \sem{\ctxt{\Delta}{\Gamma, \Gamma'} \vint{} M : A}{}
        \circ \pi
      \] where $\pi : \sem{\ctxt{\Delta}{\Gamma, x{:}C, \Gamma'}}{}
      \rightarrow \sem{\ctxt{\Delta}{\Gamma, \Gamma'}}{}$ is the
      obvious projection.
    \item
      Let $\ctxt{\Delta}{\Gamma, x{:}C, \Gamma'} \vext{} M : A$ with $x
      \not\in \fv{M}$. Then \[
        \sem{\ctxt{\Delta}{\Gamma, x{:}C, \Gamma'} \vext{} M : A}{}
          \sim
        \sem{\ctxt{\Delta}{\Gamma, \Gamma'} \vext{} M : A}{}
        \circ \pi
      \] where $\pi : \sem{\ctxt{\Delta}{\Gamma, x{:}C, \Gamma'}}{}
      \rightarrow \sem{\ctxt{\Delta}{\Gamma, \Gamma'}}{}$ is the
      obvious projection. If the iPCF model is weakly extensional
      (see \S\ref{sec:natwextmod}) then the result holds up to
      $\approx_Q$.
    \item
      Let $\ctxt{\Delta, u{:}B, \Delta'}{\Gamma} \vint{} M : A$
      with $u \not\in \fv{M}$. Then \[
        \sem{\ctxt{\Delta, u{:}B, \Delta'}{\Gamma} \vint{} M : A}{}
          \approx_Q
        \sem{\ctxt{\Delta, \Delta'}{\Gamma} \vint{} M : A}{}
        \circ \pi
      \] where $\pi : \sem{\ctxt{\Delta, u{:}B,
      \Delta'}{\Gamma}}{} \rightarrow \sem{\ctxt{\Delta,
      \Delta'}{\Gamma}}{}$ is the obvious projection.
  \item
    Let $\ctxt{\Delta, u{:}B, \Delta'}{\Gamma} \vext{} M : A$
    with $u \not\in \fv{M}$. Then \[
      \sem{\ctxt{\Delta, u{:}B, \Delta'}{\Gamma} \vext{} M : A}{}
        \sim
      \sem{\ctxt{\Delta, \Delta'}{\Gamma} \vext{} M : A}{}
      \circ \pi
    \] where $\pi : \sem{\ctxt{\Delta, u{:}B,
    \Delta'}{\Gamma}}{} \rightarrow \sem{\ctxt{\Delta,
    \Delta'}{\Gamma}}{}$ is the obvious projection. If the iPCF
    model is weakly extensional (see \S\ref{sec:natwextmod}) then
    the result holds up to $\approx_Q$.
  \end{enumerate}
\end{lem}
\begin{proof}
  By induction on the derivations. Most cases are straightforward.

  The first one holds up to $\approx_Q$ because it essentially
  consists of projecting away components, which holds
  intensionally: the fact the judgement is intensional means no
  $\lambda$'s are involved. 
  
  The second one holds only up to $\sim$, because of the occurrence
  of $\lambda(-)$'s in the semantics. If moreover the model is
  weakly extensional, $\lambda(-)$ preserves $\approx_Q$ (Cor.
  \ref{cor:idemastpresint}) so we can strengthen the inductive
  hypothesis to $\approx_Q$ and obtain the result up to
  intensional equality.

  The third one also follows easily. In the case of a
  $\ibox{(-)}$, the term within it is intensional, so we use the
  induction hypothesis and the fact\footnote{This is where
  idempotence is essential, otherwise this would hold only up to
  $\sim$ and the inductive hypothesis for $\ibox{(-)}$ would
  fail.} $(-)^\ast$ preserves $\approx_Q$. We then know that
  $(-)^\ast$ is natural for projections (Prop.
  \ref{prop:astdist}\ref{itm:astprojdist}) up to $\approx_Q$ (due
  to idempotence). There is not much to show in the case of
  $\fixlob{z}{(-)}$, as no modal variables occur freely under it.
  
  The fourth one is perhaps the most complicated, and only holds
  up to $\sim$, again because of the occurrence of $\lambda(-)$'s.
  In the case of $\ibox{(-)}$, the term within it is intensional,
  so we use the third result and the fact $(-)^\ast$ preserves
  $\approx_Q$, followed again by naturality for projections. The
  case of $\fixlob{z}{(-)}$ is again trivial.
\end{proof}

\noindent We can also show that the components of the
interpretation interact in the expected way with the corresponding
term formation rules in the language. These follows from the
analogous lemmata in \S\ref{sec:scene}, which show that the
necessary equations hold intensionally in a iPCF model, i.e. when
the exposure $Q$ is idempotent.

\begin{lem}[Double box]
  \label{lem:doublebox}
  If $\ctxt{\Delta}{\cdot} \vj{} M : A$, then \[
    \sem{\ctxt{\Delta}{\Gamma} \vj{} \ibox{(\ibox{M})} : \Box\Box A}{}
      \approx_Q
    \delta_A \circ \sem{\ctxt{\Delta}{\Gamma} \vj{} \ibox{M} : \Box A}{}
  \]
\end{lem}

\begin{proof}
  Let $f \myeq \sem{\ctxt{\Delta}{\cdot} \vj{} M : A}{}$.
  Then
  \begin{derivation}
      \sem{\ctxt{\Delta}{\Gamma}
        \vj{} \ibox{(\ibox{M})} : \Box\Box A}{}
    \since[\approx_Q]{definitions}
      \left(f^\ast\right)^\ast
        \circ \pi_\Delta^{\ctxt{\Delta}{\Gamma}}
    \since[\approx_Q]{Proposition \ref{prop:deltaepsilonast}}
      \delta_A \circ f^\ast
        \circ \pi_\Delta^{\ctxt{\Delta}{\Gamma}}
    \since[\approx_Q]{definitions}
      \delta_A
        \circ \sem{\ctxt{\Delta}{\Gamma} \vj{} \ibox{M} : \Box A}{}
  \end{derivation}
\end{proof}

\begin{lem}[Identity Lemma]
  \label{lem:identity}
  For $(u_i : B_i) \in \Delta$, \[
    \sem{\ctxt{\Delta}{\Gamma} \vj{} 
      \ibox{u_i} : \Box B_i}{}
        \approx_Q
    \pi^{\ctxt{\Delta}{\Gamma}}_{\Box B_i}
  \]
\end{lem}
\begin{proof}
  \begin{derivation}
      \sem{\ctxt{\Delta}{\Gamma} \vj{} \ibox{u_i} : \Box B_i}{}
    \since[\approx_Q]{definition}
      \left(\epsilon_{B_i}
        \circ \pi^{\ctxt{\Delta}{\cdot}}_{\Box B_i}\right)^\ast
        \circ \pi^{\ctxt{\Delta}{\Gamma}}_{\Delta}
    \since[\approx_Q]{Proposition \ref{prop:astdist}}
      \epsilon_{B_i}^\ast
        \circ \pi^{\ctxt{\Delta}{\cdot}}_{\Box B_i}
        \circ \pi^{\ctxt{\Delta}{\Gamma}}_{\Delta}
    \since[\approx_Q]{Proposition \ref{prop:astdist} \ref{itm:epsast},
                      projections}
      \pi^{\ctxt{\Delta}{\Gamma}}_{\Box B_i}
  \end{derivation}
\end{proof}

\begin{lem}[Semantics of Substitution]
  \label{lem:semsubst}
  Suppose that $\ctxt{\Delta}{\Gamma} \vj{} M_i : A_i$ for
  $i = 1, \dots, n$, and that $\ctxt{\Delta}{\cdot} \vint{} N_j : B_j$ for
  $j = 1, \dots, m$, and let \begin{align*}
    \beta_j 
      &\myeq \sem{\ctxt{\Delta}{\Gamma} \vint{} \ibox{N_j} : \Box B_j}{} \\
    \alpha_i
      &\myeq \sem{\ctxt{\Delta}{\Gamma} \vj{} M_i : A_i}{}
  \end{align*} Then,
  \begin{enumerate}
    \item
      if $\ctxt{\vec{u} : \vec{B}}{\vec{x} : \vec{A}} \vint{} P  : C$, 
      we have \[
        \sem{\ctxt{\Delta}{\Gamma} \vj{}
            P[\vec{N}/\vec{u}, \vec{M}/\vec{x}] : C}{}
          \approx_Q
        \sem{\ctxt{\vec{u} : \vec{B}}{\vec{x} : \vec{A}} \vint{}
            P : C}{}
          \circ 
            \left\langle
              \vct{\beta},
              \vct{\alpha}
            \right\rangle
      \]
    \item
      if $\ctxt{\vec{u} : \vec{B}}{\vec{x} : \vec{A}} \vext{} P  : C$, 
      we have \[
        \sem{\ctxt{\Delta}{\Gamma} \vext{}
            P[\vec{N}/\vec{u}, \vec{M}/\vec{x}] : C}{}
          \sim
        \sem{\ctxt{\vec{u} : \vec{B}}{\vec{x} : \vec{A}} \vext{}
            P : C}{}
          \circ 
            \left\langle
              \vct{\beta},
              \vct{\alpha}
            \right\rangle
      \]
  \end{enumerate}
\end{lem}

\begin{proof}
  By induction on the derivation of
  $\ctxt{\vec{u}:\vec{B}}{\vec{x}:\vec{A}} \vj{} P : C$. Most
  cases are straightforward, and use a combination of standard
  equations that hold in cartesian closed categories---see
  \cite[\S 2]{Crole1993}---in order to perform calculations very
  close the ones detailed in \cite[\S 1.6.5]{Abramsky2011a}.
  Because of the precise definitions we have used, we also need to
  make use of Lemma \ref{lem:semweak} to interpret weakening
  whenever variables in the context do not occur freely in the
  term.  We only cover the modal cases.
  \begin{indproof}
    \case{$\Box\textsf{var}$}
    Then $P \equiv u_i$ for some $u_i$ amongst the $\vec{u}$.
    Hence, the LHS is $\ctxt{\Delta}{\Gamma} \vint{} N_i : B_i$,
    whereas we calculate that the RHS, in either case, is
    \begin{derivation}
      \sem{\ctxt{\vec{u}:\vec{B}}{\vec{x}:\vec{A}} \vj{} P : C}{}
        \circ \langle \vec{\beta}, \vec{\alpha} \rangle
    \since[\approx_Q]{definition, projection}
      \epsilon_{B_i}
        \circ \sem{\ctxt{\Delta}{\Gamma} \vint{} \ibox{N_i} : \Box B_i}{}
  \since[\approx_Q]{definition}
      \epsilon_{B_i}
        \circ \sem{\ctxt{\Delta}{\cdot} \vint{} N_i : B_i}{\ast}
        \circ \pi_\Delta^{\ctxt{\Delta}{\Gamma}}
  \since[\approx_Q]{Proposition \ref{prop:deltaepsilonast}}
    \sem{\ctxt{\Delta}{\cdot} \vint{} N_i : B_i}{}
      \circ \pi_\Delta^{\ctxt{\Delta}{\Gamma}}
  \since[\approx_Q]{Semantics of Weakening (Lemma \ref{lem:semweak})}
    \sem{\ctxt{\Delta}{\Gamma} \vint{} N_i : B_i}{}
    \end{derivation}

    \case{$\Box\mathcal{I}$}
      We have that $\ctxt{\vct{u}:\vct{B}}{\vct{x}:\vct{A}} \vj{}
      \ibox{P} : \Box C$, so that $\ctxt{\vct{u}:\vct{B}}{\cdot}
      \vint{} P : C$, with the result that none of the variables
      $\vct{x}$ occur in $P$. Hence $P[\vec{N}/\vec{u},
      \vec{M}/\vec{x}] \equiv P[\vec{N}/\vec{u}]$, and we
      calculate, in either case:
      \begin{derivation}
          \sem{\ctxt{\Delta}{\Gamma} \vj{}
            \ibox{(P[\vec{N}/\vec{u}, \vec{M}/\vec{x}])} : \Box C}{}
        \since[\approx_Q]{definition, and non-occurrence of the $\vec{x}$ in $P$}
          \sem{\ctxt{\Delta}{\cdot} \vint{}
            P[\vec{N}/\vec{u}] : C}{\ast}
            \circ \pi^{\ctxt{\Delta}{\Gamma}}_\Delta
        \since[\approx_Q]{IH, $(-)^\ast$ preserves $\approx_Q$
              (Corollary \ref{cor:idemastpresint})}
          \left(
            \sem{\ctxt{\vec{u}:\vec{B}}{\cdot} \vint{} P : C}{} 
            \circ
              \left\langle 
                \vct{\sem{\ctxt{\Delta}{\cdot} \vint{} \ibox{N_i} : \Box B_i}{}}
              \right\rangle
          \right)^\ast
            \circ \pi^{\ctxt{\Delta}{\Gamma}}_\Delta
        \since[\approx_Q]{Proposition \ref{prop:astdist}}
          \sem{\ctxt{\vec{u}:\vec{B}}{\cdot} \vint{} P : C}{\bullet}
            \circ
              \left\langle 
                \vct{ 
                  \sem{\ctxt{\Delta}{\cdot} \vint{}
                    \ibox{N_i} : \Box B_i}{\ast}
                }
              \right\rangle
            \circ \pi^{\ctxt{\Delta}{\Gamma}}_\Delta
        \since[\approx_Q]{naturality of product morphism, definition}
          \sem{\ctxt{\vec{u}:\vec{B}}{\cdot}{} \vint{} P : C}{\bullet}
            \circ
              \left\langle 
                \vct{
                  \sem{\ctxt{\Delta}{\Gamma} \vint{}
                    \ibox{(\ibox{N_i})} : \Box \Box B_i}{}
                }
              \right\rangle
        \since[\approx_Q]{Double box (Theorem \ref{lem:doublebox}), 
                          $\langle \cdot, \cdot \rangle$
                          preserves $\approx_Q$}
          \sem{\ctxt{\vec{u}:\vec{B}}{\cdot} \vint{} P : C}{\bullet}
            \circ
              \left\langle
                \vct{
                  \delta_{\sem{B_i}{}}
                    \circ \sem{\ctxt{\Delta}{\Gamma} 
                            \vint{} \ibox{N_i} : \Box B_i}{}
                }
              \right\rangle
        \since[\approx_Q]{product after angled brackets}
          \sem{\ctxt{\vec{u}:\vec{B}}{\cdot} \vint{} P : C}{\bullet}
          \circ \prod_{i=1}^n \delta_{\sem{B_i}{}}
          \circ
            \left\langle 
              \vct{
                \sem{\ctxt{\Delta}{\Gamma} \vint{} \ibox{N_i} : \Box B_i}{}
              }
            \right\rangle
        \since[\approx_Q]{some projections and definition of $(-)^\ast$ 
                          and $\sem{-}{}$}
          \sem{\ctxt{\vec{u}:\vec{B}}{\vec{x}:\vec{A}} \vj{}
            \ibox{P} : \Box C}{}
          \circ
            \left\langle 
              \vct{\beta},
              \vct{\alpha}
            \right\rangle
      \end{derivation}
    \case{$\Box\textsf{fix}$}
      We have that $\ctxt{\vct{u}:\vct{B}}{\vct{x}:\vct{A}} \vj{}
      \fixlob{z}{P} : C$, so that $\ctxt{\vct{u}:\vct{B}}{z : \Box C}
      \vint{} P : C$, with the result that none of the variables
      $\vct{x}$ occur in $P$. Hence $P[\vec{N}/\vec{u},
      \vec{M}/\vec{x}] \equiv P[\vec{N}/\vec{u}]$, and we calculate:
      \begin{derivation}
          \sem{\ctxt{\Delta}{\Gamma} \vj{}
            \fixlob{z}{(P[\vec{N}/\vec{u}, \vec{M}/\vec{x}])} :  C}{}
        \since[\approx_Q]{definitions; only $z$ is free in $P$}
          \sem{\ctxt{\cdot}{z : \Box C} \vint{}
            P : C}{\dagger}
            \circ {!}
        \since[\approx_Q]{definition}
          \sem{\ctxt{\cdot}{\cdot} \vint{} \fixlob{z}{P} : C}{}
            \circ {!}
        \since[\approx_Q]{projections, definitions}
          \sem{\ctxt{\vec{u}:\vec{B}}{\vec{x} : \vec{A}}
        \vj{} \fixlob{z}{P} : C}{} 
          \circ \left\langle 
            \vct{\beta},
            \vct{\alpha}
          \right\rangle
      \end{derivation}
  \end{indproof}
\end{proof}

\begin{thm}[Soundness] \hfill
  \begin{enumerate}
    \item
If $\ctxt{\Delta}{\Gamma} \vint{} M = N : A$, then we have that \[
  \sem{\ctxt{\Delta}{\Gamma} \vint{} M : A}{}
    \approx_Q
  \sem{\ctxt{\Delta}{\Gamma} \vint{} N : A}{}
\]
    \item
If $\ctxt{\Delta}{\Gamma} \vext{} M = N : A$, then we have that \[
  \sem{\ctxt{\Delta}{\Gamma} \vext{} M : A}{}
    \sim
  \sem{\ctxt{\Delta}{\Gamma} \vext{} N : A}{}
\]
  \end{enumerate}
\end{thm}

\begin{proof} 
  By induction on the derivation of $\ctxt{\Delta}{\Gamma}
  \vdash M = N : A$. The congruence cases are clear, as is the
  majority of the ordinary clauses. All of these even hold up
  to $\approx_Q$, with the exception of $(\rightarrow\beta)$ and
  $(\rightarrow\eta)$, which only hold up to $\sim$. Only the
  modal rules remain, which we prove with direct calculation.
  
  For $(\Box\beta)$ in the case of $\mathcal{J} = \text{int.}$
  we calculate:
  \begin{derivation}
      \sem{\ctxt{\Delta}{\Gamma} \vint{} \letbox{u}{\ibox{M}}{N} : C}{}
    \since[\approx_Q]{definition}
      \sem{\ctxt{\Delta, u{:}A}{\Gamma} \vint{} N : C}{}
      \circ
        \langle
          \vct{\pi_\Delta},
          \sem{\ctxt{\Delta}{\Gamma} \vint{} \ibox{M} : \Box A}{},
          \vct{\pi_\Gamma}
        \rangle
    \since[\approx_Q]{Lemma \ref{lem:identity}}
      \sem{\ctxt{\Delta, u{:}A}{\Gamma} \vint{} N : C}{}
      \circ
        \langle
          \vct{\sem{\ctxt{\Delta}{\Gamma} 
            \vint{} \ibox{u_i} : \Box B_i}{}},
          \sem{\ctxt{\Delta}{\Gamma} \vint{} \ibox{M} : \Box A}{},
          \vct{\pi_\Gamma}
        \rangle
    \since[\approx_Q]{Lemma \ref{lem:semsubst}; 
                      $\sem{\ctxt{\Delta}{\Gamma} \vint{} x_i : A_i}{}
                      \myeq \pi^{\ctxt{\Delta}{\Gamma}}_{A_i}$}
      \sem{\ctxt{\Delta}{\Gamma} \vint{}
        N[\vec{u_i}/\vec{u_i}, M/u, \vec{x_i}/\vec{x_i}] : C}{}
  \end{derivation} In the case of $\mathcal{J} = \text{ext.}$, we
  use Lemma \ref{lem:intextsame} to write
  $\sem{\ctxt{\Delta}{\Gamma} \vext{} \ibox{M} : \Box A}{} =
  \sem{\ctxt{\Delta}{\Gamma} \vint{} \ibox{M} : \Box A}{}$, and then
  the same calculation works up to $\sim$.

  There remains the case of the fixpoint; let \[
    g \myeq 
\sem{\ctxt{\Delta}{\Gamma} \vj{}
  \fixlob{z}{M} : A}{}
  \] Then $g \myeq f^\dagger \circ {!}$, where \[
    f \myeq
\sem{\ctxt{\cdot}{z : \Box A} \vint{}
  M : A}{}
  \] But we can easily calculate that
  \begin{derivation}
f^\dagger
    \since[\sim]{definition of fixpoint (Def. \ref{def:intfp})}
f \circ \left(f^\dagger\right)^\ast
    \since[\approx_Q]{definitions}
f \circ
  \sem{\ctxt{\cdot}{\cdot}
    \vint{} \ibox{(\fixlob{z}{M})} : \Box A}{}
    \since[\approx_Q]{Lemma \ref{lem:semsubst}}
\sem{\ctxt{\cdot}{\cdot} 
  \vint{} M[\ibox{(\fixlob{z}{M})}/z] : A}{}
  \end{derivation}
  and hence $g \sim \sem{\ctxt{\Delta}{\Gamma} \vj{}
  M[\ibox{(\fixlob{z}{M})}/z] : A}{}$ by weakening (Lemma
  \ref{lem:semweak}).
%    The case of $(\Box\eta)$ is even simpler, and follows
%    immediately from Lemma \ref{lem:identity}.
\end{proof}

\section{Natural and Weakly Extensional Models}
  \label{sec:natwextmod}

In iPCF v2.0 we effected three restrictions: \begin{itemize}
  \item No free variables when taking intensional fixed points
  (except the diagonal).
  \item No $\lambda$-abstractions with free variables under boxes.
  \item IFPs only at certain types generated by $A_\textsf{fix}$.
\end{itemize}

We discussed at the end of \S\ref{sec:ipcfv2.0} the effect that
these have on the expressivity of the language, and found that it
was far too strong, so we would like to examine when these can be
lifted.

A iPCF v2.0 model must satisfy certain requirements in order for
these restrictions to be lifted. The first one can be lifted
whenever a iPCF model is \emph{natural}. The second one can be
lifted whenever a iPCF model is \emph{weakly extensional}.
Unfortunately, we are still at a loss regarding the existence of
IFPs at all objects.

\subsection{Natural iPCF v2.0 models} 

The first restriction we want to lift is the occurrence of free
variables when taking an intensional fixed point; that is, we want
to generalise the $(\Box\textsf{fix})$ rule to \[
  \begin{prooftree}
    \ctxt{\Delta}{z : \Box A_\textsf{fix}} \vint M : A_\textsf{fix}
      \justifies
    \ctxt{\Delta}{\Gamma} \vj \fixlob{z}{M} : A_\textsf{fix}
  \end{prooftree}
\] We will be able to do this in \emph{natural} models of iPCF
v2.0.

\begin{defn}
  An iPCF v2.0 model is \emph{natural} just if \begin{enumerate}
    \item
      $(-)^\dagger$ preserves $\approx_Q$; and
    \item
      $(-)^\dagger$ is natural up to $\approx_Q$, in the sense
      that for any $f : \prod_{i=1}^n QB_i \times QA \rightarrow
      A$ and $k : \prod_{j=1}^k QC_j \rightarrow \prod_{i=1}^n
      QB_i$, it is the case that \[
        \left(f \circ (k \times id)\right)^\dagger
          \approx_Q
        f^\dagger \circ k
      \]
  \end{enumerate}
\end{defn}

\noindent In this kind of iPCF model, we are free to have
parameters in our IFPs, and we can interpret the L\"ob rule by \[
  \sem{\ctxt{\Delta}{\Gamma} \vj{} \fixlob{z}{M} : A}{}
    \myeq 
    \sem{\ctxt{\Delta}{z : \Box A}
  \vint{} M : A}{\dagger}
      \circ \pi_\Delta^{\ctxt{\Delta}{\Gamma}}
\]

The lemmas for weakening (Lem. \ref{lem:semweak}) and substitution
(Lem. \ref{lem:semsubst}) directly carry over. Naturality is only
used in the appropriate cases for $(\Box\textsf{fix})$; e.g. here
is the case for substitution: \begin{derivation}
    \sem{\ctxt{\Delta}{\Gamma} \vj{}
      \fixlob{z}{(P[\vec{N}/\vec{u}, \vec{M}/\vec{x}])} :  C}{}
  \since[\approx_Q]{definition, and non-occurrence of the $\vec{x}$ in $P$}
    \sem{\ctxt{\Delta}{z : \Box C} \vint{}
      P[\vec{N}/\vec{u}] : C}{\dagger}
      \circ
  \pi^{\ctxt{\Delta}{\Gamma}}_\Delta
  \since[\approx_Q]{IH, $(-)^\dagger$ preserves $\approx_Q$, definitions}
    \left(\sem{\ctxt{\vec{u}:\vec{B}}{z : \Box C}
  \vint{} P : C}{} 
      \circ \left\langle 
        \vct{\sem{\ctxt{\Delta}{z : \Box C} \vint{}
    \ibox{N_i} : \Box B_i}{}},
        \pi^{\ctxt{\Delta}{z : \Box C}}_{z : \Box C}
      \right\rangle
    \right)^\dagger
      \circ \pi^{\ctxt{\Delta}{\Gamma}}_\Delta
  \since[\approx_Q]{weakening, $\langle \cdot, \cdot \rangle$ and
                    $(-)^\dagger$ preserve $\approx_Q$}
    \left(\sem{\ctxt{\vec{u}:\vec{B}}{z : \Box C} \vint{} P : C}{} 
      \circ
        \left\langle 
          \vct{\sem{\ctxt{\Delta}{\cdot} \vint{} \ibox{N_i} : \Box B_i}{}
                \circ \pi^{\ctxt{\Delta}{z : \Box C}}_{\ctxt{\Delta}{\cdot}}
              }, 
          \pi^{\ctxt{\Delta}{z : \Box C}}_{z : \Box C}
        \right\rangle
    \right)^\dagger
    \circ \pi^{\ctxt{\Delta}{\Gamma}}_\Delta
  \since[\approx_Q]{naturality of products, definition}
    \left(
      \sem{\ctxt{\vec{u}:\vec{B}}{z : \Box C} \vint{} P : C}{} 
        \circ 
        \left(
          \left\langle 
            \vct{\sem{\ctxt{\Delta}{\cdot} \vint{} \ibox{N_i} : \Box B_i}{}}
          \right\rangle
            \times
          id
        \right)
    \right)^\dagger
      \circ \pi^{\ctxt{\Delta}{\Gamma}}_\Delta
  \since[\approx_Q]{naturality of $(-)^\dagger$}
    \sem{\ctxt{\vec{u}:\vec{B}}{z : \Box C} \vint{} P : C}{\dagger} 
    \circ
      \left\langle 
        \vct{\sem{\ctxt{\Delta}{\cdot} \vint{} \ibox{N_i} : \Box B_i}{}}
      \right\rangle
    \circ 
      \pi^{\ctxt{\Delta}{\Gamma}}_\Delta
  \since[\approx_Q]{naturality of products, Lemma
        \ref{lem:semweak}, projections, definitions}
    \sem{\ctxt{\vec{u}:\vec{B}}{\vec{x} : \vec{A}} \vj{}
      \fixlob{z}{P} : C}{} 
    \circ
      \left\langle 
        \vct{\beta},
        \vct{\alpha}
      \right\rangle
\end{derivation} We can also calculate as usual for the fixed
point. Let $g \myeq \sem{\ctxt{\Delta}{\Gamma} \vj{} \fixlob{z}{M}
: A}{}$. Then $g \myeq f^\dagger \circ
\pi^{\ctxt{\Delta}{\Gamma}}_{\Delta}$, where \[
  f \myeq
    \sem{\ctxt{\Delta}{z : \Box A} \vint{}
      M : A}{}
\] But we can easily calculate that
\begin{derivation}
    f^\dagger
  \since[\sim]{definition of fixpoint (Def. \ref{def:intfp})}
    f
    \circ
      \langle
        id,
        \left(f^\dagger\right)^\ast
      \rangle
  \since[\approx_Q]{definitions}
    f
      \circ
        \langle
          id,
          \sem{\ctxt{\Delta}{\cdot}
            \vint{} \ibox{(\fixlob{z}{M})} : \Box A}{}
        \rangle
  \since[\approx_Q]{Lemma \ref{lem:semsubst}}
    \sem{\ctxt{\Delta}{\cdot} 
      \vint{} M[\ibox{(\fixlob{z}{M})}/z] : A}{}
\end{derivation}
and hence $g \sim \sem{\ctxt{\Delta}{\Gamma} \vj{}
M[\ibox{(\fixlob{z}{M})}/z] : A}{}$, by weakening.

\subsection{Weakly Extensional iPCF v2.0 models, or iPCF models}

In some cases, we can even rid ourselves of the distinction
between intensional and extensional judgements, eventually reaching
a language very close to the one with which we begun our
investigation in \S\ref{chap:ipcf}. The models in which this can
occur are known as \emph{weakly extensional}.

\begin{defn}[iPCF model]
  An iPCF v2.0 model is \emph{weakly extensional}---or, more
  simply, a iPCF model---just if
  \begin{enumerate}
    \item
      $\lambda(-)$ preserves $\approx_Q$, and
    \item
      $\lambda(-)$ is natural with respect to $\approx_Q$, i.e. \[
        \lambda(f \circ (g \times id)) \approx_Q \lambda(f) \circ g
      \]
  \end{enumerate}
\end{defn}

\noindent Before anything else, let us immediately remark that

\begin{lem}
  \label{lem:wextnat}
  A weakly extensional iPCF v2.0 model can be made natural.
\end{lem}
\begin{proof}
  We can use Theorem \ref{thm:ifpcs}(3) to define a strong
  intensional fixed point combinator at every object with IFPs
  given by $(-)^\ast$. Because $Q$ is idempotent, we can then use
  Theorem \ref{thm:ifpcs}(1) to yield weak fixed point combinators,
  and then Theorem \ref{thm:ifpcs}(2) to induce IFPs anew, by
  setting \[
    f^{\dagger'} \myeq \epsilon_A \circ Y \circ
      \left(\lambda(f)\right)^\ast
  \] As $Q$ is idempotent and $\lambda(-)$ preserves $\approx_Q$,
  we have---by Lemmata \ref{lem:circpresint} and
  \ref{lem:fixnat}---that $(-)^{\dagger'}$ preserves $\approx_Q$, 
  and is natural. Thus, we can replace $(-)^\dagger$ by
  $(-)^{\dagger'}$, which results in a weakly extensional and
  natural iPCF v2.0 model.
\end{proof}

We can use a weakly extensional model to interpret iPCF almost as
presented in its original form in \S\ref{chap:ipcf}: we only need
to hold back on the IFPs, and limit them to the intensional
exponential ideal $A_\textsf{fix}$. Of course, we call these
models weakly extensional as the category of assemblies
$\mathfrak{Asm}(A)$ constitutes such a model whenever $A$ is a
weakly extensional PCA: see \S\ref{sec:wextasm}.

It is not a difficult calculational exercise to show that the main
lemmata in this chapter, i.e.  weakening (Lemma \ref{lem:semweak})
and substitution (Lemma \ref{lem:semsubst}), hold up to
$\approx_Q$, with no distinction between extensional and
intensional judgements. This happens because the interpretation of
all the main language constructs, i.e.  $\lambda(-)$ and
$(-)^\ast$, preserve $\approx_Q$.

The only exception is the soundness theorem, which only holds up
to $\sim$, and it does so with good reason. Firstly, we do not
expect the equational behaviour of the constants (naturals,
booleans) to be \emph{intensional}: on account of the language
being partial, we expect both of these objects to have highly
intensional structure. But even if they did not, the cartesian
closed equations can realistically be expected to only hold
extensionally (especially $\eta$).

However, if asked to name a weakly extensional model of iPCF, we
might find ourselves at a loss. The paradigmatic example of a
weakly extensional PCA is certainly $\Lambda/=_{\beta}$, the
closed terms of the untyped $\lambda$-calculus quotiented by
$\beta$-equivalence. The construction of
$\mathfrak{Asm}(\Lambda/=_{\beta})$ and the exposure $\Box$ on it
(\S\ref{sec:realexpo}) are parametric in $A$, so all that remains
is to construct IFPs. But even if we were to construct them, we
might think that we have just engaged in an exercise in futility.
The elements of $\Box A$ would be pairs $(a, x)$ where $x \in
\dbars{a}_A$ is a realizer of a point $x \in \bars{A}$. But $x$
would be an equivalence class $[P]_{=_\beta}$ of an untyped
$\lambda$-term, which would be an object that is `too extensional'
for the level at which we have been aiming: IFPs would merely be
ordinary fixed points of $\lambda$-terms!

That leaves us with three choices: 
\begin{enumerate}[label=(\roman{*})]
  \item
    Seek the Holy Grail PCA $\mathcal{HG}$ that is weakly
    extensional, yet sufficiently intensional for IFPs w.r.t. to
    $\Box$ in $\mathfrak{Asm}(\mathcal{HG})$ to be of interest:
    this seems rather difficult, perhaps to the point of being a
    contradiction in terms.
  \item
    Try to redefine $\Box : \mathfrak{Asm}(A) \expo{}
    \mathfrak{Asm}(A)$ in the case of a weakly extensional PCA
    $A$. In the previous case we were content to use realizers as
    the `true intensions.' But how can we proceed this time? One
    attempt in the case of $A$ consisting of equivalence classes
    would be to try to `pick' a representative of each class. But
    then for $\Box$ to respect composition these would have to be
    `compatible' up to composition, which seems impossible.
  \item
    A third option would be to be in a position to accept that
    weakly extensional realizers, such as the terms of
    $\Lambda/=_{\beta}$, are sufficiently intensional. This could
    be the case of we require a lot of  extensionality at the
    assembly level, perhaps to the point where a fixed point of
    an untyped $\lambda$-term seems a rather intensional affair.
    This is again in the spirit described in the introduction
    (\S\ref{sec:intensionality}), where intensionality is argued
    to only be defined \emph{only relative to the extensional
    equality}.
\end{enumerate}

It is slowly beginning to seem that, in the most intensional of
settings, the restrictions we have demanded of iPCF v2.0 are
somehow indispensable. We will produce some further evidence for
that in the process of proposing a general method for constructing
IFPs at the end of the next section (\S\ref{sec:buildipwps}): the
simplest naturality argument we can concoct already requires weak
extensionality. This may not be proof, but nonetheless it is solid
evidence, as the method described therein is particularly useful
in constructing IFPs for very intensional PCA of classical
computability $K_1$ (\S\ref{sec:asmipcf}).

\section{Building IPWPSs categorically} 
  \label{sec:buildipwps}

In this section we shall show how to build IPWPSs from more basic
constructs. If given sufficiently many IPWPSs in a P-CCC which is
equipped with an idempotent comonadic exposure, then we can use
our Parametric Intensional Recursion Theorem (Theorem
\ref{thm:paramintrec}) to build a iPCF model.

The two main ingredients at our disposal will be a certain kind of
\emph{retraction}, and a certain kind of \emph{enumeration}. Both
of these are unlike the ones that have been considered before,
and they make deep use of the theory of exposures as developed in
this thesis. Consequently, they have a very intensional flavour.

Our enumerations will be arrows of the form $X \rightarrow A$,
where $A$ will be the object \emph{enumerated}, and $X$ the object
of `indices.' To this we will add a \emph{factorisation property},
which will be evocative of, or even directly related to, the idea
of \emph{path surjections} as briefly discussed in
\S\ref{sec:lawvere}.

To this, we will add a special notion of \emph{intensional
retraction}, which allows one to represent `code' for objects of the
form $X^{QZ}$ (for any $Z$) as a sort of retract of $X$, but only
up to extension. We will use these contraptions to formulate an
\emph{inductive argument} that constructs IPWPSs at all objects
contained in the \emph{intensional exponential ideal} $I$
generated by the `grammar' \[
    I \quad ::= \quad
                  \mathbb{B}
      \quad|\quad \mathbb{N}
      \quad|\quad I^{QZ}
\] Throughout this section, let us fix a cartesian closed P-category
$\mathfrak{C}$, and a comonadic exposure $(Q, \epsilon, \delta)$
on it.

$QX$ is an object which holds information about `codes' of objects
of type $X$. These `codes' can often be encoded as very simple
first-order data in an object $Y$; for example, $Y$ could be the
natural numbers object. Sometimes we might be able to retrieve the
original `code' in $QX$ from $Y$, making $QX$ a retract of $Y$.
But, in some cases, we might not: we will only manage to
`interpret' the data in $Y$ as data in $X$, yielding the same
extension---but not the same code---as the original one.  This
situation is precisely captured by intensional retractions.

\begin{defn}
  $X$ is an \emph{intensional retract} of $Y$ (w.r.t. $Q$)
  whenever there is a pair of arrows $s : QX \rightarrow Y$ and $r
  : Y \rightarrow X$ such that \[
    \begin{tikzcd}
      QX
        \arrow[r, "s"]
        \arrow[d, "\epsilon_X", swap]
      & Y
        \arrow[dl, "r"] \\
      X
    \end{tikzcd}
  \] commutes up to $\sim$. We call $X$ an \emph{intensional
  retract} of $Y$.
\end{defn}

The second concept that is central is that of \emph{enumeration}.
As hinted in \S\ref{sec:lawvere} in the context of various forms
of surjections, we can think of arrows $X \rightarrow A$ as
`enumerating' the elements of $A$ by `indices' in $X$. We will
require the existence of a particular type of \emph{path
surjection} (see \S\ref{sec:lawvere}), namely one whose `path'
object (denoted $N$ in \S\ref{sec:lawvere}) is an intensional
context of the form $\prod_{i=1}^n QB_i$.

\begin{defn}
  \label{def:qxenum}
  An object $A \in \mathfrak{C}$ is $(Q, X)$-enumerated by $e : X
  \rightarrow A$ just if it is a $(\prod_{i=1}^n QB_i)$-path
  surjection for every finite list of objects $\vct{B_i}$.
\end{defn}
  
\noindent That is: for every arrow $f : \prod_{i=1}^n QB_i
\rightarrow A$ there is at least one arrow $\phi_f : \prod_{i=1}^n
QB_i \rightarrow X$, not necessarily unique, that makes the
diagram \[
  \begin{tikzcd}
    \prod_{i=1}^n QB_i
      \arrow[dr, "f"]
      \arrow[d, "\phi_f", dashed, swap]
    & \\
    X
      \arrow[r, "e", swap]
    & A
  \end{tikzcd}
\] commute. We often call the arrow $e: X \rightarrow A$ a $(Q,
X)$-enumeration, and say that $A$ is $(Q, X)$-enumerable.

We are very fond of $(Q, X)$-enumerations for two reasons. 
The first reason is that they are quite easy to construct:
the domain of $\phi_f$ provides enough intensional information in
the $QB_i$'s. It would be essentially impossible to construct
something of the sort given merely a $\prod_{i=1}^n B_i$.
Intuitively, the reason is that the $B_i$'s are available
extensionally, i.e. as a kind of oracle to which we can pose a
(probably finite) number of questions. It would be unthinkable to
\emph{internally} extract an `index' for the enumeration $e : X
\rightarrow A$ in such a situation.

The second reason is simply because---under one mild
assumption---they directly give rise to IPWPSs.

\begin{lem}
  \label{lem:enumwps}
  If \begin{itemize}
    \item
      $A$ is $(Q, X)$-enumerated by $e : X \rightarrow A$; and
    \item 
      $X^{QX}$ is an intensional retract of $X$.
  \end{itemize} then there is an intensional parametric weak-point
  surjection $p : QX \times QX \rightarrow A$.
\end{lem}
\begin{proof}
  Let $(s, r)$ witness $X^{QX}$ as an intensional retract of $X$.
  Define \[
    p \myeq QX \times QX
      \xrightarrow{\epsilon \times id} X \times QX
      \xrightarrow{r \times id} X^{QX} \times QX
      \xrightarrow{\textsf{ev}} X
  \] We want to show that this is a IPWPS. Let $f : \prod_{i=1}^n
  QB_i \times QX \rightarrow A$. Then $f$ can be written as \[
    \prod_{i=1}^n QB_i \times QX
      \xrightarrow{\phi_f}
    X
      \xrightarrow{e}
    A
  \] as $e : X \rightarrow A$ is a $(Q, X)$-enumeration. We can
  $\lambda$-abstract the index, and take its co-Kleisli lifting to
  obtain $\left(\lambda(\phi_f)\right)^\ast : \prod_{i=1}^n QB_i
  \rightarrow Q(X^{QX})$. Post-composing with the lifted section
  $s^\ast : Q(X^{QX}) \rightarrow QX$ yields an `index' \[
    x_f \myeq s^\ast \circ \left(\lambda(\phi_f)\right)^\ast
      : \prod_{i=1}^n QB_i \rightarrow QX
  \] w.r.t to $p$:
  \begin{derivation}
      p 
        \circ \langle
                s^\ast \circ \left(\lambda(\phi_f)\right)^\ast,
                a
              \rangle
    \since[\sim]{definition of $p$, products after brackets}
      e
        \circ \textsf{ev}
        \circ \langle 
                r \circ \epsilon
                  \circ s^\ast
                  \circ \left(\lambda(\phi_f)\right)^\ast,
                a
              \rangle
    \since[\sim]{Prop. \ref{prop:deltaepsilonast}, int. retract.}
      e
        \circ \textsf{ev}
        \circ \langle 
                \epsilon \circ \left(\lambda(\phi_f)\right)^\ast,
                a
              \rangle
    \since[\sim]{Prop. \ref{prop:deltaepsilonast} again}
      e
        \circ \textsf{ev}
        \circ \langle 
                \lambda(\phi_f),
                a
              \rangle
    \since[\sim]{cartesian closure}
      e
        \circ \phi_f
        \circ \langle id, a \rangle
    \since[\sim]{$e$ is a $(Q, X)$-enumeration}
      f 
        \circ \langle id, a \rangle
  \end{derivation}
\end{proof}

So much for the construction of IPWPSs given enumerations. What
about higher types? In fact, the following lemma shows that, if
$X$ is sufficient to intensionally encode $X^{QZ}$, then we can
`lift' a $(Q, X)$-enumeration $e : X \rightarrow A$ to $(Q,
X)$-enumerate $A^{QZ}$. This is where our previous notion of
intensional exponential ideal comes from.

\begin{lem}
  \label{lem:enum}
  Suppose that \begin{itemize}
    \item
      $A \in \mathfrak{C}$ is $(Q, X)$-enumerated by $e : X
      \rightarrow A$, and
    \item
      $X^{QZ}$ is an intensional retract of $X$.
  \end{itemize} Then $A^{QZ}$ is $(Q, X)$-enumerable.
\end{lem}
\begin{proof}
  Take $f : \prod_{i=1}^n QB_i \rightarrow A^{QZ}$. Then $f \sim
  \lambda(g)$ for some $g : \prod_{i=1}^n QB_i \times QZ
  \rightarrow A$. By IH, we have some $\phi_g : \prod_{i=1}^n QB_i
  \times QZ \rightarrow X$ such that \[
    \begin{tikzcd}
      \prod_{i=1}^n QB_i \times QZ
        \arrow[dr, "g"]
        \arrow[d, "\phi_g", dashed, swap]
      & \\
      X
        \arrow[r, "e", swap]
      & A
    \end{tikzcd}
  \] If we apply $\lambda(-)$ to this triangle, we obtain \[
    \begin{tikzcd}
      \prod_{i=1}^n QB_i
  \arrow[dr, "f"]
  \arrow[d, "\lambda(\phi_g)", swap]
      & \\
      X^{QZ}
  \arrow[r, "e^{QZ}", swap]
      & A^{QZ}
    \end{tikzcd}
  \] We can now rewrite $\lambda(\phi_f) \sim \epsilon \circ
  \left(\lambda(\phi_f)\right)^\ast$ using Proposition 
  \ref{prop:deltaepsilonast}: \[
    \begin{tikzcd}
      \prod_{i=1}^n QB_i
        \arrow[r, "f"]
        \arrow[d, "\left(\lambda(\phi_g)\right)^\ast", swap]
      & A^{QZ} \\
      Q\left(X^{QZ}\right)
        \arrow[r, "\epsilon", swap]
      & X^{QZ}
        \arrow[u, "e^{QZ}", swap]
    \end{tikzcd}
  \] But $X^{QZ}$ is an intensional retract of $X$, so we can
  rewrite $\epsilon$ like so: \[
    \begin{tikzcd}
      \prod_{i=1}^n QB_i
        \arrow[rr, "f"]
        \arrow[d, "\left(\lambda(\phi_g)\right)^\ast", swap]
      &
      & A^{QZ} \\
      Q\left(X^{QZ}\right)
        \arrow[dr, "s", swap]
      &
      & X^{QZ}
        \arrow[u, "e^{QZ}", swap] \\
      & X
        \arrow[ur, "r", swap]
      &
    \end{tikzcd}
  \] Hence, by defining \[
    e' \myeq e^{QZ} \circ r : X \rightarrow E^{QZ}
  \] we have that $E^{QZ}$ is $(Q, X)$-enumerated by $e'$, and the
  index of $f$ is \[
    \phi_f \myeq s \circ \left(\lambda(\phi_g)\right)^\ast
  \]
\end{proof}

\begin{thm}
  \label{thm:enumifps}
  Suppose that \begin{itemize}
    \item
      $\mathbb{B}, \mathbb{N} \in \mathfrak{C}$ are $(Q,
      X)$-enumerable, and that
    \item
      $X^{QZ}$ is an intensional retract of $X$ for any $Z \in
      \mathfrak{C}$.
  \end{itemize} Then any object in the intensional exponential
  ideal generated by \[
    I \quad ::= \quad
                  \mathbb{B}_\bot
      \quad|\quad \mathbb{N}_\bot
      \quad|\quad I^{QZ}
  \] for any $Z \in \mathfrak{C}$ has parametric IFPs.
\end{thm}
\begin{proof}
  We can show by induction that every object generated by $I$ is
  $(Q, X)$-enumerable: the base cases are assumptions, and the
  inductive step is provided by Lemma \ref{lem:enum}. Then by
  Lemma \ref{lem:enumwps} there is an intensional weak-point
  surjection $p_I : QX \times QX \rightarrow I$ for every $I$.
  Finally, by the Parametric Intensional Recursion Theorem
  (Theorem \ref{thm:paramintrec}) each $I$ has IFPs.
\end{proof}

\subsubsection*{Naturality}

It is again worth asking about the necessary premises that are
sufficient for us to conclude that the IPWPSs build in this
section are \emph{natural} in the sense of
\S\ref{sec:paramlawvere}. According to our previous results, we
need $x_h \circ g \approx_Q x_{h \circ (g \times id)}$.  We can
show that, under certain assumptions, the construction of a IPWPS
from an enumeration in Lemma \ref{lem:enumwps} maintains it.
Calculating suffices to elicit the necessary assumptions:
  \begin{derivation}
      x_h \circ g
    \since[\approx_Q]{definition}
      s^\ast \circ (\lambda(\phi_h))^\ast \circ g
    \since[\approx_Q]{idempotence}
      s^\ast \circ (\lambda(\phi_h) \circ g)^\ast
    \since[\approx_Q]{$\lambda(-)$ natural up to $\approx_Q$}
      s^\ast \circ (\lambda(\phi_h \circ (g \times id)))^\ast
    \since[\approx_Q]{$\lambda(-)$ preserves $\approx_Q$,
                      $\phi_h \circ (g \times id)
                        \approx_Q \phi_{h \circ (g \times id)}$}
      s^\ast \circ (\lambda(\phi_{h \circ (g \times id)}))^\ast
  \end{derivation}
which by definition is $x_{h \circ (g \times id)}$. Hence,
\begin{cor}
  If $\lambda(-)$ preserves $\approx_Q$ and is natural up to it,
  and moreover the $(Q, X)$-enumeration $e : X \rightarrow A$
  satisfies $\phi_h \circ (g \times id) \approx_Q \phi_{h \circ (g
  \times id)}$ then the IPWPS $p : QX \times QX \rightarrow A$
  constructed in Lemma \ref{lem:enumwps} is natural in the sense
  described at the end of \S\ref{sec:paramlawvere}.
\end{cor}

\noindent In turn, an easy calculation shows that if a $(Q,
X)$-enumeration is `natural' in the above sense then the inductive
step of Lemma \ref{lem:enum} maintains this property---if
$\lambda(-)$ preserves $\approx_Q$! By induction, all the IPWPSs
constructed in Theorem \ref{thm:enumifps} are natural.

But we are---so to speak---already preaching to the choir: we have
made use of the naturality and preservation of $\lambda(-)$ up to
$\approx_Q$. Thus, the model is already \emph{weakly extensional},
and Lemma \ref{lem:wextnat} already provides a way to make it
natural.

At this point, it is beginning to seem like there is no way around
weak extensionality. On the one hand, generalising our results to
yield natural IFPs seems to already already weak extensionality.
On the other hand, we cannot conceive of a weakly extensional
model in which IFPs really are more informative than EFPs.
\emph{Intensionality and weak extensionality seem to be at odds
with each other.} If we take all of this into account, limiting
the fixed point rule to admit no free variables other than the
`diagonal' one seems almost forced in the most intensional of
settings.

\section{$\mathfrak{Asm}(K_1)$ as a model of iPCF v2.0}
  \label{sec:asmipcf}

Let us revisit the construction of the P-category of assemblies
$\mathfrak{Asm}(K_1)$ on the PCA $K_1$ of classical computability,
as described in \S\ref{sec:realexpo}. We shall prove that it is a
iPCF model, as per Definition \ref{def:ipcfmodel}. It certainly
comes with all the prerequisite structure, so all we need to check
is whether it has (natural) intensional fixed points at all the
relevant objects.  We shall construct them using Theorem
\ref{thm:enumifps}.

Before we proceed with the construction, we need to define some
objects of interest. First, we define the assembly $\mathbb{N} \in
\mathfrak{Asm}(K_1)$ of \emph{natural numbers} by \[
  \bars{\mathbb{N}} \myeq \mathbb{N},
    \qquad
  \dbars{n}_\mathbb{N} \myeq \{n\}
\] We also need to define the assembly $\mathbb{N} \in
\mathfrak{Asm}(K_1)$ of \emph{booleans} by \[
  \bars{\mathbb{B}} \myeq \{\textsf{ff}, \textsf{tt}\},
    \qquad
  \dbars{b} \myeq
    \begin{cases}
      \{ 0 \}           &\text{for } b = \textsf{ff} \\
      \{ 1, 2, \dots \} &\text{for } b = \textsf{tt}
    \end{cases}
\] At this point we urge the reader to recall the definition of
the \emph{lifted assemblies} $\mathbb{N}_\bot$ and
$\mathbb{B}_\bot$ that we defined for $K_1$ in
\S\ref{sec:passpcat}.

\begin{prop}
  \label{prop:k1intretr}
  For any assembly $Z \in \mathfrak{Asm}(K_1)$ there is an
  intensional retraction \[
    \begin{tikzcd}
      Q\left({\mathbb{N}_\bot}^{QZ}\right)
        \arrow[r, "s"]
        \arrow[d, "\epsilon_X", swap]
      & \mathbb{N}_\bot
        \arrow[dl, "r"] \\
      {\mathbb{N}_\bot}^{QZ} 
    \end{tikzcd}
  \]
\end{prop}
\begin{proof}
  For the section part, we define
  \begin{align*}
    s : \bars{Q\left(\mathbb{N}^{QZ}_\bot\right)} 
                &\longrightarrow \bars{\mathbb{N}_\bot} \\
        (f, n)  &\longmapsto     n
  \end{align*} which is realized by $\lambda^\ast nx. n$. For the
  retraction we define \begin{align*}
    r : \bars{\mathbb{N}_\bot} &\longrightarrow \bars{\mathbb{N}^{QZ}_\bot} \\
        n                      &\longmapsto     f_n \\
        \bot                   &\longmapsto     ((z, a) \mapsto \bot)
  \end{align*} where \begin{align*}
    f_n : \bars{QZ} &\longrightarrow \bars{\mathbb{N}_\bot} \\
          (z, a)    &\longmapsto 
            \begin{cases}
              m \text{ if $n \cdot a \cdot \overline{0} \simeq m$} \\
              \bot \text{ if $n \cdot a \cdot \overline{0} \uparrow$}
            \end{cases}
  \end{align*} $f_n$ is realized by $\lambda^\ast a x.\ n \cdot
  a \cdot \overline{0}$, and hence $r$ itself is realized by
  $\lambda^\ast w a  x.\ w \cdot \overline{0} \cdot a \cdot
  \overline{0}$.
  
  It remains to show that if $n \in
  \dbars{f}_{\mathbb{N}_\bot^{QZ}}$, then $f_n = f$. In order to
  show this, the first thing we have to note is that
  $\mathbb{N}_\bot$ is a \emph{modest set}: a realizer can only
  realize one element of $\bars{\mathbb{N}_\bot} \myeq \mathbb{N}
  + \{\bot\}$. Thus, if $n \cdot a \in
  \dbars{x}_{\mathbb{N}_\bot}$, then necessarily $f(z, a) = x$: if
  $f(z, a) = x'$, then we must have $n \cdot a \in
  \dbars{x'}_{\mathbb{N}_\bot}$, so $x = x'$. We can thus set up a
  chain of equivalences, \[
    f(z, a) = m\
      \Longleftrightarrow\
    n \cdot a \in \dbars{m}_{\mathbb{N}_\bot}\
      \Longleftrightarrow\
    n \cdot a \cdot \overline{0} \simeq \overline{m}\
      \Longleftrightarrow\
    f_n(z, a) = m
  \] and, similarly, \[
    f(z, a) = \bot\
      \Longleftrightarrow\
    n \cdot a \in \dbars{\bot}_{\mathbb{N}_\bot}\
      \Longleftrightarrow\
    n \cdot a \cdot \overline{0} \uparrow\
      \Longleftrightarrow\
    f_n(z, a) = \bot
  \]
\end{proof}

\begin{prop}
  \label{prop:nboxenum}
  $\mathbb{N}_\bot$ is $(\Box, \mathbb{N}_\bot)$-enumerable.
\end{prop}
\begin{proof}
  Suppose $(f, r) : \prod_{i=1}^n QB_i \rightarrow
  \mathbb{N}_\bot$. This means that, if $a_i \in
  \dbars{b_i}_{B_i}$ for all $i$ then \[
    r \cdot \langle a_1, \dots, a_n \rangle \in
      \dbars{f((b_1, a_1), \dots, (b_n, a_n))}_{\mathbb{N}_\bot}
  \] where \begin{align*}
    \langle a \rangle &\myeq a \\
    \langle a_1, \dots, a_{m+1} \rangle &\myeq
      \textsf{pair}\ \langle a_1, \dots, a_{m} \rangle\ a_{m+1}
  \end{align*} We can then define $\phi_{(f, r)} \myeq
  (g_r, s)$ where \begin{align*}
    g_r : \bars{\prod_{i=1}^n QB_i}       &\longrightarrow 
                  \bars{\mathbb{N}_\bot} \\
          ((b_1, a_1), \dots, (b_n, a_n)) &\longmapsto  
                  \langle r, a_1, \dots, a_n \rangle
  \end{align*} which is obviously realizable.  Then, define
  $e_{\mathbb{N}_\bot} : \mathbb{N}_\bot \rightarrow
  \mathbb{N}_\bot \myeq (h, v)$ by \begin{align*}
    h : \bars{\mathbb{N}_\bot} &\longrightarrow \mathbb{N}_\bot \\
         c                     &\longmapsto
            \begin{cases}
              \bot, \text{ if $(c)_1\ \cdot \langle (c)_2\ \dots
                    (c)_{n+1} \rangle \cdot \overline{0} \uparrow$} \\
              m,    \text{ if $(c)_1\ \cdot \langle (c)_2\ \dots
                    (c)_{n+1} \rangle \cdot \overline{0} \simeq m$}
            \end{cases} \\
        \bot                   &\longmapsto     \bot
  \end{align*} where $(\langle a_1, \dots, a_m \rangle)_i \myeq
  a_i$. This is realizable, and it is easy to show that the
  required diagram \[
    \begin{tikzcd}
      \prod_{i=1}^n QB_i
        \arrow[dr, "f"]
        \arrow[d, "\phi_f", dashed, swap]
      & \\
      \mathbb{N}_\bot
        \arrow[r, "e_{\mathbb{N}_\bot}", swap]
      & \mathbb{N}_\bot
    \end{tikzcd}
  \] commutes.
\end{proof}

\begin{prop}
  \label{prop:bboxenum}
  $\mathbb{B}_\bot$ is $(\Box, \mathbb{N}_\bot)$-enumerable.
\end{prop}
\begin{proof}
  The same proof as for $\mathbb{N}_\bot$ almost works: we only
  need to slightly alter the definition of $e_{\mathbb{N}_\bot} =
  (h, v) : \mathbb{N}_\bot \rightarrow \mathbb{N}_\bot$ to make it
  into an arrow $e_{\mathbb{B}_\bot} \myeq (h', v) :
  \mathbb{N}_\bot \rightarrow \mathbb{B}_\bot$ where
  \begin{align*}
    h' : \bars{\mathbb{N}_\bot} &\longrightarrow \mathbb{B}_\bot \\
         c                      &\longmapsto
            \begin{cases}
              \bot,
                \text{ if $(c)_1\ \cdot \langle (c)_2\ \dots
                (c)_{n+1} \rangle \cdot \overline{0} \uparrow$} \\
              \textsf{ff},
                \text{ if $(c)_1\ \cdot \langle (c)_2\ \dots
                (c)_{n+1} \rangle \cdot \overline{0} \simeq 0$} \\
              \textsf{tt},
                \text{ if $(c)_1\ \cdot \langle (c)_2\ \dots
                (c)_{n+1} \rangle \cdot \overline{0} \simeq m \neq 0$}
            \end{cases} \\
        \bot                   &\longmapsto     \bot
  \end{align*} The same realizer works.
\end{proof}

We thus conclude that

\begin{thm}
  $\mathfrak{Asm}(K_1)$ has intensional fixed points at the
  intensional exponential ideal generated by the `grammar' \[
    I \quad ::= \quad
                  \mathbb{B}
      \quad|\quad \mathbb{N}
      \quad|\quad I^{QZ}
  \] for any $Z \in \mathfrak{Asm}(K_1)$.
\end{thm}

\begin{proof}
  Use Propositions \ref{prop:k1intretr}, \ref{prop:bboxenum},
  \ref{prop:nboxenum} to fulfil the premises of Theorem
  \ref{thm:enumifps}.
\end{proof}

\chapter{Conclusions \& Future Work}

  \label{chap:conc}

  We briefly peruse what has been achieved in this thesis:

\begin{enumerate}[label=(\roman{*})]
  \item
    We first attempted to pin down the informal meaning of
    \emph{intensionality} as the possibility of
    \emph{non-functional operations} (\S\ref{chap:intro}), where
    non-functionality is understood in the presence of some ambient
    extensional equality.
  \item
    Then, we carefully reviewed the distinction between
    \emph{extensional and intensional recursion in computability
    theory} (\S\ref{chap:srtht}).
  \item
    This led us to the formulation of a \emph{higher-order
    intensional and reflective programming language}, in the form
    of a modal $\lambda$-calculus called Intensional PCF. This
    language included genuinely `non-functional' operations, and
    typed intensional recursion through L\"ob's rule. We showed
    that, if intensionality/`non-functionality' is limited to
    modal types, then iPCF is consistent (\S\ref{chap:ipcf}).
  \item
    In \S\ref{chap:expo} we began the search for a categorical
    semantics for that calculus. We first argued that 1-category
    theory is not the correct mathematical setting to speak of
    intensionality. As an alternative, we proposed P-category
    theory.  We then proceeded to introduce \emph{exposures}---a
    new P-categorical construct which abstracts the idea of
    intensional devices, e.g. G\"odel numberings. Then, drawing
    inspiration from comonads, we developed the theory of
    exposures.
  \item
    The claim that exposures are abstractions of intensional
    devices was substantiated by carefully constructing three
    rather different examples in \S\ref{chap:expoexamples}.

    The first one comprises an actual G\"odel numbering on Peano
    Arithmetic.
    
    The second one was drawn from higher-order
    computability/realizability. If we think of realizers as
    machine code, this main example made clear the idea that
    \emph{exposures expose the implementation}. 
    
    The third one is based on ideas from homological algebra, and
    constitutes a first attempt at recognising the occurrence of
    intensionality in fields beyond logic and computability.
  \item
    Then, in \S\ref{chap:irec}, we reformulated to intensional
    recursion, and showed that it can be captured through
    exposures. We proved abstract analogues of classic intensional
    results, like G\"odel's First Incompleteness Theorem, Tarski's
    Undefinability Theorem, and Rice's Theorem. These results lend
    credibility to the idea that exposures are a toolkit where the
    fine structure of results with an intensional flavour can be
    described.

    The culmination of this chapter was the Intensional Recursion
    Theorem (Theorem \ref{thm:intrec}), which set out conditions
    that guarantee the existence of intensional fixed points. The
    Intensional Recursion Theorem can be thought of as an abstract
    version of Kleene's Second Recursion Theorem.
  \item
    In the final two chapters (\S\ref{chap:intsem1},
    \S\ref{chap:intsem2}) we brought iPCF and exposures together.
    After some technical development, we showed that a restriction
    of iPCF, called iPCF v2.0, can be interpreted in a cartesian
    closed P-category equipped with a product-preserving comonadic
    exposure, and IFPs at appropriate objects.

    We then discussed the cases in which the restrictions that
    plague iPCF v2.0 can be waived. However, we argued that there
    are good reasons indicating that lifting those restrictions is
    at odds with intensionality, at least in the way in which we
    understand it.

    We closed the thesis by proving that $\mathfrak{Asm}(K_1)$,
    the P-category of assemblies on $K_1$, is a model of iPCF
    v2.0. As $K_1$ is a PCA based on classical computability, this
    means that iPCF v2.0 is adequate for constructing indices in
    classical computability theory: it is a typed `intensional
    metaprogramming' language for writing programs in the style of
    \cite{Jones1997}, albeit with limited expressivity.
\end{enumerate}

\noindent In the rest of this concluding chapter, we will try to
evaluate these achievements. We would like to focus on four
aspects in particular: \begin{itemize}
  \item Is intensionality really just the ability to have
  non-functional operations?
  \item Are iPCF and iPCF v2.0 adequate intensional and reflective
  languages?
  \item How do exposures compare with alternative `theories of
  intensionality'?
  \item Have we managed to elucidate the mysterious Second
  Recursion Theorem of Kleene?
\end{itemize}

\section{Is intensionality really just non-functionality?}

The formalisation of Frege's ideas of sense and reference, which
we discussed in \S\ref{sec:intensionality}, is an old problem.
Even though many have tried, there does not seem to be a
definitive account. Some even question whether such a definitive
account \emph{should} exist.

\cite{Moschovakis1993} defines the sense of logical formula as the
(possibly infinitary) algorithm that the syntax of a (first-order)
formula suggests. This is formalised using Moschovakis' own
\emph{theory of recursive algorithms}.

Occupying some middle ground, \cite{Abramsky2014} draws on a long
tradition of \emph{programming language semantics}. In a sense, he
introduced what one could call the \emph{spectrum of
intensionality}: some kinds of semantics is more
\emph{intensional} than others, in that the mathematical objects
that comprise it contain strictly more information about the
computation that is being modelled, e.g. an account of the
interactions that take place when a program is run. The gist is
that, by moving to more refined semantics, more can be captured,
even though a price might have to be paid, perhaps in the form of
\emph{quotienting the model}. However, we believe that \emph{op.
cit.} is permeated by a preliminary form of the ideas explicitly
developed in this thesis.

A third opinion is given by \cite{Girard1989}: ``the sense
contains the denotation, at least implicitly.'' Girard proposes a
study of the \emph{invariants of syntax}, in a vein inspired by
proof theory. Some interpret this statement as taking the extreme
viewpoint that `syntax = sense,' but it becomes clear in
\citep{Girard2011} that the author is simply proposing the study
of new, unorthodox proof-theoretic structures.

In this thesis we avoided this lengthy debate by \emph{defining}
intensionality to mean `anything finer than what we call
extensional equality.' We like to view this as \emph{purely
mathematical}, and \emph{philosophically agnostic}. We have merely
demonstrated in \S\ref{chap:ipcf} that modal types allow one to
treat their elements as pure syntax.

In the development of exposures in \S\ref{chap:expo}, and then in
the intensional semantics of \S\ref{chap:intsem2}, it became clear
that this view of equality is \emph{modular}. The modal types
allow one to introduce equalities in a \emph{controlled} fashion:
we began with no equations in iPCF (\S\ref{chap:ipcf}), and
gradually reached a set of equations in iPCF v2.0
(\S\ref{chap:intsem2}) which seem to mirror intensional
equality, as defined by the exposure. Whether that is a precise
reflection can be shown through a \emph{completeness theorem},
which we conjecture to hold. In turn, our flavour of exposure
(cartesian, product-preserving, weakly cartesian closed, etc.)
determines which equations intensional equality will satisfy. We
believe that the modularity of this framework is a serious
advantage that makes it adaptable to all sorts of settings, and
all levels of fine-grain intensional information.

\section{How expressive is iPCF?}
  \label{sec:expripcf}

The first two objectives of this thesis that we discussed in
\S\ref{sec:objectives} were the clarification of the
\emph{intensional and reflective programming} paradigms.

The development of iPCF elucidated the fact that we can indeed
treat terms at modal types as if they were `pure syntax' (up to
$\alpha$-equivalence), and make arbitrary decisions on them
\emph{in a typed manner}. This shows that, if we comprehend
intensional programming as entailing non-functional behaviour,
then there is a type theory in which this ability is provably
consistent. As we remarked in the introduction, there have been
many similar attempts in the past, but all were in one way or
another problematic: some led to particularly complicated
languages with unclear semantics, like those of \cite{Smith1982,
Smith1984}; others, like \cite{Gabbay2013}, were close to ours,
but proposed semantics which are---unfortunately---inconsistent. A
third class led to impossibility theorems, e.g. \cite{Wand1998}.

Our work decisively resolved many of these issues: once a modal
typing discipline is established, then we are perfectly capable of
accommodating both non-functional behaviour as well as
reflexivity. If we limit these behaviours to the modal types, and
use the typing system to stop the flow from the extensional region
of the language to the modal types, then we get a consistent
language. We also believe that we are the first to directly tie
intensionality and reflexivity together, as we think they should
be, if we are to use reflexivity in any interesting manner.

Nonetheless, as we hinted in \S\ref{sec:ipcfq}, iPCF can only be
considered a proof-of-concept. We have deliberately decided not to
concern ourselves with the task of finding good sets of
intensional primitives, but merely with the possibility of
crafting a setting in which this is possible.  We believe that
finding good intensional primitives is a particularly hard
problem, with close connections to both \emph{metaprogramming} and
\emph{higher-order computability}. Furthermore, finding
\emph{other models} is likely to prove challenging.

\subsection{Metaprogramming}

As we discussed in the introduction, metaprogramming is a
difficult task that dates back to the work of the \textsc{Lisp}
community. The area has recently witnessed a resurgence of
interest, leading to the \emph{International Summer School on
Metaprogramming} that took place at Robinson College, Cambridge
(8-12 August 2016) around the time that the author began drafting
this thesis. It is quite clear that a good foundation for
metaprogramming is still lacking: see the work of
\cite{Berger2016} for a recent discussion.

At this point we should remark that intensional operations are
\emph{not} the main subject of that area: metaprogramming is about
\emph{constructing} code dynamically from its fragments, and not
\emph{destructing} it, as we are wont to do with intensionality.

A common issue in metaprogramming is that of \emph{manipulating
code with open variables (= free variables)}. This is known to be
a rather painful limitation of the \textsf{S4}-based language of
\cite{Davies1996} on which iPCF is based. In order to overcome
this, \cite{Davies1995, Davies1996a, Davies2017} developed a
$\lambda$-calculus based on another modality that is reminiscent
of the `next' operator of Linear Temporal Logic (\textsf{LTL}).
This language has explicit annotations of the \emph{stage} at
which each computation is taking place, and is also able to handle
open code. This led to a flurry of developments, and in particular
the very influential work of \cite{Taha1997, Taha2000} on MetaML,
and then the environment classifiers of \cite{Taha2003a}, which
have recently been improved by \cite{Tsukada2010}. See \cite[\S
6]{Kavvos2016b} for further references, and for a discussion of
the modal aspects used.

In iPCF, the restriction of \emph{no open code} does not only
occur as it did before (only modal variables under boxes), but it
also vengefully reappears in the case of intensional operations:
we saw in \S\ref{sec:ipcfconfl} that, unless the terms on which we
operate intensionally are closed, we risk inconsistency. The
author's colleague, Mario Alvarez-Picallo, has begun some
preliminary work in intensionality in calculi like Davies', which
we discuss further in \S\ref{sec:altint}.

However, to achieve any meaningful notion of intensional
primitives, much more than simple modal types is required.
Consider, for example, a constant $\mb{deapp}$ that attempts to
take apart an application, in the sense that \[
  \mb{deapp}\ (\ibox{(M\,N)}) = \langle \ibox{M}, \ibox{N} \rangle
\] (where we have also assumed products in the language). This
cannot meaningfully be typed in the simple modal setting. In fact,
it seems that we need \emph{existential types} in order to
assign a type to this, as the domain of $M$ is unknown. For
example, the type would be something like \[
  \mb{deapp} : \Box B \rightarrow
    \exists A.\;\Box (A \rightarrow B) \times \Box A
\] However, this would take us deep into the waters of
\emph{second-order modal logic}, and we are not aware of any
previous work on that front.

Similar second-order type systems like this have been suggested in
the context of Barry Jay's \emph{factorisation calculi}.  These
are combinatory calculi that admit intensional operations at `face
value.' The trick by which disaster is avoided is that of having
intensional operations act only on normal forms, e.g. a partially
applied $\mb{S}$ combinator ($\mb{S}PQ$), which they are able to
take apart (\emph{factorise}) in order to reuse its immediate
subterms $P$ and $Q$: see \cite{Jay2011a}. In subsequent work by
\cite{Jay2011b} it was shown that such a system admits a
second-order Curry-style typing similar the one shown above, but
without the modalities.

\subsection{Higher-Order Computability}

This is perhaps the point of view from which the present work
originates, but also the least developed in this thesis. We
discussed some possible applications of our ideas on intensional
higher-order computation in \S\ref{sec:honfc}. Unfortunately, the
scope of this thesis could not be stretched to contain more
material in this direction. Nevertheless, the author has a
presentiment that any truly novel understanding of intensional
higher-order computation will come simultaneously with the
development of intensional primitives for a language like iPCF.

\subsection{iPCF, iPCF v2.0, and their models}

Even though we begun with the model of assemblies on $K_1$
in mind, it gradually became apparent that things were not quite
that simple. In \S\ref{chap:intsem2} we ran into severe
difficulties when trying to force $\mathfrak{Asm}(K_1)$ to be a
model of iPCF, which we related to three fundamental problems:
\begin{enumerate}
  \item free variables in intensional fixed points;
  \item free variables in abstractions of quoted terms, a.k.a
  \emph{no higher-order intensional programming}; and
  \item the construction of intensional fixed points at all types.
\end{enumerate}

Whereas the third problem is largely technical, the other two are
quite serious points against our argument that iPCF v2.0 is a good
`language of indices.' We argued extensively in
\S\ref{sec:ipcfv2.0} that, indeed, both of these problems seem
more-or-less natural from a computability theory point-of-view.
However, naturality and substitution are fundamental aspects of
any $\lambda$-calculus, and the fact that we have to disallow them
fundamentally reduces the expressiveness of what we have achieved.

It thus follows that we need to consider more candidate models of
iPCF. We can go about this task in three ways:
\begin{enumerate}[label=(\roman{*})]
  \item
    We can study $\mathfrak{Asm}(K_1)$ more closely; or
  \item
    we can look for other PCAs $A$ such that $\mathfrak{Asm}(A)$
    is an interesting \emph{natural}/\emph{weakly extensional}
    model; or
  \item
    we can look elsewhere.
\end{enumerate}

Regarding the first option, we should remark that we \emph{did not
show} that $\mathfrak{Asm}(K_1)$ is \emph{not} natural. We very
strongly believe that it is not weakly extensional, although we
have no evidence for that either. Disproving either of these
assertions is likely to be a very cumbersome exercise, and it is
perhaps best that it be automated/computer-assisted. If naturality
is shown to not be the case, then we should perhaps redefine
$\ibox{(-)}$ to enclose only \emph{completely closed} terms, with
no free variables at all, whether modal or intuitionstic. This
would directly lead us to some kind of intensional combinator
language, as also described at the end of \S\ref{sec:ipcfv2.0},
which would not be very far from what is already the case with
indices in computability theory. It may be that the `nature' of
$K_1$ simply has this shape.

However, even if $\mathfrak{Asm}(K_1)$ were to unexpectedly prove
to be natural, the key to Davies-Pfenning style staged
metaprogramming---which is an expressive improvement over indices
in computability theory---is precisely the free variables under
quoted $\lambda$-abstractions. This very urgently requires
\emph{weak extensionality}, as remarked in \S\ref{sec:natwextmod}.
In that section, we also discussed the possibility of finding some
weakly extensional PCA $A$ that can be informative in terms of
intensional programming. We concluded that the said task is likely
very difficult.

It is not inconceivable that there could be another source of
models for iPCF: perhaps a change in perspective is required, and
this change may be one that involves moving away from ideas
sourced from realizability, and more towards the emergent theory
of metaprogramming. This is also related to some ideas that we
will discuss in the next section, regarding alternative proposals
for modelling the phenomenon of intensionality at large.

Finally, as soon as more models of iPCF are identified, there are
many interesting questions that need to be answered. First,
showing a \emph{completeness theorem} for iPCF interpreted with
exposures should be a primary goal. In a sense, a completeness
theorem will demonstrate that our categorical structure indeed
corresponds to what we type-theoretically understand as
intensional types. This should be investigated as a priority.
Secondly, a more careful approach to the available intensional
operations should be taken, and some form of \emph{adequacy}
theorem for iPCF should be shown. We are not exactly sure what
form this should take, but it should certainly be much more
involved than the adequacy statements concerning PCF. This issue
is likely to be deeply intertwined with the \emph{intensional
primitives} we discussed this section. Thirdly, there must be many
interesting results concerning the mismatch between iPCF and its
various models. This is very much the case with PCF, beginning
with the work of \cite{Plotkin1977} on domain-theoretic models,
and all the way to the work of \cite{Escardo1999} on metric
models. However, the intensional case is likely to be much richer,
detailed, and tricky.

\section{Exposures vs. other theories of intensionality}
  \label{sec:altint}

Exposures attempt to abstract the general notion of intensional
devices, of which G\"odel numberings or numberings of partial
recursive functions are only particular cases, as seen in 
\S\ref{chap:expoexamples}. As such, the ambition of the work in
this thesis is slightly different when compared to previous
attempts, which are mainly concerned with presenting a categorical
account of computability theory.

We showed in \S\ref{chap:irec} that exposures are particularly
elegant settings for reproducing well-known diagonal arguments. It
has been known for some time that there are common elements
between these theorems, but the language of P-categories and
exposures allows us to capture what is needed for each argument in
very fine detail. For example, Rice's theorem necessitates an
evaluator $\epsilon : Q \natexp{} \mathsf{Id}$, whereas Tarski's
Undefinability Theorem precisely states that merely having
evaluators is catastrophic for one aspect of consistency
(fix-consistency), as this causes $\lnot$ to have a fixed point.
On the other hand, G\"odel's First Incompleteness Theorem states
that we \emph{must} have more truth values than simply true and
false. 

Contrary to the presentation of \cite{Lawvere1969, Lawvere2006},
all our theorems are in the same language of comonadic exposures.
Our development is \emph{positive}, in that it is predicated on
the existence of IFPs, and not the absence of EFPs, as in
Lawvere's paper. We have concentrated on the \emph{oughts}, and
not the \emph{ought nots}. It is for this reason that we view our
results as a refinement of those of Lawvere.

We believe that all these observations and results lend support to
the idea that exposures can become a useful toolkit situated at
the heart of a new \emph{theory of intensionality}, which would be
applicable in all sorts of settings, not necessarily related to
logic and computation. Furthermore, unlike previous work in the
same style, exposures draw inspiration from modal logic and the
Curry-Howard correspondence. It is for that reason that the
resulting framework is---unlike all its
predecessors---\emph{inherently typed}. Essentially all previous
work on the subject relied on some `universal'---in one way or
another object---which contained \emph{codes} for a whole class of
arrows.

First, we want to mention the only two generalisations of the SRT
of which we are aware. They both seem very similar, and they
concern \emph{effective Scott domains}: one is due to
\cite{Kanda1988}, and the other one due to \cite{Case2012}.

As for category-theoretic frameworks, the only previous work with
a similar flavour consists of (a) a paper of Mulry, which
uses the \emph{recursive topos}; and (b) the line of work that
culminates with Cockett and Hofstra's \emph{Turing categories}, as
well as their intellectual predecessors.

\subsection*{Mulry's Recursive Topos}

\cite{Mulry1982} constructed the \emph{recursive topos}
$\mathbf{Rec}$, which is the Grothendieck topos of canonical
sheaves on the monoid of recursive functions under composition.
Broadly speaking, the notion of (higher-order) computation in the
recursive topos is that of \emph{Banach-Mazur} computability: a
map is computable if it maps recursive sequences to recursive
sequences.\footnote{This short explanation of Banach-Mazur is due
to Andrej Bauer, and was sourced from
\href{https://mathoverflow.net/questions/21745/the-difference-between-the-recursive-and-the-effective-topos}{mathoverflow
question \#21745}.} 

In \citep{Mulry1989}, the author aims to ``contemplate a synthetic
theory of computation, i.e. an intrinsic set of categorical axioms
for recursion which captures both essential features of classical
recursion theory but is also applicable over a wide range of
applications in areas of effective mathematics and computer
science.'' 

Indeed, Mulry's work has some overlap with some of the material we
presented in \S\ref{sec:lawvere} on Lawvere's theorem. He
identifies an enumeration of the partial recursive functions as a
point surjective map, and then proceeds to interpret all the
classic results of type 2 computability (Myhill-Shepherdson,
Kreisel-Lacombe-Shoenfield, etc.) in the context of the recursive
topos. This is followed by a discussion of the connections between
the recursive topos, the \emph{effective topos} of
\cite{Hyland1982}, and \emph{effective Scott domains}. He there
points out that one can construct a $\mathbb{N}$-path surjection
$\mathbb{N} \rightarrow D$ and a point surjection $\mathbb{N}
\rightarrow D^{\mathbb{N}}$ in $\mathbf{Rec}$, for any effective
Scott domain $D$. 

Fixed points are discussed in the final section of the paper. It
is noted there that Lawvere's theorem corresponds to (a limited
version of) Kleene's FRT, simply by considering an enumeration of
(two-argument) partial recursive functions, alongside $\mathbb{N}
\cong \mathbb{N} \times \mathbb{N}$. This last isomorphism is also
used to prove versions of the FRT and SRT for effective Scott
domains, based on the observations of the previous paragraph. It
is remarked that the same can be shown in the effective topos, but
not in the category of effective domains itself, as neither of
their natural numbers objects are effective domains.

Thus, Mulry's versions of the FRT and SRT seem to depend on the
natural numbers object $\mathbb{N}$. The SRT seems to give a fixed
point for $h : \mathbb{N} \rightarrow \mathbb{N}$, in the sense
that both the fixed point $n$ and $h \circ n$ are indices for the
same element of the effective Scott domain. This is indeed a
version of the SRT for all effective Scott domains, but it lacks
the well-typed flavour of our approach.

\subsection*{The road to Turing categories}

Inspired by previous work by \cite{Heller1987}, and some of the
material in the thesis of \cite{Birkedal2000}, \cite{Cockett2008}
developed the notion of a \emph{Turing category}.

Turing categories are cartesian restriction categories: that is,
they are equipped with some structure to handle the
\emph{partiality} of their arrows, and there are cartesian
products---up to partiality. Their defining feature is that they
have a \emph{Turing object} $A$, which is a domain that contains
codes for all the arrows of the category: it has a \emph{universal
application}, \[
  \tau_{X, Y} : A \times X \rightarrow Y
\] for each pair of objects $X$ and $Y$, such that given any $f :
Z \times X \rightarrow Y$, an \emph{index} $h : Z \rightarrow A$
exists, not by any means unique, such that the following diagram
commutes: \[
  \begin{tikzcd}
    A \times X
      \arrow[r, "\tau_{X, Y}"]
    & Y \\
    Z \times X
      \arrow[ru, "f", swap]
      \arrow[u, "h \times id_X"]
    &
  \end{tikzcd}
\] So $A$ is a very weak exponential for all $X, Y$ at the same
time. In fact, every object $X$ of a Turing category is a retract
of $A$.

Whereas the work of Cockett and Hofstra is a beautiful and general
account of settings where basic recursion theory can be done, and
also give rise to an interesting theory of \emph{categorical
simulations} \citep{Cockett2010}, they are very far from the goals
of our own work: we are interested in exploring intensionality,
which is only one of the phenomena that occur in recursion theory.
We are also interested in doing so in a stringent type-theoretic
manner, and thus we perceive the reliance on Turing objects as a
problem. Cockett and Hofstra themselves point out the untyped
nature of their work: \begin{quote}
  ``A similar inherent limitation to Turing categories lies in its
  essential untypedness. Recently, Longley [...] has advocated the
  use of typed PCAs in order to clarify notions of computation at
  higher types; it is not unimaginable that the corresponding
  generalization of Turing categories can be of interest if we
  wish to handle such notions of computation in our setting. Such
  a generalization would essentially bring us back to Birkedal’s
  weakly closed partial cartesian categories.''
\end{quote} This is followed by the observation that in the Turing
category corresponding to classical computability theory no real
discussion of higher-order phenomena can occur, as one has no way
of speaking about type-2 functionals.

In conclusion, the work of Cockett and Hofstra is untyped,
fundamentally based on partiality, and mostly centred around
recursion theory. We prefer typed, total, and intensional
settings, without wanting to declare any particular allegiance to
recursion-theoretic arguments.

\subsection{An Idea for an Alternative Approach}

At this point we ought to record another candidate approach for
modelling intensionality as we see it in this thesis. This idea is
due to Martin Hyland, who examined this thesis. However, a similar
framework has also been put forward by my fellow student, Mario
Alvarez-Picallo, in his work on the semantics of metaprogramming.

In many ways, the framework of exposures can be seen as
\emph{evil} or \emph{pathological}, in that it requires a highly
non-standard property, namely that PERs are reflected instead of
preserved---as they would be for a P-functor. In that sense, it
may indeed be \emph{non-categorical} (this is another interesting
idea that needs to be investigated).

Perhaps the following proposal would be more workable. Instead of
insisting on a \emph{single} mathematical universe, i.e. a single
category, we could consider two: one that is \emph{intensional},
and one that is \emph{extensional}. The theory of exposures
already gives us two rather good candidates for these: given an
endoexposure $Q : (\mathfrak{C}, \sim) \expo{} (\mathfrak{C},
\sim)$, the intensional universe is the x-ray P-category
$(\mathfrak{C}, \approx_Q)$, wheras the extensional one is the
P-category $(\mathfrak{C}, \sim)$ itself---see \S\ref{sec:expo}
for definitions of these notions. Now, as intensional equality
implies extensional equality, it is evident that there is a
trivial P-functor, \[
  (\mathfrak{C}, \approx_Q) \longrightarrow (\mathfrak{C}, \sim)
\] which is the identity on objects, the identity on morphisms,
and full---but definitely not faithful!\footnote{The first of
these criteria is a recurrent theme of questionable categorical
status, which nonetheless occurs reasonably often, e.g. in Freyd
categories; see e.g. \cite{Staton2014}.} One could perhaps argue
that we could even move away from P-categories, and back into ordinary
categories: there could be a functor $\mathcal{C} \rightarrow
\mathcal{D}$, which is the identity on objects and full. Perhaps
$\mathcal{D}$ could be of the form $\mathcal{C}/\sim$, i.e. a
quotient category of some sort. Thus, intensionality is obtained
by having (P-)categories on \emph{two levels}, so that one is
included in the other.

Mario Alvarez has proposed some directions regarding the categorical
modelling of non-homogeneous staged metaprogramming calculi, like
the ones of \cite{Davies1995, Davies1996a, Davies2017}, which bear
a certain resemblance to the preceding idea. Put simply, he
proposes a simple metaprogramming calculus which consists of
essentially two `stages,' one `under the circle' (cf.  under the
box), and the ordinary one. Under the circle, affairs intensional,
in much the same way that they are under our boxes, but otherwise
things work up to ordinary equality. Using the syntax of this
calculus, one can construct a \emph{cartesian} category
$\mathbb{SM}$, the \emph{syntactic model}: this is a term model,
but not quotiented up to equality. A model of this is then a
cartesian closed category $\mathcal{C}$ along with a functor, \[
  F : \mathbb{SM} \longrightarrow \mathcal{C}
\] which is \emph{strictly product-preserving}. Furthermore, if
\emph{intensional operations} are to be soundly interpreted, this
functor should be \emph{faithful}. This is to be understood in the
following way: the functor $F$ \emph{encodes} syntactic
information \emph{within the semantic model}, in a manner that
does not collapse the syntax.

There is an odd tension between fullness and faithfulness in this.
Which one should we choose for intensionality? Faithfulness
guarantees that the syntax is accurately represented in the model,
whereas fullness describes the idea that for every `extensional'
view, there is at least one intensional. Nevertheless,
investigation of this idea of \emph{two level-two category}
paradigm of intensionality is likely to be a very interesting
avenue for further research.

\section{Kleene's mysterious Second Recursion Theorem}

The basic impetus behind this thesis was not intensionality
itself, but rather its r\^{o}le in Kleene's Second Recursion
Theorem, which we exhaustively studied in \S\ref{chap:srtht}.
These questions were brought to the author's attention by
\cite{Abramsky2012, Abramsky2014}, who---along with
\cite{Jones2013}---had identified the mysterious nature of this
theorem in the 1980s. \cite{Abramsky2012} also remarks that the
theorem is very powerful, and---even though very simple---its
proof is opaque, and provides no intuition. Our analysis of
intensional recursion in \S\ref{chap:irec}, and the connection we
have drawn with Lawvere's work on diagonal arguments, has shed
some light on these issues. 

First, we believe that the opacity in proof is not a real problem:
all diagonalisation proofs have a quintessentially `magical'
element that often stimulates people's imagination, bringing them
to the forefront of popular science books and scientific novels,
see e.g. \cite{Hofstadter1979, Doxiadis2001}.  What is more,
\cite{Kleene1981} explicitly states that the SRT was obtained
quite straightforwardly by translating the FRT from the
$\lambda$-calculus. He even dates this derivation before the 1st
of July 1935.

Instead, we believe that the our analysis provides a new
perspective on the \emph{structure} of the situations in which
Kleene's argument is fundamentally applicable. This amounts to a
generalisation of intensional recursion beyond first-order
computability: the Curry-Howard interpretation through L\"ob's
rule, as well as the Intensional Recursion theorem stated in the
language of comonadic exposures, reveal patterns in the statement
and the proof of the theorem that were not known before.

\section{Concluding remarks}

It is evident from the previous sections that there are many very
interesting questions that follow from our work in this thesis.
Perhaps the most interesting theoretical direction is that of
studying the expressiveness of iPCF, especially through the
related themes of metaprogramming and non-functional higher-order
computation. The present author believes that it is very likely
that we could use modalities and intensionality to obtain access
to computational power that we have both sclerotically rejected
from theoretical analysis, or left to programmers of the untyped
persuasion.

These developments---we hope---will proceed hand-in-hand with an
improved understanding of metaprogramming, intensional or not.
Progress is difficult to achieve in metaprogramming not only
because we have been lacking a theoretical foundation, but also
because industrial metaprogramming has progressed unabashedly in
the meantime. The result is that multiple hard-to-shake ad-hoc
habits have been established.

On the other hand, our intensional framework is as close to
1-category theory as one can get, and it is for this reason that
we hope that it may be more widely applicable than just in the
context of computability and logic. We have demonstrated at least
one example of this in the case of homological algebra
(\S\ref{sec:homoexpo}). However, as the applications of category
theory expand to include linguistics and philosophy---see e.g. the
forthcoming volume \citep{Landry2017}---we would like to believe
that the concept of intensionality will reappear in many different
settings, and our framework will be there to explain its
structure.

%\chapter{A Lawvere-type Theorem}
%  \label{chap:lawvere}
%
%  \subimport{./krt/}{lawvere.tex}

%\chapter{An Algebraic Approach}
%
%  \subimport{./krt/}{algsrt.tex}

%%%%%%%%%%%%%%%%%%%%
%%%% APPENDICES %%%%
%%%%%%%%%%%%%%%%%%%%

%TC:ignore
\appendix
\chapter{iPCF v2.0 in \textsc{Agda}}
  \label{app:agda}

\section{\texttt{Basics.agda}}
  \label{sec:Basics.agda}
  \begin{code}%
\>\AgdaKeyword{module} \AgdaModule{Basics} \AgdaKeyword{where}\<%
\\
\\
\>\AgdaKeyword{open} \AgdaKeyword{import} \AgdaModule{Relation.Binary.PropositionalEquality}\<%
\\
\>\AgdaKeyword{open} \AgdaKeyword{import} \AgdaModule{Data.Nat} \AgdaKeyword{using} \AgdaSymbol{(}ℕ \AgdaSymbol{;} zero \AgdaSymbol{;} suc\AgdaSymbol{)}\<%
\\
\>\AgdaKeyword{open} \AgdaKeyword{import} \AgdaModule{Data.Sum} \AgdaKeyword{renaming} \AgdaSymbol{(}\_⊎\_ \AgdaSymbol{to} \_+\_\AgdaSymbol{)}\<%
\\
\\
\>\AgdaKeyword{infixr} \AgdaNumber{1} \_=>\_\<%
\\
\>\AgdaKeyword{infixr} \AgdaNumber{5} \_∈\_\<%
\\
\>\AgdaKeyword{infixl} \AgdaNumber{1} \_⊆\_\<%
\\
\>\AgdaKeyword{infixl} \AgdaNumber{4} \_,\_\<%
\\
\>\AgdaKeyword{infixl} \AgdaNumber{3} \_++\_\<%
\\
\>\AgdaKeyword{infixl} \AgdaNumber{2} \_∧\_\<%
\\
\\
\>\AgdaComment{------------------------}\<%
\\
\>\AgdaComment{-- Types and Contexts --}\<%
\\
\>\AgdaComment{------------------------}\<%
\\
\\
\>\AgdaKeyword{data} \AgdaDatatype{Types} \AgdaSymbol{:} \AgdaPrimitiveType{Set} \AgdaKeyword{where}\<%
\\
\>[0]\AgdaIndent{2}{}\<[2]%
\>[2]\AgdaInductiveConstructor{simple} \AgdaSymbol{:} \AgdaDatatype{Types}\<%
\\
\>[0]\AgdaIndent{2}{}\<[2]%
\>[2]\AgdaInductiveConstructor{modal} \AgdaSymbol{:} \AgdaDatatype{Types}\<%
\\
\\
\>\AgdaKeyword{data} \AgdaDatatype{Ty} \AgdaSymbol{:} \AgdaDatatype{Types} \AgdaSymbol{→} \AgdaPrimitiveType{Set} \AgdaKeyword{where}\<%
\\
\>[0]\AgdaIndent{2}{}\<[2]%
\>[2]\AgdaInductiveConstructor{P} \AgdaSymbol{:} \AgdaSymbol{∀} \AgdaSymbol{\{}\AgdaBound{T}\AgdaSymbol{\}} \AgdaSymbol{→} \<[15]%
\>[15]\AgdaDatatype{ℕ} \AgdaSymbol{→} \AgdaDatatype{Ty} \AgdaBound{T}\<%
\\
\>[0]\AgdaIndent{2}{}\<[2]%
\>[2]\AgdaInductiveConstructor{\_=>\_} \AgdaSymbol{:} \AgdaSymbol{∀} \AgdaSymbol{\{}\AgdaBound{T}\AgdaSymbol{\}} \AgdaSymbol{→} \AgdaDatatype{Ty} \AgdaBound{T} \AgdaSymbol{→} \AgdaDatatype{Ty} \AgdaBound{T} \AgdaSymbol{→} \AgdaDatatype{Ty} \AgdaBound{T}\<%
\\
\>[0]\AgdaIndent{2}{}\<[2]%
\>[2]\AgdaInductiveConstructor{\_∧\_} \AgdaSymbol{:} \AgdaSymbol{∀} \AgdaSymbol{\{}\AgdaBound{T}\AgdaSymbol{\}} \AgdaSymbol{→} \AgdaDatatype{Ty} \AgdaBound{T} \AgdaSymbol{→} \AgdaDatatype{Ty} \AgdaBound{T} \AgdaSymbol{→} \AgdaDatatype{Ty} \AgdaBound{T}\<%
\\
\>[0]\AgdaIndent{2}{}\<[2]%
\>[2]\AgdaInductiveConstructor{□\_} \AgdaSymbol{:} \<[8]%
\>[8]\AgdaDatatype{Ty} \AgdaInductiveConstructor{modal} \AgdaSymbol{→} \AgdaDatatype{Ty} \AgdaInductiveConstructor{modal}\<%
\\
\\
\>\AgdaKeyword{data} \AgdaDatatype{Cx} \AgdaSymbol{:} \AgdaDatatype{Types} \AgdaSymbol{→} \AgdaPrimitiveType{Set} \AgdaKeyword{where}\<%
\\
\>[0]\AgdaIndent{2}{}\<[2]%
\>[2]\AgdaInductiveConstructor{·} \AgdaSymbol{:} \AgdaSymbol{∀} \AgdaSymbol{\{}\AgdaBound{T}\AgdaSymbol{\}} \AgdaSymbol{→} \AgdaDatatype{Cx} \AgdaBound{T}\<%
\\
\>[0]\AgdaIndent{2}{}\<[2]%
\>[2]\AgdaInductiveConstructor{\_,\_} \AgdaSymbol{:} \AgdaSymbol{∀} \AgdaSymbol{\{}\AgdaBound{T} \AgdaSymbol{:} \AgdaDatatype{Types}\AgdaSymbol{\}} \AgdaSymbol{(}\AgdaBound{Γ} \AgdaSymbol{:} \AgdaDatatype{Cx} \AgdaBound{T}\AgdaSymbol{)} \AgdaSymbol{(}\AgdaBound{A} \AgdaSymbol{:} \AgdaDatatype{Ty} \AgdaBound{T}\AgdaSymbol{)} \AgdaSymbol{→} \AgdaDatatype{Cx} \AgdaBound{T}\<%
\\
\\
\>\AgdaKeyword{data} \AgdaDatatype{\_∈\_} \AgdaSymbol{:} \AgdaSymbol{∀} \AgdaSymbol{\{}\AgdaBound{T} \AgdaSymbol{:} \AgdaDatatype{Types}\AgdaSymbol{\}} \AgdaSymbol{(}\AgdaBound{A} \AgdaSymbol{:} \AgdaDatatype{Ty} \AgdaBound{T}\AgdaSymbol{)} \AgdaSymbol{(}\AgdaBound{Γ} \AgdaSymbol{:} \AgdaDatatype{Cx} \AgdaBound{T}\AgdaSymbol{)} \AgdaSymbol{→} \AgdaPrimitiveType{Set} \AgdaKeyword{where}\<%
\\
\>[0]\AgdaIndent{2}{}\<[2]%
\>[2]\AgdaInductiveConstructor{top} \AgdaSymbol{:} \AgdaSymbol{∀} \AgdaSymbol{\{}\AgdaBound{T}\AgdaSymbol{\}} \AgdaSymbol{\{}\AgdaBound{Γ} \AgdaSymbol{:} \AgdaDatatype{Cx} \AgdaBound{T}\AgdaSymbol{\}} \AgdaSymbol{\{}\AgdaBound{A} \AgdaSymbol{:} \AgdaDatatype{Ty} \AgdaBound{T}\AgdaSymbol{\}} \AgdaSymbol{→} \AgdaBound{A} \AgdaDatatype{∈} \AgdaSymbol{(}\AgdaBound{Γ} \AgdaInductiveConstructor{,} \AgdaBound{A}\AgdaSymbol{)}\<%
\\
\>[0]\AgdaIndent{2}{}\<[2]%
\>[2]\AgdaInductiveConstructor{pop} \AgdaSymbol{:} \AgdaSymbol{∀} \AgdaSymbol{\{}\AgdaBound{T}\AgdaSymbol{\}} \AgdaSymbol{\{}\AgdaBound{Γ} \AgdaSymbol{:} \AgdaDatatype{Cx} \AgdaBound{T}\AgdaSymbol{\}} \AgdaSymbol{\{}\AgdaBound{A} \AgdaBound{B} \AgdaSymbol{:} \AgdaDatatype{Ty} \AgdaBound{T}\AgdaSymbol{\}} \AgdaSymbol{(}\AgdaBound{i} \AgdaSymbol{:} \AgdaBound{A} \AgdaDatatype{∈} \AgdaBound{Γ}\AgdaSymbol{)} \AgdaSymbol{→} \AgdaBound{A} \AgdaDatatype{∈} \AgdaSymbol{(}\AgdaBound{Γ} \AgdaInductiveConstructor{,} \AgdaBound{B}\AgdaSymbol{)}\<%
\\
\\
\>\AgdaFunction{\_⊆\_} \AgdaSymbol{:} \AgdaSymbol{∀} \AgdaSymbol{\{}\AgdaBound{T}\AgdaSymbol{\}} \AgdaSymbol{(}\AgdaBound{Γ} \AgdaBound{Δ} \AgdaSymbol{:} \AgdaDatatype{Cx} \AgdaBound{T}\AgdaSymbol{)} \AgdaSymbol{→} \AgdaPrimitiveType{Set}\<%
\\
\>\AgdaBound{Γ} \AgdaFunction{⊆} \AgdaBound{Δ} \AgdaSymbol{=} \AgdaSymbol{∀} \AgdaSymbol{\{}\AgdaBound{A}\AgdaSymbol{\}} \AgdaSymbol{→} \AgdaBound{A} \AgdaDatatype{∈} \AgdaBound{Γ} \AgdaSymbol{→} \AgdaBound{A} \AgdaDatatype{∈} \AgdaBound{Δ}\<%
\\
\>[0]\AgdaIndent{2}{}\<[2]%
\>[2]\<%
\\
\>\AgdaComment{-- Functions on contexts.}\<%
\\
\\
\>\AgdaFunction{boxcx} \AgdaSymbol{:} \AgdaDatatype{Cx} \AgdaInductiveConstructor{modal} \AgdaSymbol{→} \AgdaDatatype{Cx} \AgdaInductiveConstructor{modal}\<%
\\
\>\AgdaFunction{boxcx} \AgdaInductiveConstructor{·} \AgdaSymbol{=} \AgdaInductiveConstructor{·}\<%
\\
\>\AgdaFunction{boxcx} \AgdaSymbol{(}\AgdaBound{Γ} \AgdaInductiveConstructor{,} \AgdaBound{A}\AgdaSymbol{)} \AgdaSymbol{=} \AgdaFunction{boxcx} \AgdaBound{Γ} \AgdaInductiveConstructor{,} \AgdaInductiveConstructor{□} \AgdaBound{A}\<%
\\
\\
\>\AgdaFunction{\_++\_} \AgdaSymbol{:} \AgdaSymbol{∀} \AgdaSymbol{\{}\AgdaBound{T}\AgdaSymbol{\}} \AgdaSymbol{→} \AgdaDatatype{Cx} \AgdaBound{T} \AgdaSymbol{→} \AgdaDatatype{Cx} \AgdaBound{T} \AgdaSymbol{→} \AgdaDatatype{Cx} \AgdaBound{T}\<%
\\
\>\AgdaBound{Δ} \AgdaFunction{++} \AgdaInductiveConstructor{·} \AgdaSymbol{=} \AgdaBound{Δ}\<%
\\
\>\AgdaBound{Δ} \AgdaFunction{++} \AgdaSymbol{(}\AgdaBound{Γ} \AgdaInductiveConstructor{,} \AgdaBound{A}\AgdaSymbol{)} \AgdaSymbol{=} \AgdaSymbol{(}\AgdaBound{Δ} \AgdaFunction{++} \AgdaBound{Γ}\AgdaSymbol{)} \AgdaInductiveConstructor{,} \AgdaBound{A}\<%
\\
\\
\>\AgdaFunction{box∈cx} \AgdaSymbol{:} \AgdaSymbol{∀} \AgdaSymbol{\{}\AgdaBound{Γ} \AgdaSymbol{:} \AgdaDatatype{Cx} \AgdaInductiveConstructor{modal}\AgdaSymbol{\}} \AgdaSymbol{\{}\AgdaBound{A} \AgdaSymbol{:} \AgdaDatatype{Ty} \AgdaInductiveConstructor{modal}\AgdaSymbol{\}} \AgdaSymbol{→} \AgdaBound{A} \AgdaDatatype{∈} \AgdaBound{Γ} \AgdaSymbol{→} \AgdaInductiveConstructor{□} \AgdaBound{A} \AgdaDatatype{∈} \AgdaFunction{boxcx} \AgdaBound{Γ}\<%
\\
\>\AgdaFunction{box∈cx} \AgdaInductiveConstructor{top} \AgdaSymbol{=} \AgdaInductiveConstructor{top}\<%
\\
\>\AgdaFunction{box∈cx} \AgdaSymbol{(}\AgdaInductiveConstructor{pop} \AgdaBound{d}\AgdaSymbol{)} \AgdaSymbol{=} \AgdaInductiveConstructor{pop} \AgdaSymbol{(}\AgdaFunction{box∈cx} \AgdaBound{d}\AgdaSymbol{)}\<%
\\
\\
\>\AgdaFunction{subsetdef} \AgdaSymbol{:} \<[13]%
\>[13]\AgdaSymbol{∀} \AgdaSymbol{\{}\AgdaBound{T}\AgdaSymbol{\}} \AgdaSymbol{\{}\AgdaBound{Γ} \AgdaBound{Δ} \AgdaSymbol{:} \AgdaDatatype{Cx} \AgdaBound{T}\AgdaSymbol{\}} \AgdaSymbol{\{}\AgdaBound{A}\AgdaSymbol{\}} \AgdaSymbol{→} \AgdaBound{A} \AgdaDatatype{∈} \AgdaBound{Γ} \AgdaSymbol{→} \AgdaBound{Γ} \AgdaFunction{⊆} \AgdaBound{Δ} \AgdaSymbol{→} \AgdaBound{A} \AgdaDatatype{∈} \AgdaBound{Δ}\<%
\\
\>\AgdaFunction{subsetdef} \AgdaBound{d} \AgdaBound{f} \AgdaSymbol{=} \AgdaBound{f} \AgdaBound{d}\<%
\\
\\
\>\AgdaFunction{subsetempty} \AgdaSymbol{:} \AgdaSymbol{∀} \AgdaSymbol{\{}\AgdaBound{T}\AgdaSymbol{\}} \AgdaSymbol{\{}\AgdaBound{Γ} \AgdaSymbol{:} \AgdaDatatype{Cx} \AgdaBound{T}\AgdaSymbol{\}} \AgdaSymbol{→} \AgdaInductiveConstructor{·} \AgdaFunction{⊆} \AgdaBound{Γ}\<%
\\
\>\AgdaFunction{subsetempty} \AgdaSymbol{()}\<%
\\
\\
\>\AgdaFunction{subsetid} \AgdaSymbol{:} \AgdaSymbol{∀} \AgdaSymbol{\{}\AgdaBound{T}\AgdaSymbol{\}} \AgdaSymbol{\{}\AgdaBound{Γ} \AgdaSymbol{:} \AgdaDatatype{Cx} \AgdaBound{T}\AgdaSymbol{\}} \AgdaSymbol{→} \AgdaBound{Γ} \AgdaFunction{⊆} \AgdaBound{Γ}\<%
\\
\>\AgdaFunction{subsetid} \AgdaSymbol{=} \AgdaSymbol{λ} \AgdaSymbol{\{}\AgdaBound{Γ}\AgdaSymbol{\}} \AgdaSymbol{\{}\AgdaBound{A}\AgdaSymbol{\}} \AgdaBound{z} \AgdaSymbol{→} \AgdaBound{z}\<%
\\
\\
\>\AgdaFunction{weakone} \AgdaSymbol{:} \AgdaSymbol{∀} \AgdaSymbol{\{}\AgdaBound{T}\AgdaSymbol{\}} \AgdaSymbol{\{}\AgdaBound{Γ} \AgdaBound{Δ} \AgdaSymbol{:} \AgdaDatatype{Cx} \AgdaBound{T}\AgdaSymbol{\}} \AgdaSymbol{\{}\AgdaBound{A}\AgdaSymbol{\}} \AgdaSymbol{→} \AgdaBound{Γ} \AgdaFunction{⊆} \AgdaBound{Δ} \AgdaSymbol{→} \AgdaBound{Γ} \AgdaFunction{⊆} \AgdaSymbol{(}\AgdaBound{Δ} \AgdaInductiveConstructor{,} \AgdaBound{A}\AgdaSymbol{)}\<%
\\
\>\AgdaFunction{weakone} \AgdaBound{p} \AgdaSymbol{=} \AgdaSymbol{λ} \AgdaSymbol{\{}\AgdaBound{A}\AgdaSymbol{\}} \AgdaBound{z} \AgdaSymbol{→} \AgdaInductiveConstructor{pop} \AgdaSymbol{(}\AgdaBound{p} \AgdaBound{z}\AgdaSymbol{)}\<%
\\
\\
\>\AgdaFunction{weakboth} \AgdaSymbol{:} \AgdaSymbol{∀} \AgdaSymbol{\{}\AgdaBound{T}\AgdaSymbol{\}} \AgdaSymbol{\{}\AgdaBound{Γ} \AgdaBound{Δ} \AgdaSymbol{:} \AgdaDatatype{Cx} \AgdaBound{T}\AgdaSymbol{\}} \AgdaSymbol{\{}\AgdaBound{A}\AgdaSymbol{\}} \AgdaSymbol{→} \AgdaBound{Γ} \AgdaFunction{⊆} \AgdaBound{Δ} \AgdaSymbol{→} \AgdaBound{Γ} \AgdaInductiveConstructor{,} \AgdaBound{A} \AgdaFunction{⊆} \AgdaBound{Δ} \AgdaInductiveConstructor{,} \AgdaBound{A}\<%
\\
\>\AgdaFunction{weakboth} \AgdaBound{p} \AgdaInductiveConstructor{top} \AgdaSymbol{=} \AgdaInductiveConstructor{top}\<%
\\
\>\AgdaFunction{weakboth} \AgdaBound{p} \AgdaSymbol{(}\AgdaInductiveConstructor{pop} \AgdaBound{x}\AgdaSymbol{)} \AgdaSymbol{=} \AgdaFunction{subsetdef} \AgdaBound{x} \AgdaSymbol{(}\AgdaFunction{weakone} \AgdaBound{p}\AgdaSymbol{)}\<%
\\
\\
\>\AgdaFunction{weakmany} \AgdaSymbol{:} \AgdaSymbol{∀} \AgdaSymbol{\{}\AgdaBound{T}\AgdaSymbol{\}} \AgdaSymbol{(}\AgdaBound{Γ} \AgdaBound{Δ} \AgdaSymbol{:} \AgdaDatatype{Cx} \AgdaBound{T}\AgdaSymbol{)} \AgdaSymbol{→} \AgdaBound{Γ} \AgdaFunction{⊆} \AgdaBound{Γ} \AgdaFunction{++} \AgdaBound{Δ}\<%
\\
\>\AgdaFunction{weakmany} \AgdaBound{Γ} \AgdaInductiveConstructor{·} \AgdaBound{x} \AgdaSymbol{=} \AgdaBound{x}\<%
\\
\>\AgdaFunction{weakmany} \AgdaBound{Γ} \AgdaSymbol{(}\AgdaBound{Δ} \AgdaInductiveConstructor{,} \AgdaBound{A}\AgdaSymbol{)} \AgdaBound{x} \AgdaSymbol{=} \AgdaInductiveConstructor{pop} \AgdaSymbol{(}\AgdaFunction{weakmany} \AgdaBound{Γ} \AgdaBound{Δ} \AgdaBound{x}\AgdaSymbol{)}\<%
\\
\\
\>\AgdaFunction{concat-subset-1} \AgdaSymbol{:} \AgdaSymbol{∀} \AgdaSymbol{\{}\AgdaBound{T}\AgdaSymbol{\}} \AgdaSymbol{(}\AgdaBound{Γ} \AgdaBound{Δ} \AgdaSymbol{:} \AgdaDatatype{Cx} \AgdaBound{T}\AgdaSymbol{)} \AgdaSymbol{→} \AgdaBound{Γ} \AgdaFunction{⊆} \AgdaBound{Γ} \AgdaFunction{++} \AgdaBound{Δ}\<%
\\
\>\AgdaFunction{concat-subset-1} \AgdaBound{Γ} \AgdaInductiveConstructor{·} \AgdaBound{x} \AgdaSymbol{=} \AgdaBound{x}\<%
\\
\>\AgdaFunction{concat-subset-1} \AgdaBound{Γ} \AgdaSymbol{(}\AgdaBound{Δ} \AgdaInductiveConstructor{,} \AgdaBound{A}\AgdaSymbol{)} \AgdaBound{x} \AgdaSymbol{=} \AgdaFunction{subsetdef} \AgdaBound{x} \AgdaSymbol{(}\AgdaFunction{weakone} \AgdaSymbol{(}\AgdaFunction{concat-subset-1} \AgdaBound{Γ} \AgdaBound{Δ}\AgdaSymbol{))}\<%
\\
\\
\>\AgdaFunction{concat-subset-2} \AgdaSymbol{:} \AgdaSymbol{∀} \AgdaSymbol{\{}\AgdaBound{T}\AgdaSymbol{\}} \AgdaSymbol{(}\AgdaBound{Γ} \AgdaBound{Δ} \AgdaSymbol{:} \AgdaDatatype{Cx} \AgdaBound{T}\AgdaSymbol{)} \AgdaSymbol{→} \AgdaBound{Δ} \AgdaFunction{⊆} \AgdaBound{Γ} \AgdaFunction{++} \AgdaBound{Δ}\<%
\\
\>\AgdaFunction{concat-subset-2} \AgdaBound{Γ} \AgdaInductiveConstructor{·} \AgdaSymbol{()}\<%
\\
\>\AgdaFunction{concat-subset-2} \AgdaBound{Γ} \AgdaSymbol{(}\AgdaBound{Δ} \AgdaInductiveConstructor{,} \AgdaBound{A}\AgdaSymbol{)} \AgdaBound{x} \AgdaSymbol{=} \AgdaFunction{subsetdef} \AgdaBound{x} \AgdaSymbol{(}\AgdaFunction{weakboth} \AgdaSymbol{(}\AgdaFunction{concat-subset-2} \AgdaBound{Γ} \AgdaBound{Δ}\AgdaSymbol{))}\<%
\\
\\
\>\AgdaFunction{incl-trans} \AgdaSymbol{:} \AgdaSymbol{∀} \AgdaSymbol{\{}\AgdaBound{T}\AgdaSymbol{\}} \AgdaSymbol{\{}\AgdaBound{Γ} \AgdaBound{Γ'} \AgdaBound{Γ''} \AgdaSymbol{:} \AgdaDatatype{Cx} \AgdaBound{T}\AgdaSymbol{\}} \AgdaSymbol{→} \AgdaBound{Γ} \AgdaFunction{⊆} \AgdaBound{Γ'} \AgdaSymbol{→} \AgdaBound{Γ'} \AgdaFunction{⊆} \AgdaBound{Γ''} \AgdaSymbol{→} \AgdaBound{Γ} \AgdaFunction{⊆} \AgdaBound{Γ''}\<%
\\
\>\AgdaFunction{incl-trans} \AgdaBound{p} \AgdaBound{q} \AgdaBound{x} \AgdaSymbol{=} \AgdaBound{q} \AgdaSymbol{(}\AgdaBound{p} \AgdaBound{x}\AgdaSymbol{)}\<%
\\
\\
\>\AgdaFunction{swap-last} \AgdaSymbol{:} \AgdaSymbol{∀} \AgdaSymbol{\{}\AgdaBound{T}\AgdaSymbol{\}} \AgdaSymbol{\{}\AgdaBound{Γ} \AgdaSymbol{:} \AgdaDatatype{Cx} \AgdaBound{T}\AgdaSymbol{\}} \AgdaSymbol{\{}\AgdaBound{A} \AgdaBound{B}\AgdaSymbol{\}} \AgdaSymbol{→} \AgdaBound{Γ} \AgdaInductiveConstructor{,} \AgdaBound{A} \AgdaInductiveConstructor{,} \AgdaBound{B} \AgdaFunction{⊆} \AgdaBound{Γ} \AgdaInductiveConstructor{,} \AgdaBound{B} \AgdaInductiveConstructor{,} \AgdaBound{A}\<%
\\
\>\AgdaFunction{swap-last} \AgdaSymbol{\{\_\}} \AgdaSymbol{\{}\AgdaInductiveConstructor{·}\AgdaSymbol{\}} \AgdaInductiveConstructor{top} \AgdaSymbol{=} \AgdaInductiveConstructor{pop} \AgdaInductiveConstructor{top}\<%
\\
\>\AgdaFunction{swap-last} \AgdaSymbol{\{\_\}} \AgdaSymbol{\{}\AgdaInductiveConstructor{·}\AgdaSymbol{\}} \AgdaSymbol{(}\AgdaInductiveConstructor{pop} \AgdaInductiveConstructor{top}\AgdaSymbol{)} \AgdaSymbol{=} \AgdaInductiveConstructor{top}\<%
\\
\>\AgdaFunction{swap-last} \AgdaSymbol{\{\_\}} \AgdaSymbol{\{}\AgdaInductiveConstructor{·}\AgdaSymbol{\}} \AgdaSymbol{(}\AgdaInductiveConstructor{pop} \AgdaSymbol{(}\AgdaInductiveConstructor{pop} \AgdaBound{x}\AgdaSymbol{))} \AgdaSymbol{=} \AgdaInductiveConstructor{pop} \AgdaSymbol{(}\AgdaInductiveConstructor{pop} \AgdaBound{x}\AgdaSymbol{)}\<%
\\
\>\AgdaFunction{swap-last} \AgdaSymbol{\{\_\}} \AgdaSymbol{\{}\AgdaBound{Γ} \AgdaInductiveConstructor{,} \AgdaBound{A}\AgdaSymbol{\}} \AgdaInductiveConstructor{top} \AgdaSymbol{=} \AgdaInductiveConstructor{pop} \AgdaInductiveConstructor{top}\<%
\\
\>\AgdaFunction{swap-last} \AgdaSymbol{\{\_\}} \AgdaSymbol{\{}\AgdaBound{Γ} \AgdaInductiveConstructor{,} \AgdaBound{A}\AgdaSymbol{\}} \AgdaSymbol{(}\AgdaInductiveConstructor{pop} \AgdaInductiveConstructor{top}\AgdaSymbol{)} \AgdaSymbol{=} \AgdaInductiveConstructor{top}\<%
\\
\>\AgdaFunction{swap-last} \AgdaSymbol{\{\_\}} \AgdaSymbol{\{}\AgdaBound{Γ} \AgdaInductiveConstructor{,} \AgdaBound{A}\AgdaSymbol{\}} \AgdaSymbol{(}\AgdaInductiveConstructor{pop} \AgdaSymbol{(}\AgdaInductiveConstructor{pop} \AgdaBound{x}\AgdaSymbol{))} \AgdaSymbol{=} \AgdaInductiveConstructor{pop} \AgdaSymbol{(}\AgdaInductiveConstructor{pop} \AgdaBound{x}\AgdaSymbol{)}\<%
\\
\\
\>\AgdaFunction{cx-exch} \AgdaSymbol{:} \AgdaSymbol{∀} \AgdaSymbol{\{}\AgdaBound{T}\AgdaSymbol{\}} \AgdaSymbol{\{}\AgdaBound{Γ} \AgdaBound{Δ} \AgdaSymbol{:} \AgdaDatatype{Cx} \AgdaBound{T}\AgdaSymbol{\}} \AgdaSymbol{\{}\AgdaBound{A} \AgdaBound{B}\AgdaSymbol{\}} \AgdaSymbol{→} \AgdaSymbol{(}\AgdaBound{Γ} \AgdaInductiveConstructor{,} \AgdaBound{A} \AgdaInductiveConstructor{,} \AgdaBound{B}\AgdaSymbol{)} \AgdaFunction{++} \AgdaBound{Δ} \AgdaFunction{⊆} \AgdaSymbol{(}\AgdaBound{Γ} \AgdaInductiveConstructor{,} \AgdaBound{B} \AgdaInductiveConstructor{,} \AgdaBound{A}\AgdaSymbol{)} \AgdaFunction{++} \AgdaBound{Δ}\<%
\\
\>\AgdaFunction{cx-exch} \AgdaSymbol{\{}\AgdaArgument{Δ} \AgdaSymbol{=} \AgdaInductiveConstructor{·}\AgdaSymbol{\}} \AgdaBound{d} \AgdaSymbol{=} \AgdaFunction{swap-last} \AgdaBound{d}\<%
\\
\>\AgdaFunction{cx-exch} \AgdaSymbol{\{}\AgdaArgument{Δ} \AgdaSymbol{=} \AgdaBound{Δ} \AgdaInductiveConstructor{,} \AgdaBound{A₁}\AgdaSymbol{\}} \AgdaInductiveConstructor{top} \AgdaSymbol{=} \AgdaInductiveConstructor{top}\<%
\\
\>\AgdaFunction{cx-exch} \AgdaSymbol{\{}\AgdaArgument{Δ} \AgdaSymbol{=} \AgdaBound{Δ} \AgdaInductiveConstructor{,} \AgdaBound{A₁}\AgdaSymbol{\}} \AgdaSymbol{(}\AgdaInductiveConstructor{pop} \AgdaBound{d}\AgdaSymbol{)} \AgdaSymbol{=} \AgdaFunction{subsetdef} \AgdaBound{d} \AgdaSymbol{(}\AgdaFunction{weakone} \AgdaSymbol{(}\AgdaFunction{cx-exch} \AgdaSymbol{\{}\AgdaArgument{Δ} \AgdaSymbol{=} \AgdaBound{Δ}\AgdaSymbol{\}))}\<%
\\
\\
\>\AgdaFunction{cx-contr} \AgdaSymbol{:} \AgdaSymbol{∀} \AgdaSymbol{\{}\AgdaBound{T}\AgdaSymbol{\}} \AgdaSymbol{\{}\AgdaBound{Γ} \AgdaBound{Δ} \AgdaSymbol{:} \AgdaDatatype{Cx} \AgdaBound{T}\AgdaSymbol{\}} \AgdaSymbol{\{}\AgdaBound{A}\AgdaSymbol{\}} \AgdaSymbol{→} \AgdaSymbol{(}\AgdaBound{Γ} \AgdaInductiveConstructor{,} \AgdaBound{A} \AgdaInductiveConstructor{,} \AgdaBound{A}\AgdaSymbol{)} \AgdaFunction{++} \AgdaBound{Δ} \AgdaFunction{⊆} \AgdaSymbol{(}\AgdaBound{Γ} \AgdaInductiveConstructor{,} \AgdaBound{A}\AgdaSymbol{)} \AgdaFunction{++} \AgdaBound{Δ}\<%
\\
\>\AgdaFunction{cx-contr} \AgdaSymbol{\{}\AgdaArgument{Δ} \AgdaSymbol{=} \AgdaInductiveConstructor{·}\AgdaSymbol{\}} \AgdaInductiveConstructor{top} \AgdaSymbol{=} \AgdaInductiveConstructor{top}\<%
\\
\>\AgdaFunction{cx-contr} \AgdaSymbol{\{}\AgdaArgument{Δ} \AgdaSymbol{=} \AgdaInductiveConstructor{·}\AgdaSymbol{\}} \AgdaSymbol{(}\AgdaInductiveConstructor{pop} \AgdaBound{d}\AgdaSymbol{)} \AgdaSymbol{=} \AgdaBound{d}\<%
\\
\>\AgdaFunction{cx-contr} \AgdaSymbol{\{}\AgdaArgument{Δ} \AgdaSymbol{=} \AgdaBound{Δ} \AgdaInductiveConstructor{,} \AgdaBound{A₁}\AgdaSymbol{\}} \AgdaInductiveConstructor{top} \AgdaSymbol{=} \AgdaInductiveConstructor{top}\<%
\\
\>\AgdaFunction{cx-contr} \AgdaSymbol{\{}\AgdaArgument{Δ} \AgdaSymbol{=} \AgdaBound{Δ} \AgdaInductiveConstructor{,} \AgdaBound{A₁}\AgdaSymbol{\}} \AgdaSymbol{(}\AgdaInductiveConstructor{pop} \AgdaBound{d}\AgdaSymbol{)} \AgdaSymbol{=} \AgdaFunction{subsetdef} \AgdaBound{d} \AgdaSymbol{(}\AgdaFunction{weakone} \AgdaSymbol{(}\AgdaFunction{cx-contr} \AgdaSymbol{\{}\AgdaArgument{Δ} \AgdaSymbol{=} \AgdaBound{Δ}\AgdaSymbol{\}))}\<%
\\
\\
\>\AgdaFunction{is-in} \AgdaSymbol{:} \AgdaSymbol{∀} \AgdaSymbol{\{}\AgdaBound{T}\AgdaSymbol{\}} \AgdaSymbol{(}\AgdaBound{Γ} \AgdaBound{Γ'} \AgdaSymbol{:} \AgdaDatatype{Cx} \AgdaBound{T}\AgdaSymbol{)} \AgdaSymbol{(}\AgdaBound{A} \AgdaSymbol{:} \AgdaDatatype{Ty} \AgdaBound{T}\AgdaSymbol{)} \AgdaSymbol{→} \AgdaBound{A} \AgdaDatatype{∈} \AgdaSymbol{(}\AgdaBound{Γ} \AgdaInductiveConstructor{,} \AgdaBound{A} \AgdaFunction{++} \AgdaBound{Γ'}\AgdaSymbol{)}\<%
\\
\>\AgdaFunction{is-in} \AgdaBound{Γ} \AgdaInductiveConstructor{·} \AgdaBound{A} \AgdaSymbol{=} \AgdaInductiveConstructor{top}\<%
\\
\>\AgdaFunction{is-in} \AgdaBound{Γ} \AgdaSymbol{(}\AgdaBound{Γ'} \AgdaInductiveConstructor{,} \AgdaBound{A'}\AgdaSymbol{)} \AgdaBound{A} \AgdaSymbol{=} \AgdaInductiveConstructor{pop} \AgdaSymbol{(}\AgdaFunction{is-in} \AgdaBound{Γ} \AgdaBound{Γ'} \AgdaBound{A}\AgdaSymbol{)}\<%
\\
\\
\>\AgdaFunction{ctxt-disj} \AgdaSymbol{:} \AgdaSymbol{∀} \AgdaSymbol{\{}\AgdaBound{T}\AgdaSymbol{\}} \AgdaSymbol{(}\AgdaBound{Γ} \AgdaBound{Γ'} \AgdaSymbol{:} \AgdaDatatype{Cx} \AgdaBound{T}\AgdaSymbol{)} \AgdaSymbol{(}\AgdaBound{A} \AgdaSymbol{:} \AgdaDatatype{Ty} \AgdaBound{T}\AgdaSymbol{)} \AgdaSymbol{→} \AgdaBound{A} \AgdaDatatype{∈} \AgdaSymbol{(}\AgdaBound{Γ} \AgdaFunction{++} \AgdaBound{Γ'}\AgdaSymbol{)} \AgdaSymbol{→} \AgdaBound{A} \AgdaDatatype{∈} \AgdaBound{Γ} \AgdaDatatype{+} \AgdaBound{A} \AgdaDatatype{∈} \AgdaBound{Γ'}\<%
\\
\>\AgdaFunction{ctxt-disj} \AgdaBound{Γ} \AgdaInductiveConstructor{·} \AgdaBound{A} \AgdaBound{x} \AgdaSymbol{=} \AgdaInductiveConstructor{inj₁} \AgdaBound{x}\<%
\\
\>\AgdaFunction{ctxt-disj} \AgdaBound{Γ} \AgdaSymbol{(}\AgdaBound{Γ'} \AgdaInductiveConstructor{,} \AgdaBound{A'}\AgdaSymbol{)} \AgdaSymbol{.}\AgdaBound{A'} \AgdaInductiveConstructor{top} \AgdaSymbol{=} \AgdaInductiveConstructor{inj₂} \AgdaInductiveConstructor{top}\<%
\\
\>\AgdaFunction{ctxt-disj} \AgdaBound{Γ} \AgdaSymbol{(}\AgdaBound{Γ'} \AgdaInductiveConstructor{,} \AgdaBound{A'}\AgdaSymbol{)} \AgdaBound{A} \AgdaSymbol{(}\AgdaInductiveConstructor{pop} \AgdaBound{x}\AgdaSymbol{)}\<%
\\
\>[0]\AgdaIndent{2}{}\<[2]%
\>[2]\AgdaKeyword{with} \AgdaFunction{ctxt-disj} \AgdaBound{Γ} \AgdaBound{Γ'} \AgdaBound{A} \AgdaBound{x}\<%
\\
\>\AgdaFunction{ctxt-disj} \AgdaBound{Γ} \AgdaSymbol{(}\AgdaBound{Γ'} \AgdaInductiveConstructor{,} \AgdaBound{A'}\AgdaSymbol{)} \AgdaBound{A} \AgdaSymbol{(}\AgdaInductiveConstructor{pop} \AgdaBound{x}\AgdaSymbol{)} \AgdaSymbol{|} \AgdaInductiveConstructor{inj₁} \AgdaBound{z} \AgdaSymbol{=} \AgdaInductiveConstructor{inj₁} \AgdaBound{z}\<%
\\
\>\AgdaFunction{ctxt-disj} \AgdaBound{Γ} \AgdaSymbol{(}\AgdaBound{Γ'} \AgdaInductiveConstructor{,} \AgdaBound{A'}\AgdaSymbol{)} \AgdaBound{A} \AgdaSymbol{(}\AgdaInductiveConstructor{pop} \AgdaBound{x}\AgdaSymbol{)} \AgdaSymbol{|} \AgdaInductiveConstructor{inj₂} \AgdaBound{z} \AgdaSymbol{=} \AgdaInductiveConstructor{inj₂} \AgdaSymbol{(}\AgdaInductiveConstructor{pop} \AgdaBound{z}\AgdaSymbol{)}\<%
\\
\\
\>\AgdaFunction{swap-out} \AgdaSymbol{:} \AgdaSymbol{∀} \AgdaSymbol{\{}\AgdaBound{T}\AgdaSymbol{\}} \AgdaSymbol{(}\AgdaBound{Δ} \AgdaBound{Γ} \AgdaSymbol{:} \AgdaDatatype{Cx} \AgdaBound{T}\AgdaSymbol{)} \AgdaSymbol{(}\AgdaBound{A} \AgdaSymbol{:} \AgdaDatatype{Ty} \AgdaBound{T}\AgdaSymbol{)} \AgdaSymbol{→} \AgdaSymbol{(}\AgdaBound{Δ} \AgdaInductiveConstructor{,} \AgdaBound{A}\AgdaSymbol{)} \AgdaFunction{++} \AgdaBound{Γ} \AgdaFunction{⊆} \AgdaSymbol{(}\AgdaBound{Δ} \AgdaFunction{++} \AgdaBound{Γ}\AgdaSymbol{)} \AgdaInductiveConstructor{,} \AgdaBound{A}\<%
\\
\>\AgdaFunction{swap-out} \AgdaBound{Δ} \AgdaInductiveConstructor{·} \AgdaBound{A} \AgdaBound{x} \AgdaSymbol{=} \AgdaBound{x}\<%
\\
\>\AgdaFunction{swap-out} \AgdaBound{Δ} \AgdaSymbol{(}\AgdaBound{Γ} \AgdaInductiveConstructor{,} \AgdaBound{B}\AgdaSymbol{)} \AgdaBound{A} \AgdaBound{x} \AgdaSymbol{=} \AgdaFunction{swap-last} \AgdaSymbol{(}\AgdaFunction{subsetdef} \AgdaBound{x} \AgdaSymbol{(}\AgdaFunction{weakboth} \AgdaSymbol{(}\AgdaFunction{swap-out} \AgdaBound{Δ} \AgdaBound{Γ} \AgdaBound{A}\AgdaSymbol{)))}\<%
\\
\\
\>\AgdaFunction{swap-in} \AgdaSymbol{:} \AgdaSymbol{∀} \AgdaSymbol{\{}\AgdaBound{T}\AgdaSymbol{\}} \AgdaSymbol{(}\AgdaBound{Δ} \AgdaBound{Γ} \AgdaSymbol{:} \AgdaDatatype{Cx} \AgdaBound{T}\AgdaSymbol{)} \AgdaSymbol{(}\AgdaBound{A} \AgdaSymbol{:} \AgdaDatatype{Ty} \AgdaBound{T}\AgdaSymbol{)} \AgdaSymbol{→} \AgdaSymbol{(}\AgdaBound{Δ} \AgdaFunction{++} \AgdaBound{Γ}\AgdaSymbol{)} \AgdaInductiveConstructor{,} \AgdaBound{A} \AgdaFunction{⊆} \AgdaSymbol{(}\AgdaBound{Δ} \AgdaInductiveConstructor{,} \AgdaBound{A}\AgdaSymbol{)} \AgdaFunction{++} \AgdaBound{Γ}\<%
\\
\>\AgdaFunction{swap-in} \AgdaBound{Δ} \AgdaBound{Γ} \AgdaBound{A} \AgdaInductiveConstructor{top} \AgdaSymbol{=} \AgdaFunction{is-in} \AgdaBound{Δ} \AgdaBound{Γ} \AgdaBound{A}\<%
\\
\>\AgdaFunction{swap-in} \AgdaBound{Δ} \AgdaBound{Γ} \AgdaBound{A} \AgdaSymbol{(}\AgdaInductiveConstructor{pop} \AgdaBound{x}\AgdaSymbol{)}\<%
\\
\>[0]\AgdaIndent{2}{}\<[2]%
\>[2]\AgdaKeyword{with} \AgdaFunction{ctxt-disj} \AgdaBound{Δ} \AgdaBound{Γ} \AgdaSymbol{\_} \AgdaBound{x}\<%
\\
\>\AgdaFunction{swap-in} \AgdaBound{Δ} \AgdaBound{Γ} \AgdaBound{A} \AgdaSymbol{(}\AgdaInductiveConstructor{pop} \AgdaBound{x}\AgdaSymbol{)} \AgdaSymbol{|} \AgdaInductiveConstructor{inj₁} \AgdaBound{y} \AgdaSymbol{=} \AgdaFunction{concat-subset-1} \AgdaSymbol{(}\AgdaBound{Δ} \AgdaInductiveConstructor{,} \AgdaBound{A}\AgdaSymbol{)} \AgdaBound{Γ} \AgdaSymbol{(}\AgdaInductiveConstructor{pop} \AgdaBound{y}\AgdaSymbol{)}\<%
\\
\>\AgdaFunction{swap-in} \AgdaBound{Δ} \AgdaBound{Γ} \AgdaBound{A} \AgdaSymbol{(}\AgdaInductiveConstructor{pop} \AgdaBound{x}\AgdaSymbol{)} \AgdaSymbol{|} \AgdaInductiveConstructor{inj₂} \AgdaBound{y} \AgdaSymbol{=} \AgdaFunction{concat-subset-2} \AgdaSymbol{(}\AgdaBound{Δ} \AgdaInductiveConstructor{,} \AgdaBound{A}\AgdaSymbol{)} \AgdaBound{Γ} \AgdaBound{y}\<%
\end{code}

\pagebreak

\section{\texttt{iPCF.agda}}
  \label{sec:iPCF.agda}
  \begin{code}%
\>\AgdaKeyword{module} \AgdaModule{iPCF} \AgdaKeyword{where}\<%
\\
\\
\>\AgdaKeyword{infixl} \AgdaNumber{0} \_/\_⊢\_\<%
\\
\\
\>\AgdaKeyword{open} \AgdaKeyword{import} \AgdaModule{Basics}\<%
\\
\\
\>\AgdaComment{-- Definition}\<%
\\
\\
\>\AgdaKeyword{data} \AgdaDatatype{\_/\_⊢\_} \AgdaSymbol{(}\AgdaBound{Δ} \AgdaBound{Γ} \AgdaSymbol{:} \AgdaDatatype{Cx} \AgdaInductiveConstructor{modal}\AgdaSymbol{)} \AgdaSymbol{:} \<[31]%
\>[31]\AgdaDatatype{Ty} \AgdaInductiveConstructor{modal} \AgdaSymbol{→} \AgdaPrimitiveType{Set} \AgdaKeyword{where}\<%
\\
\\
\>[0]\AgdaIndent{2}{}\<[2]%
\>[2]\AgdaInductiveConstructor{iPCF-var} \AgdaSymbol{:} \AgdaSymbol{∀} \AgdaSymbol{\{}\AgdaBound{A}\AgdaSymbol{\}}\<%
\\
\>[0]\AgdaIndent{2}{}\<[2]%
\>[2]\<%
\\
\>[2]\AgdaIndent{5}{}\<[5]%
\>[5]\AgdaSymbol{→} \AgdaBound{A} \AgdaDatatype{∈} \AgdaBound{Γ}\<%
\\
\>[0]\AgdaIndent{4}{}\<[4]%
\>[4]\AgdaComment{-------------}\<%
\\
\>[0]\AgdaIndent{4}{}\<[4]%
\>[4]\AgdaSymbol{→} \AgdaBound{Δ} \AgdaDatatype{/} \AgdaBound{Γ} \AgdaDatatype{⊢} \AgdaBound{A}\<%
\\
\\
\>[0]\AgdaIndent{2}{}\<[2]%
\>[2]\AgdaInductiveConstructor{iPCF-modal-var} \AgdaSymbol{:} \AgdaSymbol{∀} \AgdaSymbol{\{}\AgdaBound{A}\AgdaSymbol{\}}\<%
\\
\>[0]\AgdaIndent{2}{}\<[2]%
\>[2]\<%
\\
\>[2]\AgdaIndent{5}{}\<[5]%
\>[5]\AgdaSymbol{→} \AgdaBound{A} \AgdaDatatype{∈} \AgdaBound{Δ}\<%
\\
\>[0]\AgdaIndent{4}{}\<[4]%
\>[4]\AgdaComment{------------}\<%
\\
\>[0]\AgdaIndent{4}{}\<[4]%
\>[4]\AgdaSymbol{→} \AgdaBound{Δ} \AgdaDatatype{/} \AgdaBound{Γ} \AgdaDatatype{⊢} \AgdaBound{A}\<%
\\
\>[4]\AgdaIndent{15}{}\<[15]%
\>[15]\<%
\\
\>[0]\AgdaIndent{2}{}\<[2]%
\>[2]\AgdaInductiveConstructor{iPCF-app} \AgdaSymbol{:} \AgdaSymbol{∀} \AgdaSymbol{\{}\AgdaBound{A} \AgdaBound{B}\AgdaSymbol{\}}\<%
\\
\>[0]\AgdaIndent{2}{}\<[2]%
\>[2]\<%
\\
\>[2]\AgdaIndent{4}{}\<[4]%
\>[4]\AgdaSymbol{→} \AgdaBound{Δ} \AgdaDatatype{/} \AgdaBound{Γ} \AgdaDatatype{⊢} \AgdaBound{A} \AgdaInductiveConstructor{=>} \AgdaBound{B} \<[24]%
\>[24]\AgdaSymbol{→} \AgdaBound{Δ} \AgdaDatatype{/} \AgdaBound{Γ} \AgdaDatatype{⊢} \AgdaBound{A}\<%
\\
\>[2]\AgdaIndent{4}{}\<[4]%
\>[4]\AgdaComment{---------------------------------}\<%
\\
\>[4]\AgdaIndent{10}{}\<[10]%
\>[10]\AgdaSymbol{→} \AgdaBound{Δ} \AgdaDatatype{/} \AgdaBound{Γ} \AgdaDatatype{⊢} \AgdaBound{B}\<%
\\
\>[10]\AgdaIndent{26}{}\<[26]%
\>[26]\<%
\\
\>[0]\AgdaIndent{2}{}\<[2]%
\>[2]\AgdaInductiveConstructor{iPCF-lam} \AgdaSymbol{:} \AgdaSymbol{∀} \AgdaSymbol{\{}\AgdaBound{A} \AgdaBound{B}\AgdaSymbol{\}}\<%
\\
\>[0]\AgdaIndent{2}{}\<[2]%
\>[2]\<%
\\
\>[2]\AgdaIndent{5}{}\<[5]%
\>[5]\AgdaSymbol{→} \AgdaBound{Δ} \AgdaDatatype{/} \AgdaSymbol{(}\AgdaBound{Γ} \AgdaInductiveConstructor{,} \AgdaBound{A}\AgdaSymbol{)} \AgdaDatatype{⊢} \AgdaBound{B}\<%
\\
\>[2]\AgdaIndent{5}{}\<[5]%
\>[5]\AgdaComment{------------------}\<%
\\
\>[2]\AgdaIndent{5}{}\<[5]%
\>[5]\AgdaSymbol{→} \AgdaBound{Δ} \AgdaDatatype{/} \AgdaBound{Γ} \AgdaDatatype{⊢} \AgdaBound{A} \AgdaInductiveConstructor{=>} \AgdaBound{B}\<%
\\
\>[5]\AgdaIndent{15}{}\<[15]%
\>[15]\<%
\\
\>[0]\AgdaIndent{2}{}\<[2]%
\>[2]\AgdaInductiveConstructor{iPCF-prod} \AgdaSymbol{:} \AgdaSymbol{∀} \AgdaSymbol{\{}\AgdaBound{A} \AgdaBound{B}\AgdaSymbol{\}}\<%
\\
\>[0]\AgdaIndent{2}{}\<[2]%
\>[2]\<%
\\
\>[2]\AgdaIndent{4}{}\<[4]%
\>[4]\AgdaSymbol{→} \AgdaBound{Δ} \AgdaDatatype{/} \AgdaBound{Γ} \AgdaDatatype{⊢} \AgdaBound{A} \<[19]%
\>[19]\AgdaSymbol{→} \AgdaBound{Δ} \AgdaDatatype{/} \AgdaBound{Γ} \AgdaDatatype{⊢} \AgdaBound{B}\<%
\\
\>[2]\AgdaIndent{4}{}\<[4]%
\>[4]\AgdaComment{----------------------------}\<%
\\
\>[4]\AgdaIndent{7}{}\<[7]%
\>[7]\AgdaSymbol{→} \AgdaBound{Δ} \AgdaDatatype{/} \AgdaBound{Γ} \AgdaDatatype{⊢} \AgdaBound{A} \AgdaInductiveConstructor{∧} \AgdaBound{B}\<%
\\
\>[7]\AgdaIndent{21}{}\<[21]%
\>[21]\<%
\\
\>[0]\AgdaIndent{2}{}\<[2]%
\>[2]\AgdaInductiveConstructor{iPCF-fst} \AgdaSymbol{:} \AgdaSymbol{∀} \AgdaSymbol{\{}\AgdaBound{A} \AgdaBound{B}\AgdaSymbol{\}}\<%
\\
\>[0]\AgdaIndent{2}{}\<[2]%
\>[2]\<%
\\
\>[2]\AgdaIndent{4}{}\<[4]%
\>[4]\AgdaSymbol{→} \AgdaBound{Δ} \AgdaDatatype{/} \AgdaBound{Γ} \AgdaDatatype{⊢} \AgdaBound{A} \AgdaInductiveConstructor{∧} \AgdaBound{B}\<%
\\
\>[2]\AgdaIndent{4}{}\<[4]%
\>[4]\AgdaComment{-----------------}\<%
\\
\>[4]\AgdaIndent{6}{}\<[6]%
\>[6]\AgdaSymbol{→} \AgdaBound{Δ} \AgdaDatatype{/} \AgdaBound{Γ} \AgdaDatatype{⊢} \AgdaBound{A}\<%
\\
\>[6]\AgdaIndent{17}{}\<[17]%
\>[17]\<%
\\
\>[0]\AgdaIndent{2}{}\<[2]%
\>[2]\AgdaInductiveConstructor{iPCF-snd} \AgdaSymbol{:} \AgdaSymbol{∀} \AgdaSymbol{\{}\AgdaBound{A} \AgdaBound{B}\AgdaSymbol{\}}\<%
\\
\>[0]\AgdaIndent{2}{}\<[2]%
\>[2]\<%
\\
\>[2]\AgdaIndent{4}{}\<[4]%
\>[4]\AgdaSymbol{→} \AgdaBound{Δ} \AgdaDatatype{/} \AgdaBound{Γ} \AgdaDatatype{⊢} \AgdaBound{A} \AgdaInductiveConstructor{∧} \AgdaBound{B}\<%
\\
\>[2]\AgdaIndent{4}{}\<[4]%
\>[4]\AgdaComment{----------------}\<%
\\
\>[4]\AgdaIndent{6}{}\<[6]%
\>[6]\AgdaSymbol{→} \AgdaBound{Δ} \AgdaDatatype{/} \AgdaBound{Γ} \AgdaDatatype{⊢} \AgdaBound{B}\<%
\\
\>[6]\AgdaIndent{17}{}\<[17]%
\>[17]\<%
\\
\>[0]\AgdaIndent{2}{}\<[2]%
\>[2]\AgdaInductiveConstructor{iPCF-boxI} \AgdaSymbol{:} \AgdaSymbol{∀} \AgdaSymbol{\{}\AgdaBound{A}\AgdaSymbol{\}}\<%
\\
\>[0]\AgdaIndent{2}{}\<[2]%
\>[2]\<%
\\
\>[2]\AgdaIndent{5}{}\<[5]%
\>[5]\AgdaSymbol{→} \AgdaBound{Δ} \AgdaDatatype{/} \AgdaInductiveConstructor{·} \AgdaDatatype{⊢} \AgdaBound{A}\<%
\\
\>[0]\AgdaIndent{4}{}\<[4]%
\>[4]\AgdaComment{---------------}\<%
\\
\>[0]\AgdaIndent{4}{}\<[4]%
\>[4]\AgdaSymbol{→} \AgdaBound{Δ} \AgdaDatatype{/} \AgdaBound{Γ} \AgdaDatatype{⊢} \AgdaInductiveConstructor{□} \AgdaBound{A}\<%
\\
\>[4]\AgdaIndent{15}{}\<[15]%
\>[15]\<%
\\
\>[0]\AgdaIndent{2}{}\<[2]%
\>[2]\AgdaInductiveConstructor{iPCF-boxE} \AgdaSymbol{:} \AgdaSymbol{∀} \AgdaSymbol{\{}\AgdaBound{A} \AgdaBound{C}\AgdaSymbol{\}}\<%
\\
\>[0]\AgdaIndent{2}{}\<[2]%
\>[2]\<%
\\
\>[2]\AgdaIndent{4}{}\<[4]%
\>[4]\AgdaSymbol{→} \AgdaBound{Δ} \AgdaDatatype{/} \AgdaBound{Γ} \AgdaDatatype{⊢} \AgdaInductiveConstructor{□} \AgdaBound{A} \<[21]%
\>[21]\AgdaSymbol{→} \AgdaSymbol{(}\AgdaBound{Δ} \AgdaInductiveConstructor{,} \AgdaBound{A}\AgdaSymbol{)} \AgdaDatatype{/} \AgdaBound{Γ} \AgdaDatatype{⊢} \AgdaBound{C}\<%
\\
\>[2]\AgdaIndent{4}{}\<[4]%
\>[4]\AgdaComment{------------------------------------}\<%
\\
\>[4]\AgdaIndent{14}{}\<[14]%
\>[14]\AgdaSymbol{→} \AgdaBound{Δ} \AgdaDatatype{/} \AgdaBound{Γ} \AgdaDatatype{⊢} \AgdaBound{C}\<%
\\
\\
\>[0]\AgdaIndent{2}{}\<[2]%
\>[2]\AgdaInductiveConstructor{iPCF-fix} \AgdaSymbol{:} \AgdaSymbol{∀} \AgdaSymbol{\{}\AgdaBound{A}\AgdaSymbol{\}}\<%
\\
\>[0]\AgdaIndent{2}{}\<[2]%
\>[2]\<%
\\
\>[2]\AgdaIndent{5}{}\<[5]%
\>[5]\AgdaSymbol{→} \AgdaBound{Δ} \AgdaDatatype{/} \AgdaSymbol{(}\AgdaInductiveConstructor{·} \AgdaInductiveConstructor{,} \AgdaInductiveConstructor{□} \AgdaBound{A}\AgdaSymbol{)} \AgdaDatatype{⊢} \AgdaBound{A}\<%
\\
\>[0]\AgdaIndent{4}{}\<[4]%
\>[4]\AgdaComment{---------------}\<%
\\
\>[0]\AgdaIndent{4}{}\<[4]%
\>[4]\AgdaSymbol{→} \AgdaBound{Δ} \AgdaDatatype{/} \AgdaBound{Γ} \AgdaDatatype{⊢} \AgdaBound{A}\<%
\\
\\
\\
\>\AgdaComment{-- Weakening and exchange.}\<%
\\
\\
\\
\>\AgdaFunction{exch} \AgdaSymbol{:} \AgdaSymbol{∀} \AgdaSymbol{\{}\AgdaBound{Δ} \AgdaBound{Γ} \AgdaBound{A} \AgdaBound{B} \AgdaBound{C}\AgdaSymbol{\}} \AgdaSymbol{(}\AgdaBound{Γ'} \AgdaSymbol{:} \AgdaDatatype{Cx} \AgdaInductiveConstructor{modal}\AgdaSymbol{)}\<%
\\
\\
\>[0]\AgdaIndent{4}{}\<[4]%
\>[4]\AgdaSymbol{→} \AgdaBound{Δ} \AgdaDatatype{/} \AgdaSymbol{(}\AgdaBound{Γ} \AgdaInductiveConstructor{,} \AgdaBound{A} \AgdaInductiveConstructor{,} \AgdaBound{B}\AgdaSymbol{)} \AgdaFunction{++} \AgdaBound{Γ'} \AgdaDatatype{⊢} \AgdaBound{C}\<%
\\
\>[0]\AgdaIndent{4}{}\<[4]%
\>[4]\AgdaComment{-----------------------------}\<%
\\
\>[0]\AgdaIndent{4}{}\<[4]%
\>[4]\AgdaSymbol{→} \AgdaBound{Δ} \AgdaDatatype{/} \AgdaSymbol{(}\AgdaBound{Γ} \AgdaInductiveConstructor{,} \AgdaBound{B} \AgdaInductiveConstructor{,} \AgdaBound{A}\AgdaSymbol{)} \AgdaFunction{++} \AgdaBound{Γ'} \AgdaDatatype{⊢} \AgdaBound{C}\<%
\\
\\
\>\AgdaFunction{exch} \AgdaBound{Γ'} \AgdaSymbol{(}\AgdaInductiveConstructor{iPCF-var} \AgdaBound{x}\AgdaSymbol{)} \AgdaSymbol{=} \AgdaInductiveConstructor{iPCF-var} \AgdaSymbol{(}\AgdaFunction{cx-exch} \AgdaSymbol{\{}\AgdaArgument{Δ} \AgdaSymbol{=} \AgdaBound{Γ'}\AgdaSymbol{\}} \AgdaBound{x}\AgdaSymbol{)}\<%
\\
\>\AgdaFunction{exch} \AgdaBound{Γ'} \AgdaSymbol{(}\AgdaInductiveConstructor{iPCF-modal-var} \AgdaBound{x}\AgdaSymbol{)} \AgdaSymbol{=} \AgdaInductiveConstructor{iPCF-modal-var} \AgdaBound{x}\<%
\\
\>\AgdaFunction{exch} \AgdaBound{Γ'} \AgdaSymbol{(}\AgdaInductiveConstructor{iPCF-app} \AgdaBound{d} \AgdaBound{d₁}\AgdaSymbol{)} \AgdaSymbol{=} \AgdaInductiveConstructor{iPCF-app} \AgdaSymbol{(}\AgdaFunction{exch} \AgdaBound{Γ'} \AgdaBound{d}\AgdaSymbol{)} \AgdaSymbol{(}\AgdaFunction{exch} \AgdaBound{Γ'} \AgdaBound{d₁}\AgdaSymbol{)}\<%
\\
\>\AgdaFunction{exch} \AgdaSymbol{\{}\AgdaArgument{C} \AgdaSymbol{=} \AgdaBound{A} \AgdaInductiveConstructor{=>} \AgdaBound{B}\AgdaSymbol{\}} \AgdaBound{Γ'} \AgdaSymbol{(}\AgdaInductiveConstructor{iPCF-lam} \AgdaBound{d}\AgdaSymbol{)} \AgdaSymbol{=} \AgdaInductiveConstructor{iPCF-lam} \AgdaSymbol{(}\AgdaFunction{exch} \AgdaSymbol{(}\AgdaBound{Γ'} \AgdaInductiveConstructor{,} \AgdaBound{A}\AgdaSymbol{)} \AgdaBound{d}\AgdaSymbol{)}\<%
\\
\>\AgdaFunction{exch} \AgdaBound{Γ'} \AgdaSymbol{(}\AgdaInductiveConstructor{iPCF-prod} \AgdaBound{d} \AgdaBound{e}\AgdaSymbol{)} \AgdaSymbol{=} \AgdaInductiveConstructor{iPCF-prod} \AgdaSymbol{(}\AgdaFunction{exch} \AgdaBound{Γ'} \AgdaBound{d}\AgdaSymbol{)} \AgdaSymbol{(}\AgdaFunction{exch} \AgdaBound{Γ'} \AgdaBound{e}\AgdaSymbol{)}\<%
\\
\>\AgdaFunction{exch} \AgdaBound{Γ'} \AgdaSymbol{(}\AgdaInductiveConstructor{iPCF-fst} \AgdaBound{d}\AgdaSymbol{)} \AgdaSymbol{=} \AgdaInductiveConstructor{iPCF-fst} \AgdaSymbol{(}\AgdaFunction{exch} \AgdaBound{Γ'} \AgdaBound{d}\AgdaSymbol{)}\<%
\\
\>\AgdaFunction{exch} \AgdaBound{Γ'} \AgdaSymbol{(}\AgdaInductiveConstructor{iPCF-snd} \AgdaBound{d}\AgdaSymbol{)} \AgdaSymbol{=} \AgdaInductiveConstructor{iPCF-snd} \AgdaSymbol{(}\AgdaFunction{exch} \AgdaBound{Γ'} \AgdaBound{d}\AgdaSymbol{)}\<%
\\
\>\AgdaFunction{exch} \AgdaBound{Γ'} \AgdaSymbol{(}\AgdaInductiveConstructor{iPCF-boxI} \AgdaBound{d}\AgdaSymbol{)} \AgdaSymbol{=} \AgdaInductiveConstructor{iPCF-boxI} \AgdaBound{d}\<%
\\
\>\AgdaFunction{exch} \AgdaBound{Γ'} \AgdaSymbol{(}\AgdaInductiveConstructor{iPCF-boxE} \AgdaBound{d} \AgdaBound{e}\AgdaSymbol{)} \AgdaSymbol{=} \AgdaInductiveConstructor{iPCF-boxE} \AgdaSymbol{(}\AgdaFunction{exch} \AgdaBound{Γ'} \AgdaBound{d}\AgdaSymbol{)} \AgdaSymbol{(}\AgdaFunction{exch} \AgdaBound{Γ'} \AgdaBound{e}\AgdaSymbol{)}\<%
\\
\>\AgdaFunction{exch} \AgdaBound{Γ'} \AgdaSymbol{(}\AgdaInductiveConstructor{iPCF-fix} \AgdaBound{d}\AgdaSymbol{)} \AgdaSymbol{=} \AgdaInductiveConstructor{iPCF-fix} \AgdaBound{d}\<%
\\
\\
\\
\>\AgdaFunction{exch-modal} \AgdaSymbol{:} \AgdaSymbol{∀} \AgdaSymbol{\{}\AgdaBound{Δ} \AgdaBound{Γ} \AgdaBound{A} \AgdaBound{B} \AgdaBound{C}\AgdaSymbol{\}} \AgdaSymbol{(}\AgdaBound{Δ'} \AgdaSymbol{:} \AgdaDatatype{Cx} \AgdaInductiveConstructor{modal}\AgdaSymbol{)}\<%
\\
\\
\>[0]\AgdaIndent{4}{}\<[4]%
\>[4]\AgdaSymbol{→} \AgdaSymbol{(}\AgdaBound{Δ} \AgdaInductiveConstructor{,} \AgdaBound{A} \AgdaInductiveConstructor{,} \AgdaBound{B}\AgdaSymbol{)} \AgdaFunction{++} \AgdaBound{Δ'} \AgdaDatatype{/} \AgdaBound{Γ} \<[29]%
\>[29]\AgdaDatatype{⊢} \AgdaBound{C}\<%
\\
\>[0]\AgdaIndent{4}{}\<[4]%
\>[4]\AgdaComment{------------------------------}\<%
\\
\>[0]\AgdaIndent{4}{}\<[4]%
\>[4]\AgdaSymbol{→} \AgdaSymbol{(}\AgdaBound{Δ} \AgdaInductiveConstructor{,} \AgdaBound{B} \AgdaInductiveConstructor{,} \AgdaBound{A}\AgdaSymbol{)} \AgdaFunction{++} \AgdaBound{Δ'} \AgdaDatatype{/} \AgdaBound{Γ} \AgdaDatatype{⊢} \AgdaBound{C}\<%
\\
\>[4]\AgdaIndent{20}{}\<[20]%
\>[20]\<%
\\
\>\AgdaFunction{exch-modal} \AgdaBound{Δ'} \AgdaSymbol{(}\AgdaInductiveConstructor{iPCF-var} \AgdaBound{x}\AgdaSymbol{)} \AgdaSymbol{=} \AgdaInductiveConstructor{iPCF-var} \AgdaBound{x}\<%
\\
\>\AgdaFunction{exch-modal} \AgdaBound{Δ'} \AgdaSymbol{(}\AgdaInductiveConstructor{iPCF-modal-var} \AgdaBound{x}\AgdaSymbol{)} \AgdaSymbol{=}\<%
\\
\>[0]\AgdaIndent{2}{}\<[2]%
\>[2]\AgdaInductiveConstructor{iPCF-modal-var} \AgdaSymbol{(}\AgdaFunction{subsetdef} \AgdaBound{x} \AgdaSymbol{(}\AgdaFunction{cx-exch} \AgdaSymbol{\{}\AgdaArgument{Δ} \AgdaSymbol{=} \AgdaBound{Δ'}\AgdaSymbol{\}))}\<%
\\
\>\AgdaFunction{exch-modal} \AgdaBound{Δ'} \AgdaSymbol{(}\AgdaInductiveConstructor{iPCF-app} \AgdaBound{d} \AgdaBound{e}\AgdaSymbol{)} \AgdaSymbol{=}\<%
\\
\>[0]\AgdaIndent{2}{}\<[2]%
\>[2]\AgdaInductiveConstructor{iPCF-app} \AgdaSymbol{(}\AgdaFunction{exch-modal} \AgdaBound{Δ'} \AgdaBound{d}\AgdaSymbol{)} \AgdaSymbol{(}\AgdaFunction{exch-modal} \AgdaBound{Δ'} \AgdaBound{e}\AgdaSymbol{)}\<%
\\
\>\AgdaFunction{exch-modal} \AgdaBound{Δ'} \AgdaSymbol{(}\AgdaInductiveConstructor{iPCF-lam} \AgdaBound{d}\AgdaSymbol{)} \AgdaSymbol{=} \AgdaInductiveConstructor{iPCF-lam} \AgdaSymbol{(}\AgdaFunction{exch-modal} \AgdaBound{Δ'} \AgdaBound{d}\AgdaSymbol{)}\<%
\\
\>\AgdaFunction{exch-modal} \AgdaBound{Δ'} \AgdaSymbol{(}\AgdaInductiveConstructor{iPCF-prod} \AgdaBound{d} \AgdaBound{e}\AgdaSymbol{)} \AgdaSymbol{=}\<%
\\
\>[0]\AgdaIndent{2}{}\<[2]%
\>[2]\AgdaInductiveConstructor{iPCF-prod} \AgdaSymbol{(}\AgdaFunction{exch-modal} \AgdaBound{Δ'} \AgdaBound{d}\AgdaSymbol{)} \AgdaSymbol{(}\AgdaFunction{exch-modal} \AgdaBound{Δ'} \AgdaBound{e}\AgdaSymbol{)}\<%
\\
\>\AgdaFunction{exch-modal} \AgdaBound{Δ'} \AgdaSymbol{(}\AgdaInductiveConstructor{iPCF-fst} \AgdaBound{d}\AgdaSymbol{)} \AgdaSymbol{=} \AgdaInductiveConstructor{iPCF-fst} \AgdaSymbol{(}\AgdaFunction{exch-modal} \AgdaBound{Δ'} \AgdaBound{d}\AgdaSymbol{)}\<%
\\
\>\AgdaFunction{exch-modal} \AgdaBound{Δ'} \AgdaSymbol{(}\AgdaInductiveConstructor{iPCF-snd} \AgdaBound{d}\AgdaSymbol{)} \AgdaSymbol{=} \AgdaInductiveConstructor{iPCF-snd} \AgdaSymbol{(}\AgdaFunction{exch-modal} \AgdaBound{Δ'} \AgdaBound{d}\AgdaSymbol{)}\<%
\\
\>\AgdaFunction{exch-modal} \AgdaBound{Δ'} \AgdaSymbol{(}\AgdaInductiveConstructor{iPCF-boxI} \AgdaBound{d}\AgdaSymbol{)} \AgdaSymbol{=} \AgdaInductiveConstructor{iPCF-boxI} \AgdaSymbol{(}\AgdaFunction{exch-modal} \AgdaBound{Δ'} \AgdaBound{d}\AgdaSymbol{)}\<%
\\
\>\AgdaFunction{exch-modal} \AgdaBound{Δ'} \AgdaSymbol{(}\AgdaInductiveConstructor{iPCF-boxE} \AgdaBound{d} \AgdaBound{e}\AgdaSymbol{)} \AgdaSymbol{=}\<%
\\
\>[0]\AgdaIndent{2}{}\<[2]%
\>[2]\AgdaInductiveConstructor{iPCF-boxE} \AgdaSymbol{(}\AgdaFunction{exch-modal} \AgdaBound{Δ'} \AgdaBound{d}\AgdaSymbol{)} \AgdaSymbol{(}\AgdaFunction{exch-modal} \AgdaSymbol{(}\AgdaBound{Δ'} \AgdaInductiveConstructor{,} \AgdaSymbol{\_)} \AgdaBound{e}\AgdaSymbol{)}\<%
\\
\>\AgdaFunction{exch-modal} \AgdaBound{Δ'} \AgdaSymbol{(}\AgdaInductiveConstructor{iPCF-fix} \AgdaBound{d}\AgdaSymbol{)} \AgdaSymbol{=} \AgdaInductiveConstructor{iPCF-fix} \AgdaSymbol{(}\AgdaFunction{exch-modal} \AgdaBound{Δ'} \AgdaBound{d}\AgdaSymbol{)}\<%
\\
\\
\\
\>\AgdaFunction{weak} \AgdaSymbol{:} \AgdaSymbol{∀} \AgdaSymbol{\{}\AgdaBound{Δ} \AgdaBound{Γ} \AgdaBound{Γ'} \AgdaBound{A}\AgdaSymbol{\}}\<%
\\
\\
\>[2]\AgdaIndent{4}{}\<[4]%
\>[4]\AgdaSymbol{→} \AgdaBound{Δ} \AgdaDatatype{/} \AgdaBound{Γ} \AgdaDatatype{⊢} \AgdaBound{A} \<[19]%
\>[19]\AgdaSymbol{→} \AgdaBound{Γ} \AgdaFunction{⊆} \AgdaBound{Γ'}\<%
\\
\>[2]\AgdaIndent{4}{}\<[4]%
\>[4]\AgdaComment{-------------------------}\<%
\\
\>[4]\AgdaIndent{8}{}\<[8]%
\>[8]\AgdaSymbol{→} \AgdaSymbol{(}\AgdaBound{Δ} \AgdaDatatype{/} \AgdaBound{Γ'} \AgdaDatatype{⊢} \AgdaBound{A}\AgdaSymbol{)}\<%
\\
\\
\>\AgdaFunction{weak} \AgdaSymbol{(}\AgdaInductiveConstructor{iPCF-var} \AgdaBound{x}\AgdaSymbol{)} \AgdaBound{f} \AgdaSymbol{=} \AgdaInductiveConstructor{iPCF-var} \AgdaSymbol{(}\AgdaBound{f} \AgdaBound{x}\AgdaSymbol{)}\<%
\\
\>\AgdaFunction{weak} \AgdaSymbol{(}\AgdaInductiveConstructor{iPCF-modal-var} \AgdaBound{x}\AgdaSymbol{)} \AgdaBound{f} \AgdaSymbol{=} \AgdaInductiveConstructor{iPCF-modal-var} \AgdaBound{x}\<%
\\
\>\AgdaFunction{weak} \AgdaSymbol{(}\AgdaInductiveConstructor{iPCF-app} \AgdaBound{d} \AgdaBound{e}\AgdaSymbol{)} \AgdaBound{f} \AgdaSymbol{=} \AgdaInductiveConstructor{iPCF-app} \AgdaSymbol{(}\AgdaFunction{weak} \AgdaBound{d} \AgdaBound{f}\AgdaSymbol{)} \AgdaSymbol{(}\AgdaFunction{weak} \AgdaBound{e} \AgdaBound{f}\AgdaSymbol{)}\<%
\\
\>\AgdaFunction{weak} \AgdaSymbol{(}\AgdaInductiveConstructor{iPCF-lam} \AgdaBound{d}\AgdaSymbol{)} \AgdaBound{f} \AgdaSymbol{=} \AgdaInductiveConstructor{iPCF-lam} \AgdaSymbol{(}\AgdaFunction{weak} \AgdaBound{d} \AgdaSymbol{(}\AgdaFunction{weakboth} \AgdaBound{f}\AgdaSymbol{))}\<%
\\
\>\AgdaFunction{weak} \AgdaSymbol{(}\AgdaInductiveConstructor{iPCF-prod} \AgdaBound{d} \AgdaBound{e}\AgdaSymbol{)} \AgdaBound{f} \AgdaSymbol{=} \AgdaInductiveConstructor{iPCF-prod} \AgdaSymbol{(}\AgdaFunction{weak} \AgdaBound{d} \AgdaBound{f}\AgdaSymbol{)} \AgdaSymbol{(}\AgdaFunction{weak} \AgdaBound{e} \AgdaBound{f}\AgdaSymbol{)}\<%
\\
\>\AgdaFunction{weak} \AgdaSymbol{(}\AgdaInductiveConstructor{iPCF-fst} \AgdaBound{d}\AgdaSymbol{)} \AgdaBound{f} \AgdaSymbol{=} \AgdaInductiveConstructor{iPCF-fst} \AgdaSymbol{(}\AgdaFunction{weak} \AgdaBound{d} \AgdaBound{f}\AgdaSymbol{)}\<%
\\
\>\AgdaFunction{weak} \AgdaSymbol{(}\AgdaInductiveConstructor{iPCF-snd} \AgdaBound{d}\AgdaSymbol{)} \AgdaBound{f} \AgdaSymbol{=} \AgdaInductiveConstructor{iPCF-snd} \AgdaSymbol{(}\AgdaFunction{weak} \AgdaBound{d} \AgdaBound{f}\AgdaSymbol{)}\<%
\\
\>\AgdaFunction{weak} \AgdaSymbol{(}\AgdaInductiveConstructor{iPCF-boxI} \AgdaBound{d}\AgdaSymbol{)} \AgdaBound{f} \AgdaSymbol{=} \AgdaInductiveConstructor{iPCF-boxI} \AgdaBound{d}\<%
\\
\>\AgdaFunction{weak} \AgdaSymbol{(}\AgdaInductiveConstructor{iPCF-boxE} \AgdaBound{d} \AgdaBound{e}\AgdaSymbol{)} \AgdaBound{f} \AgdaSymbol{=}\<%
\\
\>[0]\AgdaIndent{2}{}\<[2]%
\>[2]\AgdaInductiveConstructor{iPCF-boxE} \AgdaSymbol{(}\AgdaFunction{weak} \AgdaBound{d} \AgdaBound{f}\AgdaSymbol{)} \AgdaSymbol{(}\AgdaFunction{weak} \AgdaBound{e} \AgdaBound{f}\AgdaSymbol{)}\<%
\\
\>\AgdaFunction{weak} \AgdaSymbol{(}\AgdaInductiveConstructor{iPCF-fix} \AgdaBound{d}\AgdaSymbol{)} \AgdaBound{f} \AgdaSymbol{=} \AgdaInductiveConstructor{iPCF-fix} \AgdaBound{d}\<%
\\
\\
\\
\>\AgdaFunction{weak-modal} \AgdaSymbol{:} \AgdaSymbol{∀} \AgdaSymbol{\{}\AgdaBound{Δ} \AgdaBound{Δ'} \AgdaBound{Γ} \AgdaBound{A}\AgdaSymbol{\}}\<%
\\
\\
\>[2]\AgdaIndent{3}{}\<[3]%
\>[3]\AgdaSymbol{→} \AgdaBound{Δ} \AgdaDatatype{/} \AgdaBound{Γ} \AgdaDatatype{⊢} \AgdaBound{A} \<[18]%
\>[18]\AgdaSymbol{→} \AgdaBound{Δ} \AgdaFunction{⊆} \AgdaBound{Δ'}\<%
\\
\>[2]\AgdaIndent{3}{}\<[3]%
\>[3]\AgdaComment{-------------------------}\<%
\\
\>[3]\AgdaIndent{9}{}\<[9]%
\>[9]\AgdaSymbol{→} \AgdaBound{Δ'} \AgdaDatatype{/} \AgdaBound{Γ} \AgdaDatatype{⊢} \AgdaBound{A}\<%
\\
\\
\>\AgdaFunction{weak-modal} \AgdaSymbol{(}\AgdaInductiveConstructor{iPCF-var} \AgdaBound{p}\AgdaSymbol{)} \AgdaBound{x} \AgdaSymbol{=} \AgdaInductiveConstructor{iPCF-var} \AgdaBound{p}\<%
\\
\>\AgdaFunction{weak-modal} \AgdaSymbol{(}\AgdaInductiveConstructor{iPCF-modal-var} \AgdaBound{p}\AgdaSymbol{)} \AgdaBound{x} \AgdaSymbol{=} \AgdaInductiveConstructor{iPCF-modal-var} \AgdaSymbol{(}\AgdaBound{x} \AgdaBound{p}\AgdaSymbol{)}\<%
\\
\>\AgdaFunction{weak-modal} \AgdaSymbol{(}\AgdaInductiveConstructor{iPCF-app} \AgdaBound{t} \AgdaBound{u}\AgdaSymbol{)} \AgdaBound{x} \AgdaSymbol{=} \AgdaInductiveConstructor{iPCF-app} \AgdaSymbol{(}\AgdaFunction{weak-modal} \AgdaBound{t} \AgdaBound{x}\AgdaSymbol{)}\<%
\\
\>[9]\AgdaIndent{37}{}\<[37]%
\>[37]\AgdaSymbol{(}\AgdaFunction{weak-modal} \AgdaBound{u} \AgdaBound{x}\AgdaSymbol{)}\<%
\\
\>\AgdaFunction{weak-modal} \AgdaSymbol{(}\AgdaInductiveConstructor{iPCF-lam} \AgdaBound{t}\AgdaSymbol{)} \AgdaBound{x} \AgdaSymbol{=} \AgdaInductiveConstructor{iPCF-lam} \AgdaSymbol{(}\AgdaFunction{weak-modal} \AgdaBound{t} \AgdaBound{x}\AgdaSymbol{)}\<%
\\
\>\AgdaFunction{weak-modal} \AgdaSymbol{(}\AgdaInductiveConstructor{iPCF-prod} \AgdaBound{t} \AgdaBound{u}\AgdaSymbol{)} \AgdaBound{x} \AgdaSymbol{=} \AgdaInductiveConstructor{iPCF-prod} \AgdaSymbol{(}\AgdaFunction{weak-modal} \AgdaBound{t} \AgdaBound{x}\AgdaSymbol{)}\<%
\\
\>[37]\AgdaIndent{39}{}\<[39]%
\>[39]\AgdaSymbol{(}\AgdaFunction{weak-modal} \AgdaBound{u} \AgdaBound{x}\AgdaSymbol{)}\<%
\\
\>\AgdaFunction{weak-modal} \AgdaSymbol{(}\AgdaInductiveConstructor{iPCF-fst} \AgdaBound{t}\AgdaSymbol{)} \AgdaBound{x} \AgdaSymbol{=} \AgdaInductiveConstructor{iPCF-fst} \AgdaSymbol{(}\AgdaFunction{weak-modal} \AgdaBound{t} \AgdaBound{x}\AgdaSymbol{)}\<%
\\
\>\AgdaFunction{weak-modal} \AgdaSymbol{(}\AgdaInductiveConstructor{iPCF-snd} \AgdaBound{t}\AgdaSymbol{)} \AgdaBound{x} \AgdaSymbol{=} \AgdaInductiveConstructor{iPCF-snd} \AgdaSymbol{(}\AgdaFunction{weak-modal} \AgdaBound{t} \AgdaBound{x}\AgdaSymbol{)}\<%
\\
\>\AgdaFunction{weak-modal} \AgdaSymbol{(}\AgdaInductiveConstructor{iPCF-boxI} \AgdaBound{t}\AgdaSymbol{)} \AgdaBound{x} \AgdaSymbol{=} \AgdaInductiveConstructor{iPCF-boxI} \AgdaSymbol{(}\AgdaFunction{weak-modal} \AgdaBound{t} \AgdaBound{x}\AgdaSymbol{)}\<%
\\
\>\AgdaFunction{weak-modal} \AgdaSymbol{(}\AgdaInductiveConstructor{iPCF-boxE} \AgdaBound{t} \AgdaBound{u}\AgdaSymbol{)} \AgdaBound{x} \AgdaSymbol{=}\<%
\\
\>[0]\AgdaIndent{2}{}\<[2]%
\>[2]\AgdaInductiveConstructor{iPCF-boxE} \AgdaSymbol{(}\AgdaFunction{weak-modal} \AgdaBound{t} \AgdaBound{x}\AgdaSymbol{)}\<%
\\
\>[2]\AgdaIndent{11}{}\<[11]%
\>[11]\AgdaSymbol{(}\AgdaFunction{weak-modal} \AgdaBound{u} \AgdaSymbol{(}\AgdaFunction{weakboth} \AgdaBound{x}\AgdaSymbol{))}\<%
\\
\>\AgdaFunction{weak-modal} \AgdaSymbol{(}\AgdaInductiveConstructor{iPCF-fix} \AgdaBound{t}\AgdaSymbol{)} \AgdaBound{x} \AgdaSymbol{=} \AgdaInductiveConstructor{iPCF-fix} \AgdaSymbol{(}\AgdaFunction{weak-modal} \AgdaBound{t} \AgdaBound{x}\AgdaSymbol{)}\<%
\\
\\
\\
\>\AgdaComment{-- Cut.}\<%
\\
\\
\>\AgdaFunction{cut} \AgdaSymbol{:} \AgdaSymbol{∀} \AgdaSymbol{\{}\AgdaBound{Δ} \AgdaBound{Γ} \AgdaBound{A} \AgdaBound{B}\AgdaSymbol{\}} \AgdaSymbol{→} \AgdaSymbol{(}\AgdaBound{Γ'} \AgdaSymbol{:} \AgdaDatatype{Cx} \AgdaInductiveConstructor{modal}\AgdaSymbol{)}\<%
\\
\\
\>[0]\AgdaIndent{4}{}\<[4]%
\>[4]\AgdaSymbol{→} \AgdaBound{Δ} \AgdaDatatype{/} \AgdaBound{Γ} \AgdaDatatype{⊢} \AgdaBound{A} \<[19]%
\>[19]\AgdaSymbol{→} \AgdaBound{Δ} \AgdaDatatype{/} \AgdaBound{Γ} \AgdaInductiveConstructor{,} \AgdaBound{A} \AgdaFunction{++} \AgdaBound{Γ'} \AgdaDatatype{⊢} \AgdaBound{B}\<%
\\
\>[0]\AgdaIndent{4}{}\<[4]%
\>[4]\AgdaComment{---------------------------------------}\<%
\\
\>[4]\AgdaIndent{14}{}\<[14]%
\>[14]\AgdaSymbol{→} \AgdaBound{Δ} \AgdaDatatype{/} \AgdaBound{Γ} \AgdaFunction{++} \AgdaBound{Γ'} \AgdaDatatype{⊢} \AgdaBound{B}\<%
\\
\\
\>\AgdaFunction{cut} \AgdaInductiveConstructor{·} \AgdaBound{d} \AgdaSymbol{(}\AgdaInductiveConstructor{iPCF-var} \AgdaInductiveConstructor{top}\AgdaSymbol{)} \AgdaSymbol{=} \AgdaBound{d}\<%
\\
\>\AgdaFunction{cut} \AgdaInductiveConstructor{·} \AgdaBound{d} \AgdaSymbol{(}\AgdaInductiveConstructor{iPCF-var} \AgdaSymbol{(}\AgdaInductiveConstructor{pop} \AgdaBound{x}\AgdaSymbol{))} \AgdaSymbol{=} \AgdaInductiveConstructor{iPCF-var} \AgdaBound{x}\<%
\\
\>\AgdaFunction{cut} \AgdaSymbol{(}\AgdaBound{Γ'} \AgdaInductiveConstructor{,} \AgdaBound{B}\AgdaSymbol{)} \AgdaBound{d} \AgdaSymbol{(}\AgdaInductiveConstructor{iPCF-var} \AgdaInductiveConstructor{top}\AgdaSymbol{)} \AgdaSymbol{=} \AgdaInductiveConstructor{iPCF-var} \AgdaInductiveConstructor{top}\<%
\\
\>\AgdaFunction{cut} \AgdaSymbol{(}\AgdaBound{Γ'} \AgdaInductiveConstructor{,} \AgdaBound{A'}\AgdaSymbol{)} \AgdaBound{d} \AgdaSymbol{(}\AgdaInductiveConstructor{iPCF-var} \AgdaSymbol{(}\AgdaInductiveConstructor{pop} \AgdaBound{x}\AgdaSymbol{))} \AgdaSymbol{=}\<%
\\
\>[0]\AgdaIndent{2}{}\<[2]%
\>[2]\AgdaFunction{weak} \AgdaSymbol{(}\AgdaFunction{cut} \AgdaBound{Γ'} \AgdaBound{d} \AgdaSymbol{(}\AgdaInductiveConstructor{iPCF-var} \AgdaBound{x}\AgdaSymbol{))} \AgdaSymbol{(}\AgdaFunction{weakone} \AgdaFunction{subsetid}\AgdaSymbol{)}\<%
\\
\>\AgdaFunction{cut} \AgdaBound{Γ'} \AgdaBound{d} \AgdaSymbol{(}\AgdaInductiveConstructor{iPCF-modal-var} \AgdaBound{p}\AgdaSymbol{)} \AgdaSymbol{=} \AgdaInductiveConstructor{iPCF-modal-var} \AgdaBound{p}\<%
\\
\>\AgdaFunction{cut} \AgdaBound{Γ'} \AgdaBound{d} \AgdaSymbol{(}\AgdaInductiveConstructor{iPCF-app} \AgdaBound{t} \AgdaBound{u}\AgdaSymbol{)} \AgdaSymbol{=} \AgdaInductiveConstructor{iPCF-app} \AgdaSymbol{(}\AgdaFunction{cut} \AgdaBound{Γ'} \AgdaBound{d} \AgdaBound{t}\AgdaSymbol{)} \AgdaSymbol{(}\AgdaFunction{cut} \AgdaBound{Γ'} \AgdaBound{d} \AgdaBound{u}\AgdaSymbol{)}\<%
\\
\>\AgdaFunction{cut} \AgdaBound{Γ'} \AgdaBound{d} \AgdaSymbol{(}\AgdaInductiveConstructor{iPCF-lam} \AgdaBound{e}\AgdaSymbol{)} \AgdaSymbol{=} \AgdaInductiveConstructor{iPCF-lam} \AgdaSymbol{(}\AgdaFunction{cut} \AgdaSymbol{(}\AgdaBound{Γ'} \AgdaInductiveConstructor{,} \AgdaSymbol{\_)} \AgdaBound{d} \AgdaBound{e}\AgdaSymbol{)}\<%
\\
\>\AgdaFunction{cut} \AgdaBound{Γ'} \AgdaBound{d} \AgdaSymbol{(}\AgdaInductiveConstructor{iPCF-prod} \AgdaBound{t} \AgdaBound{u}\AgdaSymbol{)} \AgdaSymbol{=} \AgdaInductiveConstructor{iPCF-prod} \AgdaSymbol{(}\AgdaFunction{cut} \AgdaBound{Γ'} \AgdaBound{d} \AgdaBound{t}\AgdaSymbol{)} \AgdaSymbol{(}\AgdaFunction{cut} \AgdaBound{Γ'} \AgdaBound{d} \AgdaBound{u}\AgdaSymbol{)}\<%
\\
\>\AgdaFunction{cut} \AgdaBound{Γ'} \AgdaBound{d} \AgdaSymbol{(}\AgdaInductiveConstructor{iPCF-fst} \AgdaBound{e}\AgdaSymbol{)} \AgdaSymbol{=} \AgdaInductiveConstructor{iPCF-fst} \AgdaSymbol{(}\AgdaFunction{cut} \AgdaBound{Γ'} \AgdaBound{d} \AgdaBound{e}\AgdaSymbol{)}\<%
\\
\>\AgdaFunction{cut} \AgdaBound{Γ'} \AgdaBound{d} \AgdaSymbol{(}\AgdaInductiveConstructor{iPCF-snd} \AgdaBound{e}\AgdaSymbol{)} \AgdaSymbol{=} \AgdaInductiveConstructor{iPCF-snd} \AgdaSymbol{(}\AgdaFunction{cut} \AgdaBound{Γ'} \AgdaBound{d} \AgdaBound{e}\AgdaSymbol{)}\<%
\\
\>\AgdaFunction{cut} \AgdaBound{Γ'} \AgdaBound{d} \AgdaSymbol{(}\AgdaInductiveConstructor{iPCF-boxI} \AgdaBound{e}\AgdaSymbol{)} \AgdaSymbol{=} \AgdaInductiveConstructor{iPCF-boxI} \AgdaBound{e}\<%
\\
\>\AgdaFunction{cut} \AgdaBound{Γ'} \AgdaBound{d} \AgdaSymbol{(}\AgdaInductiveConstructor{iPCF-boxE} \AgdaBound{t} \AgdaBound{u}\AgdaSymbol{)} \AgdaSymbol{=}\<%
\\
\>[0]\AgdaIndent{2}{}\<[2]%
\>[2]\AgdaInductiveConstructor{iPCF-boxE} \AgdaSymbol{(}\AgdaFunction{cut} \AgdaBound{Γ'} \AgdaBound{d} \AgdaBound{t}\AgdaSymbol{)}\<%
\\
\>[2]\AgdaIndent{12}{}\<[12]%
\>[12]\AgdaSymbol{(}\AgdaFunction{cut} \AgdaBound{Γ'} \AgdaSymbol{(}\AgdaFunction{weak-modal} \AgdaBound{d} \AgdaSymbol{(}\AgdaFunction{weakone} \AgdaSymbol{(}\AgdaFunction{subsetid}\AgdaSymbol{)))} \AgdaBound{u}\AgdaSymbol{)}\<%
\\
\>\AgdaFunction{cut} \AgdaBound{Γ'} \AgdaBound{d} \AgdaSymbol{(}\AgdaInductiveConstructor{iPCF-fix} \AgdaBound{t}\AgdaSymbol{)} \AgdaSymbol{=} \AgdaInductiveConstructor{iPCF-fix} \AgdaBound{t}\<%
\\
\\
\\
\>\AgdaFunction{cut-modal} \AgdaSymbol{:} \AgdaSymbol{∀} \AgdaSymbol{\{}\AgdaBound{Δ} \AgdaBound{Γ} \AgdaBound{A} \AgdaBound{B}\AgdaSymbol{\}} \AgdaSymbol{→} \AgdaSymbol{(}\AgdaBound{Δ'} \AgdaSymbol{:} \AgdaDatatype{Cx} \AgdaInductiveConstructor{modal}\AgdaSymbol{)}\<%
\\
\\
\>[0]\AgdaIndent{4}{}\<[4]%
\>[4]\AgdaSymbol{→} \AgdaBound{Δ} \AgdaDatatype{/} \AgdaInductiveConstructor{·} \AgdaDatatype{⊢} \AgdaBound{A} \<[19]%
\>[19]\AgdaSymbol{→} \AgdaBound{Δ} \AgdaInductiveConstructor{,} \AgdaBound{A} \AgdaFunction{++} \AgdaBound{Δ'} \AgdaDatatype{/} \AgdaBound{Γ} \<[38]%
\>[38]\AgdaDatatype{⊢} \AgdaBound{B}\<%
\\
\>[0]\AgdaIndent{4}{}\<[4]%
\>[4]\AgdaComment{---------------------------------------}\<%
\\
\>[4]\AgdaIndent{13}{}\<[13]%
\>[13]\AgdaSymbol{→} \AgdaBound{Δ} \AgdaFunction{++} \AgdaBound{Δ'} \AgdaDatatype{/} \AgdaBound{Γ} \AgdaDatatype{⊢} \AgdaBound{B}\<%
\\
\\
\>\AgdaFunction{cut-modal} \AgdaBound{Δ'} \AgdaBound{d} \AgdaSymbol{(}\AgdaInductiveConstructor{iPCF-var} \AgdaBound{x}\AgdaSymbol{)} \AgdaSymbol{=} \AgdaInductiveConstructor{iPCF-var} \AgdaBound{x}\<%
\\
\>\AgdaFunction{cut-modal} \AgdaInductiveConstructor{·} \AgdaBound{d} \AgdaSymbol{(}\AgdaInductiveConstructor{iPCF-modal-var} \AgdaInductiveConstructor{top}\AgdaSymbol{)} \AgdaSymbol{=} \AgdaFunction{weak} \AgdaBound{d} \AgdaFunction{subsetempty}\<%
\\
\>\AgdaFunction{cut-modal} \AgdaInductiveConstructor{·} \AgdaBound{d} \AgdaSymbol{(}\AgdaInductiveConstructor{iPCF-modal-var} \AgdaSymbol{(}\AgdaInductiveConstructor{pop} \AgdaBound{x}\AgdaSymbol{))} \AgdaSymbol{=} \AgdaInductiveConstructor{iPCF-modal-var} \AgdaBound{x}\<%
\\
\>\AgdaFunction{cut-modal} \AgdaSymbol{(}\AgdaBound{Δ'} \AgdaInductiveConstructor{,} \AgdaBound{B}\AgdaSymbol{)} \AgdaBound{d} \AgdaSymbol{(}\AgdaInductiveConstructor{iPCF-modal-var} \AgdaInductiveConstructor{top}\AgdaSymbol{)} \AgdaSymbol{=} \AgdaInductiveConstructor{iPCF-modal-var} \AgdaInductiveConstructor{top}\<%
\\
\>\AgdaFunction{cut-modal} \AgdaSymbol{(}\AgdaBound{Δ'} \AgdaInductiveConstructor{,} \AgdaBound{A'}\AgdaSymbol{)} \AgdaBound{d} \AgdaSymbol{(}\AgdaInductiveConstructor{iPCF-modal-var} \AgdaSymbol{(}\AgdaInductiveConstructor{pop} \AgdaBound{x}\AgdaSymbol{))} \AgdaSymbol{=}\<%
\\
\>[0]\AgdaIndent{2}{}\<[2]%
\>[2]\AgdaFunction{weak-modal} \AgdaSymbol{(}\AgdaFunction{cut-modal} \AgdaBound{Δ'} \AgdaBound{d} \AgdaSymbol{(}\AgdaInductiveConstructor{iPCF-modal-var} \AgdaBound{x}\AgdaSymbol{))} \AgdaSymbol{(}\AgdaFunction{weakone} \AgdaFunction{subsetid}\AgdaSymbol{)}\<%
\\
\>\AgdaFunction{cut-modal} \AgdaBound{Δ'} \AgdaBound{d} \AgdaSymbol{(}\AgdaInductiveConstructor{iPCF-app} \AgdaBound{p} \AgdaBound{q}\AgdaSymbol{)} \AgdaSymbol{=}\<%
\\
\>[0]\AgdaIndent{2}{}\<[2]%
\>[2]\AgdaInductiveConstructor{iPCF-app} \AgdaSymbol{(}\AgdaFunction{cut-modal} \AgdaBound{Δ'} \AgdaBound{d} \AgdaBound{p}\AgdaSymbol{)} \AgdaSymbol{(}\AgdaFunction{cut-modal} \AgdaBound{Δ'} \AgdaBound{d} \AgdaBound{q}\AgdaSymbol{)}\<%
\\
\>\AgdaFunction{cut-modal} \AgdaBound{Δ'} \AgdaBound{d} \AgdaSymbol{(}\AgdaInductiveConstructor{iPCF-lam} \AgdaBound{e}\AgdaSymbol{)} \AgdaSymbol{=} \AgdaInductiveConstructor{iPCF-lam} \AgdaSymbol{(}\AgdaFunction{cut-modal} \AgdaBound{Δ'} \AgdaBound{d} \AgdaBound{e}\AgdaSymbol{)}\<%
\\
\>\AgdaFunction{cut-modal} \AgdaBound{Δ'} \AgdaBound{d} \AgdaSymbol{(}\AgdaInductiveConstructor{iPCF-prod} \AgdaBound{p} \AgdaBound{q}\AgdaSymbol{)} \AgdaSymbol{=}\<%
\\
\>[0]\AgdaIndent{2}{}\<[2]%
\>[2]\AgdaInductiveConstructor{iPCF-prod} \AgdaSymbol{(}\AgdaFunction{cut-modal} \AgdaBound{Δ'} \AgdaBound{d} \AgdaBound{p}\AgdaSymbol{)} \AgdaSymbol{(}\AgdaFunction{cut-modal} \AgdaBound{Δ'} \AgdaBound{d} \AgdaBound{q}\AgdaSymbol{)}\<%
\\
\>\AgdaFunction{cut-modal} \AgdaBound{Δ'} \AgdaBound{d} \AgdaSymbol{(}\AgdaInductiveConstructor{iPCF-fst} \AgdaBound{e}\AgdaSymbol{)} \AgdaSymbol{=} \AgdaInductiveConstructor{iPCF-fst} \AgdaSymbol{(}\AgdaFunction{cut-modal} \AgdaBound{Δ'} \AgdaBound{d} \AgdaBound{e}\AgdaSymbol{)}\<%
\\
\>\AgdaFunction{cut-modal} \AgdaBound{Δ'} \AgdaBound{d} \AgdaSymbol{(}\AgdaInductiveConstructor{iPCF-snd} \AgdaBound{e}\AgdaSymbol{)} \AgdaSymbol{=} \AgdaInductiveConstructor{iPCF-snd} \AgdaSymbol{(}\AgdaFunction{cut-modal} \AgdaBound{Δ'} \AgdaBound{d} \AgdaBound{e}\AgdaSymbol{)}\<%
\\
\>\AgdaFunction{cut-modal} \AgdaBound{Δ'} \AgdaBound{d} \AgdaSymbol{(}\AgdaInductiveConstructor{iPCF-boxI} \AgdaBound{e}\AgdaSymbol{)} \AgdaSymbol{=} \AgdaInductiveConstructor{iPCF-boxI} \AgdaSymbol{(}\AgdaFunction{cut-modal} \AgdaBound{Δ'} \AgdaBound{d} \AgdaBound{e}\AgdaSymbol{)}\<%
\\
\>\AgdaFunction{cut-modal} \AgdaBound{Δ'} \AgdaBound{d} \AgdaSymbol{(}\AgdaInductiveConstructor{iPCF-boxE} \AgdaBound{p} \AgdaBound{q}\AgdaSymbol{)} \AgdaSymbol{=}\<%
\\
\>[0]\AgdaIndent{2}{}\<[2]%
\>[2]\AgdaInductiveConstructor{iPCF-boxE} \AgdaSymbol{(}\AgdaFunction{cut-modal} \AgdaBound{Δ'} \AgdaBound{d} \AgdaBound{p}\AgdaSymbol{)} \AgdaSymbol{(}\AgdaFunction{cut-modal} \AgdaSymbol{(}\AgdaBound{Δ'} \AgdaInductiveConstructor{,} \AgdaSymbol{\_)} \AgdaBound{d} \AgdaBound{q}\AgdaSymbol{)}\<%
\\
\>\AgdaFunction{cut-modal} \AgdaBound{Δ'} \AgdaBound{d} \AgdaSymbol{(}\AgdaInductiveConstructor{iPCF-fix} \AgdaBound{e}\AgdaSymbol{)} \AgdaSymbol{=} \AgdaInductiveConstructor{iPCF-fix} \AgdaSymbol{(}\AgdaFunction{cut-modal} \AgdaBound{Δ'} \AgdaBound{d} \AgdaBound{e}\AgdaSymbol{)}\<%
\end{code}

\pagebreak

\section{\texttt{iPCF2.agda}}
  \label{sec:iPCF2.agda}
  \begin{code}%
\>\AgdaKeyword{module} \AgdaModule{iPCF2} \AgdaKeyword{where}\<%
\\
\\
\>\AgdaKeyword{infixl} \AgdaNumber{0} \_/\_⊢\_::\_\<%
\\
\\
\>\AgdaKeyword{open} \AgdaKeyword{import} \AgdaModule{Basics}\<%
\\
\\
\>\AgdaComment{-- Definition}\<%
\\
\\
\>\AgdaKeyword{data} \AgdaDatatype{Judgement} \AgdaSymbol{:} \AgdaPrimitiveType{Set} \AgdaKeyword{where}\<%
\\
\>[0]\AgdaIndent{2}{}\<[2]%
\>[2]\AgdaInductiveConstructor{int} \AgdaSymbol{:} \AgdaDatatype{Judgement}\<%
\\
\>[0]\AgdaIndent{2}{}\<[2]%
\>[2]\AgdaInductiveConstructor{ext} \AgdaSymbol{:} \AgdaDatatype{Judgement}\<%
\\
\>[0]\AgdaIndent{2}{}\<[2]%
\>[2]\<%
\\
\>\AgdaKeyword{data} \AgdaDatatype{\_/\_⊢\_::\_} \AgdaSymbol{(}\AgdaBound{Δ} \AgdaBound{Γ} \AgdaSymbol{:} \AgdaDatatype{Cx} \AgdaInductiveConstructor{modal}\AgdaSymbol{)} \AgdaSymbol{:} \AgdaDatatype{Judgement} \AgdaSymbol{→} \AgdaDatatype{Ty} \AgdaInductiveConstructor{modal} \AgdaSymbol{→} \AgdaPrimitiveType{Set} \AgdaKeyword{where}\<%
\\
\\
\>[0]\AgdaIndent{2}{}\<[2]%
\>[2]\AgdaInductiveConstructor{iPCF-var} \AgdaSymbol{:} \AgdaSymbol{∀} \AgdaSymbol{\{}\AgdaBound{J} \AgdaBound{A}\AgdaSymbol{\}}\<%
\\
\>[0]\AgdaIndent{2}{}\<[2]%
\>[2]\<%
\\
\>[2]\AgdaIndent{8}{}\<[8]%
\>[8]\AgdaSymbol{→} \AgdaBound{A} \AgdaDatatype{∈} \AgdaBound{Γ}\<%
\\
\>[0]\AgdaIndent{4}{}\<[4]%
\>[4]\AgdaComment{---------------------------}\<%
\\
\>[0]\AgdaIndent{4}{}\<[4]%
\>[4]\AgdaSymbol{→} \AgdaBound{Δ} \AgdaDatatype{/} \AgdaBound{Γ} \AgdaDatatype{⊢} \AgdaBound{J} \AgdaDatatype{::} \AgdaBound{A}\<%
\\
\\
\>[0]\AgdaIndent{2}{}\<[2]%
\>[2]\AgdaInductiveConstructor{iPCF-modal-var} \AgdaSymbol{:} \AgdaSymbol{∀} \AgdaSymbol{\{}\AgdaBound{J} \AgdaBound{A}\AgdaSymbol{\}}\<%
\\
\>[0]\AgdaIndent{2}{}\<[2]%
\>[2]\<%
\\
\>[2]\AgdaIndent{8}{}\<[8]%
\>[8]\AgdaSymbol{→} \AgdaBound{A} \AgdaDatatype{∈} \AgdaBound{Δ}\<%
\\
\>[0]\AgdaIndent{4}{}\<[4]%
\>[4]\AgdaComment{---------------------------}\<%
\\
\>[0]\AgdaIndent{4}{}\<[4]%
\>[4]\AgdaSymbol{→} \AgdaBound{Δ} \AgdaDatatype{/} \AgdaBound{Γ} \AgdaDatatype{⊢} \AgdaBound{J} \AgdaDatatype{::} \AgdaBound{A}\<%
\\
\>[4]\AgdaIndent{15}{}\<[15]%
\>[15]\<%
\\
\>[0]\AgdaIndent{2}{}\<[2]%
\>[2]\AgdaInductiveConstructor{iPCF-app} \AgdaSymbol{:} \AgdaSymbol{∀} \AgdaSymbol{\{}\AgdaBound{J} \AgdaBound{A} \AgdaBound{B}\AgdaSymbol{\}}\<%
\\
\>[0]\AgdaIndent{2}{}\<[2]%
\>[2]\<%
\\
\>[2]\AgdaIndent{4}{}\<[4]%
\>[4]\AgdaSymbol{→} \AgdaBound{Δ} \AgdaDatatype{/} \AgdaBound{Γ} \AgdaDatatype{⊢} \AgdaBound{J} \AgdaDatatype{::} \AgdaBound{A} \AgdaInductiveConstructor{=>} \AgdaBound{B} \<[29]%
\>[29]\AgdaSymbol{→} \AgdaBound{Δ} \AgdaDatatype{/} \AgdaBound{Γ} \AgdaDatatype{⊢} \AgdaBound{J} \AgdaDatatype{::} \AgdaBound{A}\<%
\\
\>[2]\AgdaIndent{4}{}\<[4]%
\>[4]\AgdaComment{----------------------------------------------------------}\<%
\\
\>[4]\AgdaIndent{10}{}\<[10]%
\>[10]\AgdaSymbol{→} \AgdaBound{Δ} \AgdaDatatype{/} \AgdaBound{Γ} \AgdaDatatype{⊢} \AgdaBound{J} \AgdaDatatype{::} \AgdaBound{B}\<%
\\
\>[10]\AgdaIndent{26}{}\<[26]%
\>[26]\<%
\\
\>[0]\AgdaIndent{2}{}\<[2]%
\>[2]\AgdaInductiveConstructor{iPCF-lam-ext} \AgdaSymbol{:} \AgdaSymbol{∀} \AgdaSymbol{\{}\AgdaBound{A} \AgdaBound{B}\AgdaSymbol{\}}\<%
\\
\>[0]\AgdaIndent{2}{}\<[2]%
\>[2]\<%
\\
\>[2]\AgdaIndent{5}{}\<[5]%
\>[5]\AgdaSymbol{→} \AgdaBound{Δ} \AgdaDatatype{/} \AgdaSymbol{(}\AgdaBound{Γ} \AgdaInductiveConstructor{,} \AgdaBound{A}\AgdaSymbol{)} \AgdaDatatype{⊢} \AgdaInductiveConstructor{ext} \AgdaDatatype{::} \AgdaBound{B}\<%
\\
\>[2]\AgdaIndent{5}{}\<[5]%
\>[5]\AgdaComment{--------------------------------------}\<%
\\
\>[2]\AgdaIndent{5}{}\<[5]%
\>[5]\AgdaSymbol{→} \AgdaBound{Δ} \AgdaDatatype{/} \AgdaBound{Γ} \AgdaDatatype{⊢} \AgdaInductiveConstructor{ext} \AgdaDatatype{::} \AgdaBound{A} \AgdaInductiveConstructor{=>} \AgdaBound{B}\<%
\\
\\
\>[0]\AgdaIndent{2}{}\<[2]%
\>[2]\AgdaInductiveConstructor{iPCF-lam-int} \AgdaSymbol{:} \AgdaSymbol{∀} \AgdaSymbol{\{}\AgdaBound{J} \AgdaBound{A} \AgdaBound{B}\AgdaSymbol{\}}\<%
\\
\>[0]\AgdaIndent{2}{}\<[2]%
\>[2]\<%
\\
\>[2]\AgdaIndent{5}{}\<[5]%
\>[5]\AgdaSymbol{→} \AgdaInductiveConstructor{·} \AgdaDatatype{/} \AgdaSymbol{(}\AgdaInductiveConstructor{·} \AgdaInductiveConstructor{,} \AgdaBound{A}\AgdaSymbol{)} \AgdaDatatype{⊢} \AgdaBound{J} \AgdaDatatype{::} \AgdaBound{B}\<%
\\
\>[2]\AgdaIndent{5}{}\<[5]%
\>[5]\AgdaComment{----------------------------------}\<%
\\
\>[2]\AgdaIndent{5}{}\<[5]%
\>[5]\AgdaSymbol{→} \AgdaBound{Δ} \AgdaDatatype{/} \AgdaBound{Γ} \AgdaDatatype{⊢} \AgdaInductiveConstructor{int} \AgdaDatatype{::} \AgdaBound{A} \AgdaInductiveConstructor{=>} \AgdaBound{B}\<%
\\
\>[5]\AgdaIndent{17}{}\<[17]%
\>[17]\<%
\\
\>[0]\AgdaIndent{2}{}\<[2]%
\>[2]\AgdaInductiveConstructor{iPCF-boxI} \AgdaSymbol{:} \AgdaSymbol{∀} \AgdaSymbol{\{}\AgdaBound{J} \AgdaBound{A}\AgdaSymbol{\}}\<%
\\
\>[0]\AgdaIndent{2}{}\<[2]%
\>[2]\<%
\\
\>[2]\AgdaIndent{5}{}\<[5]%
\>[5]\AgdaSymbol{→} \AgdaBound{Δ} \AgdaDatatype{/} \AgdaInductiveConstructor{·} \AgdaDatatype{⊢} \AgdaInductiveConstructor{int} \AgdaDatatype{::} \AgdaBound{A}\<%
\\
\>[0]\AgdaIndent{4}{}\<[4]%
\>[4]\AgdaComment{-------------------------------}\<%
\\
\>[0]\AgdaIndent{4}{}\<[4]%
\>[4]\AgdaSymbol{→} \AgdaBound{Δ} \AgdaDatatype{/} \AgdaBound{Γ} \AgdaDatatype{⊢} \AgdaBound{J} \AgdaDatatype{::} \AgdaInductiveConstructor{□} \AgdaBound{A}\<%
\\
\>[4]\AgdaIndent{15}{}\<[15]%
\>[15]\<%
\\
\>[0]\AgdaIndent{2}{}\<[2]%
\>[2]\AgdaInductiveConstructor{iPCF-boxE} \AgdaSymbol{:} \AgdaSymbol{∀} \AgdaSymbol{\{}\AgdaBound{J} \AgdaBound{A} \AgdaBound{C}\AgdaSymbol{\}}\<%
\\
\>[0]\AgdaIndent{2}{}\<[2]%
\>[2]\<%
\\
\>[2]\AgdaIndent{4}{}\<[4]%
\>[4]\AgdaSymbol{→} \AgdaBound{Δ} \AgdaDatatype{/} \AgdaBound{Γ} \AgdaDatatype{⊢} \AgdaBound{J} \AgdaDatatype{::} \AgdaInductiveConstructor{□} \AgdaBound{A} \<[26]%
\>[26]\AgdaSymbol{→} \AgdaSymbol{(}\AgdaBound{Δ} \AgdaInductiveConstructor{,} \AgdaBound{A}\AgdaSymbol{)} \AgdaDatatype{/} \AgdaBound{Γ} \AgdaDatatype{⊢} \AgdaBound{J} \AgdaDatatype{::} \AgdaBound{C}\<%
\\
\>[2]\AgdaIndent{4}{}\<[4]%
\>[4]\AgdaComment{------------------------------------------------------------------}\<%
\\
\>[4]\AgdaIndent{14}{}\<[14]%
\>[14]\AgdaSymbol{→} \AgdaBound{Δ} \AgdaDatatype{/} \AgdaBound{Γ} \AgdaDatatype{⊢} \AgdaBound{J} \AgdaDatatype{::} \AgdaBound{C}\<%
\\
\\
\>[0]\AgdaIndent{2}{}\<[2]%
\>[2]\AgdaInductiveConstructor{iPCF-fix} \AgdaSymbol{:} \AgdaSymbol{∀} \AgdaSymbol{\{}\AgdaBound{J} \AgdaBound{A}\AgdaSymbol{\}}\<%
\\
\>[0]\AgdaIndent{2}{}\<[2]%
\>[2]\<%
\\
\>[2]\AgdaIndent{5}{}\<[5]%
\>[5]\AgdaSymbol{→} \AgdaInductiveConstructor{·} \AgdaDatatype{/} \AgdaSymbol{(}\AgdaInductiveConstructor{·} \AgdaInductiveConstructor{,} \AgdaInductiveConstructor{□} \AgdaBound{A}\AgdaSymbol{)} \AgdaDatatype{⊢} \AgdaInductiveConstructor{int} \AgdaDatatype{::} \AgdaBound{A}\<%
\\
\>[0]\AgdaIndent{4}{}\<[4]%
\>[4]\AgdaComment{--------------------------------------}\<%
\\
\>[0]\AgdaIndent{4}{}\<[4]%
\>[4]\AgdaSymbol{→} \AgdaBound{Δ} \AgdaDatatype{/} \AgdaBound{Γ} \AgdaDatatype{⊢} \AgdaBound{J} \AgdaDatatype{::} \AgdaBound{A}\<%
\\
\\
\\
\>\AgdaComment{-- Weakening and exchange.}\<%
\\
\\
\\
\>\AgdaFunction{exch} \AgdaSymbol{:} \AgdaSymbol{∀} \AgdaSymbol{\{}\AgdaBound{Δ} \AgdaBound{Γ} \AgdaBound{J} \AgdaBound{A} \AgdaBound{B} \AgdaBound{C}\AgdaSymbol{\}} \AgdaSymbol{(}\AgdaBound{Γ'} \AgdaSymbol{:} \AgdaDatatype{Cx} \AgdaInductiveConstructor{modal}\AgdaSymbol{)}\<%
\\
\\
\>[0]\AgdaIndent{4}{}\<[4]%
\>[4]\AgdaSymbol{→} \AgdaBound{Δ} \AgdaDatatype{/} \AgdaSymbol{(}\AgdaBound{Γ} \AgdaInductiveConstructor{,} \AgdaBound{A} \AgdaInductiveConstructor{,} \AgdaBound{B}\AgdaSymbol{)} \AgdaFunction{++} \AgdaBound{Γ'} \AgdaDatatype{⊢} \AgdaBound{J} \AgdaDatatype{::} \AgdaBound{C}\<%
\\
\>[0]\AgdaIndent{4}{}\<[4]%
\>[4]\AgdaComment{-----------------------------}\<%
\\
\>[0]\AgdaIndent{4}{}\<[4]%
\>[4]\AgdaSymbol{→} \AgdaBound{Δ} \AgdaDatatype{/} \AgdaSymbol{(}\AgdaBound{Γ} \AgdaInductiveConstructor{,} \AgdaBound{B} \AgdaInductiveConstructor{,} \AgdaBound{A}\AgdaSymbol{)} \AgdaFunction{++} \AgdaBound{Γ'} \AgdaDatatype{⊢} \AgdaBound{J} \AgdaDatatype{::} \AgdaBound{C}\<%
\\
\\
\>\AgdaFunction{exch} \AgdaBound{Γ'} \AgdaSymbol{(}\AgdaInductiveConstructor{iPCF-var} \AgdaBound{x}\AgdaSymbol{)} \AgdaSymbol{=} \AgdaInductiveConstructor{iPCF-var} \AgdaSymbol{(}\AgdaFunction{cx-exch} \AgdaSymbol{\{}\AgdaArgument{Δ} \AgdaSymbol{=} \AgdaBound{Γ'}\AgdaSymbol{\}} \AgdaBound{x}\AgdaSymbol{)}\<%
\\
\>\AgdaFunction{exch} \AgdaBound{Γ'} \AgdaSymbol{(}\AgdaInductiveConstructor{iPCF-modal-var} \AgdaBound{x}\AgdaSymbol{)} \AgdaSymbol{=} \AgdaInductiveConstructor{iPCF-modal-var} \AgdaBound{x}\<%
\\
\>\AgdaFunction{exch} \AgdaBound{Γ'} \AgdaSymbol{(}\AgdaInductiveConstructor{iPCF-app} \AgdaBound{d} \AgdaBound{d₁}\AgdaSymbol{)} \AgdaSymbol{=} \AgdaInductiveConstructor{iPCF-app} \AgdaSymbol{(}\AgdaFunction{exch} \AgdaBound{Γ'} \AgdaBound{d}\AgdaSymbol{)} \AgdaSymbol{(}\AgdaFunction{exch} \AgdaBound{Γ'} \AgdaBound{d₁}\AgdaSymbol{)}\<%
\\
\>\AgdaFunction{exch} \AgdaSymbol{\{}\AgdaArgument{C} \AgdaSymbol{=} \AgdaBound{A} \AgdaInductiveConstructor{=>} \AgdaBound{B}\AgdaSymbol{\}} \AgdaBound{Γ'} \AgdaSymbol{(}\AgdaInductiveConstructor{iPCF-lam-ext} \AgdaBound{d}\AgdaSymbol{)} \AgdaSymbol{=} \AgdaInductiveConstructor{iPCF-lam-ext} \AgdaSymbol{(}\AgdaFunction{exch} \AgdaSymbol{(}\AgdaBound{Γ'} \AgdaInductiveConstructor{,} \AgdaBound{A}\AgdaSymbol{)} \AgdaBound{d}\AgdaSymbol{)}\<%
\\
\>\AgdaFunction{exch} \AgdaBound{Γ'} \AgdaSymbol{(}\AgdaInductiveConstructor{iPCF-lam-int} \AgdaBound{d}\AgdaSymbol{)} \AgdaSymbol{=} \AgdaInductiveConstructor{iPCF-lam-int} \AgdaBound{d}\<%
\\
\>\AgdaFunction{exch} \AgdaBound{Γ'} \AgdaSymbol{(}\AgdaInductiveConstructor{iPCF-boxI} \AgdaBound{d}\AgdaSymbol{)} \AgdaSymbol{=} \AgdaInductiveConstructor{iPCF-boxI} \AgdaBound{d}\<%
\\
\>\AgdaFunction{exch} \AgdaBound{Γ'} \AgdaSymbol{(}\AgdaInductiveConstructor{iPCF-boxE} \AgdaBound{d} \AgdaBound{e}\AgdaSymbol{)} \AgdaSymbol{=} \AgdaInductiveConstructor{iPCF-boxE} \AgdaSymbol{(}\AgdaFunction{exch} \AgdaBound{Γ'} \AgdaBound{d}\AgdaSymbol{)} \AgdaSymbol{(}\AgdaFunction{exch} \AgdaBound{Γ'} \AgdaBound{e}\AgdaSymbol{)}\<%
\\
\>\AgdaFunction{exch} \AgdaBound{Γ'} \AgdaSymbol{(}\AgdaInductiveConstructor{iPCF-fix} \AgdaBound{f}\AgdaSymbol{)} \AgdaSymbol{=} \AgdaInductiveConstructor{iPCF-fix} \AgdaBound{f}\<%
\\
\\
\\
\>\AgdaFunction{exch-modal} \AgdaSymbol{:} \AgdaSymbol{∀} \AgdaSymbol{\{}\AgdaBound{Δ} \AgdaBound{Γ} \AgdaBound{J} \AgdaBound{A} \AgdaBound{B} \AgdaBound{C}\AgdaSymbol{\}} \AgdaSymbol{(}\AgdaBound{Δ'} \AgdaSymbol{:} \AgdaDatatype{Cx} \AgdaInductiveConstructor{modal}\AgdaSymbol{)}\<%
\\
\\
\>[0]\AgdaIndent{4}{}\<[4]%
\>[4]\AgdaSymbol{→} \AgdaSymbol{(}\AgdaBound{Δ} \AgdaInductiveConstructor{,} \AgdaBound{A} \AgdaInductiveConstructor{,} \AgdaBound{B}\AgdaSymbol{)} \AgdaFunction{++} \AgdaBound{Δ'} \AgdaDatatype{/} \AgdaBound{Γ} \<[29]%
\>[29]\AgdaDatatype{⊢} \AgdaBound{J} \AgdaDatatype{::} \AgdaBound{C}\<%
\\
\>[0]\AgdaIndent{4}{}\<[4]%
\>[4]\AgdaComment{------------------------------}\<%
\\
\>[0]\AgdaIndent{4}{}\<[4]%
\>[4]\AgdaSymbol{→} \AgdaSymbol{(}\AgdaBound{Δ} \AgdaInductiveConstructor{,} \AgdaBound{B} \AgdaInductiveConstructor{,} \AgdaBound{A}\AgdaSymbol{)} \AgdaFunction{++} \AgdaBound{Δ'} \AgdaDatatype{/} \AgdaBound{Γ} \AgdaDatatype{⊢} \AgdaBound{J} \AgdaDatatype{::} \AgdaBound{C}\<%
\\
\>[4]\AgdaIndent{20}{}\<[20]%
\>[20]\<%
\\
\>\AgdaFunction{exch-modal} \AgdaBound{Δ'} \AgdaSymbol{(}\AgdaInductiveConstructor{iPCF-var} \AgdaBound{x}\AgdaSymbol{)} \AgdaSymbol{=} \AgdaInductiveConstructor{iPCF-var} \AgdaBound{x}\<%
\\
\>\AgdaFunction{exch-modal} \AgdaBound{Δ'} \AgdaSymbol{(}\AgdaInductiveConstructor{iPCF-modal-var} \AgdaBound{x}\AgdaSymbol{)} \AgdaSymbol{=}\<%
\\
\>[0]\AgdaIndent{2}{}\<[2]%
\>[2]\AgdaInductiveConstructor{iPCF-modal-var} \AgdaSymbol{(}\AgdaFunction{subsetdef} \AgdaBound{x} \AgdaSymbol{(}\AgdaFunction{cx-exch} \AgdaSymbol{\{}\AgdaArgument{Δ} \AgdaSymbol{=} \AgdaBound{Δ'}\AgdaSymbol{\}))}\<%
\\
\>\AgdaFunction{exch-modal} \AgdaBound{Δ'} \AgdaSymbol{(}\AgdaInductiveConstructor{iPCF-app} \AgdaBound{d} \AgdaBound{e}\AgdaSymbol{)} \AgdaSymbol{=}\<%
\\
\>[0]\AgdaIndent{2}{}\<[2]%
\>[2]\AgdaInductiveConstructor{iPCF-app} \AgdaSymbol{(}\AgdaFunction{exch-modal} \AgdaBound{Δ'} \AgdaBound{d}\AgdaSymbol{)} \AgdaSymbol{(}\AgdaFunction{exch-modal} \AgdaBound{Δ'} \AgdaBound{e}\AgdaSymbol{)}\<%
\\
\>\AgdaFunction{exch-modal} \AgdaBound{Δ'} \AgdaSymbol{(}\AgdaInductiveConstructor{iPCF-lam-ext} \AgdaBound{d}\AgdaSymbol{)} \AgdaSymbol{=} \AgdaInductiveConstructor{iPCF-lam-ext} \AgdaSymbol{(}\AgdaFunction{exch-modal} \AgdaBound{Δ'} \AgdaBound{d}\AgdaSymbol{)}\<%
\\
\>\AgdaFunction{exch-modal} \AgdaBound{Δ'} \AgdaSymbol{(}\AgdaInductiveConstructor{iPCF-lam-int} \AgdaBound{d}\AgdaSymbol{)} \AgdaSymbol{=} \AgdaInductiveConstructor{iPCF-lam-int} \AgdaBound{d}\<%
\\
\>\AgdaFunction{exch-modal} \AgdaBound{Δ'} \AgdaSymbol{(}\AgdaInductiveConstructor{iPCF-boxI} \AgdaBound{d}\AgdaSymbol{)} \AgdaSymbol{=} \AgdaInductiveConstructor{iPCF-boxI} \AgdaSymbol{(}\AgdaFunction{exch-modal} \AgdaBound{Δ'} \AgdaBound{d}\AgdaSymbol{)}\<%
\\
\>\AgdaFunction{exch-modal} \AgdaBound{Δ'} \AgdaSymbol{(}\AgdaInductiveConstructor{iPCF-boxE} \AgdaBound{d} \AgdaBound{e}\AgdaSymbol{)} \AgdaSymbol{=}\<%
\\
\>[0]\AgdaIndent{2}{}\<[2]%
\>[2]\AgdaInductiveConstructor{iPCF-boxE} \AgdaSymbol{(}\AgdaFunction{exch-modal} \AgdaBound{Δ'} \AgdaBound{d}\AgdaSymbol{)} \AgdaSymbol{(}\AgdaFunction{exch-modal} \AgdaSymbol{(}\AgdaBound{Δ'} \AgdaInductiveConstructor{,} \AgdaSymbol{\_)} \AgdaBound{e}\AgdaSymbol{)}\<%
\\
\>\AgdaFunction{exch-modal} \AgdaBound{Δ'} \AgdaSymbol{(}\AgdaInductiveConstructor{iPCF-fix} \AgdaBound{f}\AgdaSymbol{)} \AgdaSymbol{=} \AgdaInductiveConstructor{iPCF-fix} \AgdaBound{f}\<%
\\
\\
\\
\>\AgdaFunction{weak} \AgdaSymbol{:} \AgdaSymbol{∀} \AgdaSymbol{\{}\AgdaBound{Δ} \AgdaBound{Γ} \AgdaBound{Γ'} \AgdaBound{J} \AgdaBound{A}\AgdaSymbol{\}}\<%
\\
\\
\>[2]\AgdaIndent{4}{}\<[4]%
\>[4]\AgdaSymbol{→} \AgdaBound{Δ} \AgdaDatatype{/} \AgdaBound{Γ} \AgdaDatatype{⊢} \AgdaBound{J} \AgdaDatatype{::} \<[20]%
\>[20]\AgdaBound{A} \<[25]%
\>[25]\AgdaSymbol{→} \AgdaBound{Γ} \AgdaFunction{⊆} \AgdaBound{Γ'}\<%
\\
\>[2]\AgdaIndent{4}{}\<[4]%
\>[4]\AgdaComment{-------------------------------}\<%
\\
\>[4]\AgdaIndent{8}{}\<[8]%
\>[8]\AgdaSymbol{→} \AgdaBound{Δ} \AgdaDatatype{/} \AgdaBound{Γ'} \AgdaDatatype{⊢} \AgdaBound{J} \AgdaDatatype{::} \AgdaBound{A}\<%
\\
\\
\>\AgdaFunction{weak} \AgdaSymbol{(}\AgdaInductiveConstructor{iPCF-var} \AgdaBound{x}\AgdaSymbol{)} \AgdaBound{f} \AgdaSymbol{=} \AgdaInductiveConstructor{iPCF-var} \AgdaSymbol{(}\AgdaBound{f} \AgdaBound{x}\AgdaSymbol{)}\<%
\\
\>\AgdaFunction{weak} \AgdaSymbol{(}\AgdaInductiveConstructor{iPCF-modal-var} \AgdaBound{x}\AgdaSymbol{)} \AgdaBound{f} \AgdaSymbol{=} \AgdaInductiveConstructor{iPCF-modal-var} \AgdaBound{x}\<%
\\
\>\AgdaFunction{weak} \AgdaSymbol{(}\AgdaInductiveConstructor{iPCF-app} \AgdaBound{d} \AgdaBound{e}\AgdaSymbol{)} \AgdaBound{f} \AgdaSymbol{=} \AgdaInductiveConstructor{iPCF-app} \AgdaSymbol{(}\AgdaFunction{weak} \AgdaBound{d} \AgdaBound{f}\AgdaSymbol{)} \AgdaSymbol{(}\AgdaFunction{weak} \AgdaBound{e} \AgdaBound{f}\AgdaSymbol{)}\<%
\\
\>\AgdaFunction{weak} \AgdaSymbol{(}\AgdaInductiveConstructor{iPCF-lam-int} \AgdaBound{d}\AgdaSymbol{)} \AgdaBound{f} \AgdaSymbol{=} \AgdaInductiveConstructor{iPCF-lam-int} \AgdaBound{d}\<%
\\
\>\AgdaFunction{weak} \AgdaSymbol{(}\AgdaInductiveConstructor{iPCF-lam-ext} \AgdaBound{d}\AgdaSymbol{)} \AgdaBound{f} \AgdaSymbol{=} \AgdaInductiveConstructor{iPCF-lam-ext} \AgdaSymbol{(}\AgdaFunction{weak} \AgdaBound{d} \AgdaSymbol{(}\AgdaFunction{weakboth} \AgdaBound{f}\AgdaSymbol{))}\<%
\\
\>\AgdaFunction{weak} \AgdaSymbol{(}\AgdaInductiveConstructor{iPCF-boxI} \AgdaBound{d}\AgdaSymbol{)} \AgdaBound{f} \AgdaSymbol{=} \AgdaInductiveConstructor{iPCF-boxI} \AgdaBound{d}\<%
\\
\>\AgdaFunction{weak} \AgdaSymbol{(}\AgdaInductiveConstructor{iPCF-boxE} \AgdaBound{d} \AgdaBound{e}\AgdaSymbol{)} \AgdaBound{f} \AgdaSymbol{=}\<%
\\
\>[0]\AgdaIndent{2}{}\<[2]%
\>[2]\AgdaInductiveConstructor{iPCF-boxE} \AgdaSymbol{(}\AgdaFunction{weak} \AgdaBound{d} \AgdaBound{f}\AgdaSymbol{)} \AgdaSymbol{(}\AgdaFunction{weak} \AgdaBound{e} \AgdaBound{f}\AgdaSymbol{)}\<%
\\
\>\AgdaFunction{weak} \AgdaSymbol{(}\AgdaInductiveConstructor{iPCF-fix} \AgdaBound{d}\AgdaSymbol{)} \AgdaBound{f} \AgdaSymbol{=} \AgdaInductiveConstructor{iPCF-fix} \AgdaBound{d}\<%
\\
\\
\\
\>\AgdaFunction{weak-modal} \AgdaSymbol{:} \AgdaSymbol{∀} \AgdaSymbol{\{}\AgdaBound{Δ} \AgdaBound{Δ'} \AgdaBound{Γ} \AgdaBound{J} \AgdaBound{A}\AgdaSymbol{\}}\<%
\\
\\
\>[2]\AgdaIndent{3}{}\<[3]%
\>[3]\AgdaSymbol{→} \AgdaBound{Δ} \AgdaDatatype{/} \AgdaBound{Γ} \AgdaDatatype{⊢} \AgdaBound{J} \AgdaDatatype{::} \AgdaBound{A} \<[23]%
\>[23]\AgdaSymbol{→} \AgdaBound{Δ} \AgdaFunction{⊆} \AgdaBound{Δ'}\<%
\\
\>[2]\AgdaIndent{3}{}\<[3]%
\>[3]\AgdaComment{-------------------------}\<%
\\
\>[3]\AgdaIndent{9}{}\<[9]%
\>[9]\AgdaSymbol{→} \AgdaBound{Δ'} \AgdaDatatype{/} \AgdaBound{Γ} \AgdaDatatype{⊢} \AgdaBound{J} \AgdaDatatype{::} \AgdaBound{A}\<%
\\
\\
\>\AgdaFunction{weak-modal} \AgdaSymbol{(}\AgdaInductiveConstructor{iPCF-var} \AgdaBound{p}\AgdaSymbol{)} \AgdaBound{x} \AgdaSymbol{=} \AgdaInductiveConstructor{iPCF-var} \AgdaBound{p}\<%
\\
\>\AgdaFunction{weak-modal} \AgdaSymbol{(}\AgdaInductiveConstructor{iPCF-modal-var} \AgdaBound{p}\AgdaSymbol{)} \AgdaBound{x} \AgdaSymbol{=} \AgdaInductiveConstructor{iPCF-modal-var} \AgdaSymbol{(}\AgdaBound{x} \AgdaBound{p}\AgdaSymbol{)}\<%
\\
\>\AgdaFunction{weak-modal} \AgdaSymbol{(}\AgdaInductiveConstructor{iPCF-app} \AgdaBound{t} \AgdaBound{u}\AgdaSymbol{)} \AgdaBound{x} \AgdaSymbol{=} \AgdaInductiveConstructor{iPCF-app} \AgdaSymbol{(}\AgdaFunction{weak-modal} \AgdaBound{t} \AgdaBound{x}\AgdaSymbol{)}\<%
\\
\>[9]\AgdaIndent{37}{}\<[37]%
\>[37]\AgdaSymbol{(}\AgdaFunction{weak-modal} \AgdaBound{u} \AgdaBound{x}\AgdaSymbol{)}\<%
\\
\>\AgdaFunction{weak-modal} \AgdaSymbol{(}\AgdaInductiveConstructor{iPCF-lam-int} \AgdaBound{t}\AgdaSymbol{)} \AgdaBound{x} \AgdaSymbol{=} \AgdaInductiveConstructor{iPCF-lam-int} \AgdaBound{t}\<%
\\
\>\AgdaFunction{weak-modal} \AgdaSymbol{(}\AgdaInductiveConstructor{iPCF-lam-ext} \AgdaBound{t}\AgdaSymbol{)} \AgdaBound{x} \AgdaSymbol{=} \AgdaInductiveConstructor{iPCF-lam-ext} \AgdaSymbol{(}\AgdaFunction{weak-modal} \AgdaBound{t} \AgdaBound{x}\AgdaSymbol{)}\<%
\\
\>\AgdaFunction{weak-modal} \AgdaSymbol{(}\AgdaInductiveConstructor{iPCF-boxI} \AgdaBound{t}\AgdaSymbol{)} \AgdaBound{x} \AgdaSymbol{=} \AgdaInductiveConstructor{iPCF-boxI} \AgdaSymbol{(}\AgdaFunction{weak-modal} \AgdaBound{t} \AgdaBound{x}\AgdaSymbol{)}\<%
\\
\>\AgdaFunction{weak-modal} \AgdaSymbol{(}\AgdaInductiveConstructor{iPCF-boxE} \AgdaBound{t} \AgdaBound{u}\AgdaSymbol{)} \AgdaBound{x} \AgdaSymbol{=}\<%
\\
\>[0]\AgdaIndent{2}{}\<[2]%
\>[2]\AgdaInductiveConstructor{iPCF-boxE} \AgdaSymbol{(}\AgdaFunction{weak-modal} \AgdaBound{t} \AgdaBound{x}\AgdaSymbol{)}\<%
\\
\>[2]\AgdaIndent{11}{}\<[11]%
\>[11]\AgdaSymbol{(}\AgdaFunction{weak-modal} \AgdaBound{u} \AgdaSymbol{(}\AgdaFunction{weakboth} \AgdaBound{x}\AgdaSymbol{))}\<%
\\
\>\AgdaFunction{weak-modal} \AgdaSymbol{(}\AgdaInductiveConstructor{iPCF-fix} \AgdaBound{f}\AgdaSymbol{)} \AgdaBound{x} \AgdaSymbol{=} \AgdaInductiveConstructor{iPCF-fix} \AgdaBound{f}\<%
\\
\\
\\
\>\AgdaComment{-- Including intensional into extensional.}\<%
\\
\\
\>\AgdaFunction{incl} \AgdaSymbol{:} \AgdaSymbol{∀} \AgdaSymbol{\{}\AgdaBound{Δ} \AgdaBound{Γ} \AgdaBound{A}\AgdaSymbol{\}}\<%
\\
\\
\>[0]\AgdaIndent{4}{}\<[4]%
\>[4]\AgdaSymbol{→} \AgdaBound{Δ} \AgdaDatatype{/} \AgdaBound{Γ} \AgdaDatatype{⊢} \AgdaInductiveConstructor{int} \AgdaDatatype{::} \AgdaBound{A}\<%
\\
\>[0]\AgdaIndent{4}{}\<[4]%
\>[4]\AgdaComment{-------------------}\<%
\\
\>[0]\AgdaIndent{4}{}\<[4]%
\>[4]\AgdaSymbol{→} \AgdaBound{Δ} \AgdaDatatype{/} \AgdaBound{Γ} \AgdaDatatype{⊢} \AgdaInductiveConstructor{ext} \AgdaDatatype{::} \AgdaBound{A}\<%
\\
\\
\>\AgdaFunction{incl} \AgdaSymbol{(}\AgdaInductiveConstructor{iPCF-var} \AgdaBound{x}\AgdaSymbol{)} \AgdaSymbol{=} \AgdaInductiveConstructor{iPCF-var} \AgdaBound{x}\<%
\\
\>\AgdaFunction{incl} \AgdaSymbol{(}\AgdaInductiveConstructor{iPCF-modal-var} \AgdaBound{x}\AgdaSymbol{)} \AgdaSymbol{=} \AgdaInductiveConstructor{iPCF-modal-var} \AgdaBound{x}\<%
\\
\>\AgdaFunction{incl} \AgdaSymbol{(}\AgdaInductiveConstructor{iPCF-app} \AgdaBound{d} \AgdaBound{e}\AgdaSymbol{)} \AgdaSymbol{=} \AgdaInductiveConstructor{iPCF-app} \AgdaSymbol{(}\AgdaFunction{incl} \AgdaBound{d}\AgdaSymbol{)} \AgdaSymbol{(}\AgdaFunction{incl} \AgdaBound{e}\AgdaSymbol{)}\<%
\\
\>\AgdaFunction{incl} \AgdaSymbol{(}\AgdaInductiveConstructor{iPCF-lam-int} \AgdaSymbol{\{}\AgdaInductiveConstructor{int}\AgdaSymbol{\}} \AgdaBound{d}\AgdaSymbol{)} \AgdaSymbol{=}\<%
\\
\>[0]\AgdaIndent{2}{}\<[2]%
\>[2]\AgdaInductiveConstructor{iPCF-lam-ext} \AgdaSymbol{(}\AgdaFunction{weak} \AgdaSymbol{(}\AgdaFunction{weak-modal} \AgdaSymbol{(}\AgdaFunction{incl} \AgdaBound{d}\AgdaSymbol{)} \AgdaFunction{subsetempty}\AgdaSymbol{)} \AgdaSymbol{(}\AgdaFunction{weakboth} \AgdaFunction{subsetempty}\AgdaSymbol{))}\<%
\\
\>\AgdaFunction{incl} \AgdaSymbol{(}\AgdaInductiveConstructor{iPCF-lam-int} \AgdaSymbol{\{}\AgdaInductiveConstructor{ext}\AgdaSymbol{\}} \AgdaBound{d}\AgdaSymbol{)} \AgdaSymbol{=}\<%
\\
\>[0]\AgdaIndent{2}{}\<[2]%
\>[2]\AgdaInductiveConstructor{iPCF-lam-ext} \AgdaSymbol{(}\AgdaFunction{weak} \AgdaSymbol{(}\AgdaFunction{weak-modal} \AgdaBound{d} \AgdaFunction{subsetempty}\AgdaSymbol{)} \AgdaSymbol{(}\AgdaFunction{weakboth} \AgdaFunction{subsetempty}\AgdaSymbol{))}\<%
\\
\>\AgdaFunction{incl} \AgdaSymbol{(}\AgdaInductiveConstructor{iPCF-boxI} \AgdaBound{d}\AgdaSymbol{)} \AgdaSymbol{=} \AgdaInductiveConstructor{iPCF-boxI} \AgdaBound{d}\<%
\\
\>\AgdaFunction{incl} \AgdaSymbol{(}\AgdaInductiveConstructor{iPCF-boxE} \AgdaBound{d} \AgdaBound{e}\AgdaSymbol{)} \AgdaSymbol{=} \AgdaInductiveConstructor{iPCF-boxE} \AgdaSymbol{(}\AgdaFunction{incl} \AgdaBound{d}\AgdaSymbol{)} \AgdaSymbol{(}\AgdaFunction{incl} \AgdaBound{e}\AgdaSymbol{)}\<%
\\
\>\AgdaFunction{incl} \AgdaSymbol{(}\AgdaInductiveConstructor{iPCF-fix} \AgdaBound{f}\AgdaSymbol{)} \AgdaSymbol{=} \AgdaInductiveConstructor{iPCF-fix} \AgdaBound{f}\<%
\\
\\
\\
\>\AgdaFunction{incl-either-ext} \AgdaSymbol{:} \AgdaSymbol{∀} \AgdaSymbol{\{}\AgdaBound{Δ} \AgdaBound{J} \AgdaBound{Γ} \AgdaBound{A}\AgdaSymbol{\}}\<%
\\
\\
\>[2]\AgdaIndent{4}{}\<[4]%
\>[4]\AgdaSymbol{→} \AgdaBound{Δ} \AgdaDatatype{/} \AgdaBound{Γ} \AgdaDatatype{⊢} \AgdaBound{J} \AgdaDatatype{::} \AgdaBound{A}\<%
\\
\>[2]\AgdaIndent{4}{}\<[4]%
\>[4]\AgdaComment{-------------------}\<%
\\
\>[2]\AgdaIndent{4}{}\<[4]%
\>[4]\AgdaSymbol{→} \AgdaBound{Δ} \AgdaDatatype{/} \AgdaBound{Γ} \AgdaDatatype{⊢} \AgdaInductiveConstructor{ext} \AgdaDatatype{::} \AgdaBound{A}\<%
\\
\\
\>\AgdaFunction{incl-either-ext} \AgdaSymbol{\{}\AgdaArgument{J} \AgdaSymbol{=} \AgdaInductiveConstructor{int}\AgdaSymbol{\}} \AgdaBound{d} \AgdaSymbol{=} \AgdaFunction{incl} \AgdaBound{d}\<%
\\
\>\AgdaFunction{incl-either-ext} \AgdaSymbol{\{}\AgdaArgument{J} \AgdaSymbol{=} \AgdaInductiveConstructor{ext}\AgdaSymbol{\}} \AgdaBound{d} \AgdaSymbol{=} \AgdaBound{d}\<%
\\
\\
\\
\>\AgdaFunction{incl-either-int} \AgdaSymbol{:} \AgdaSymbol{∀} \AgdaSymbol{\{}\AgdaBound{Δ} \AgdaBound{J} \AgdaBound{Γ} \AgdaBound{A}\AgdaSymbol{\}}\<%
\\
\\
\>[2]\AgdaIndent{4}{}\<[4]%
\>[4]\AgdaSymbol{→} \AgdaBound{Δ} \AgdaDatatype{/} \AgdaBound{Γ} \AgdaDatatype{⊢} \AgdaInductiveConstructor{int} \AgdaDatatype{::} \AgdaBound{A}\<%
\\
\>[2]\AgdaIndent{4}{}\<[4]%
\>[4]\AgdaComment{-------------------}\<%
\\
\>[2]\AgdaIndent{4}{}\<[4]%
\>[4]\AgdaSymbol{→} \AgdaBound{Δ} \AgdaDatatype{/} \AgdaBound{Γ} \AgdaDatatype{⊢} \AgdaBound{J} \AgdaDatatype{::} \AgdaBound{A}\<%
\\
\\
\>\AgdaFunction{incl-either-int} \AgdaSymbol{\{}\AgdaArgument{J} \AgdaSymbol{=} \AgdaInductiveConstructor{int}\AgdaSymbol{\}} \AgdaBound{d} \AgdaSymbol{=} \AgdaBound{d}\<%
\\
\>\AgdaFunction{incl-either-int} \AgdaSymbol{\{}\AgdaArgument{J} \AgdaSymbol{=} \AgdaInductiveConstructor{ext}\AgdaSymbol{\}} \AgdaBound{d} \AgdaSymbol{=} \AgdaFunction{incl} \AgdaBound{d}\<%
\\
\\
\>\AgdaComment{-- Cut.}\<%
\\
\\
\\
\>\AgdaFunction{cut-ext} \AgdaSymbol{:} \AgdaSymbol{∀} \AgdaSymbol{\{}\AgdaBound{Δ} \AgdaBound{Γ} \AgdaBound{J} \AgdaBound{A} \AgdaBound{B}\AgdaSymbol{\}} \AgdaSymbol{→} \AgdaSymbol{(}\AgdaBound{Γ'} \AgdaSymbol{:} \AgdaDatatype{Cx} \AgdaInductiveConstructor{modal}\AgdaSymbol{)}\<%
\\
\\
\>[2]\AgdaIndent{4}{}\<[4]%
\>[4]\AgdaSymbol{→} \AgdaBound{Δ} \AgdaDatatype{/} \AgdaBound{Γ} \AgdaDatatype{⊢} \AgdaInductiveConstructor{ext} \AgdaDatatype{::} \AgdaBound{A} \<[26]%
\>[26]\AgdaSymbol{→} \AgdaBound{Δ} \AgdaDatatype{/} \AgdaBound{Γ} \AgdaInductiveConstructor{,} \AgdaBound{A} \AgdaFunction{++} \AgdaBound{Γ'} \AgdaDatatype{⊢} \AgdaBound{J} \AgdaDatatype{::} \AgdaBound{B}\<%
\\
\>[2]\AgdaIndent{4}{}\<[4]%
\>[4]\AgdaComment{------------------------------------------------}\<%
\\
\>[4]\AgdaIndent{14}{}\<[14]%
\>[14]\AgdaSymbol{→} \AgdaBound{Δ} \AgdaDatatype{/} \AgdaBound{Γ} \AgdaFunction{++} \AgdaBound{Γ'} \AgdaDatatype{⊢} \AgdaInductiveConstructor{ext} \AgdaDatatype{::} \AgdaBound{B}\<%
\\
\\
\>\AgdaFunction{cut-ext} \AgdaInductiveConstructor{·} \AgdaBound{d} \AgdaSymbol{(}\AgdaInductiveConstructor{iPCF-var} \AgdaInductiveConstructor{top}\AgdaSymbol{)} \AgdaSymbol{=} \AgdaBound{d}\<%
\\
\>\AgdaFunction{cut-ext} \AgdaInductiveConstructor{·} \AgdaBound{d} \AgdaSymbol{(}\AgdaInductiveConstructor{iPCF-var} \AgdaSymbol{(}\AgdaInductiveConstructor{pop} \AgdaBound{x}\AgdaSymbol{))} \AgdaSymbol{=} \AgdaInductiveConstructor{iPCF-var} \AgdaBound{x}\<%
\\
\>\AgdaFunction{cut-ext} \AgdaSymbol{(}\AgdaBound{Γ'} \AgdaInductiveConstructor{,} \AgdaBound{B}\AgdaSymbol{)} \AgdaBound{d} \AgdaSymbol{(}\AgdaInductiveConstructor{iPCF-var} \AgdaInductiveConstructor{top}\AgdaSymbol{)} \AgdaSymbol{=} \AgdaInductiveConstructor{iPCF-var} \AgdaInductiveConstructor{top}\<%
\\
\>\AgdaFunction{cut-ext} \AgdaSymbol{(}\AgdaBound{Γ'} \AgdaInductiveConstructor{,} \AgdaBound{A'}\AgdaSymbol{)} \AgdaBound{d} \AgdaSymbol{(}\AgdaInductiveConstructor{iPCF-var} \AgdaSymbol{(}\AgdaInductiveConstructor{pop} \AgdaBound{x}\AgdaSymbol{))} \AgdaSymbol{=}\<%
\\
\>[0]\AgdaIndent{2}{}\<[2]%
\>[2]\AgdaFunction{weak} \AgdaSymbol{(}\AgdaFunction{cut-ext} \AgdaSymbol{\{}\AgdaArgument{J} \AgdaSymbol{=} \AgdaInductiveConstructor{ext}\AgdaSymbol{\}} \AgdaBound{Γ'} \AgdaBound{d} \AgdaSymbol{(}\AgdaInductiveConstructor{iPCF-var} \AgdaBound{x}\AgdaSymbol{))} \AgdaSymbol{(}\AgdaFunction{weakone} \AgdaFunction{subsetid}\AgdaSymbol{)}\<%
\\
\>\AgdaFunction{cut-ext} \AgdaBound{Γ'} \AgdaBound{d} \AgdaSymbol{(}\AgdaInductiveConstructor{iPCF-modal-var} \AgdaBound{p}\AgdaSymbol{)} \AgdaSymbol{=} \AgdaInductiveConstructor{iPCF-modal-var} \AgdaBound{p}\<%
\\
\>\AgdaFunction{cut-ext} \AgdaBound{Γ'} \AgdaBound{d} \AgdaSymbol{(}\AgdaInductiveConstructor{iPCF-app} \AgdaBound{t} \AgdaBound{u}\AgdaSymbol{)} \AgdaSymbol{=} \AgdaInductiveConstructor{iPCF-app} \AgdaSymbol{(}\AgdaFunction{cut-ext} \AgdaBound{Γ'} \AgdaBound{d} \AgdaBound{t}\AgdaSymbol{)} \AgdaSymbol{(}\AgdaFunction{cut-ext} \AgdaBound{Γ'} \AgdaBound{d} \AgdaBound{u}\AgdaSymbol{)}\<%
\\
\>\AgdaFunction{cut-ext} \AgdaBound{Γ'} \AgdaBound{d} \AgdaSymbol{(}\AgdaInductiveConstructor{iPCF-lam-int} \AgdaBound{e}\AgdaSymbol{)} \AgdaSymbol{=} \AgdaFunction{incl} \AgdaSymbol{(}\AgdaInductiveConstructor{iPCF-lam-int} \AgdaBound{e}\AgdaSymbol{)}\<%
\\
\>\AgdaFunction{cut-ext} \AgdaBound{Γ'} \AgdaBound{d} \AgdaSymbol{(}\AgdaInductiveConstructor{iPCF-lam-ext} \AgdaBound{e}\AgdaSymbol{)} \AgdaSymbol{=} \AgdaInductiveConstructor{iPCF-lam-ext} \AgdaSymbol{(}\AgdaFunction{cut-ext} \AgdaSymbol{(}\AgdaBound{Γ'} \AgdaInductiveConstructor{,} \AgdaSymbol{\_)} \AgdaBound{d} \AgdaBound{e}\AgdaSymbol{)}\<%
\\
\>\AgdaFunction{cut-ext} \AgdaBound{Γ'} \AgdaBound{d} \AgdaSymbol{(}\AgdaInductiveConstructor{iPCF-boxI} \AgdaBound{e}\AgdaSymbol{)} \AgdaSymbol{=} \AgdaInductiveConstructor{iPCF-boxI} \AgdaBound{e}\<%
\\
\>\AgdaFunction{cut-ext} \AgdaBound{Γ'} \AgdaBound{d} \AgdaSymbol{(}\AgdaInductiveConstructor{iPCF-boxE} \AgdaBound{t} \AgdaBound{u}\AgdaSymbol{)} \AgdaSymbol{=}\<%
\\
\>[0]\AgdaIndent{2}{}\<[2]%
\>[2]\AgdaInductiveConstructor{iPCF-boxE} \AgdaSymbol{(}\AgdaFunction{cut-ext} \AgdaBound{Γ'} \AgdaBound{d} \AgdaBound{t}\AgdaSymbol{)}\<%
\\
\>[2]\AgdaIndent{12}{}\<[12]%
\>[12]\AgdaSymbol{(}\AgdaFunction{cut-ext} \AgdaBound{Γ'} \AgdaSymbol{(}\AgdaFunction{weak-modal} \AgdaBound{d} \AgdaSymbol{(}\AgdaFunction{weakone} \AgdaSymbol{(}\AgdaFunction{subsetid}\AgdaSymbol{)))} \AgdaBound{u}\AgdaSymbol{)}\<%
\\
\>\AgdaFunction{cut-ext} \AgdaBound{Γ'} \AgdaBound{d} \AgdaSymbol{(}\AgdaInductiveConstructor{iPCF-fix} \AgdaBound{f}\AgdaSymbol{)} \AgdaSymbol{=} \AgdaInductiveConstructor{iPCF-fix} \AgdaBound{f} \<[39]%
\>[39]\<%
\\
\\
\\
\>\AgdaFunction{cut-int} \AgdaSymbol{:} \AgdaSymbol{∀} \AgdaSymbol{\{}\AgdaBound{Δ} \AgdaBound{Γ} \AgdaBound{J} \AgdaBound{A} \AgdaBound{B}\AgdaSymbol{\}} \AgdaSymbol{→} \AgdaSymbol{(}\AgdaBound{Γ'} \AgdaSymbol{:} \AgdaDatatype{Cx} \AgdaInductiveConstructor{modal}\AgdaSymbol{)}\<%
\\
\\
\>[0]\AgdaIndent{4}{}\<[4]%
\>[4]\AgdaSymbol{→} \AgdaBound{Δ} \AgdaDatatype{/} \AgdaBound{Γ} \AgdaDatatype{⊢} \AgdaInductiveConstructor{int} \AgdaDatatype{::} \AgdaBound{A} \<[26]%
\>[26]\AgdaSymbol{→} \AgdaBound{Δ} \AgdaDatatype{/} \AgdaBound{Γ} \AgdaInductiveConstructor{,} \AgdaBound{A} \AgdaFunction{++} \AgdaBound{Γ'} \AgdaDatatype{⊢} \AgdaBound{J} \AgdaDatatype{::} \AgdaBound{B}\<%
\\
\>[0]\AgdaIndent{4}{}\<[4]%
\>[4]\AgdaComment{------------------------------------------------}\<%
\\
\>[4]\AgdaIndent{14}{}\<[14]%
\>[14]\AgdaSymbol{→} \AgdaBound{Δ} \AgdaDatatype{/} \AgdaBound{Γ} \AgdaFunction{++} \AgdaBound{Γ'} \AgdaDatatype{⊢} \AgdaBound{J} \AgdaDatatype{::} \AgdaBound{B}\<%
\\
\\
\>\AgdaFunction{cut-int} \AgdaInductiveConstructor{·} \AgdaBound{d} \AgdaSymbol{(}\AgdaInductiveConstructor{iPCF-var} \AgdaInductiveConstructor{top}\AgdaSymbol{)} \AgdaSymbol{=} \AgdaFunction{incl-either-int} \AgdaBound{d}\<%
\\
\>\AgdaFunction{cut-int} \AgdaInductiveConstructor{·} \AgdaBound{d} \AgdaSymbol{(}\AgdaInductiveConstructor{iPCF-var} \AgdaSymbol{(}\AgdaInductiveConstructor{pop} \AgdaBound{x}\AgdaSymbol{))} \AgdaSymbol{=} \AgdaInductiveConstructor{iPCF-var} \AgdaBound{x}\<%
\\
\>\AgdaFunction{cut-int} \AgdaSymbol{(}\AgdaBound{Γ'} \AgdaInductiveConstructor{,} \AgdaBound{B}\AgdaSymbol{)} \AgdaBound{d} \AgdaSymbol{(}\AgdaInductiveConstructor{iPCF-var} \AgdaInductiveConstructor{top}\AgdaSymbol{)} \AgdaSymbol{=} \AgdaInductiveConstructor{iPCF-var} \AgdaInductiveConstructor{top}\<%
\\
\>\AgdaFunction{cut-int} \AgdaSymbol{(}\AgdaBound{Γ'} \AgdaInductiveConstructor{,} \AgdaBound{A'}\AgdaSymbol{)} \AgdaBound{d} \AgdaSymbol{(}\AgdaInductiveConstructor{iPCF-var} \AgdaSymbol{(}\AgdaInductiveConstructor{pop} \AgdaBound{x}\AgdaSymbol{))} \AgdaSymbol{=}\<%
\\
\>[0]\AgdaIndent{2}{}\<[2]%
\>[2]\AgdaFunction{weak} \AgdaSymbol{(}\AgdaFunction{cut-int} \AgdaBound{Γ'} \AgdaBound{d} \AgdaSymbol{(}\AgdaInductiveConstructor{iPCF-var} \AgdaBound{x}\AgdaSymbol{))} \AgdaSymbol{(}\AgdaFunction{weakone} \AgdaFunction{subsetid}\AgdaSymbol{)}\<%
\\
\>\AgdaFunction{cut-int} \AgdaBound{Γ'} \AgdaBound{d} \AgdaSymbol{(}\AgdaInductiveConstructor{iPCF-modal-var} \AgdaBound{p}\AgdaSymbol{)} \AgdaSymbol{=} \AgdaInductiveConstructor{iPCF-modal-var} \AgdaBound{p}\<%
\\
\>\AgdaFunction{cut-int} \AgdaBound{Γ'} \AgdaBound{d} \AgdaSymbol{(}\AgdaInductiveConstructor{iPCF-app} \AgdaBound{t} \AgdaBound{u}\AgdaSymbol{)} \AgdaSymbol{=} \AgdaInductiveConstructor{iPCF-app} \AgdaSymbol{(}\AgdaFunction{cut-int} \AgdaBound{Γ'} \AgdaBound{d} \AgdaBound{t}\AgdaSymbol{)} \AgdaSymbol{(}\AgdaFunction{cut-int} \AgdaBound{Γ'} \AgdaBound{d} \AgdaBound{u}\AgdaSymbol{)}\<%
\\
\>\AgdaFunction{cut-int} \AgdaBound{Γ'} \AgdaBound{d} \AgdaSymbol{(}\AgdaInductiveConstructor{iPCF-lam-int} \AgdaBound{e}\AgdaSymbol{)} \AgdaSymbol{=} \AgdaInductiveConstructor{iPCF-lam-int} \AgdaBound{e}\<%
\\
\>\AgdaFunction{cut-int} \AgdaBound{Γ'} \AgdaBound{d} \AgdaSymbol{(}\AgdaInductiveConstructor{iPCF-lam-ext} \AgdaBound{e}\AgdaSymbol{)} \AgdaSymbol{=} \AgdaInductiveConstructor{iPCF-lam-ext} \AgdaSymbol{(}\AgdaFunction{cut-ext} \AgdaSymbol{(}\AgdaBound{Γ'} \AgdaInductiveConstructor{,} \AgdaSymbol{\_)} \AgdaSymbol{(}\AgdaFunction{incl} \AgdaBound{d}\AgdaSymbol{)} \AgdaBound{e}\AgdaSymbol{)}\<%
\\
\>\AgdaFunction{cut-int} \AgdaBound{Γ'} \AgdaBound{d} \AgdaSymbol{(}\AgdaInductiveConstructor{iPCF-boxI} \AgdaBound{e}\AgdaSymbol{)} \AgdaSymbol{=} \AgdaInductiveConstructor{iPCF-boxI} \AgdaBound{e}\<%
\\
\>\AgdaFunction{cut-int} \AgdaBound{Γ'} \AgdaBound{d} \AgdaSymbol{(}\AgdaInductiveConstructor{iPCF-boxE} \AgdaBound{t} \AgdaBound{u}\AgdaSymbol{)} \AgdaSymbol{=}\<%
\\
\>[0]\AgdaIndent{2}{}\<[2]%
\>[2]\AgdaInductiveConstructor{iPCF-boxE} \AgdaSymbol{(}\AgdaFunction{cut-int} \AgdaBound{Γ'} \AgdaBound{d} \AgdaBound{t}\AgdaSymbol{)}\<%
\\
\>[2]\AgdaIndent{12}{}\<[12]%
\>[12]\AgdaSymbol{(}\AgdaFunction{cut-int} \AgdaBound{Γ'} \AgdaSymbol{(}\AgdaFunction{weak-modal} \AgdaBound{d} \AgdaSymbol{(}\AgdaFunction{weakone} \AgdaSymbol{(}\AgdaFunction{subsetid}\AgdaSymbol{)))} \AgdaBound{u}\AgdaSymbol{)}\<%
\\
\>\AgdaFunction{cut-int} \AgdaBound{Γ'} \AgdaBound{d} \AgdaSymbol{(}\AgdaInductiveConstructor{iPCF-fix} \AgdaBound{e}\AgdaSymbol{)} \AgdaSymbol{=} \AgdaInductiveConstructor{iPCF-fix} \AgdaBound{e}\<%
\\
\\
\\
\>\AgdaFunction{cut-modal} \AgdaSymbol{:} \AgdaSymbol{∀} \AgdaSymbol{\{}\AgdaBound{Δ} \AgdaBound{Γ} \AgdaBound{J} \AgdaBound{A} \AgdaBound{B}\AgdaSymbol{\}} \AgdaSymbol{→} \AgdaSymbol{(}\AgdaBound{Δ'} \AgdaSymbol{:} \AgdaDatatype{Cx} \AgdaInductiveConstructor{modal}\AgdaSymbol{)}\<%
\\
\\
\>[0]\AgdaIndent{4}{}\<[4]%
\>[4]\AgdaSymbol{→} \AgdaBound{Δ} \AgdaDatatype{/} \AgdaInductiveConstructor{·} \AgdaDatatype{⊢} \AgdaInductiveConstructor{int} \AgdaDatatype{::} \AgdaBound{A} \<[26]%
\>[26]\AgdaSymbol{→} \AgdaBound{Δ} \AgdaInductiveConstructor{,} \AgdaBound{A} \AgdaFunction{++} \AgdaBound{Δ'} \AgdaDatatype{/} \AgdaBound{Γ} \<[45]%
\>[45]\AgdaDatatype{⊢} \AgdaBound{J} \AgdaDatatype{::} \AgdaBound{B}\<%
\\
\>[0]\AgdaIndent{4}{}\<[4]%
\>[4]\AgdaComment{--------------------------------------------------}\<%
\\
\>[4]\AgdaIndent{13}{}\<[13]%
\>[13]\AgdaSymbol{→} \AgdaBound{Δ} \AgdaFunction{++} \AgdaBound{Δ'} \AgdaDatatype{/} \AgdaBound{Γ} \AgdaDatatype{⊢} \AgdaBound{J} \AgdaDatatype{::} \AgdaBound{B}\<%
\\
\\
\>\AgdaFunction{cut-modal} \AgdaBound{Δ'} \AgdaBound{d} \AgdaSymbol{(}\AgdaInductiveConstructor{iPCF-var} \AgdaBound{x}\AgdaSymbol{)} \AgdaSymbol{=} \AgdaInductiveConstructor{iPCF-var} \AgdaBound{x}\<%
\\
\>\AgdaFunction{cut-modal} \AgdaInductiveConstructor{·} \AgdaBound{d} \AgdaSymbol{(}\AgdaInductiveConstructor{iPCF-modal-var} \AgdaInductiveConstructor{top}\AgdaSymbol{)} \AgdaSymbol{=} \AgdaFunction{incl-either-int} \AgdaSymbol{(}\AgdaFunction{weak} \AgdaBound{d} \AgdaFunction{subsetempty}\AgdaSymbol{)}\<%
\\
\>\AgdaFunction{cut-modal} \AgdaInductiveConstructor{·} \AgdaBound{d} \AgdaSymbol{(}\AgdaInductiveConstructor{iPCF-modal-var} \AgdaSymbol{(}\AgdaInductiveConstructor{pop} \AgdaBound{x}\AgdaSymbol{))} \AgdaSymbol{=} \AgdaInductiveConstructor{iPCF-modal-var} \AgdaBound{x}\<%
\\
\>\AgdaFunction{cut-modal} \AgdaSymbol{(}\AgdaBound{Δ'} \AgdaInductiveConstructor{,} \AgdaBound{B}\AgdaSymbol{)} \AgdaBound{d} \AgdaSymbol{(}\AgdaInductiveConstructor{iPCF-modal-var} \AgdaInductiveConstructor{top}\AgdaSymbol{)} \AgdaSymbol{=} \AgdaInductiveConstructor{iPCF-modal-var} \AgdaInductiveConstructor{top}\<%
\\
\>\AgdaFunction{cut-modal} \AgdaSymbol{(}\AgdaBound{Δ'} \AgdaInductiveConstructor{,} \AgdaBound{A'}\AgdaSymbol{)} \AgdaBound{d} \AgdaSymbol{(}\AgdaInductiveConstructor{iPCF-modal-var} \AgdaSymbol{(}\AgdaInductiveConstructor{pop} \AgdaBound{x}\AgdaSymbol{))} \AgdaSymbol{=}\<%
\\
\>[0]\AgdaIndent{2}{}\<[2]%
\>[2]\AgdaFunction{weak-modal} \AgdaSymbol{(}\AgdaFunction{cut-modal} \AgdaBound{Δ'} \AgdaBound{d} \AgdaSymbol{(}\AgdaInductiveConstructor{iPCF-modal-var} \AgdaBound{x}\AgdaSymbol{))} \AgdaSymbol{(}\AgdaFunction{weakone} \AgdaFunction{subsetid}\AgdaSymbol{)}\<%
\\
\>\AgdaFunction{cut-modal} \AgdaBound{Δ'} \AgdaBound{d} \AgdaSymbol{(}\AgdaInductiveConstructor{iPCF-app} \AgdaBound{p} \AgdaBound{q}\AgdaSymbol{)} \AgdaSymbol{=}\<%
\\
\>[0]\AgdaIndent{2}{}\<[2]%
\>[2]\AgdaInductiveConstructor{iPCF-app} \AgdaSymbol{(}\AgdaFunction{cut-modal} \AgdaBound{Δ'} \AgdaBound{d} \AgdaBound{p}\AgdaSymbol{)} \AgdaSymbol{(}\AgdaFunction{cut-modal} \AgdaBound{Δ'} \AgdaBound{d} \AgdaBound{q}\AgdaSymbol{)}\<%
\\
\>\AgdaFunction{cut-modal} \AgdaBound{Δ'} \AgdaBound{d} \AgdaSymbol{(}\AgdaInductiveConstructor{iPCF-lam-ext} \AgdaBound{e}\AgdaSymbol{)} \AgdaSymbol{=} \AgdaInductiveConstructor{iPCF-lam-ext} \AgdaSymbol{(}\AgdaFunction{cut-modal} \AgdaBound{Δ'} \AgdaBound{d} \AgdaBound{e}\AgdaSymbol{)}\<%
\\
\>\AgdaFunction{cut-modal} \AgdaBound{Δ'} \AgdaBound{d} \AgdaSymbol{(}\AgdaInductiveConstructor{iPCF-lam-int} \AgdaBound{e}\AgdaSymbol{)} \AgdaSymbol{=} \AgdaInductiveConstructor{iPCF-lam-int} \AgdaBound{e}\<%
\\
\>\AgdaFunction{cut-modal} \AgdaBound{Δ'} \AgdaBound{d} \AgdaSymbol{(}\AgdaInductiveConstructor{iPCF-boxI} \AgdaBound{e}\AgdaSymbol{)} \AgdaSymbol{=} \AgdaInductiveConstructor{iPCF-boxI} \AgdaSymbol{(}\AgdaFunction{cut-modal} \AgdaBound{Δ'} \AgdaBound{d} \AgdaBound{e}\AgdaSymbol{)} \<[60]%
\>[60]\<%
\\
\>\AgdaFunction{cut-modal} \AgdaBound{Δ'} \AgdaBound{d} \AgdaSymbol{(}\AgdaInductiveConstructor{iPCF-boxE} \AgdaBound{p} \AgdaBound{q}\AgdaSymbol{)} \AgdaSymbol{=}\<%
\\
\>[0]\AgdaIndent{2}{}\<[2]%
\>[2]\AgdaInductiveConstructor{iPCF-boxE} \AgdaSymbol{(}\AgdaFunction{cut-modal} \AgdaBound{Δ'} \AgdaBound{d} \AgdaBound{p}\AgdaSymbol{)} \AgdaSymbol{(}\AgdaFunction{cut-modal} \AgdaSymbol{(}\AgdaBound{Δ'} \AgdaInductiveConstructor{,} \AgdaSymbol{\_)} \AgdaBound{d} \AgdaBound{q}\AgdaSymbol{)}\<%
\\
\>\AgdaFunction{cut-modal} \AgdaBound{Δ'} \AgdaBound{d} \AgdaSymbol{(}\AgdaInductiveConstructor{iPCF-fix} \AgdaBound{f}\AgdaSymbol{)} \AgdaSymbol{=} \AgdaInductiveConstructor{iPCF-fix} \AgdaBound{f}\<%
\end{code}

%TC:endignore

%%%%%%%%%%%%%%%%%%%%%%
%%%% BIBLIOGRAPHY %%%%
%%%%%%%%%%%%%%%%%%%%%%

% Add bibliography to table of contents.
\addcontentsline{toc}{chapter}{Bibliography}

%\bibliography{library}

\begin{thebibliography}{179}
\providecommand{\natexlab}[1]{#1}
\providecommand{\url}[1]{\texttt{#1}}
\expandafter\ifx\csname urlstyle\endcsname\relax
  \providecommand{\doi}[1]{doi: #1}\else
  \providecommand{\doi}{doi: \begingroup \urlstyle{rm}\Url}\fi

\bibitem[Abramsky(1990)]{Abramsky1990b}
Samson Abramsky.
\newblock {The Lazy Lambda Calculus}.
\newblock In \emph{Research Topics in Functional Programming}, pages 65--117.
  Addison Wesley, 1990.
\newblock URL \url{https://www.cs.ox.ac.uk/files/293/lazy.pdf}.

\bibitem[Abramsky(2006)]{Abramsky2006}
Samson Abramsky.
\newblock {What are the Fundamental Structures of Concurrency?}
\newblock \emph{Electronic Notes in Theoretical Computer Science},
  162:\penalty0 37--41, 2006.
\newblock ISSN 15710661.
\newblock \doi{10.1016/j.entcs.2005.12.075}.
\newblock URL
  \url{http://linkinghub.elsevier.com/retrieve/pii/S1571066106004105}.

\bibitem[Abramsky(2012)]{Abramsky2012}
Samson Abramsky.
\newblock {Notes on Intensional Recursion}, 2012.

\bibitem[Abramsky(2014)]{Abramsky2014}
Samson Abramsky.
\newblock {Intensionality, Definability and Computation}.
\newblock In Alexandru Baltag and Sonja Smets, editors, \emph{Johan van Benthem
  on Logic and Information Dynamics}, pages 121--142. Springer-Verlag, 2014.
\newblock \doi{10.1007/978-3-319-06025-5_5}.
\newblock URL \url{https://dx.doi.org/10.1007/978-3-319-06025-5_5}.

\bibitem[Abramsky and Jung(1994)]{Abramsky1994}
Samson Abramsky and Achim Jung.
\newblock {Domain Theory}.
\newblock \emph{Handbook of Logic in Computer Science}, 3:\penalty0 1--168,
  1994.
\newblock URL \url{https://www.cs.bham.ac.uk/$\sim$axj/pub/papers/handy1.pdf}.

\bibitem[Abramsky and McCusker(1996)]{Abramsky1996a}
Samson Abramsky and Guy McCusker.
\newblock {Linearity, Sharing and State: a fully abstract game semantics for
  Idealized Algol with active expressions}.
\newblock \emph{Electronic Notes in Theoretical Computer Science}, 3:\penalty0
  2--14, 1996.
\newblock ISSN 15710661.
\newblock \doi{10.1016/S1571-0661(05)80398-6}.
\newblock URL
  \url{http://linkinghub.elsevier.com/retrieve/pii/S1571066105803986}.

\bibitem[Abramsky and Tzevelekos(2011)]{Abramsky2011a}
Samson Abramsky and Nikos Tzevelekos.
\newblock {Introduction to Categories and Categorical Logic}.
\newblock In Bob Coecke, editor, \emph{New Structures for Physics}, pages
  3--94. Springer-Verlag, 2011.
\newblock \doi{10.1007/978-3-642-12821-9_1}.
\newblock URL \url{http://arxiv.org/abs/1102.1313}.

\bibitem[Abramsky et~al.(1996)Abramsky, Jagadeesan, and
  Malacaria]{Abramsky1996}
Samson Abramsky, Radha Jagadeesan, and Pasquale Malacaria.
\newblock {Full Abstraction for PCF}.
\newblock \emph{Information and Computation}, 163:\penalty0 409--470, 1996.

\bibitem[Abramsky et~al.(1998)Abramsky, Honda, and McCusker]{Abramsky1998}
Samson Abramsky, Kohei Honda, and G~McCusker.
\newblock {A fully abstract game semantics for general references}.
\newblock In \emph{Proceedings of theThirteenth Annual IEEE Symposium on Logic
  in Computer Science}. IEEE Comput. Soc, 1998.
\newblock ISBN 0-8186-8506-9.
\newblock \doi{10.1109/LICS.1998.705669}.
\newblock URL
  \url{http://ieeexplore.ieee.org/lpdocs/epic03/wrapper.htm?arnumber=705669}.

\bibitem[Adleman(1990)]{Adleman1990}
Leonard~M. Adleman.
\newblock {An Abstract Theory of Computer Viruses}.
\newblock In \emph{Advances in Cryptology - CRYPTO' 88}, volume 403 of
  \emph{Lecture Notes in Computer Science}, pages 354--374. Springer New York,
  New York, NY, 1990.
\newblock ISBN 3-540-97196-3.
\newblock \doi{10.1007/0-387-34799-2_28}.
\newblock URL \url{https://dx.doi.org/10.1007/0-387-34799-2_28}.

\bibitem[Altenkirch et~al.(2017)Altenkirch, Danielsson, and
  Kraus]{Altenkirch2017}
Thorsten Altenkirch, Nils~Anders Danielsson, and Nicolai Kraus.
\newblock {Partiality, Revisited: The Partiality Monad as a Quotient
  Inductive-Inductive Type}.
\newblock In \emph{Proceedings of the 20th International Conference on
  Foundations of Software Science and Computation Structures (FoSSaCS)}, oct
  2017.
\newblock URL \url{http://arxiv.org/abs/1610.09254}.

\bibitem[Awodey(2010)]{Awodey2010}
Steve Awodey.
\newblock \emph{{Category Theory}}.
\newblock Oxford Logic Guides. Oxford University Press, 2010.
\newblock ISBN 9780191612558.
\newblock URL \url{https://books.google.co.uk/books?id=zLs8BAAAQBAJ}.

\bibitem[Barber(1996)]{Barber1996}
Andrew~Graham Barber.
\newblock {Dual Intuitionistic Linear Logic}.
\newblock Technical report, ECS-LFCS-96-347, Laboratory for Foundations of
  Computer Science, University of Edinburgh, 1996.
\newblock URL \url{http://www.lfcs.inf.ed.ac.uk/reports/96/ECS-LFCS-96-347/}.

\bibitem[Barendregt(1984)]{Barendregt1984}
Henk Barendregt.
\newblock \emph{{Lambda Calculus: Its Syntax and Semantics}}.
\newblock North-Holland, Amsterdam, 1984.
\newblock ISBN 978-0444875082.

\bibitem[Barendregt(1991)]{Barendregt1991}
Henk Barendregt.
\newblock {Self-Interpretation in Lambda Calculus}.
\newblock \emph{Journal of Functional Programming}, 1\penalty0 (2):\penalty0
  229--233, 1991.
\newblock URL \url{https://dx.doi.org/10.1017/S0956796800020062}.

\bibitem[Bauer(2011)]{Bauer2011}
Andrej Bauer.
\newblock {Definability and extensionality of the modulus of continuity
  functional}.
\newblock \emph{Mathematics and Computation (blog)}, 2011.
\newblock URL
  \url{http://math.andrej.com/2011/07/27/definability-and-extensionality-of-the-modulus-of-continuity-functional/}.

\bibitem[Bawden(1999)]{Bawden1999}
Alan Bawden.
\newblock {Quasiquotation in LISP}.
\newblock In \emph{Proceedings of the 6th ACM SIGPLAN Workshop on Partial
  Evaluation and Semantics-Based Program Manipulation (PEPM '99)}, 1999.
\newblock URL
  \url{http://repository.readscheme.org/ftp/papers/pepm99/bawden.pdf}.

\bibitem[Beeson(1985)]{Beeson1985}
Michael~J. Beeson.
\newblock \emph{{Foundations of Constructive Mathematics}}.
\newblock Springer Berlin Heidelberg, 1985.
\newblock ISBN 978-3-642-68954-3.
\newblock \doi{10.1007/978-3-642-68952-9}.
\newblock URL \url{https://dx.doi.org/10.1007/978-3-642-68952-9}.

\bibitem[Berger(2016)]{Berger2016}
Martin Berger.
\newblock {Foundations of meta-programming}, 2016.
\newblock URL
  \url{http://users.sussex.ac.uk/$\sim$mfb21/publications/mp-slides/slides.pdf}.

\bibitem[Bierman(2000)]{Bierman2000}
Gavin~M. Bierman.
\newblock {Program equivalence in a linear functional language}.
\newblock \emph{Journal of Functional Programming}, 10\penalty0 (2):\penalty0
  167--190, 2000.
\newblock ISSN 09567968.
\newblock \doi{10.1017/S0956796899003639}.

\bibitem[Bierman and de~Paiva(2000)]{Bierman2000a}
Gavin~M. Bierman and Valeria de~Paiva.
\newblock {On an Intuitionistic Modal Logic}.
\newblock \emph{Studia Logica}, 65\penalty0 (3):\penalty0 383--416, 2000.
\newblock \doi{10.1023/A:1005291931660}.
\newblock URL \url{https://dx.doi.org/10.1023/A:1005291931660}.

\bibitem[Birkedal(2000)]{Birkedal2000}
Lars Birkedal.
\newblock {Developing theories of types and computability via realizability}.
\newblock \emph{Electronic Notes in Theoretical Computer Science}, 34:\penalty0
  2, 2000.
\newblock ISSN 15710661.
\newblock \doi{10.1016/S1571-0661(05)80642-5}.
\newblock URL \url{http://cs.au.dk/$\sim$birke/papers/devttc.pdf}.

\bibitem[Bishop(1967)]{Bishop1967}
Errett Bishop.
\newblock \emph{{Foundations of Constructive Analysis}}.
\newblock McGraw-Hill, 1967.

\bibitem[Bloom and Riecke(1989)]{Bloom1989}
B~Bloom and J~G Riecke.
\newblock {LCF Should Be Lifted}.
\newblock 1989.

\bibitem[Blum(1967)]{Blum1967}
Manuel Blum.
\newblock {On the size of machines}.
\newblock \emph{Information and Control}, 11\penalty0 (3):\penalty0 257--265,
  sep 1967.
\newblock ISSN 00199958.
\newblock \doi{10.1016/S0019-9958(67)90546-3}.
\newblock URL
  \url{http://linkinghub.elsevier.com/retrieve/pii/S0019995867905463}.

\bibitem[Blumensath and Winschel(2013)]{Blumensath2013}
Achim Blumensath and Viktor Winschel.
\newblock {A Coalgebraic Framework for Games in Economics}.
\newblock 2013.

\bibitem[Bonfante et~al.(2005)Bonfante, Kaczmarek, and Marion]{Bonfante2005}
Guillaume Bonfante, Matthieu Kaczmarek, and Jean-Yves Marion.
\newblock {Toward an Abstract Computer Virology}.
\newblock In Dan {Van Hung} and Martin Wirsing, editors, \emph{Theoretical
  Aspects of Computing – ICTAC 2005: Second International Colloquium, Hanoi,
  Vietnam, October 17-21, 2005. Proceedings}, volume 3722 of \emph{Lecture
  Notes in Computer Science}, pages 579--593. Springer Berlin Heidelberg, 2005.
\newblock ISBN 3540291075.
\newblock \doi{10.1007/11560647_38}.
\newblock URL \url{http://link.springer.com/10.1007/11560647_38}.

\bibitem[Bonfante et~al.(2006)Bonfante, Kaczmarek, and Marion]{Bonfante2006}
Guillaume Bonfante, Matthieu Kaczmarek, and J.-Y. Marion.
\newblock {On Abstract Computer Virology from a Recursion Theoretic
  Perspective}.
\newblock \emph{Journal in Computer Virology}, 1\penalty0 (3):\penalty0 45--54,
  2006.
\newblock ISSN 1772-9890.
\newblock \doi{10.1007/s11416-005-0007-4}.
\newblock URL \url{http://link.springer.com/10.1007/s11416-005-0007-4}.

\bibitem[Bonfante et~al.(2007)Bonfante, Kaczmarek, and Marion]{Bonfante2007}
Guillaume Bonfante, Matthieu Kaczmarek, and Jean-Yves Marion.
\newblock {A Classification of Viruses Through Recursion Theorems}.
\newblock In S~Barry Cooper, Benedikt L{\"{o}}we, and Andrea Sorbi, editors,
  \emph{Computation and Logic in the Real World: Third Conference on
  Computability in Europe, CiE 2007, Siena, Italy, June 18-23, 2007.
  Proceedings}, volume 4497 of \emph{Lecture Notes in Computer Science}, pages
  73--82. Springer Berlin Heidelberg, Berlin, Heidelberg, 2007.
\newblock \doi{10.1007/978-3-540-73001-9_8}.
\newblock URL \url{http://link.springer.com/10.1007/978-3-540-73001-9_8}.

\bibitem[Boolos(1994)]{Boolos1994}
George~S. Boolos.
\newblock \emph{{The Logic of Provability}}.
\newblock Cambridge University Press, Cambridge, 1994.
\newblock ISBN 9780511625183.
\newblock \doi{10.1017/CBO9780511625183}.
\newblock URL \url{https://dx.doi.org/10.1017/CBO9780511625183}.

\bibitem[Bove and Capretta(2005)]{Bove2005}
Ana Bove and Venanzio Capretta.
\newblock {Modelling general recursion in type theory}.
\newblock \emph{Mathematical Structures in Computer Science}, 15\penalty0
  (4):\penalty0 671--708, aug 2005.

\bibitem[Bra{\"{u}}ner(1995)]{Brauner1995}
Torben Bra{\"{u}}ner.
\newblock {The Girard Translation Extended with Recursion}.
\newblock In Leszek Pacholski and Jerzy Tiuryn, editors, \emph{Computer Science
  Logic, 8th International Workshop, CSL '94, Kazimierz, Poland, September
  25-30, 1994, Selected Papers}, number Lecture Notes in Computer Science 933,
  pages 31--45, 1995.
\newblock ISBN 3-540-60017-5.
\newblock \doi{10.1007/BFb0022245}.

\bibitem[Bra{\"{u}}ner(1997)]{Brauner1997b}
Torben Bra{\"{u}}ner.
\newblock {A general adequacy result for a linear functional language}.
\newblock \emph{Theoretical Computer Science}, 177\penalty0 (1):\penalty0
  27--58, 1997.
\newblock ISSN 03043975.
\newblock \doi{10.1016/S0304-3975(96)00233-2}.

\bibitem[Capretta(2005)]{Capretta2005}
Venanzio Capretta.
\newblock {General recursion via coinductive types}.
\newblock \emph{Logical Methods in Computer Science}, 1\penalty0 (2):\penalty0
  1--28, jul 2005.
\newblock ISSN 18605974.
\newblock \doi{10.2168/LMCS-1(2:1)2005}.
\newblock URL \url{http://www.lmcs-online.org/ojs/viewarticle.php?id=55}.

\bibitem[Case and Moelius(2007)]{Case2007}
John Case and Samuel~E. Moelius.
\newblock {Properties Complementary to Program Self-reference}.
\newblock In Lud{\v{e}}k Ku{\v{c}}era and Anton{\'{i}}n Ku{\v{c}}era, editors,
  \emph{Proceedings of the 32nd International Symposium on Mathematical
  Foundations of Computer Science 2007 (MFCS'07)}, volume 4708 of \emph{Lecture
  Notes in Computer Science}, pages 253--263, Berlin, Heidelberg, 2007.
  Springer Berlin Heidelberg.
\newblock \doi{10.1007/978-3-540-74456-6_24}.
\newblock URL \url{https://dx.doi.org/10.1007/978-3-540-74456-6_24}.

\bibitem[Case and Moelius(2009{\natexlab{a}})]{Case2009}
John Case and Samuel~E. Moelius.
\newblock {Characterizing programming systems allowing program self-reference}.
\newblock \emph{Theory of Computing Systems}, 45\penalty0 (4):\penalty0
  756--772, 2009{\natexlab{a}}.
\newblock ISSN 14324350.
\newblock \doi{10.1007/s00224-009-9168-8}.
\newblock URL \url{https://dx.doi.org/10.1007/s00224-009-9168-8}.

\bibitem[Case and Moelius(2009{\natexlab{b}})]{Case2009a}
John Case and Samuel~E. Moelius.
\newblock {Independence Results for n-Ary Recursion Theorems}.
\newblock In Miroslaw Kutylowski, Witold Charatonik, and Maciej Gebala,
  editors, \emph{Fundamentals of Computation Theory: Proceedings of the 17th
  International Symposium, FCT 2009, Wroc{\l}aw, Poland, September 2-4, 2009},
  volume 5699 of \emph{Lecture Notes in Computer Science}, pages 38--49.
  Springer, 2009{\natexlab{b}}.
\newblock ISBN 978-3-642-03408-4.
\newblock \doi{10.1007/978-3-642-03409-1_5}.
\newblock URL \url{https://dx.doi.org/10.1007/978-3-642-03409-1_5}.

\bibitem[Case and Moelius(2012)]{Case2012}
John Case and Samuel~E. Moelius.
\newblock {Program Self-Reference in Constructive Scott Subdomains}.
\newblock \emph{Theory of Computing Systems}, 51\penalty0 (1):\penalty0 22--49,
  2012.
\newblock ISSN 14324350.
\newblock \doi{10.1007/s00224-011-9372-1}.
\newblock URL \url{https://dx.doi.org/10.1007/s00224-011-9372-1}.

\bibitem[Cockett and Hofstra(2008)]{Cockett2008}
J.~R.~B. Cockett and Pieter J.~W. Hofstra.
\newblock {Introduction to Turing categories}.
\newblock \emph{Annals of Pure and Applied Logic}, 156\penalty0 (2-3):\penalty0
  183--209, 2008.
\newblock \doi{10.1016/j.apal.2008.04.005}.
\newblock URL \url{http://dx.doi.org/10.1016/j.apal.2008.04.005}.

\bibitem[Cockett and Hofstra(2010)]{Cockett2010}
J.~R.~B. Cockett and Pieter J.~W. Hofstra.
\newblock {Categorical simulations}.
\newblock \emph{Journal of Pure and Applied Algebra}, 214\penalty0
  (10):\penalty0 1835--1853, 2010.
\newblock ISSN 00224049.
\newblock \doi{10.1016/j.jpaa.2009.12.028}.
\newblock URL \url{http://dx.doi.org/10.1016/j.jpaa.2009.12.028}.

\bibitem[Cohen(1989)]{Cohen1989}
Fred Cohen.
\newblock {Computational aspects of computer viruses}.
\newblock \emph{Computers and Security}, 8\penalty0 (4):\penalty0 325--344,
  1989.
\newblock ISSN 01674048.
\newblock \doi{10.1016/0167-4048(89)90094-1}.

\bibitem[Constable and Smith(1993)]{Constable1993}
Robert~L. Constable and Scott~F. Smith.
\newblock {Computational foundations of basic recursive function theory}.
\newblock \emph{Theoretical Computer Science}, 121\penalty0 (1-2):\penalty0
  89--112, dec 1993.
\newblock ISSN 03043975.
\newblock \doi{10.1016/0304-3975(93)90085-8}.
\newblock URL \url{https://dx.doi.org/10.1016/0304-3975(93)90085-8}.

\bibitem[Crole(1993)]{Crole1993}
Roy~L. Crole.
\newblock \emph{{Categories for Types}}.
\newblock Cambridge University Press, 1993.
\newblock ISBN 0 521 45701 7.

\bibitem[{\v{C}}ubri{\'{c}} et~al.(1998){\v{C}}ubri{\'{c}}, Dybjer, and
  Scott]{Cubric1998}
Djordje {\v{C}}ubri{\'{c}}, Peter Dybjer, and Philip~J. Scott.
\newblock {Normalization and the Yoneda embedding}.
\newblock \emph{Mathematical Structures in Computer Science}, 8\penalty0
  (2):\penalty0 153--192, 1998.
\newblock \doi{10.1017/s0960129597002508}.
\newblock URL \url{https://dx.doi.org/10.1017/s0960129597002508}.

\bibitem[Cutland(1980)]{Cutland1980}
Nigel Cutland.
\newblock \emph{{Computability: An Introduction to Recursive Function Theory}}.
\newblock Cambridge University Press, 1980.
\newblock ISBN 9780521294652.

\bibitem[Danvy and Malmkjaer(1988)]{Danvy1988}
Olivier Danvy and Karoline Malmkjaer.
\newblock {Intensions and extensions in a reflective tower}.
\newblock In \emph{Proceedings of the 1988 ACM conference on LISP and
  functional programming (LFP '88)}, pages 327--341, New York, New York, USA,
  1988. ACM Press.
\newblock ISBN 089791273X.
\newblock \doi{10.1145/62678.62725}.
\newblock URL \url{https://dx.doi.org/10.1145/62678.62725}.

\bibitem[Davies(1995)]{Davies1995}
Rowan Davies.
\newblock {A Temporal-Logic Approach to Binding-Time Analysis}.
\newblock Technical report, BRICS Report Series RS-95-51, 1995.

\bibitem[Davies(1996)]{Davies1996a}
Rowan Davies.
\newblock {A temporal-logic approach to binding-time analysis}.
\newblock \emph{Proceedings 11th Annual IEEE Symposium on Logic in Computer
  Science}, pages 184--195, 1996.
\newblock ISSN 1043-6871.
\newblock \doi{10.1109/LICS.1996.561317}.

\bibitem[Davies(2017)]{Davies2017}
Rowan Davies.
\newblock {A Temporal Logic Approach to Binding-Time Analysis}.
\newblock \emph{Journal of the ACM}, 64\penalty0 (1):\penalty0 1--45, mar 2017.
\newblock ISSN 00045411.
\newblock \doi{10.1145/3011069}.
\newblock URL \url{http://dx.doi.org/10.1145/3011069}.

\bibitem[Davies and Pfenning(1996)]{Davies1996}
Rowan Davies and Frank Pfenning.
\newblock {A modal analysis of staged computation}.
\newblock In \emph{Proceedings of the 23rd ACM SIGPLAN-SIGACT Symposium on
  Principles of Programming Languages (POPL'96)}, pages 258--270, 1996.
\newblock ISBN 0897917693.
\newblock \doi{10.1145/382780.382785}.
\newblock URL \url{https://dx.doi.org/10.1145/382780.382785}.

\bibitem[Davies and Pfenning(2001)]{Davies2001a}
Rowan Davies and Frank Pfenning.
\newblock {A modal analysis of staged computation}.
\newblock \emph{Journal of the ACM}, 48\penalty0 (3):\penalty0 555--604, 2001.
\newblock ISSN 00045411.
\newblock \doi{10.1145/382780.382785}.
\newblock URL \url{https://dx.doi.org/10.1145/382780.382785}.

\bibitem[Demers and Malenfant(1995)]{Malenfant1995}
Fran{\c{c}}ois-Nicola Demers and Jacques Malenfant.
\newblock {Reflection in logic, functional and object-oriented programming: a
  Short Comparative Study}.
\newblock In \emph{Proceedings of the IJCAI '95 Workshop on Reflection and
  Metalevel Architectures and their Applications in AI}, pages 29--38, 1995.

\bibitem[Doxiadis(2001)]{Doxiadis2001}
Apostolos Doxiadis.
\newblock \emph{{Uncle Petros and Goldbach's Conjecture}}.
\newblock Faber and Faber, 2001.

\bibitem[Eilenberg and Kelly(1966)]{Eilenberg1966}
Samuel Eilenberg and G.~Max Kelly.
\newblock {Closed Categories}.
\newblock In \emph{Proceedings of the Conference on Categorical Algebra}, pages
  421--562. Springer Berlin Heidelberg, Berlin, Heidelberg, 1966.
\newblock \doi{10.1007/978-3-642-99902-4_22}.
\newblock URL
  \url{http://www.springerlink.com/index/10.1007/978-3-642-99902-4_22}.

\bibitem[Ershov(1977)]{Ershov1977}
A.~P. Ershov.
\newblock {On the partial computation principle}.
\newblock \emph{Information Processing Letters}, 6\penalty0 (2):\penalty0
  38--41, apr 1977.
\newblock ISSN 00200190.
\newblock \doi{10.1016/0020-0190(77)90078-3}.
\newblock URL \url{https://dx.doi.org/10.1016/0020-0190(77)90078-3}.

\bibitem[Ershov(1982)]{Ershov1982}
A.~P. Ershov.
\newblock {Mixed computation: potential applications and problems for study}.
\newblock \emph{Theoretical Computer Science}, 18\penalty0 (1):\penalty0
  41--67, apr 1982.
\newblock ISSN 03043975.
\newblock \doi{10.1016/0304-3975(82)90111-6}.
\newblock URL
  \url{http://linkinghub.elsevier.com/retrieve/pii/0304397582901116}.

\bibitem[Escard{\'{o}}(1999)]{Escardo1999}
Mart{\'{i}}n~H{\"{o}}tzel Escard{\'{o}}.
\newblock {A metric model of PCF}, 1999.
\newblock URL \url{http://cs.bham.ac.uk/$\sim$mhe/papers/metricpcf.pdf}.

\bibitem[Fitting(2015)]{Fitting2015}
Melvin Fitting.
\newblock {Intensional Logic}.
\newblock In Edward~N Zalta, editor, \emph{The Stanford Encyclopedia of
  Philosophy}. Metaphysics Research Lab, Stanford University, summer '15
  edition, 2015.
\newblock URL
  \url{https://plato.stanford.edu/archives/sum2015/entries/logic-intensional/}.

\bibitem[Friedman and Wand(1984)]{Friedman1984}
Daniel~P. Friedman and Mitchell Wand.
\newblock {Reification: Reflection without metaphysics}.
\newblock In \emph{Proceedings of the 1984 ACM Symposium on LISP and functional
  programming (LFP '84)}, pages 348--355, New York, New York, USA, 1984. ACM
  Press.
\newblock ISBN 0897911423.
\newblock \doi{10.1145/800055.802051}.
\newblock URL \url{https://dx.doi.org/10.1145/800055.802051}.

\bibitem[Friedman(1998)]{Friedman1998}
Harvey Friedman.
\newblock {FOM: renaming recursion theory}, 1998.
\newblock URL
  \url{http://www.cs.nyu.edu/pipermail/fom/1998-August/002017.html}.

\bibitem[Futamura(1999)]{Futamura1999}
Yoshihiko Futamura.
\newblock {Partial evaluation of computation process--an approach to a
  compiler-compiler}.
\newblock \emph{Higher-Order and Symbolic Computation}, 12\penalty0
  (4):\penalty0 381--391, 1999.
\newblock ISSN 13883690.
\newblock \doi{10.1023/A:1010095604496}.
\newblock URL \url{http://link.springer.com/article/10.1023/A:1010095604496}.

\bibitem[Gabbay and Nanevski(2013)]{Gabbay2013}
Murdoch~J. Gabbay and Aleksandar Nanevski.
\newblock {Denotation of contextual modal type theory (CMTT): Syntax and
  meta-programming}.
\newblock \emph{Journal of Applied Logic}, 11\penalty0 (1):\penalty0 1--29, mar
  2013.
\newblock ISSN 15708683.
\newblock \doi{10.1016/j.jal.2012.07.002}.
\newblock URL \url{http://dx.doi.org/10.1016/j.jal.2012.07.002}.

\bibitem[Gandy(1988)]{Gandy1988}
R~Gandy.
\newblock {The Confluence of Ideas in 1936}.
\newblock In \emph{A Half-century Survey on The Universal Turing Machine},
  pages 55--111, New York, NY, USA, 1988. Oxford University Press, Inc.
\newblock ISBN 0-19-853741-7.
\newblock URL \url{http://dl.acm.org/citation.cfm?id=57249.57252}.

\bibitem[Gandy(1956)]{Gandy1956}
R.~O. Gandy.
\newblock {On the axiom of extensionality – Part I}.
\newblock \emph{The Journal of Symbolic Logic}, 21\penalty0 (01):\penalty0
  36--48, mar 1956.
\newblock ISSN 0022-4812.
\newblock \doi{10.2307/2268484}.
\newblock URL
  \url{https://www.cambridge.org/core/product/identifier/S0022481200085352/type/journal_article}.

\bibitem[Gandy(1959)]{Gandy1959}
R.~O. Gandy.
\newblock {On the axiom of extensionality, Part II}.
\newblock \emph{The Journal of Symbolic Logic}, 24\penalty0 (04):\penalty0
  287--300, dec 1959.
\newblock ISSN 0022-4812.
\newblock \doi{10.2307/2963897}.
\newblock URL
  \url{https://www.cambridge.org/core/product/identifier/S0022481200123266/type/journal_article}.

\bibitem[Girard(1972)]{Girard1972}
Jean-Yves Girard.
\newblock \emph{{Interpr{\'{e}}tation fonctionelle et {\'{e}}limination des
  coupures de l'arithm{\'{e}}tique d'ordre sup{\'{e}}rieur}}.
\newblock PhD thesis, Universit{\'{e}} Paris VII, 1972.

\bibitem[Girard(2011)]{Girard2011}
Jean-Yves Girard.
\newblock \emph{{The Blind Spot: Lectures on Logic}}.
\newblock European Mathematical Society, 2011.
\newblock ISBN 978-3037190883.

\bibitem[Girard et~al.(1989)Girard, Lafont, and Taylor]{Girard1989}
Jean-Yves Girard, Yves Lafont, and Paul Taylor.
\newblock \emph{{Proofs and Types}}.
\newblock Cambridge University Press, 1989.

\bibitem[Graham(1993)]{Graham1993}
Paul Graham.
\newblock \emph{{On LISP: Advanced Techniques for Common LISP}}.
\newblock Prentice Hall, 1993.
\newblock ISBN 0130305529.

\bibitem[Gunter(1992)]{Gunter1992}
Carl~A. Gunter.
\newblock \emph{{Semantics of programming languages: structures and
  techniques}}.
\newblock Foundations of Computing. The MIT Press, 1992.

\bibitem[Hansen et~al.(1989)Hansen, Nikolajsen, Tr{\"{a}}ff, and
  Jones]{Hansen1989}
Torben~Amtoft Hansen, Thomas Nikolajsen, Jesper~Larsson Tr{\"{a}}ff, and
  Neil~D. Jones.
\newblock {Experiments with Implementations of Two Theoretical Constructions}.
\newblock In \emph{Proceedings of the Symposium on Logical Foundations of
  Computer Science: Logic at Botik '89}, pages 119--133, London, UK, 1989.
  Springer-Verlag.
\newblock ISBN 3-540-51237-3.
\newblock URL \url{http://dl.acm.org/citation.cfm?id=646798.759646}.

\bibitem[Harper(2014)]{Harper2014}
Robert Harper.
\newblock {Old Neglected Theorems Are Still Theorems}, 2014.
\newblock URL
  \url{https://existentialtype.wordpress.com/2014/03/20/old-neglected-theorems-are-still-theorems/}.

\bibitem[Hofstadter(1979)]{Hofstadter1979}
Douglas~R Hofstadter.
\newblock \emph{{Godel, Escher, Bach: An Eternal Golden Braid}}.
\newblock Basic Books, Inc., New York, NY, USA, 1979.
\newblock ISBN 0465026850.

\bibitem[Huwig and Poign{\'{e}}(1990)]{Huwig1990}
Hagen Huwig and Axel Poign{\'{e}}.
\newblock {A note on inconsistencies caused by fixpoints in a cartesian closed
  category}.
\newblock \emph{Theoretical Computer Science}, 73\penalty0 (1):\penalty0
  101--112, jun 1990.
\newblock ISSN 03043975.
\newblock \doi{10.1016/0304-3975(90)90165-E}.
\newblock URL
  \url{http://linkinghub.elsevier.com/retrieve/pii/030439759090165E}.

\bibitem[Hyland(1982)]{Hyland1982}
J.~M.~E. Hyland.
\newblock {The Effective Topos}.
\newblock In Anne~S. Troelstra and Dick van Dalen, editors, \emph{The L. E. J.
  Brouwer Centenary Symposium}, pages 165--216. North-Holland, 1982.
\newblock URL
  \url{https://webdpmms.maths.cam.ac.uk/$\sim$martin/Research/Oldpapers/hyland-effectivetopos.pdf}.

\bibitem[Hyland(2016)]{Hyland2016}
J.M.E. Hyland.
\newblock {The Forgotten Turing}.
\newblock In S.~Barry Cooper and Andrew Hodges, editors, \emph{The Once and
  Future Turing}, pages 20--33. Cambridge University Press, Cambridge, 2016.
\newblock \doi{10.1017/CBO9780511863196.005}.
\newblock URL \url{http://ebooks.cambridge.org/ref/id/CBO9780511863196A012}.

\bibitem[Hyland and Ong(2000)]{Hyland2000}
J.M.E. Hyland and C.-H.L. Ong.
\newblock {On Full Abstraction for PCF: I, II, and III}.
\newblock \emph{Information and Computation}, 163\penalty0 (2):\penalty0
  285--408, 2000.
\newblock \doi{10.1006/inco.2000.2917}.
\newblock URL
  \url{http://linkinghub.elsevier.com/retrieve/pii/S0890540100929171}.

\bibitem[Hyland and Power(2007)]{Hyland2007}
Martin Hyland and John Power.
\newblock {The Category Theoretic Understanding of Universal Algebra: Lawvere
  Theories and Monads}.
\newblock \emph{Electronic Notes in Theoretical Computer Science},
  172:\penalty0 437--458, 2007.
\newblock ISSN 15710661.
\newblock \doi{10.1016/j.entcs.2007.02.019}.

\bibitem[Jacobs(1999)]{Jacobs1999}
Bart Jacobs.
\newblock \emph{{Categorical Logic and Type Theory}}.
\newblock Number 141 in Studies in Logic and the Foundations of Mathematics.
  North Holland, Amsterdam, 1999.

\bibitem[Jay(2016)]{Jay2016}
Barry Jay.
\newblock {Programs as Data Structures in $\lambda$SF-Calculus}.
\newblock \emph{Electronic Notes in Theoretical Computer Science},
  325:\penalty0 221--236, 2016.
\newblock ISSN 15710661.
\newblock \doi{10.1016/j.entcs.2016.09.040}.
\newblock URL \url{http://dx.doi.org/10.1016/j.entcs.2016.09.040}.

\bibitem[Jay and Given-Wilson(2011)]{Jay2011a}
Barry Jay and Thomas Given-Wilson.
\newblock {A Combinatory Account of Internal Structure}.
\newblock \emph{The Journal of Symbolic Logic}, 76\penalty0 (3):\penalty0
  807--826, 2011.
\newblock URL \url{http://www.jstor.org/stable/23041848}.

\bibitem[Jay and Palsberg(2011)]{Jay2011b}
Barry Jay and Jens Palsberg.
\newblock {Typed self-interpretation by pattern matching}.
\newblock \emph{ACM SIGPLAN Notices}, 46\penalty0 (9):\penalty0 247, sep 2011.
\newblock ISSN 03621340.
\newblock \doi{10.1145/2034574.2034808}.
\newblock URL \url{http://dl.acm.org/citation.cfm?doid=2034574.2034808}.

\bibitem[Jones(1992)]{Jones1992}
Neil~D. Jones.
\newblock {Computer Implementation and Applications of Kleene's S-M-N and
  Recursion Theorems}.
\newblock In Yiannis~N Moschovakis, editor, \emph{Logic from Computer Science:
  Proceedings of a Workshop Held November 13-17, 1989 [at MSRI]}, volume~21 of
  \emph{Mathematical Sciences Research Institute Publications}, pages 243--263.
  Springer New York, 1992.
\newblock \doi{10.1007/978-1-4612-2822-6_9}.
\newblock URL \url{https://dx.doi.org/10.1007/978-1-4612-2822-6_9}.

\bibitem[Jones(1996)]{Jones1996}
Neil~D. Jones.
\newblock {An introduction to partial evaluation}.
\newblock \emph{ACM Computing Surveys}, 28\penalty0 (3):\penalty0 480--503,
  1996.
\newblock ISSN 03600300.
\newblock \doi{10.1145/243439.243447}.
\newblock URL \url{http://doi.acm.org/10.1145/243439.243447}.

\bibitem[Jones(1997)]{Jones1997}
Neil~D. Jones.
\newblock \emph{{Computability and Complexity: From a Programming
  Perspective}}.
\newblock Foundations of Computing. MIT Press, 1997.

\bibitem[Jones(2013)]{Jones2013}
Neil~D. Jones.
\newblock {A Swiss Pocket Knife for Computability}.
\newblock \emph{Electronic Proceedings in Theoretical Computer Science},
  129:\penalty0 1--17, sep 2013.
\newblock ISSN 2075-2180.
\newblock \doi{10.4204/EPTCS.129.1}.
\newblock URL \url{http://arxiv.org/abs/1309.5128v1}.

\bibitem[Jones et~al.(1993)Jones, Gomard, and Sestoft]{Jones1993}
Neil~D. Jones, Carsten~K. Gomard, and Peter Sestoft.
\newblock \emph{{Partial Evaluation and Automatic Program Generation}}.
\newblock Prentice Hall International, 1993.
\newblock ISBN 0-13-020249-5.

\bibitem[Kanda(1988)]{Kanda1988}
Akira Kanda.
\newblock {Recursion theorems and effective domains}.
\newblock \emph{Annals of Pure and Applied Logic}, 38\penalty0 (3):\penalty0
  289--300, 1988.
\newblock ISSN 01680072.
\newblock \doi{10.1016/0168-0072(88)90029-2}.

\bibitem[Kavvos(2016)]{Kavvos2016b}
G.~A. Kavvos.
\newblock {The Many Worlds of Modal $\lambda$-calculi: I. Curry-Howard for
  Necessity, Possibility and Time}.
\newblock \emph{CoRR}, 2016.

\bibitem[Kavvos(2017{\natexlab{a}})]{Kavvos2017a}
G.~A. Kavvos.
\newblock {On the Semantics of Intensionality}.
\newblock In Javier Esparza and Andrzej~S. Murawski, editors, \emph{Proceedings
  of the 20th International Conference on Foundations of Software Science and
  Computation Structures (FoSSaCS)}, volume 10203 of \emph{Lecture Notes in
  Computer Science}, pages 550--566. Springer-Verlag Berlin Heidelberg,
  2017{\natexlab{a}}.
\newblock \doi{10.1007/978-3-662-54458-7_32}.
\newblock URL \url{https://dx.doi.org/10.1007/978-3-662-54458-7_32}.

\bibitem[Kavvos(2017{\natexlab{b}})]{Kavvos2017b}
G.~A. Kavvos.
\newblock {Dual-context calculi for modal logic}.
\newblock In \emph{2017 32nd Annual ACM/IEEE Symposium on Logic in Computer
  Science (LICS)}. IEEE, 2017{\natexlab{b}}.
\newblock ISBN 978-1-5090-3018-7.
\newblock \doi{10.1109/LICS.2017.8005089}.

\bibitem[Kavvos(2017{\natexlab{c}})]{Kavvos2017c}
G.~A. Kavvos.
\newblock {Dual-context calculi for modal logic (technical report)}.
\newblock Technical report, University of Oxford, 2017{\natexlab{c}}.
\newblock URL \url{http://www.lambdabetaeta.eu/papers/dualcalc.pdf}.

\bibitem[Kavvos(2017{\natexlab{d}})]{Kavvos2017d}
G.~A. Kavvos.
\newblock {Intensionality, Intensional Recursion, and the G{\"{o}}del-L{\"{o}}b
  axiom}.
\newblock In \emph{Proceedings of 7th Workshop on Intuitionistic Modal Logic
  and Applications (IMLA 2017)}, 2017{\natexlab{d}}.

\bibitem[Kiselyov(2015)]{Kiselyov2015}
Oleg Kiselyov.
\newblock {In'y\=o to ichiokubai k\=osokuka no monogatari: Kans\=u
  puroguramingu ni yoru Kleene daini saiki teiri no sh\=omei [A story of
  quotation and $10^8$-fold speed-up: A proof of Kleene's second recursion
  theorem by functional programming]}.
\newblock In \emph{The 17th Programming and Programming Language Workshop (PPL
  2015)}, 2015.
\newblock URL \url{http://okmij.org/ftp/Computation/Kleene.pdf}.

\bibitem[Kleene(1936)]{Kleene1936}
S.~C. Kleene.
\newblock $\lambda$-definability and recursiveness.
\newblock \emph{Duke Mathematical Journal}, 2\penalty0 (2):\penalty0 340--353,
  1936.
\newblock ISSN 0012-7094.
\newblock \doi{10.1215/S0012-7094-36-00227-2}.
\newblock URL \url{https://dx.doi.org/10.1215/S0012-7094-36-00227-2}.

\bibitem[Kleene(1938)]{Kleene1938}
Stephen~C. Kleene.
\newblock {On notation for ordinal numbers}.
\newblock \emph{The Journal of Symbolic Logic}, 3\penalty0 (04):\penalty0
  150--155, 1938.
\newblock ISSN 0022-4812.
\newblock \doi{10.2307/2267778}.
\newblock URL \url{https://dx.doi.org/10.2307/2267778}.

\bibitem[Kleene(1952)]{Kleene1952}
Stephen~C. Kleene.
\newblock \emph{{Introduction to Metamathematics}}.
\newblock North-Holland, Amsterdam, 1952.

\bibitem[Kleene(1981)]{Kleene1981}
Stephen~C. Kleene.
\newblock {Origins of Recursive Function Theory}.
\newblock \emph{IEEE Annals of the History of Computing}, 3\penalty0
  (1):\penalty0 52--67, 1981.
\newblock ISSN 1058-6180.
\newblock \doi{10.1109/MAHC.1981.10004}.
\newblock URL \url{https://dx.doi.org/10.1109/MAHC.1981.10004}.

\bibitem[Kobayashi(1997)]{Kobayashi1997}
Satoshi Kobayashi.
\newblock {Monad as modality}.
\newblock \emph{Theoretical Computer Science}, 175\penalty0 (1):\penalty0
  29--74, 1997.
\newblock \doi{10.1016/S0304-3975(96)00169-7}.

\bibitem[Lambek and Scott(1988)]{Lambek1988}
Joachim Lambek and Philip~J. Scott.
\newblock \emph{{Introduction to Higher-Order Categorical Logic}}.
\newblock Cambridge University Press, 1988.
\newblock ISBN 9780521356534.

\bibitem[Landry(2017)]{Landry2017}
Elaine Landry, editor.
\newblock \emph{{Categories for the Working Philosopher}}.
\newblock Oxford University Press, 2017.
\newblock ISBN 9780198748991.

\bibitem[Lassez et~al.(1982)Lassez, Nguyen, and Sonenberg]{Lassez1982}
J.-L. Lassez, V.L. Nguyen, and E.a. Sonenberg.
\newblock {Fixed point theorems and semantics: a folk tale}.
\newblock \emph{Information Processing Letters}, 14\penalty0 (3):\penalty0
  112--116, 1982.
\newblock ISSN 00200190.
\newblock \doi{10.1016/0020-0190(82)90065-5}.

\bibitem[Lawvere(1969)]{Lawvere1969}
F.~William Lawvere.
\newblock {Diagonal arguments and cartesian closed categories}.
\newblock In \emph{Category Theory, Homology Theory and their Applications II},
  number~15, pages 134--145. Springer Berlin Heidelberg, 1969.
\newblock ISBN 978-3-540-04611-0.
\newblock \doi{10.1007/BFb0080769}.
\newblock URL
  \url{http://link.springer.com/content/pdf/10.1007/BFb0080769.pdf}.

\bibitem[Lawvere(2006)]{Lawvere2006}
F.~William Lawvere.
\newblock {Diagonal arguments and cartesian closed categories}.
\newblock \emph{Reprints in Theory and Applications of Categories},
  15:\penalty0 1--13, 2006.
\newblock URL
  \url{http://www.tac.mta.ca/tac/reprints/articles/15/tr15abs.html}.

\bibitem[Levy(2003)]{Levy2003}
Paul~Blain Levy.
\newblock \emph{{Call-by-Push-Value: A Functional-Imperative Synthesis}}.
\newblock Semantic Structures in Computation. Springer, 2003.
\newblock ISBN 978-94-010-3752-5.
\newblock \doi{10.1007/978-94-007-0954-6}.

\bibitem[Lewis and Papadimitriou(1997)]{Lewis1997}
Harry~R Lewis and Christos~H Papadimitriou.
\newblock \emph{{Elements of the Theory of Computation}}.
\newblock Prentice Hall PTR, Upper Saddle River, NJ, USA, 2nd edition, 1997.
\newblock ISBN 0132624788.

\bibitem[Litak(2014)]{Litak2014}
Tadeusz Litak.
\newblock {Constructive Modalities with Provability Smack}.
\newblock In Guram Bezhanishvili, editor, \emph{Leo Esakia on duality in modal
  and intuitionistic logics}, pages 179--208. Springer, 2014.
\newblock \doi{10.1007/978-94-017-8860-1_8}.
\newblock URL \url{https://www8.cs.fau.de/ext/litak/esakiaarxivfull.pdf}.

\bibitem[Longley(1995)]{Longley1995}
John~R. Longley.
\newblock \emph{{Realizability Toposes and Language Semantics}}.
\newblock PhD thesis, University of Edinburgh. College of Science and
  Engineering. School of Informatics., 1995.
\newblock URL \url{http://www.lfcs.inf.ed.ac.uk/reports/95/ECS-LFCS-95-332/}.

\bibitem[Longley(1999{\natexlab{a}})]{Longley1999}
John~R. Longley.
\newblock {Matching typed and untyped realizability}.
\newblock \emph{Electronic Notes in Theoretical Computer Science}, 23\penalty0
  (1):\penalty0 74--100, 1999{\natexlab{a}}.
\newblock ISSN 15710661.
\newblock \doi{10.1016/S1571-0661(04)00105-7}.
\newblock URL
  \url{http://linkinghub.elsevier.com/retrieve/pii/S1571066104001057}.

\bibitem[Longley(1999{\natexlab{b}})]{Longley1999b}
John~R. Longley.
\newblock {Unifying Typed and Untyped Realizability}, 1999{\natexlab{b}}.
\newblock URL \url{http://homepages.inf.ed.ac.uk/jrl/Research/unifying.txt}.

\bibitem[Longley(2005)]{Longley2000}
John~R. Longley.
\newblock {Notions of computability at higher types I}.
\newblock In \emph{Logic Colloquium 2000: Proceedings of the Annual European
  Summer Meeting of the Association for Symbolic Logic, held in Paris, France,
  July 23-31, 2000}, volume~19 of \emph{Lecture Notes in Logic}, pages 32--142.
  A. K. Peters, 2005.

\bibitem[Longley and Normann(2015)]{Longley2015}
John~R. Longley and Dag Normann.
\newblock \emph{{Higher-Order Computability}}.
\newblock Theory and Applications of Computability. Springer Berlin Heidelberg,
  Berlin, Heidelberg, 2015.
\newblock ISBN 978-3-662-47991-9.
\newblock \doi{10.1007/978-3-662-47992-6}.
\newblock URL \url{https://dx.doi.org/10.1007/978-3-662-47992-6}.

\bibitem[Longley and Simpson(1997)]{Longley1997}
John~R. Longley and Alex~K. Simpson.
\newblock {A uniform approach to domain theory in realizability models}.
\newblock \emph{Mathematical Structures in Computer Science}, 7:\penalty0
  469--505, 1997.
\newblock \doi{10.1017/S0960129597002387}.
\newblock URL \url{https://dx.doi.org/10.1017/S0960129597002387}.

\bibitem[Longo and Moggi(1991)]{Longo1991}
Giuseppe Longo and Eugenio Moggi.
\newblock {Constructive natural deduction and its ‘omega-set'
  interpretation}.
\newblock \emph{Mathematical Structures in Computer Science}, 1\penalty0
  (02):\penalty0 215, 1991.
\newblock ISSN 0960-1295.
\newblock \doi{10.1017/S0960129500001298}.
\newblock URL
  \url{http://www.journals.cambridge.org/abstract_S0960129500001298}.

\bibitem[{Mac Lane}(1978)]{MacLane1978}
Saunders {Mac Lane}.
\newblock \emph{{Categories for the Working Mathematician}}, volume~5 of
  \emph{Graduate Texts in Mathematics}.
\newblock Springer New York, New York, NY, 1978.
\newblock ISBN 978-1-4419-3123-8.
\newblock \doi{10.1007/978-1-4757-4721-8}.
\newblock URL \url{http://link.springer.com/10.1007/978-1-4757-4721-8}.

\bibitem[Machtey and Young(1978)]{MY:1978}
Michael Machtey and Paul Young.
\newblock \emph{{An introduction to the general theory of algorithms}}.
\newblock Theory of Computation Series. Elsevier North-Holland, New York, 1978.
\newblock ISBN 9780444002266.
\newblock URL \url{http://books.google.co.uk/books?id=qncEAQAAIAAJ}.

\bibitem[Machtey et~al.(1978)Machtey, Winklmann, and Young]{MWY:1978}
Michael Machtey, Karl Winklmann, and Paul Young.
\newblock {Simple G{\"{o}}del Numberings, Isomorphisms, and Programming
  Properties}.
\newblock \emph{SIAM Journal on Computing}, 7\penalty0 (1):\penalty0 39--60,
  feb 1978.
\newblock ISSN 0097-5397.
\newblock \doi{10.1137/0207003}.
\newblock URL \url{https://dx.doi.org/10.1137/0207003}.

\bibitem[Maraist et~al.(1995)Maraist, Odersky, Turner, and Wadler]{Maraist1995}
John Maraist, Martin Odersky, David~N Turner, and Philip Wadler.
\newblock {Call-by-name, call-by-value, call-by-need, and the linear lambda
  calculus}.
\newblock \emph{Electronic Notes in Theoretical Computer Science}, 1:\penalty0
  370--392, 1995.

\bibitem[Marion(2012)]{Marion2012}
Jean-Yves Marion.
\newblock {From Turing machines to computer viruses}.
\newblock \emph{Philosophical Transactions of the Royal Society A:
  Mathematical, Physical and Engineering Sciences}, 370\penalty0
  (1971):\penalty0 3319--3339, 2012.
\newblock ISSN 1364-503X.
\newblock \doi{10.1098/rsta.2011.0332}.
\newblock URL
  \url{http://rsta.royalsocietypublishing.org/cgi/doi/10.1098/rsta.2011.0332}.

\bibitem[Martin-L{\"{o}}f(1984)]{MartinLof1984}
Per Martin-L{\"{o}}f.
\newblock \emph{{Intuitionistic type theory}}, volume~1 of \emph{Studies in
  Proof Theory}.
\newblock Bibliopolis, 1984.
\newblock ISBN 88-7088-105-9.

\bibitem[Martin-L{\"{o}}f(1998)]{MartinLof1972}
Per Martin-L{\"{o}}f.
\newblock {An intuitionistic theory of types}.
\newblock In Giovanni Sambin and Jan~M Smith, editors, \emph{Twenty-five years
  of constructive type theory ({V}enice, 1995)}, volume~36 of \emph{Oxford
  Logic Guides}, pages 127--172. Oxford University Press, 1998.

\bibitem[Melli{\`{e}}s(2009)]{Mellies2009}
Paul-Andr{\'{e}} Melli{\`{e}}s.
\newblock {Categorical Semantics of Linear Logic}.
\newblock In Pierre-Louis Curien, Hugo Herbelin, Jean-Louis Krivine, and
  Paul-Andr{\'{e}} Melli{\`{e}}s, editors, \emph{Panoramas et synth{\`{e}}ses
  27: Interactive models of computation and program behaviour}.
  Soci{\'{e}}t{\'{e}} Math{\'{e}}matique de France, 2009.
\newblock ISBN 978-2-85629-273-0.
\newblock URL
  \url{http://www.pps.univ-paris-diderot.fr/$\sim$mellies/papers/panorama.pdf}.

\bibitem[Mitchell(1996)]{Mitchell1996}
John~C. Mitchell.
\newblock \emph{{Foundations for programming languages}}.
\newblock Foundations of Computing. The MIT Press, 1996.
\newblock ISBN 9780262133210.

\bibitem[Moelius(2009)]{Moelius2009}
Samuel~E. Moelius.
\newblock \emph{{Program Self-Reference}}.
\newblock PhD thesis, University of Delaware, 2009.

\bibitem[Mogensen(1992)]{Mogensen1992}
Torben~{\AE}. Mogensen.
\newblock {Efficient self-interpretation in lambda calculus}.
\newblock \emph{Journal of Functional Programming}, 2\penalty0 (03):\penalty0
  345--364, jul 1992.
\newblock ISSN 0956-7968.
\newblock \doi{10.1017/S0956796800000423}.
\newblock URL
  \url{http://www.journals.cambridge.org/abstract_S0956796800000423}.

\bibitem[Moggi(1989)]{Moggi1989}
Eugenio Moggi.
\newblock {Computational lambda-calculus and monads}.
\newblock In \emph{[1989] Proceedings. Fourth Annual Symposium on Logic in
  Computer Science}, pages 14--23. IEEE Comput. Soc. Press, 1989.
\newblock ISBN 0-8186-1954-6.
\newblock \doi{10.1109/LICS.1989.39155}.

\bibitem[Moggi(1991)]{Moggi1991}
Eugenio Moggi.
\newblock {Notions of computation and monads}.
\newblock \emph{Information and Computation}, 93\penalty0 (1):\penalty0 55--92,
  1991.
\newblock ISSN 08905401.
\newblock \doi{10.1016/0890-5401(91)90052-4}.
\newblock URL \url{https://dx.doi.org/10.1016/0890-5401(91)90052-4}.

\bibitem[Moschovakis(1993)]{Moschovakis1993}
Yiannis~N. Moschovakis.
\newblock {Sense and Denotation as Algorithm and Value}.
\newblock \emph{Logic Colloquium '90: ASL Summer Meeting in Helsinki},
  2:\penalty0 210--249, 1993.
\newblock URL \url{http://www.math.ucla.edu/$\sim$ynm/papers/frege.pdf}.

\bibitem[Moschovakis(2010)]{Moschovakis2010}
Yiannis~N Moschovakis.
\newblock {Kleene's Amazing Second Recursion Theorem}.
\newblock \emph{Bulletin of Symbolic Logic}, 16\penalty0 (2):\penalty0
  189--239, 2010.

\bibitem[Mulry(1982)]{Mulry1982}
Philip~S. Mulry.
\newblock {Generalized Banach-Mazur functionals in the topos of recursive
  sets}.
\newblock \emph{Journal of Pure and Applied Algebra}, 26\penalty0 (1):\penalty0
  71--83, oct 1982.
\newblock ISSN 00224049.
\newblock \doi{10.1016/0022-4049(82)90030-5}.
\newblock URL
  \url{http://linkinghub.elsevier.com/retrieve/pii/0022404982900305}.

\bibitem[Mulry(1989)]{Mulry1989}
Philip~S. Mulry.
\newblock {A categorical approach to the theory of computation}.
\newblock \emph{Annals of Pure and Applied Logic}, 43\penalty0 (3):\penalty0
  293--305, aug 1989.
\newblock ISSN 01680072.
\newblock \doi{10.1016/0168-0072(89)90072-9}.
\newblock URL
  \url{http://linkinghub.elsevier.com/retrieve/pii/0168007289900729}.

\bibitem[Myhill and Shepherdson(1955)]{Myhill1955}
J.~Myhill and J.~C. Shepherdson.
\newblock {Effective operations on partial recursive functions}.
\newblock \emph{Zeitschrift f{\"{u}}r Mathematische Logik und Grundlagen der
  Mathematik}, 1\penalty0 (4):\penalty0 310--317, 1955.
\newblock ISSN 00443050.
\newblock \doi{10.1002/malq.19550010407}.
\newblock URL \url{http://doi.wiley.com/10.1002/malq.19550010407}.

\bibitem[Nanevski(2002)]{Nanevski2002}
Aleksandar Nanevski.
\newblock {Meta-programming with names and necessity}.
\newblock \emph{ACM SIGPLAN Notices}, 37:\penalty0 206--217, 2002.
\newblock ISSN 03621340.
\newblock \doi{10.1145/583852.581498}.

\bibitem[Nanevski and Pfenning(2005)]{Nanevski2005}
Aleksandar Nanevski and Frank Pfenning.
\newblock {Staged computation with names and necessity}.
\newblock \emph{Journal of Functional Programming}, 15\penalty0 (06):\penalty0
  893, 2005.
\newblock ISSN 0956-7968.
\newblock \doi{10.1017/S095679680500568X}.

\bibitem[Nielson and Nielson(1992)]{Nielson1992}
Flemming Nielson and Hanne~Riis Nielson.
\newblock \emph{{Two-Level Functional Languages}}.
\newblock Cambridge Tracts in Theoretical Computer Science. Cambridge
  University Press, 1992.
\newblock ISBN 9780521018470.

\bibitem[Nordstr{\"{o}}m et~al.(1990)Nordstr{\"{o}}m, Petersson, and
  Smith]{Nordstrom1990}
Bengt Nordstr{\"{o}}m, Kent Petersson, and Jan~M. Smith.
\newblock \emph{{Programming in Martin-L{\"{o}}f's Type Theory: an
  Introduction}}.
\newblock Oxford University Press, 1990.
\newblock ISBN 0-19-853814-6.
\newblock \doi{10.1016/0377-0427(91)90052-L}.

\bibitem[Odifreddi(1992)]{Odifreddi1992}
Piergiorgio Odifreddi.
\newblock \emph{{Classical recursion theory: The theory of functions and sets
  of natural numbers}}.
\newblock Elsevier, 1992.
\newblock ISBN 9780444894830.

\bibitem[Paola and Heller(1987)]{Heller1987}
Robert A.~Di Paola and Alex Heller.
\newblock {Dominical Categories: Recursion Theory without Elements}.
\newblock \emph{The Journal of Symbolic Logic}, 52\penalty0 (3):\penalty0 594,
  sep 1987.
\newblock ISSN 00224812.
\newblock \doi{10.2307/2274352}.
\newblock URL \url{http://www.jstor.org/stable/2274352?origin=crossref}.

\bibitem[Pfenning and Davies(2001)]{Davies2001}
Frank Pfenning and Rowan Davies.
\newblock {A judgmental reconstruction of modal logic}.
\newblock \emph{Mathematical Structures in Computer Science}, 11\penalty0
  (4):\penalty0 511--540, 2001.
\newblock ISSN 0960-1295.
\newblock \doi{10.1017/S0960129501003322}.
\newblock URL \url{https://dx.doi.org/10.1017/S0960129501003322}.

\bibitem[Platek(1966)]{Platek1966}
Richard~Alan Platek.
\newblock \emph{{Foundations of Recursion Theory}}.
\newblock PhD thesis, Stanford University, 1966.

\bibitem[Plotkin and Power(2004)]{Plotkin2004b}
Gordon Plotkin and John Power.
\newblock {Computational Effects and Operations: An Overview}.
\newblock \emph{Electronic Notes in Theoretical Computer Science}, 73\penalty0
  (March 2002):\penalty0 149--163, oct 2004.
\newblock ISSN 15710661.
\newblock \doi{10.1016/j.entcs.2004.08.008}.
\newblock URL
  \url{http://linkinghub.elsevier.com/retrieve/pii/S1571066104050893}.

\bibitem[Plotkin(1977)]{Plotkin1977}
Gordon~D. Plotkin.
\newblock {LCF considered as a programming language}.
\newblock \emph{Theoretical Computer Science}, 5\penalty0 (3):\penalty0
  223--255, 1977.
\newblock ISSN 03043975.
\newblock \doi{10.1016/0304-3975(77)90044-5}.

\bibitem[Plotkin(1993)]{Plotkin1993}
Gordon~D. Plotkin.
\newblock {Type theory and recursion}.
\newblock In \emph{Proceedings Eighth Annual IEEE Symposium on Logic in
  Computer Science}, page 374. IEEE Comput. Soc. Press, 1993.
\newblock ISBN 0-8186-3140-6.
\newblock \doi{10.1109/LICS.1993.287571}.
\newblock URL \url{https://dx.doi.org/10.1109/LICS.1993.287571}.

\bibitem[Poign{\'{e}}(1992)]{Poigne1992}
Axel Poign{\'{e}}.
\newblock {Basic category theory}.
\newblock In \emph{Handbook of Logic in Computer Science}. Clarendon Press,
  1992.
\newblock ISBN 9780198537359.

\bibitem[Polonsky(2011)]{Polonsky2011}
Andrew Polonsky.
\newblock {Axiomatizing the Quote}.
\newblock In Marc Bezem, editor, \emph{Computer Science Logic (CSL'11) - 25th
  International Workshop/20th Annual Conference of the EACSL}, volume~12 of
  \emph{Leibniz International Proceedings in Informatics (LIPIcs)}, pages
  458--469. Schloss Dagstuhl - Leibniz-Zentrum fuer Informatik, 2011.
\newblock \doi{10.4230/LIPIcs.CSL.2011.458}.
\newblock URL \url{https://dx.doi.org/10.4230/LIPIcs.CSL.2011.458}.

\bibitem[Riccardi(1980)]{Riccardi1980}
Gregory~A. Riccardi.
\newblock \emph{{The Independence of Control Structures in Abstract Programming
  Systems}}.
\newblock PhD thesis, State University of New York at Buffalo, 1980.

\bibitem[Riccardi(1981)]{Riccardi1981}
Gregory~A. Riccardi.
\newblock {The independence of control structures in abstract programming
  systems}.
\newblock \emph{Journal of Computer and System Sciences}, 22\penalty0
  (2):\penalty0 107--143, apr 1981.
\newblock ISSN 00220000.
\newblock \doi{10.1016/0022-0000(81)90024-6}.
\newblock URL \url{https://dx.doi.org/10.1016/0022-0000(81)90024-6}.

\bibitem[Rogers(1958)]{Rogers1958}
Hartley Rogers.
\newblock {G{\"{o}}del numberings of partial recursive functions}.
\newblock \emph{The Journal of Symbolic Logic}, 23\penalty0 (03):\penalty0
  331--341, sep 1958.
\newblock ISSN 0022-4812.
\newblock \doi{10.2307/2964292}.
\newblock URL
  \url{http://www.journals.cambridge.org/abstract_S0022481200058011}.

\bibitem[Rogers(1987)]{Rogers1987}
Hartley Rogers.
\newblock \emph{{Theory of recursive functions and effective computability}}.
\newblock MIT Press, Cambridge, MA, USA, 1987.
\newblock ISBN 0-262-68052-1.

\bibitem[Royer(1987)]{Royer1987}
James~S. Royer.
\newblock \emph{{A Connotational Theory of Program Structure}}, volume 273 of
  \emph{Lecture Notes in Computer Science}.
\newblock Springer Berlin Heidelberg, Berlin, Heidelberg, 1987.
\newblock ISBN 978-3-540-18253-5.
\newblock \doi{10.1007/3-540-18253-5}.
\newblock URL \url{http://link.springer.com/10.1007/3-540-18253-5
  www.cis.syr.edu/$\sim$royer/archive/ctps.ps}.

\bibitem[Royer and Case(1994)]{Royer1994}
James~S. Royer and John Case.
\newblock \emph{{Subrecursive Programming Systems}}.
\newblock Birkh{\"{a}}user Boston, Boston, MA, 1994.
\newblock ISBN 978-1-4612-6680-8.
\newblock \doi{10.1007/978-1-4612-0249-3}.
\newblock URL \url{http://link.springer.com/10.1007/978-1-4612-0249-3}.

\bibitem[Scott(1975)]{Scott1975}
Dana Scott.
\newblock {Lambda Calculus and Recursion Theory (Preliminary Version)}.
\newblock In Stig Kanger, editor, \emph{Proceedings of the Third Scandinavian
  Logic Symposium}, volume~82 of \emph{Studies in Logic and the Foundations of
  Mathematics}, pages 154--193. North-Holland, 1975.
\newblock \doi{10.1016/S0049-237X(08)70730-4}.
\newblock URL
  \url{http://linkinghub.elsevier.com/retrieve/pii/S0049237X08707304}.

\bibitem[Scott(1976)]{Scott1976}
Dana~S. Scott.
\newblock {Data Types as Lattices}.
\newblock \emph{SIAM Journal on Computing}, 5\penalty0 (3):\penalty0 522--587,
  1976.
\newblock \doi{10.1137/0205037}.

\bibitem[Scott(1993)]{Scott1993}
Dana~S. Scott.
\newblock {A type-theoretical alternative to ISWIM, CUCH, OWHY}.
\newblock \emph{Theoretical Computer Science}, 121\penalty0 (1-2):\penalty0
  411--440, 1993.
\newblock ISSN 03043975.
\newblock \doi{10.1016/0304-3975(93)90095-B}.

\bibitem[Simpson and Plotkin(2000)]{Simpson2000}
Alex~K. Simpson and Gordon~D. Plotkin.
\newblock {Complete axioms for categorical fixed-point operators}.
\newblock In \emph{Proceedings of the 15th Annual IEEE Symposium on Logic in
  Computer Science (LICS 2000)}, pages 30--41. IEEE Comput. Soc, 2000.
\newblock ISBN 0-7695-0725-5.
\newblock \doi{10.1109/LICS.2000.855753}.
\newblock URL \url{http://ieeexplore.ieee.org/document/855753/}.

\bibitem[Smith(1982)]{Smith1982}
Brian~Cantwell Smith.
\newblock \emph{{Procedural reflection in programming languages}}.
\newblock PhD thesis, Massachusetts Institute of Technology, 1982.
\newblock URL \url{http://hdl.handle.net/1721.1/15961}.

\bibitem[Smith(1984)]{Smith1984}
Brian~Cantwell Smith.
\newblock {Reflection and Semantics in LISP}.
\newblock In \emph{Proceedings of the 11th ACM SIGACT-SIGPLAN Symposium on
  Principles of Programming Languages (POPL '84)}, pages 23--35, New York, New
  York, USA, 1984. ACM Press.
\newblock ISBN 0897911253.
\newblock \doi{10.1145/800017.800513}.
\newblock URL \url{https://dx.doi.org/10.1145/800017.800513}.

\bibitem[Smullyan(1992)]{Smullyan1992}
Raymond~M. Smullyan.
\newblock \emph{{G{\"{o}}del's Incompleteness Theorems}}.
\newblock Oxford University Press, 1992.

\bibitem[Staton(2014)]{Staton2014}
Sam Staton.
\newblock {Freyd categories are Enriched Lawvere Theories}.
\newblock \emph{Electronic Notes in Theoretical Computer Science},
  303:\penalty0 197--206, 2014.
\newblock ISSN 15710661.
\newblock \doi{10.1016/j.entcs.2014.02.010}.
\newblock URL \url{http://dx.doi.org/10.1016/j.entcs.2014.02.010}.

\bibitem[Strachey(2000)]{Strachey2000}
Christopher Strachey.
\newblock {Fundamental Concepts in Programming Languages}.
\newblock \emph{Higher-Order and Symbolic Computation}, 13:\penalty0 11--49,
  2000.
\newblock ISSN 13883690.
\newblock \doi{10.1023/A:1010000313106}.
\newblock URL \url{http://dx.doi.org/10.1023/A:1010000313106}.

\bibitem[Streicher(2006)]{Streicher2006}
Thomas Streicher.
\newblock \emph{{Domain-theoretic Foundations of Functional Programming}}.
\newblock World Scientific, 2006.

\bibitem[Streicher(2015)]{Streicher2013}
Thomas Streicher.
\newblock {How Intensional Is Homotopy Type Theory?}
\newblock In Maria del~Mar Gonz{\'{a}}lez, Paul~C. Yang, Nicola Gambino, and
  Joachim Kock, editors, \emph{Extended Abstracts Fall 2013: Geometrical
  Analysis; Type Theory, Homotopy Theory and Univalent Foundations}, number
  September in Research Perspectives CRM Barcelona, pages 105--110.
  Birkh{\"{a}}user Basel, Cham, 2015.
\newblock ISBN 978-3-319-21283-8.
\newblock \doi{10.1007/978-3-319-21284-5_20}.
\newblock URL \url{https://dx.doi.org/10.1007/978-3-319-21284-5_20}.

\bibitem[Taha and Nielsen(2003)]{Taha2003a}
Walid Taha and Michael~Florentin Nielsen.
\newblock {Environment classifiers}.
\newblock \emph{ACM SIGPLAN Notices}, 38:\penalty0 26--37, 2003.
\newblock ISSN 03621340.
\newblock \doi{10.1145/640128.604134}.

\bibitem[Taha and Sheard(1997)]{Taha1997}
Walid Taha and Tim Sheard.
\newblock {Multi-stage programming with explicit annotations}.
\newblock In \emph{Proceedings of the 1997 ACM SIGPLAN symposium on Partial
  evaluation and semantics-based program manipulation (PEPM '97)}, pages
  203--217, New York, New York, USA, 1997. ACM Press.
\newblock ISBN 0897919173.
\newblock \doi{10.1145/258993.259019}.
\newblock URL \url{https://dx.doi.org/10.1145/258993.259019}.

\bibitem[Taha and Sheard(2000)]{Taha2000}
Walid Taha and Tim Sheard.
\newblock {MetaML and multi-stage programming with explicit annotations}.
\newblock \emph{Theoretical Computer Science}, 248\penalty0 (1-2):\penalty0
  211--242, 2000.
\newblock ISSN 03043975.
\newblock \doi{10.1016/S0304-3975(00)00053-0}.
\newblock URL \url{https://dx.doi.org/10.1016/S0304-3975(00)00053-0}.

\bibitem[Takahashi(1995)]{Takahashi1995}
M.~Takahashi.
\newblock {Parallel Reductions in $\lambda$-Calculus}.
\newblock \emph{Information and Computation}, 118\penalty0 (1):\penalty0
  120--127, apr 1995.
\newblock ISSN 08905401.
\newblock \doi{10.1006/inco.1995.1057}.
\newblock URL \url{https://dx.doi.org/10.1006/inco.1995.1057}.

\bibitem[Tsukada and Igarashi(2010)]{Tsukada2010}
Takeshi Tsukada and Atsushi Igarashi.
\newblock {A logical foundation for environment classifiers}.
\newblock \emph{Logical Methods in Computer Science}, 6\penalty0 (4):\penalty0
  1--43, 2010.
\newblock ISSN 18605974.
\newblock \doi{10.2168/LMCS-6(4:8)2010}.
\newblock URL \url{https://dx.doi.org/10.2168/LMCS-6(4:8)2010}.

\bibitem[Turing(1937)]{Turing1937}
Alan~M Turing.
\newblock {On Computable Numbers, with an Application to the
  Entscheidungsproblem}.
\newblock \emph{Proceedings of the London Mathematical Society}, s2-42\penalty0
  (1):\penalty0 230--265, jan 1937.
\newblock ISSN 0024-6115.
\newblock \doi{10.1112/plms/s2-42.1.230}.
\newblock URL
  \url{http://plms.oxfordjournals.org/cgi/doi/10.1112/plms/s2-42.1.230}.

\bibitem[Turing(1939)]{Turing1939}
Alan~M Turing.
\newblock {Systems of logic based on ordinals}.
\newblock \emph{Proceedings of the London Mathematical Society}, 2\penalty0
  (1):\penalty0 161--228, 1939.

\bibitem[Ursini(1979{\natexlab{a}})]{Ursini1979}
Aldo Ursini.
\newblock {Intuitionistic diagonalizable algebras}.
\newblock \emph{Algebra Universalis}, 9\penalty0 (1):\penalty0 229--237,
  1979{\natexlab{a}}.
\newblock ISSN 00025240.
\newblock \doi{10.1007/BF02488034}.

\bibitem[Ursini(1979{\natexlab{b}})]{Ursini1979a}
Aldo Ursini.
\newblock {A modal calculus analogous to K4W, based on intuitionistic
  propositional logic}.
\newblock \emph{Studia Logica}, 38\penalty0 (3):\penalty0 297--311,
  1979{\natexlab{b}}.
\newblock ISSN 0039-3215.
\newblock \doi{10.1007/BF00405387}.
\newblock URL \url{https://dx.doi.org/10.1007/BF00405387}.

\bibitem[van Oosten(2008)]{VanOosten2008}
Jaap van Oosten.
\newblock \emph{{Realizability: An Introduction to its Categorical Side}},
  volume 152.
\newblock Elsevier, 2008.
\newblock ISBN 978-0-444-51584-1.
\newblock URL
  \url{http://www.sciencedirect.com/science/bookseries/0049237X/152}.

\bibitem[Vassilakis(1989)]{Vassilakis1989}
Spyros Vassilakis.
\newblock {Economic Data Types}.
\newblock 1989.

\bibitem[Vassilakis(1992)]{Vassilakis1992}
Spyros Vassilakis.
\newblock {Some economic applications of Scott domains}.
\newblock \emph{Mathematical Social Sciences}, 24\penalty0 (2-3):\penalty0
  173--208, nov 1992.
\newblock ISSN 01654896.
\newblock \doi{10.1016/0165-4896(92)90061-9}.
\newblock URL
  \url{http://linkinghub.elsevier.com/retrieve/pii/0165489692900619}.

\bibitem[Velupillai(2000)]{Velupillai2000}
Kumaraswamy Velupillai.
\newblock \emph{{Computable economics}}.
\newblock The Arne Ryde Memorial Lectures. Oxford University Press, 2000.
\newblock ISBN 978 1 84376 239 3.

\bibitem[Wadler(1987)]{Wadler1987}
P~Wadler.
\newblock {A critique of Abelson and Sussman or why calculating is better than
  scheming}.
\newblock \emph{ACM SIGPLAN Notices}, 22\penalty0 (3):\penalty0 83--94, mar
  1987.
\newblock ISSN 03621340.
\newblock \doi{10.1145/24697.24706}.
\newblock URL \url{https://dx.doi.org/10.1145/24697.24706}.

\bibitem[Wand(1998)]{Wand1998}
Mitchell Wand.
\newblock {The Theory of Fexprs is Trivial}.
\newblock \emph{LISP and Symbolic Computation}, 10\penalty0 (3):\penalty0
  189--199, 1998.
\newblock ISSN 08924635.
\newblock \doi{10.1023/A:1007720632734}.
\newblock URL \url{https://dx.doi.org/10.1023/A:1007720632734}.

\bibitem[Wand and Friedman(1988)]{Wand1988}
Mitchell Wand and Daniel~P. Friedman.
\newblock {The mystery of the tower revealed: A nonreflective description of
  the reflective tower}.
\newblock \emph{Lisp and Symbolic Computation}, 1\penalty0 (1):\penalty0
  11--38, jun 1988.
\newblock ISSN 0892-4635.
\newblock \doi{10.1007/BF01806174}.
\newblock URL \url{https://dx.doi.org/10.1007/BF01806174}.

\bibitem[Winrich(1984)]{Winrich1984}
Steven~J Winrich.
\newblock {Self-Reference and the Incomplete Structure of Neoclassical
  Economics}.
\newblock \emph{Journal of Economic Issues}, 18\penalty0 (4):\penalty0
  987--1005, 1984.

\end{thebibliography}

\end{document}